\documentclass[oneside,english]{extbook}
\usepackage[T1]{fontenc}
\usepackage[latin9]{inputenc}
\usepackage{color}
\usepackage{babel}
\usepackage{array}
\usepackage{calc}
\usepackage{units}
\usepackage{textcomp}
\usepackage{mathrsfs}
\usepackage{enumitem}
\usepackage{multirow}
\usepackage{amsmath}
\usepackage{amsthm}
\usepackage{amssymb}
\usepackage{cancel}
\usepackage{stmaryrd}
\usepackage{makeidx}
\makeindex
\usepackage{graphicx}
\usepackage{wasysym}
\PassOptionsToPackage{normalem}{ulem}
\usepackage{ulem}
\usepackage[unicode=true]
 {hyperref}

\makeatletter

\newcommand{\lyxmathsym}[1]{\ifmmode\begingroup\def\b@ld{bold}
  \text{\ifx\math@version\b@ld\bfseries\fi#1}\endgroup\else#1\fi}

\providecommand{\tabularnewline}{\\}

\numberwithin{figure}{section}
  \theoremstyle{definition}
    \ifx\thechapter\undefined
      \newtheorem{defn}{\protect\definitionname}
    \else
      \newtheorem{defn}{\protect\definitionname}[chapter]
    \fi
  \theoremstyle{definition}
    \ifx\thechapter\undefined
      \newtheorem{example}{\protect\examplename}
    \else
      \newtheorem{example}{\protect\examplename}[chapter]
    \fi
\theoremstyle{plain}
    \ifx\thechapter\undefined
    \newtheorem{thm}{\protect\theoremname}
  \else
      \newtheorem{thm}{\protect\theoremname}[chapter]
    \fi
  \theoremstyle{remark}
    \ifx\thechapter\undefined
      \newtheorem{rem}{\protect\remarkname}
    \else
      \newtheorem{rem}{\protect\remarkname}[chapter]
    \fi
  \theoremstyle{plain}
    \ifx\thechapter\undefined
      \newtheorem{conjecture}{\protect\conjecturename}
    \else
      \newtheorem{conjecture}{\protect\conjecturename}[chapter]
    \fi
  \theoremstyle{plain}
    \ifx\thechapter\undefined
      \newtheorem{lem}{\protect\lemmaname}
    \else
      \newtheorem{lem}{\protect\lemmaname}[chapter]
    \fi
  \theoremstyle{plain}
    \ifx\thechapter\undefined
    \newtheorem{cor}{\protect\corollaryname}
  \else
      \newtheorem{cor}{\protect\corollaryname}[chapter]
    \fi
  \theoremstyle{plain}
    \ifx\thechapter\undefined
      \newtheorem{prop}{\protect\propositionname}
    \else
      \newtheorem{prop}{\protect\propositionname}[chapter]
    \fi

\newif\ifS
\Sfalse
\newcommand{\exo}{}
\usepackage{tikz}
\usepackage{old-arrows}
\newcommand\encircle[1]{%
  \tikz[baseline=(X.base)] 
    \node (X) [draw, shape=circle, inner sep=0] {\strut #1};}
\usetikzlibrary{shapes.geometric,decorations.pathreplacing,matrix,positioning,calc}
\newcommand{\warningsign}{\tikz[baseline=-.75ex] \node[shape=regular polygon, regular polygon sides=3, inner sep=0pt, draw, thick] {\textbf{!}};}
            \usepackage{relsize}
\numberwithin{equation}{chapter}
\hypersetup{
  pdfinfo = {
    Author = {Morgan Le Delliou},
    Title = {GR2 Course}
  },
  colorlinks = true,
  breaklinks = true,
 linkcolor = blue,
 urlcolor = blue,
 citecolor = blue,
}

\usepackage{abraces}
\newcommand{\mrot}[1]{\rotatebox{90}{$#1\mathstrut$}}

\tikzset{
mybrace/.style={
  decorate,
  decoration={brace,aspect=#1},
  line width=1pt
  }
}

\newcommand{\lboxed}[1]{\begin{array}{|l}\hline#1\\\hline\end{array}}
\newcommand{\rboxed}[1]{\begin{array}{r|}\hline#1\\\hline\end{array}}
\makeatletter
\newcommand{\llboxed}[1]{{%
  \let\@frameb@x\l@frameb@x\fbox{#1}%
}}
\newcommand{\rrboxed}[1]{{%
  \let\@frameb@x\r@frameb@x\fbox{#1}%
}}

\def\l@frameb@x#1{%
  \@tempdima\fboxrule
  \advance\@tempdima\fboxsep
  \advance\@tempdima\dp\@tempboxa
  \hbox{%
    \lower\@tempdima\hbox{%
      \vbox{%
        \hrule\@height\fboxrule
        \hbox{%
          \vrule\@width\fboxrule
          #1%
          \vbox{%
            \vskip\fboxsep
            \box\@tempboxa
            \vskip\fboxsep}%
          #1%
                    }%
        \hrule\@height\fboxrule}%
                          }%
        }%
}
\def\r@frameb@x#1{%
  \@tempdima\fboxrule
  \advance\@tempdima\fboxsep
  \advance\@tempdima\dp\@tempboxa
  \hbox{%
    \lower\@tempdima\hbox{%
      \vbox{%
        \hrule\@height\fboxrule
        \hbox{%
          #1%
          \vbox{%
            \vskip\fboxsep
            \box\@tempboxa
            \vskip\fboxsep}%
          #1%
          \vrule\@width\fboxrule}%
        \hrule\@height\fboxrule}%
                          }%
        }%
}
\newcommand{\pushright}[1]{\ifmeasuring@#1\else\omit\hfill$\displaystyle#1$\fi\ignorespaces}
\newcommand{\pushleft}[1]{\ifmeasuring@#1\else\omit$\displaystyle#1$\hfill\fi\ignorespaces}

\@ifundefined{showcaptionsetup}{}{%
 \PassOptionsToPackage{caption=false}{subfig}}
\usepackage{subfig}
\makeatother

  \providecommand{\conjecturename}{Conjecture}
  \providecommand{\definitionname}{Definition}
  \providecommand{\examplename}{Example}
  \providecommand{\lemmaname}{Lemma}
  \providecommand{\propositionname}{Proposition}
  \providecommand{\remarkname}{Remark}
\providecommand{\corollaryname}{Corollary}
\providecommand{\theoremname}{Theorem}

\begin{document}

\title{Advanced General Relativity\ifS\else ~Notes\fi}

\author{M. Le Delliou}

\maketitle
In this class is implied the knowledge of a GR1 course.

\paragraph{Conventions: }

In most of the course we will use
\begin{description}
\item [{$c=1$}]~
\end{description}
In some parts we will set 
\begin{description}
\item [{$G=1$}]~
\end{description}
Following \cite{EllisMaartensMacCallum2012}, we chose the conventions
for
\begin{itemize}
\item the metric signature in 4D as $\begin{array}{cccc}
- & + & + & +\end{array}$
\item the sign of the Riemann tensor, as set in the Ricci identities
\begin{align*}
2u_{c[;b;a]}= & R_{abcd}u^{d}
\end{align*}
\item the sign of the Ricci tensor, from the choice of contraction
\begin{align*}
R_{ab}= & R_{\:acb}^{c}
\end{align*}
\item the Einstein summation
\begin{align*}
X_{\:a}^{a}\equiv & \sum_{a=0}^{3}X_{\:a}^{a}
\end{align*}
\item the covariant derivative ($\pm\Gamma$ being the connection, with
the sign depending on the indices of $X$)
\begin{align*}
X_{;a}= & \partial_{a}X+\sum\left(\pm1\right)\Gamma X
\end{align*}
\item the Lie derivative
\begin{align*}
\mathcal{L}_{v}X= & v^{a}X_{;a}+\sum_{i,j}\left(-v_{;b}^{a_{i}}X^{\cdots b\cdots}+v_{;b_{j}}^{a}X_{\cdots a\cdots}\right)
\end{align*}
\item the indices (anti)symmetrisations (with $\sigma$ permutation among
indices of signature $\epsilon_{\sigma}$)
\begin{align*}
X_{\left(\begin{array}{ccc}
a_{1} & \cdots & a_{n}\end{array}\right)}= & \frac{1}{n!}\sum_{\sigma}X_{\begin{array}{ccc}
\sigma\left(a_{1}\right) & \cdots & \sigma\left(a_{n}\right)\end{array}}\\
X_{\left[\begin{array}{ccc}
a_{1} & \cdots & a_{n}\end{array}\right]}= & \frac{1}{n!}\sum_{\sigma}\left(-1\right)^{\epsilon_{\sigma}}X_{\begin{array}{ccc}
\sigma\left(a_{1}\right) & \cdots & \sigma\left(a_{n}\right)\end{array}}
\end{align*}
\\
and in particular, we have
\begin{align*}
X_{\left(ab\right)}= & \frac{1}{2}\left(X_{ab}+X_{ba}\right)\\
X_{\left[ab\right]}= & \frac{1}{2}\left(X_{ab}-X_{ba}\right)
\end{align*}
\end{itemize}
Beware that those conventions vary from authors or book to authors
or books!

A short bibliography for the class can be given as \cite{carroll-2004,EllisMaartensMacCallum2012,Stephani:2004ud,Chandrasekhar:1985kt,dInverno92,Plebanski:2006sd,wald-book,hawking,misner-thorne-wheeler,weinberg}

\ifS\else see at the end of the document for full reference\fi\tableofcontents{}

\listoffigures

\chapter[The many forms of the Schwarzschild BH ]{The many forms of the Schwarzschild Black Hole }
\begin{defn}
Black Hole

Black Holes are vacuum, non-trivial solutions
of the General Relativity\ifS
\else\\ \fi results of primordial extreme density
fluctuations or of end of life process of massive stars.

The simplest types are
\begin{itemize}
\item spherically symmetric
\item static
\end{itemize}
This is the Schwarzschild solution.
\end{defn}

\section{Standard Schwarzschild solution}

\subsection{Characterisation}

Published first in \cite{Schwarzschild:1916uq}, it is a
\begin{itemize}
\item spherically symmetric solution,
\end{itemize}
$(t,r,\theta,\varphi)$
independently of the angular variables. \ifS\else\\\fi
metric on a 2-sphere
\begin{align*}
d\Omega^{2}= & d\theta^{2}+\sin^{2}\theta\,d\varphi^{2},\ifS\else\fi
\end{align*}

\begin{itemize}
\item vacuum solution,
\end{itemize}
\begin{align*}
R_{ab}= & 0,\ifS\else\fi
\end{align*}

\begin{itemize}
\item static solution,
\end{itemize}
\begin{align*}
g_{ab}= & g_{ab}(r),\ifS\else\fi\\
g_{0i}= & 0.\ifS\else\fi
\end{align*}

\begin{example}
spherical~Minkowski

\begin{align*}
ds_{M}^{2}= & -dt^{2}+dr^{2}+r^{2}d\Omega^{2},\ifS\else\fi
\end{align*}
\begin{defn}

$r$ areal radius
\begin{align*}
\mathscr{S}_{2}= & \int_{\mathscr{S}_{2}}r^{2}d\Omega^{2}=4\pi r^{2}.\ifS\else\fi
\end{align*}
\end{defn}
\end{example}

\subsection{static, spherical~metric
}

\begin{align*}
ds^{2}= & -e^{2\alpha(r)}dt^{2}+e^{2\beta(r)}dr^{2}+e^{2\gamma(r)}r^{2}d\Omega^{2}.\ifS\else\fi
\end{align*}
Without loss of generality, coordinate change 
\begin{align*}
\bar{r}= & e^{\gamma}r\\
\Rightarrow d\bar{r}= & e^{\gamma}dr+e^{\gamma}r\,d\gamma=\left(1+\frac{d\gamma}{d\ln r}\right)e^{\gamma}dr,\ifS\else\fi
\end{align*}
so the metric reads
\begin{align*}
ds^{2}= & -e^{2\alpha}dt^{2}+\left(1+r\gamma^{\prime}\right)^{-2}e^{2\left(\beta-\gamma\right)}d\bar{r}^{2}+\bar{r}^{2}d\Omega^{2},\ifS\else\fi
\end{align*}
relabel 
\begin{align*}
\bar{r}\rightarrow & r,\ifS\else\fi\\
\left(1+r\gamma^{\prime}\right)^{-2}e^{2\left(\beta-\gamma\right)}\rightarrow & e^{2\beta},\ifS\else\fi
\end{align*}
\begin{align*}
ds^{2}= & -e^{2\alpha}dt^{2}+e^{2\beta}dr^{2}+r^{2}d\Omega^{2}\\
\Leftrightarrow g_{ab}= & \left(\begin{array}{cccc}
-e^{2\alpha} &  &  & 0\\
 & e^{2\beta}\\
 &  & r^{2}\\
0 &  &  & r^{2}\sin^{2}\theta
\end{array}\right).\ifS\else\fi
\end{align*}

\subsection{EFE solutions\label{subsec:EFE-solutions}}

\subsubsection{Geometry of static, spherically symmetric metric}

Christofel ()
connection, no torsion~$\Gamma_{bc}^{a}=\frac{g^{ad}}{2}\left(\partial_{b}g_{cd}+\partial_{c}g_{bd}-\partial_{d}g_{bc}\right)$,
\ifS\else
only $r$ dependence, \fi  $\partial_{r}\equiv^{\prime}$, 
\begin{gather}
\begin{array}{rlrlrl}
\Gamma_{01}^{0}= & \alpha^{\prime}, & \Gamma_{00}^{1}= & e^{2\left(\alpha-\beta\right)}\alpha^{\prime}, & \Gamma_{11}^{1}= & \beta^{\prime},\\
\Gamma_{12}^{2}= & \frac{1}{r}, & \Gamma_{22}^{1}= & -re^{-2\beta}, & \Gamma_{13}^{3}= & \frac{1}{r},\ifS\else\nonumber\fi\\
\Gamma_{33}^{1}= & -re^{-2\beta}\sin^{2}\theta, & \Gamma_{33}^{2}= & -\sin\theta\cos\theta, & \Gamma_{23}^{3}= & \cot\theta,
\end{array}
\end{gather}
\ifS\else(exercise)\exo rest \fi are $0$
or symmetric.

\ifS
\else (From\fi 
\begin{align*}
2\left[\nabla_{c},\nabla_{d}\right]V^{a}= & R_{\:bcd}^{a}V^{b}-T_{\:cd}^{e}\nabla_{e}V^{a},\ifS\else\fi
\end{align*}
\begin{align*}
T_{\:cd}^{e}= & 2\Gamma_{\left[cd\right]}^{e},\ifS\else\fi\\
R_{\:bcd}^{a}= & \partial_{c}\Gamma_{db}^{a}-\partial_{d}\Gamma_{cb}^{a}+\Gamma_{ce}^{a}\Gamma_{db}^{e}-\Gamma_{de}^{a}\Gamma_{cb}^{e}\\
= & 2\Gamma_{b\left[d\,,c\right]}^{a}+2\Gamma_{[c|e}^{a}\Gamma_{|d]b}^{e}\ifS\else\fi
\end{align*}
\ifS \else)\fi  Then \ifS
\else Riemann \fi 
\begin{gather}
\begin{array}{rlrl}
R_{\:101}^{0}= & \alpha^{\prime}\beta^{\prime}-\alpha^{\prime\prime}-\left(\alpha^{\prime}\right)^{2}, & R_{\:202}^{0}= & -re^{-2\beta}\alpha^{\prime},\\
R_{\:303}^{0}= & -re^{-2\beta}\sin^{2}\theta\alpha^{\prime}, & R_{\:212}^{1}= & re^{-2\beta}\beta^{\prime},\ifS\else\nonumber\fi\\
R_{\:313}^{1}= & re^{-2\beta}\sin^{2}\theta\beta^{\prime}, & R_{\:323}^{2}= & \left(1-e^{-2\beta}\right)\sin^{2}\theta,
\end{array}
\end{gather}
\ifS\else (exercise)\exo rest \fi are $0$
or symmetric.

\ifS\else Ricci \& Ricci
scalar \fi 
\begin{align*}
R_{00}= & R_{\:010}^{1}+R_{\:020}^{2}+R_{\:030}^{3}=g^{11}g_{00}R_{\:101}^{0}+g^{22}g_{00}R_{\:202}^{0}+g^{33}g_{00}R_{\:303}^{0}\\
= & e^{2\left(\alpha-\beta\right)}\left[\alpha^{\prime\prime}+\left(\alpha^{\prime}\right)^{2}-\alpha^{\prime}\beta^{\prime}+\frac{2}{r}\alpha^{\prime}\right],\ifS\else\fi\\
R_{11}= & -\alpha^{\prime\prime}-\left(\alpha^{\prime}\right)^{2}+\alpha^{\prime}\beta^{\prime}+\frac{2}{r}\beta^{\prime},\ifS\else\fi\\
R_{22}= & e^{-2\beta}\left[r\left(\beta^{\prime}-\alpha^{\prime}\right)-1\right]+1,\ifS\else\fi\\
R_{33}= & R_{22}\sin^{2}\theta,\ifS\else\fi
\end{align*}
\ifS\else (exercise)\exo rest \fi are $0$
or symmetric,
\begin{align*}
\Rightarrow R= & e^{-2\beta}\left[\alpha^{\prime\prime}+\left(\alpha^{\prime}\right)^{2}-\alpha^{\prime}\beta^{\prime}+\frac{2}{r}\left(\alpha^{\prime}-\beta^{\prime}\right)+\frac{1}{r^{2}}\left(1-e^{2\beta}\right)\right].\ifS\else\fi
\end{align*}

\subsubsection{Solutions of EFE}

\ifS\else Vacuum: \fi 
$R_{ab}=0$. 
\begin{itemize}
\item 
\begin{align*}
0= & e^{2\left(\beta-\alpha\right)}R_{00}+R_{11}=\frac{2}{r}\left(\alpha+\beta\right)^{\prime}\\
\Rightarrow\alpha= & -\beta+C,\ifS\else\fi
\end{align*}
so 
\begin{align*}
g_{00}dt^{2}= & e^{-2\beta}\left(e^{C}dt\right)^{2},\ifS\else\fi
\end{align*}
rescale $t\to e^{-C}t$ set $C$ to $0$ 
\begin{align*}
\Rightarrow\alpha= & -\beta,\ifS\else\fi
\end{align*}
\item 
\begin{align*}
R_{22}= & 0\\
\Leftrightarrow1= & e^{2\alpha}\left[1+2r\alpha^{\prime}\right]=\left(re^{2\alpha}\right)^{\prime}\\
\Rightarrow e^{2\alpha}= & 1+\frac{A}{r}=1-\frac{R_{s}}{r},\ifS\else\fi
\end{align*}
\ifS
\else chosing $A$\fi .
\end{itemize}
\ifS\else  Metric
becomes\fi 
\begin{align*}
ds^{2}= & -\left(1-\frac{R_{s}}{r}\right)dt^{2}+\left(1-\frac{R_{s}}{r}\right)^{-1}dr^{2}+r^{2}d\Omega^{2}.\ifS\else\fi
\end{align*}
It is easy to verify \ifS\else (exercise)\exo
$R_{s}$ Schwarzschild radius \fi 
\begin{align*}
R_{00}\propto & \left(re^{2\alpha}\right)^{\prime\prime}=0,\ifS\else\fi\\
R_{11}\propto & R_{00}=0.\ifS\else\fi
\end{align*}

\paragraph{\ifS\else Recall \fi  }

Weak field limit:

To fix $R_{s}$\ifS\else$\to$ \fi  Newtonian gravity.

Static field $\partial_{0}g_{ab}=0$

Slowly moving objects : $\frac{dx^{i}}{d\tau}\ll\frac{dt}{d\tau}$

Geodesic equations :
\begin{align*}
\frac{d^{2}x^{a}}{d\tau^{2}}+\Gamma_{00}^{a}\left(\frac{dt}{d\tau}\right)^{2}= & 0.\ifS\else\fi
\end{align*}
\begin{align*}
\partial_{0}g_{ab}=0\Rightarrow\Gamma_{00}^{a}= & \frac{g^{ad}}{2}\left(\partial_{0}g_{0d}+\partial_{0}g_{0d}-\partial_{d}g_{00}\right)\\
= & -\frac{1}{2}g^{ab}\partial_{b}g_{00}.\ifS\else\fi
\end{align*}
\ifS
\else Decompose metric into small perturbations
around inertial Minkowski\fi 
\begin{align*}
g_{ab}= & \eta_{ab}+h_{ab},\,\left|h_{ab}\right|\ll1\ifS\else\fi\\
g_{ac}g^{cb}= & \delta_{a}^{\,b}\\
\Rightarrow g^{ab}= & \eta^{ab}-h^{ab}\textrm{ to 1st order in }h\\
\textrm{with }h^{ab}= & \eta^{ac}\eta^{bd}h_{cd}\ifS\else\fi
\end{align*}
\ifS\else To \fi  first order
\begin{align*}
\Gamma_{00}^{a}= & -\frac{1}{2}\eta^{ab}\partial_{b}h_{00}\textrm{ and}\ifS\else\fi\\
\frac{d^{2}x^{a}}{d\tau^{2}}= & \frac{1}{2}\eta^{ab}\partial_{b}h_{00}\left(\frac{dt}{d\tau}\right)^{2}.\ifS\else\fi
\end{align*}
From $\partial_{0}h_{00}=0$,
we get
\begin{align*}
\frac{d^{2}x^{0}}{d\tau^{2}}= & 0\Leftrightarrow\frac{dt}{d\tau}=cst\ifS\else\fi
\end{align*}
Thus 
\begin{align*}
\frac{d^{2}x^{\mu}}{d\tau^{2}}= & \frac{1}{2}\partial_{\mu}h_{00}\left(\frac{dt}{d\tau}\right)^{2}=\frac{dt}{d\tau}\frac{d}{dt}\left(\frac{dt}{d\tau}\frac{dx^{\mu}}{dt}\right)=\left(\frac{dt}{d\tau}\right)^{2}\frac{d^{2}x^{\mu}}{dt^{2}}\\
\Leftrightarrow\frac{d^{2}x^{\mu}}{dt^{2}}= & \frac{1}{2}\partial_{\mu}h_{00},\ifS\else\fi
\end{align*}
Newton's equation for $h_{00}=-2\Phi$ 
\begin{align*}
g_{00}\underset{r\gg R_{s}}{=} & -\left(1+2\Phi\right)=-\left(1-\frac{R_{s}}{r}\right).\ifS\else\fi
\end{align*}
\ifS
\else In our case we have vacuum
and \fi  a central mass
\begin{align*}
\Rightarrow\Phi= & -\frac{GM}{r} & \Rightarrow R_{s}= & 2GM.\ifS\else\fi
\end{align*}
M \ifS
\else parameter:
Newtonian mass in weak field limit \fi  $r\gg2GM$ \ifS
\else Schwarzschild
metric \fi 
\begin{align*}
ds^{2}= & -\left(1-\frac{2GM}{r}\right)dt^{2}+\left(1-\frac{2GM}{r}\right)^{-1}dr^{2}+r^{2}d\Omega^{2}.\ifS\else\fi
\end{align*}
\ifS
\else Recover Minkowski \fi 
\begin{align*}
M\to & 0\\
r\to & \infty
\end{align*}
\ifS\else $\to$\fi 
asymptotic flatness. \ifS
\else $G$ set to $1$: natural units \fi 
\begin{example}
For the Sun
\end{example}
\begin{align*}
R_{\odot}= & 10^{6}GM_{\odot}\gg R_{s_{\odot}}.
\end{align*}

\section{Spatially conformally flat Schwarzschild metric}

\ifS
\else  A.K.A. \fi  Isotropic
radial coordinate metric \cite[Sec. 23.6]{Stephani:2004ud}, first
proposed by Eddington \cite[p93]{Eddington1923}. We want to write
\ifS
\else Schwarzschild
in the Euclidean form\fi 
\begin{align*}
ds^{2}= & -A^{2}(\bar{r})dt^{2}+B^{2}(\bar{r})d\Sigma^{2}\textrm{ with }d\Sigma^{2}=d\bar{r}^{2}+\bar{r}^{2}d\Omega^{2}\\
= & -A^{2}dt^{2}+B^{2}d\bar{r}^{2}+B^{2}\bar{r}^{2}d\Omega^{2}\\
= & -\left(1-\frac{2M}{r}\right)dt^{2}+\left(1-\frac{2M}{r}\right)^{-1}dr^{2}+r^{2}d\Omega^{2},\ifS\else\fi
\end{align*}
\ifS
\else  equate with Schwarzschild \fi  
\begin{align*}
r^{2}= & B^{2}\bar{r}^{2}\ifS\else\fi\\
\frac{dr^{2}}{1-\frac{2M}{r}}= & B^{2}d\bar{r}^{2}\ifS\else\fi
\end{align*}
\ifS\else  Divide:\fi  
\begin{align*}
\frac{dr^{2}}{\left(r-2M\right)r}= & \left(\frac{d\bar{r}}{\bar{r}}\right)^{2},\ifS\else\fi
\end{align*}
\ifS
\else  solution: (exercise: verify
solution)\exo\fi 
\begin{align*}
r= & \bar{r}\left(1+\frac{M}{2\bar{r}}\right)^{2}\ifS\else\fi\\
\Rightarrow B^{2}= & \left(\frac{r}{\bar{r}}\right)^{2}=\left(1+\frac{M}{2\bar{r}}\right)^{4},\ifS\else\fi\\
A^{2}= & 1-\frac{2M}{r}=\frac{\bar{r}\left(1+\frac{2M}{2\bar{r}}+\frac{M^{2}}{4\bar{r}^{2}}\right)-2M}{\bar{r}\left(1+\frac{M}{2\bar{r}}\right)^{2}}=\left(\frac{1-\frac{M}{2\bar{r}}}{1+\frac{M}{2\bar{r}}}\right)^{2}.\ifS\else\fi
\end{align*}
\begin{align*}
ds^{2}= & -\left(\frac{1-\frac{M}{2\bar{r}}}{1+\frac{M}{2\bar{r}}}\right)^{2}dt^{2}+\left(1+\frac{M}{2\bar{r}}\right)^{4}\left[d\bar{r}^{2}+\bar{r}^{2}d\Omega^{2}\right].\ifS\else\fi
\end{align*}

\section{Lemaître free-falling Schwarzschild metric}

\ifS
\else  Transforming to Lemaître coordinates\fi 
\begin{align*}
dT= & dt+\sqrt{\frac{2GM}{r}}\frac{dr}{1-\frac{2GM}{r}}& dR= & dt+\sqrt{\frac{r}{2GM}}\frac{dr}{1-\frac{2GM}{r}}
\end{align*}
\ifS
\else yields
from Schwarzschild to the Lemaître metric, removing the $r=2GM$ singularity,
a synchronous metric as all static observers admit the same proper
time ($g_{tt}=cst$)\fi 
\begin{align*}
ds^{2}= & -dT^{2}+\frac{2GM}{r}dR^{2}+r^{2}d\Omega^{2}, & r= & r\left(T,R\right).\ifS\else\fi
\end{align*}
\ifS
\else From
coordinates transformation, $dR-dT=\sqrt{\frac{r}{2GM}}dr$, which
integrates into $r=\left[\frac{3}{2}\left(R-T\right)\right]^{\frac{2}{3}}\left(2GM\right)^{\frac{1}{3}}$
\fi .

\ifS
\else Radial free fall can be found
for static Lemaître observers:
\begin{align*}
dR= & 0 & \Rightarrow dt= & -\sqrt{\frac{r}{2GM}}\frac{dr}{1-\frac{2GM}{r}}\;\underset{r\to\left(2GM\right)^{+}}{\sim}-\frac{2GM}{r-2GM}dr & \Rightarrow\frac{r-2GM}{r_{i}-2GM}= & e^{-\frac{t-t_{i}}{2GM}}
\end{align*}
Infinite time $t$ to get to Schwarzschild radius while\fi 

\begin{align*}
\Delta T_{i\to Sch.}= & \int_{r_{i}}^{2GM}dT=\int_{r_{i}}^{2GM}\sqrt{\frac{r}{2GM}}\left(\frac{2GM}{r}-1\right)\frac{dr}{1-\frac{2GM}{r}}\\
= & -\int_{r_{i}}^{2GM}\sqrt{\frac{r}{2GM}}dr=\frac{2}{3}\frac{r_{i}^{\nicefrac{3}{2}}-\left(2GM\right)^{\nicefrac{3}{2}}}{\sqrt{2GM}}<\infty\ifS\else\fi
\end{align*}
\ifS
\else Samely \fi 
\begin{align*}
\Delta T_{Sch.\to0}= & \int_{2GM}^{0}dT=\frac{4}{3}GM.\ifS\else\fi
\end{align*}

\section{Eddington-Filkenstein metric}

\subsection{Singularity of Schwarzschild}

\begin{figure}
\includegraphics[width=1\columnwidth]{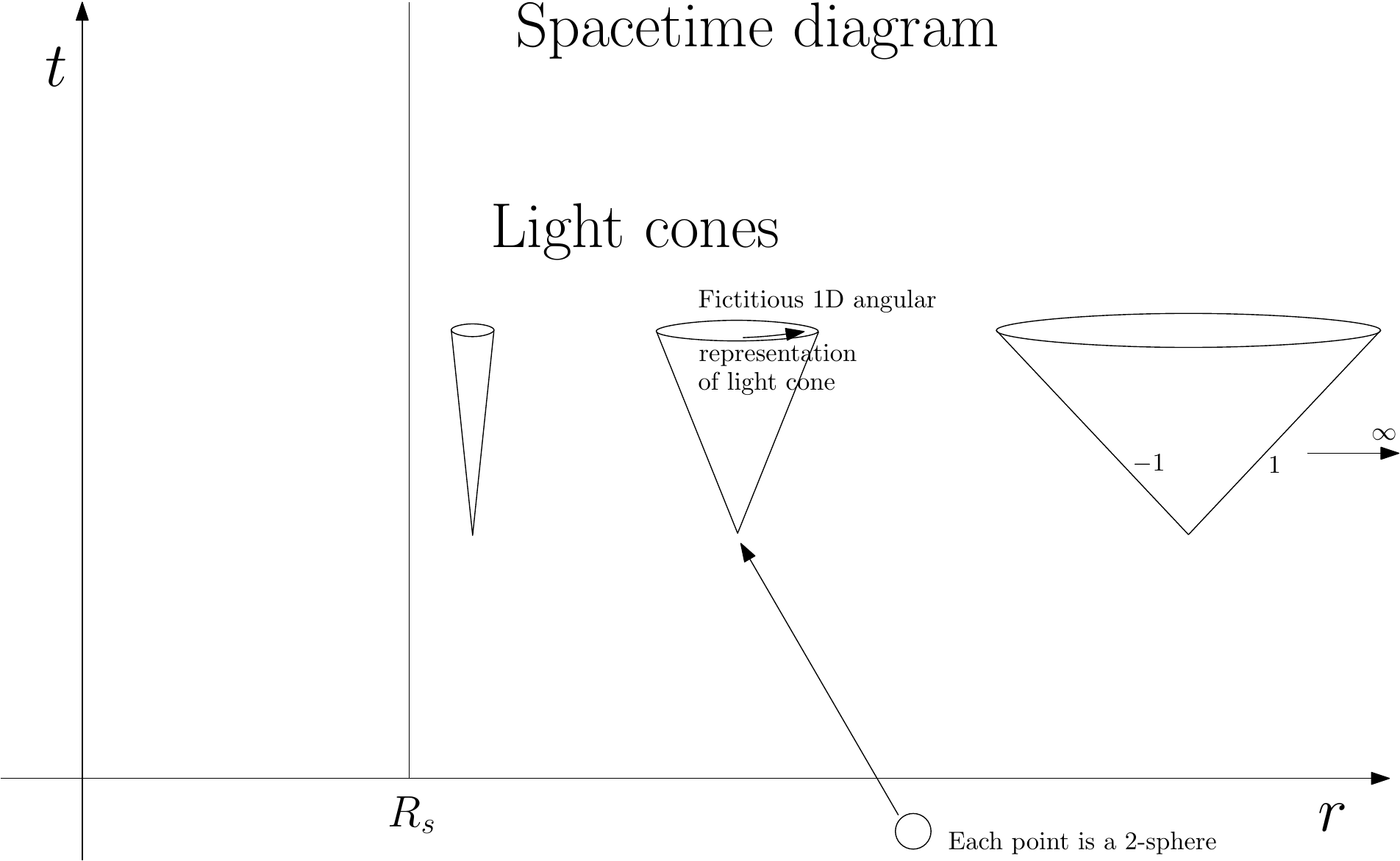}

\caption{\label{fig:Spacetime-diagram-of}\ifS
\else Schwarzschild radial spacetime diagram\fi  }
\end{figure}
\ifS
\else The Sch. metric admits
2 coordinate singularities at $r=0$ (term $\frac{R_{s}}{r}$) and
$r=2GM$ ($g_{rr}$ term). 

Computing the scalar\fi 
\begin{align*}
R_{abcd}R^{abcd}= & \frac{48G^{2}M^{2}}{r^{6}},\ifS\else\fi
\end{align*}
\ifS
\else we see 0 is probably
a true singularity.

$2GM$ may be a mere coordinate singularity\fi 

\subsection{Light curves in Schwarzschild}

\ifS
\else We look at light curves crossing $r=2GM$ or not.

From metric, radial null curves $d\Omega=0=ds^{2}$
\begin{align*}
ds^{2}=0= & -\left(1-\frac{2GM}{r}\right)dt^{2}+\left(1-\frac{2GM}{r}\right)^{-1}dr^{2}\\
\Rightarrow\frac{dr}{dt}= & \pm\left(1-\frac{2GM}{r}\right)\left\{ \begin{array}{cr}
\underset{r\to R_{s}}{\longrightarrow} & 0\\
\underset{r\to\infty}{\longrightarrow} & \pm1
\end{array}\right.\\
\textrm{or }\frac{dt}{dr}= & \pm\left(1-\frac{2GM}{r}\right)^{-1}\left\{ \begin{array}{cr}
\underset{r\to R_{s}^{+}}{\longrightarrow} & \pm\infty\\
\underset{r\to\infty}{\longrightarrow} & \pm1
\end{array}\right.
\end{align*}
thus light cone aperture$\to0$\fi 

\subsection{Tortoise coordinates\label{subsec:Tortoise-coordinates}}

\begin{figure}
\includegraphics[width=1\columnwidth]{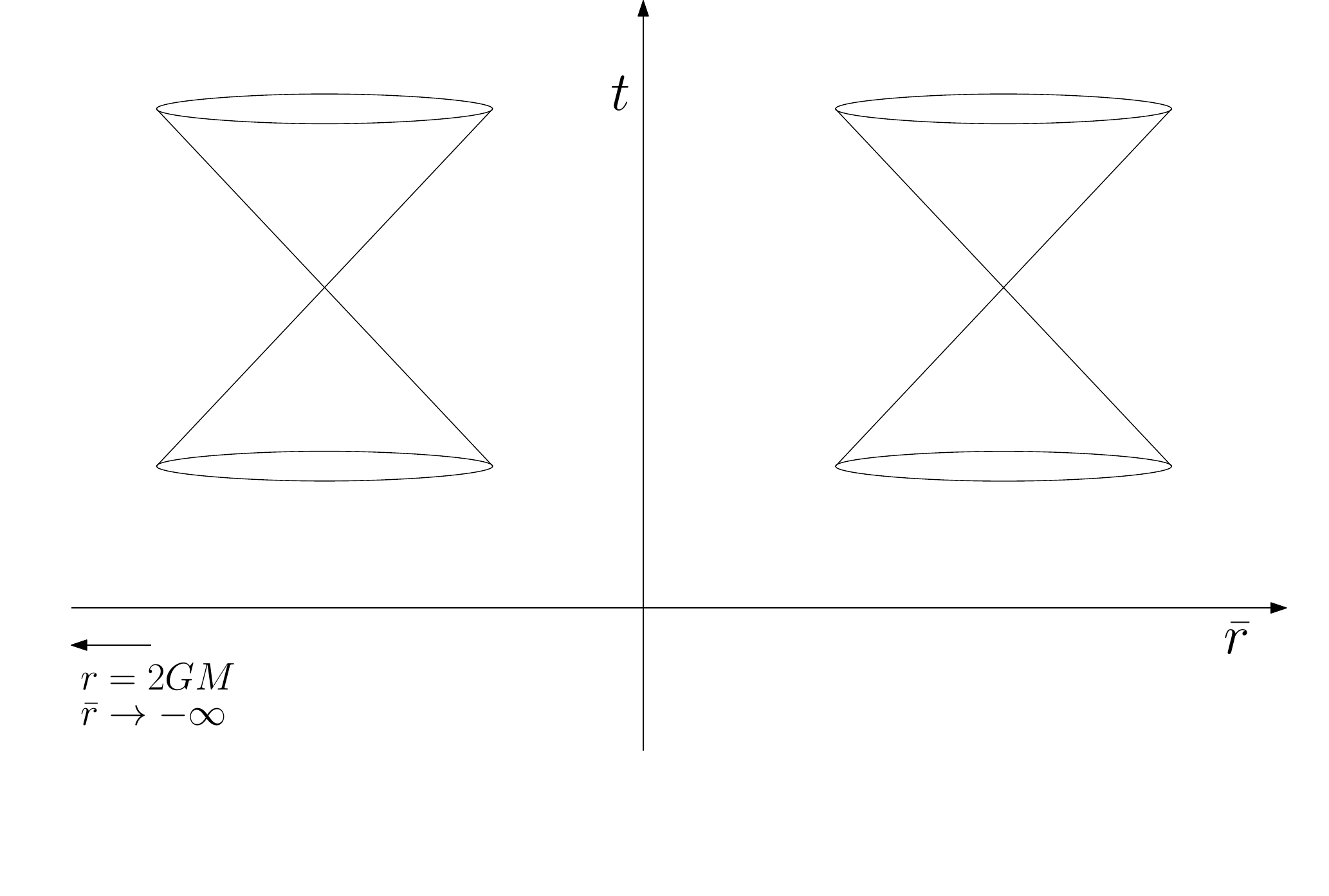}

\caption{\label{fig:-Outer-tortoise}\ifS
\else Outer tortoise coordinates spacetime
diagram\fi }
\end{figure}
\ifS
\else Solve $\frac{dt}{dr}$
to get new radial coordinates with constant light cone\fi 
\begin{align*}
t= & \pm\bar{r}+cst.\ifS\else\fi
\end{align*}
\ifS
\else 
\begin{align*}
\Rightarrow dt=\pm\frac{dr}{1-\frac{2GM}{r}}=\pm\frac{rdr}{r-2GM}= & \pm\left(dr+\frac{2GMdr}{r-2GM}\right)\\
= & \left\{ \begin{array}{ll}
\pm\left(dr+2GMd\ln\left(\frac{r}{2GM}-1\right)\right), & r>2GM\\
\pm\left(dr+2GMd\ln\left(1-\frac{r}{2GM}\right)\right), & r<2GM
\end{array}\right\} \\
\Rightarrow\bar{r}= & r+2GM\ln\left(\frac{r}{2GM}-1\right),\;r>2GM
\end{align*}
Metric becomes\fi 
\begin{align*}
ds^{2}= & -\left(1-\frac{2GM}{r}\right)\left(-dt^{2}+dr^{2}\right)+r^{2}d\Omega^{2}
\end{align*}
\ifS
\else Light cones constant but \fi $r=2GM$ corresponds to $\bar{r}\to-\infty$.

\subsection{Eddington-Finkelstein coordinates\label{subsec:Eddington-Finkelstein-coordinate}}

\begin{figure}
\begin{centering}
\includegraphics[width=0.5\columnwidth]{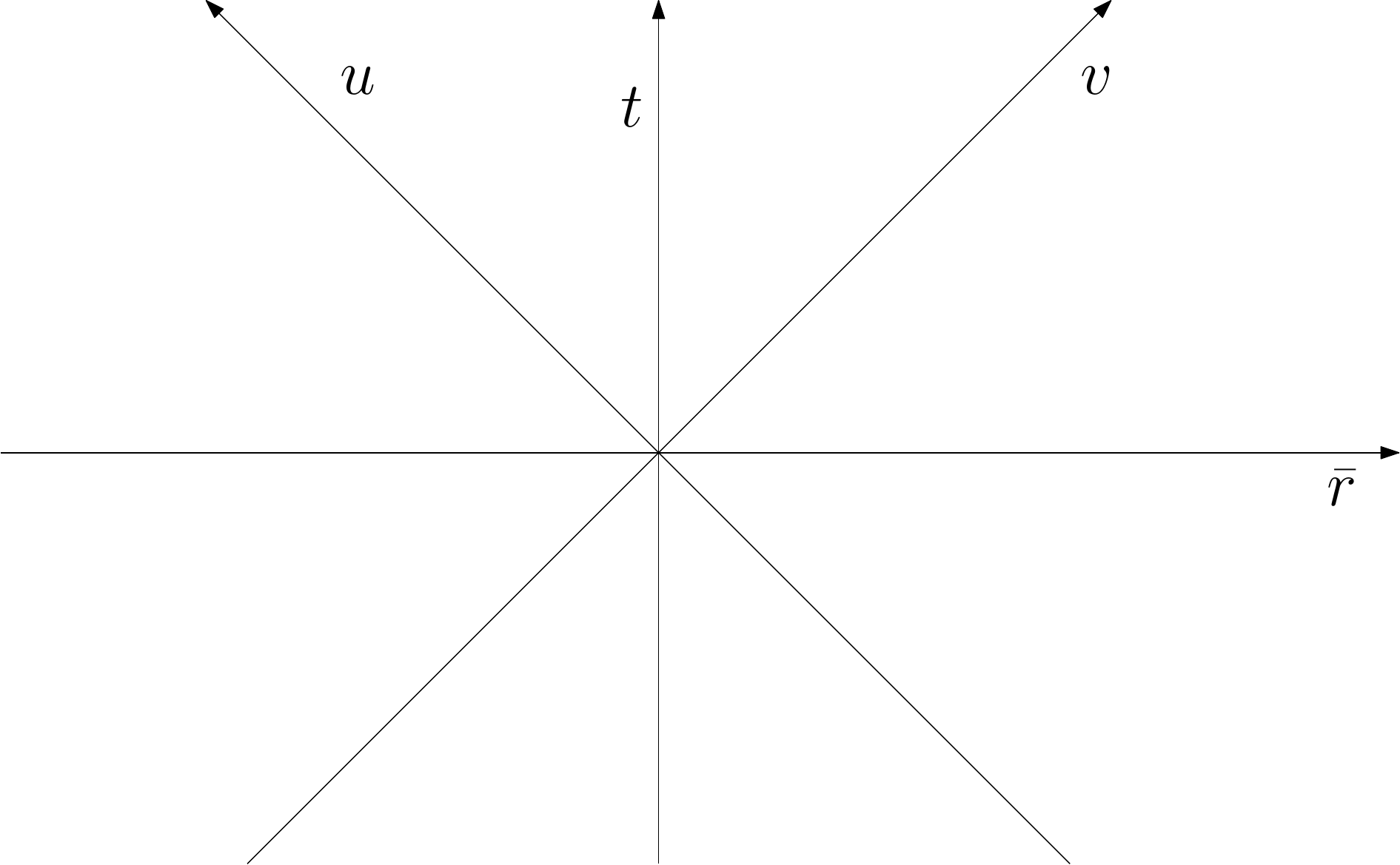}
\par\end{centering}
\caption{\label{fig:Tortoise-null-coordinates}Tortoise null coordinates}

\end{figure}
\ifS
\else Natural null coordinates: along tortoise light cones: \fi 
\begin{align*}
v= & t+\bar{r}, & \textrm{ingoing curves: }v= & cst,\ifS\else\fi\\
u= & t-\bar{r}, & \textrm{outgoing curves: }u= & cst.\ifS\else\fi
\end{align*}
\ifS
\else Interested
in ingoing light curves: choose $v$ and $r$. Metric becomes Eddington-Finkelstein:(exercise\exo :
find E.F. from S.)\fi 
\begin{align*}
ds^{2}= & -\left(1-\frac{2GM}{r}\right)dv^{2}+2dvdr+r^{2}d\Omega^{2}.\ifS\else\fi
\end{align*}
\begin{figure}
\includegraphics[width=1\columnwidth]{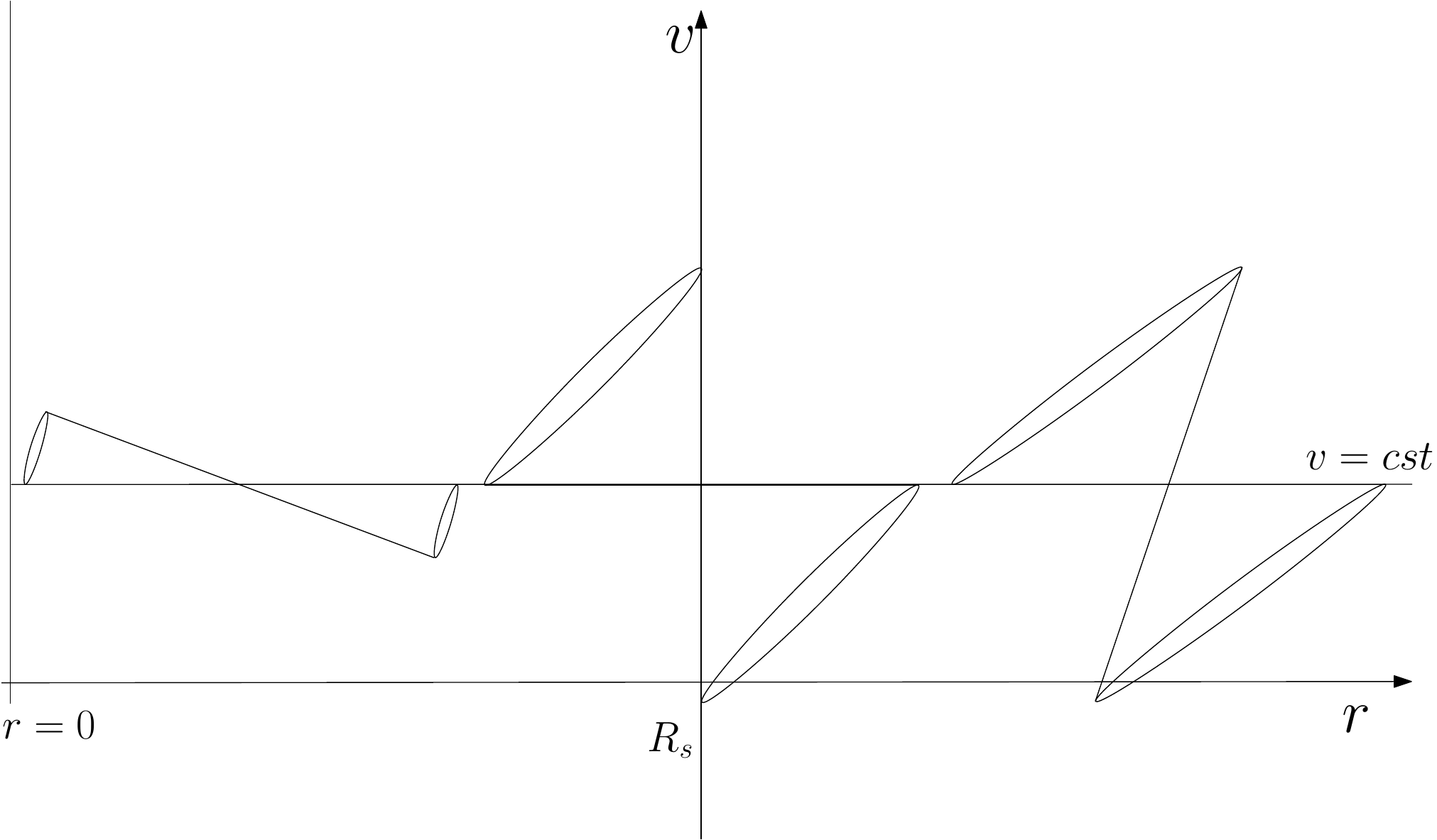}

\caption{\label{fig:Eddigton-Finkelstein-light-cones}Eddigton-Finkelstein
light cones}
\end{figure}
\ifS
\else At $r=2GM$, $\det g=-r^{4}\sin^{2}\theta\ne0$
so $g$ invertible and $R_{s}$ not singular.

Null curves (radial)
\begin{align*}
ds^{2}=0= & -\left(1-\frac{2GM}{r}\right)dv^{2}+2dvdr\\
\Rightarrow & \begin{array}{l}
\left\{ \begin{array}{ll}
\frac{dv}{dr}=0 & :\textrm{ ingoing}\\
\frac{dv}{dr}=\frac{2}{1-\frac{2GM}{r}} & :\textrm{ outgoing}
\end{array}\right.\\
\hfill\underset{r\to2GM}{\longrightarrow}\infty
\end{array}
\end{align*}
so outgoing l.c. becomes ingoing too at $R_{s}:$ 

Event horizon: particle cannot go to bigger $r$\fi 

\section{Kruskal metric}

\ifS
\else In E.F., $v=cst$ and $r\searrow R_{s}^{+}\Rightarrow t\to+\infty$
($v=t+\underset{\overset{\downarrow R_{s}}{-\infty}}{\bar{r}}$) 

If $u$ is chosen, we get a symmetric picture but $u=cst$ and $r\searrow R_{s}^{+}\Rightarrow t\to-\infty$
($u=t-\underset{\overset{\downarrow R_{s}}{-\infty}}{\bar{r}}$) (exercise:
find this version\exo )\fi 
\begin{align*}
ds^{2}= & -\left(1-\frac{2GM}{r}\right)du^{2}-2dudr+r^{2}d\Omega^{2}.\ifS\else\fi
\end{align*}
 
\begin{figure}
\includegraphics[width=1\columnwidth]{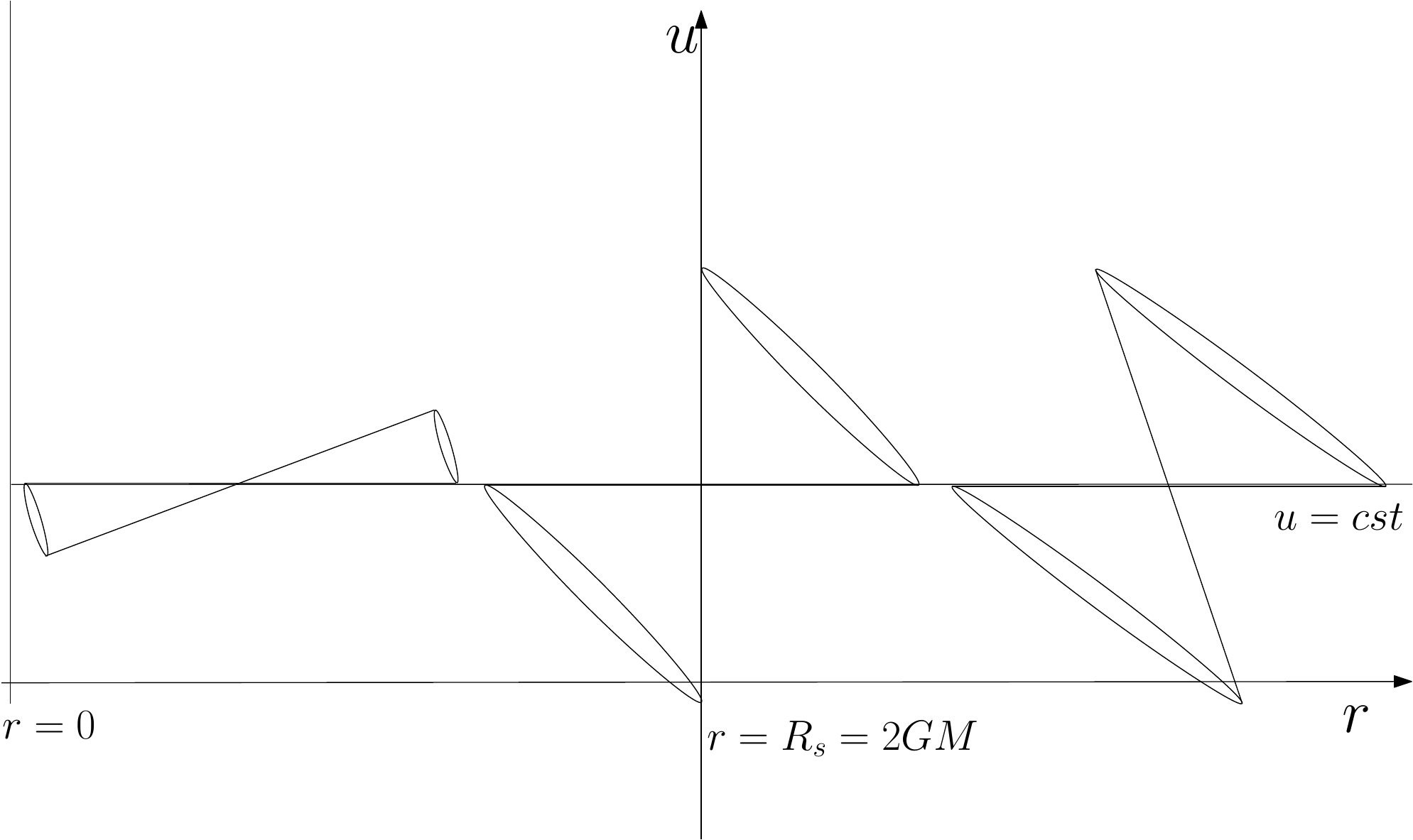}

\caption{\label{fig:Eddigton-Finkelstein-light}Eddigton-Finkelstein past light
cones}
\end{figure}
\ifS
\else To
cover all light curves, use $u$ and $v$. However $r$ spatial localisation
implicit
\begin{align*}
\left.\begin{array}{rl}
dv= & dt+d\bar{r}\\
du= & dt-d\bar{r}
\end{array}\right\} \Rightarrow-dt^{2}+d\bar{r}^{2}= & -dudv
\end{align*}
so
\begin{align*}
ds^{2}= & -\left(1-\frac{2GM}{r}\right)dudv+r^{2}d\Omega^{2}
\end{align*}
but
\begin{align*}
\frac{1}{2}\left(v-u\right)=\bar{r}= & r+2GM\ln\left(\frac{r}{2GM}-1\right)\underset{r\to2GM}{\longrightarrow}-\infty
\end{align*}
so$u\to+\infty$ or $v\to-\infty$\fi 

We need to compactify the null directions\ifS
\else Tortoise radius in null
coordinates yields\fi 
\begin{align*}
1-\frac{2GM}{r}= & \frac{2GM}{r}\left(\frac{r}{2GM}-1\right)=\frac{2GM}{r}e^{-\frac{r}{2GM}}e^{\frac{v-u}{4GM}}\epsilon_{s},\ifS\else\fi
\end{align*}
\ifS
\else so\fi 
\begin{align*}
-\left(1-\frac{2GM}{r}\right)dudv= & -\frac{2GM}{r}e^{-\frac{r}{2GM}}\epsilon_{s}e^{-\frac{u}{4GM}}du\,e^{\frac{v}{4GM}}dv,\ifS\else\fi
\end{align*}
\ifS
\else from which we can
extract\fi 
\begin{align*}
e^{-\frac{u}{4GM}}du= & 4GM\epsilon_{U}dU\\
e^{\frac{v}{4GM}}dv= & 4GM\epsilon_{V}dV
\end{align*}
\begin{figure}
\begin{centering}
\includegraphics[width=0.5\columnwidth]{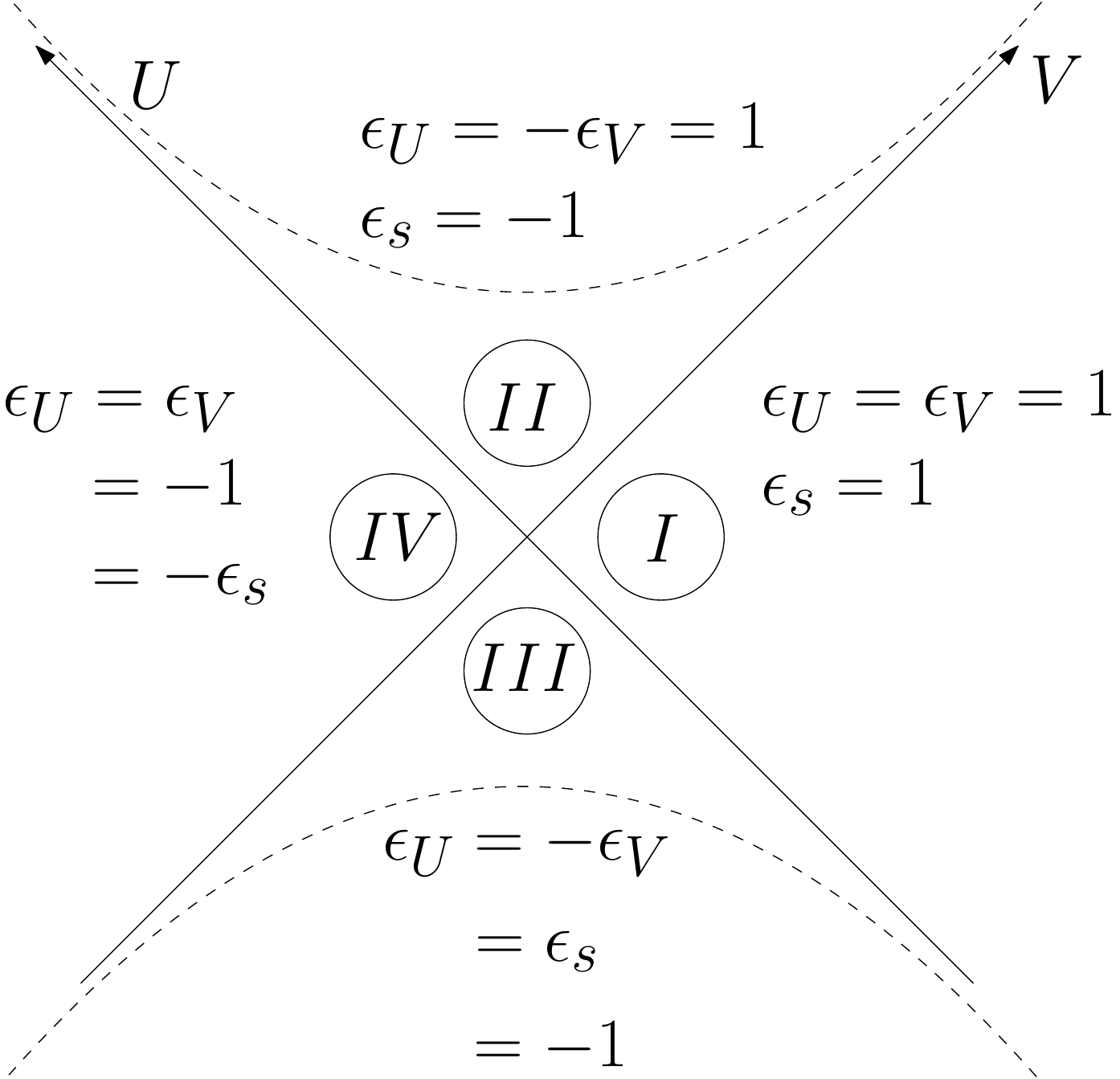}
\par\end{centering}
\caption{\label{fig:Compact-null-coordinates}Compact null coordinates sign
chart. The $U-V$ plane is divided into four sections numbered I-IV.}

\end{figure}
\ifS
\else $\Rightarrow$
\begin{align*}
U= & -\epsilon_{U}e^{-\frac{u}{4GM}} & \underset{u\to\infty}{\longrightarrow}0\\
V= & \epsilon_{V}e^{\frac{v}{4GM}} & \underset{v\to-\infty}{\longrightarrow}0
\end{align*}
$(r\to R_{s})$, also given in Schwarzschild coordinates by
\begin{align*}
V= & \sqrt{\epsilon_{s}\left(\frac{r}{2GM}-1\right)}\epsilon_{V}e^{\frac{r+t}{4GM}}\\
U= & -\sqrt{\epsilon_{s}\left(\frac{r}{2GM}-1\right)}\epsilon_{U}e^{\frac{r-t}{4GM}}
\end{align*}
(exercise: from tortoise\exo ) and the metric reads (Fig.~\ref{fig:Compact-null-coordinates}$\Rightarrow\epsilon_{s}=\epsilon_{U}\epsilon_{V}$)
\begin{align*}
ds^{2}= & -\frac{1}{2}\frac{\left(4GM\right)^{3}}{r}e^{-\frac{r}{2GM}}dUdV+r^{2}d\Omega^{2}
\end{align*}
Go from dual null $\left(\begin{array}{cccc}
U, & V, & \theta & \varphi\end{array}\right)$ system to $\left(\begin{array}{cccc}
T, & R, & \theta & \varphi\end{array}\right)$:

Define 
\begin{align*}
T=\frac{1}{2}\left(U+V\right)= & \sqrt{\epsilon_{s}\frac{r}{2GM}-1}e^{\frac{r}{4GM}}\epsilon_{V}\left[\frac{1+\epsilon_{s}}{2}\sinh\frac{t}{4GM}+\frac{1-\epsilon_{s}}{2}\cosh\frac{t}{4GM}\right]\\
R=\frac{1}{2}\left(V-U\right)= & \sqrt{\epsilon_{s}\frac{r}{2GM}-1}e^{\frac{r}{4GM}}\epsilon_{V}\left[\frac{1+\epsilon_{s}}{2}\cosh\frac{t}{4GM}+\frac{1-\epsilon_{s}}{2}\sinh\frac{t}{4GM}\right]\\
\Leftrightarrow & \left\{ \begin{array}{rl}
U= & T-R\\
V= & T+R
\end{array}\right.
\end{align*}
Kruskal\fi 

\begin{figure}
\begin{centering}
\includegraphics[width=1\columnwidth]{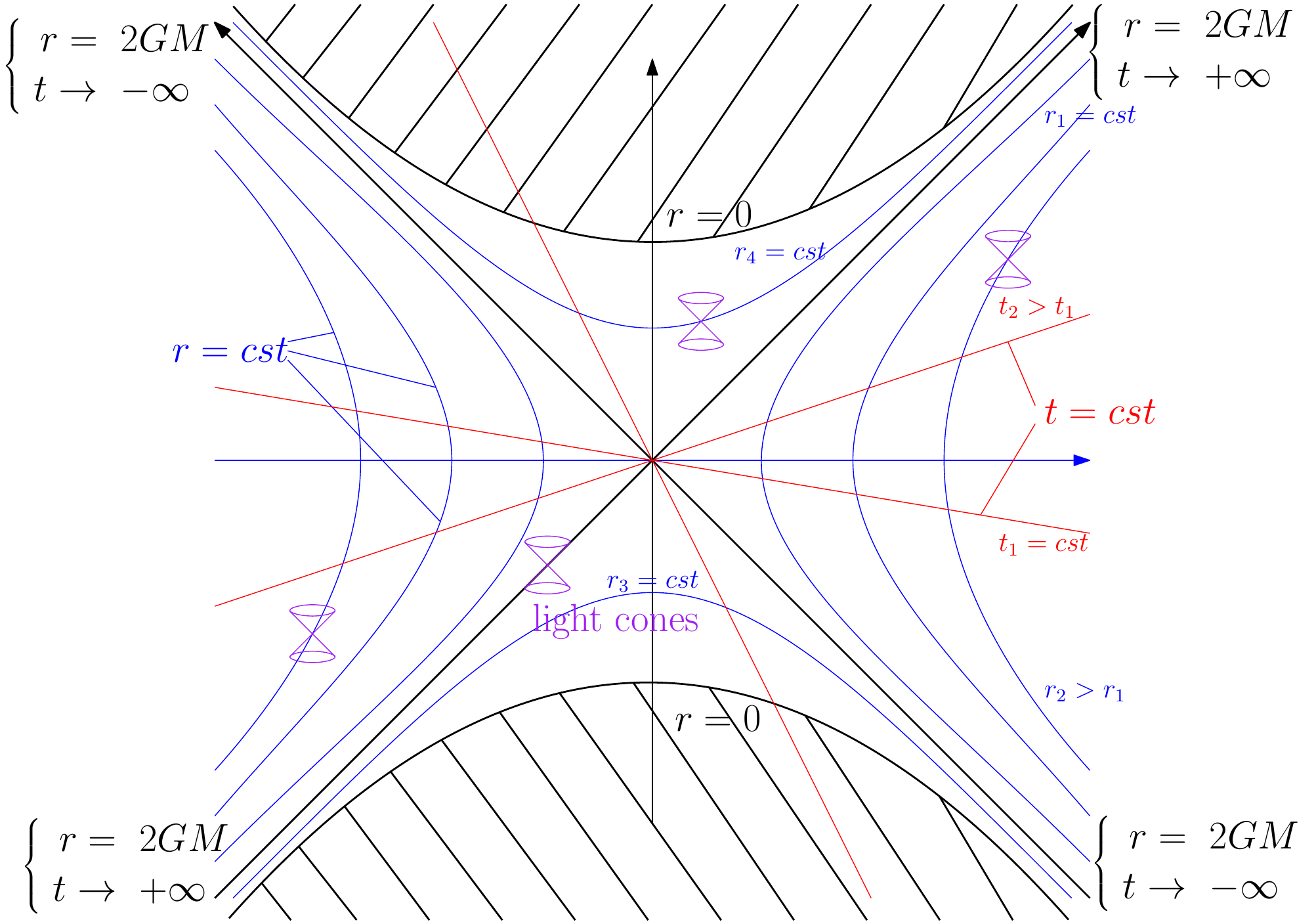}
\par\end{centering}
\caption{\label{fig:Kruskal-spacetime-diagram}Kruskal spacetime diagram}

\end{figure}

\begin{figure}
\begin{centering}
\begin{tabular}{cc}
\multirow{5}{*}{\includegraphics[width=0.6\columnwidth]{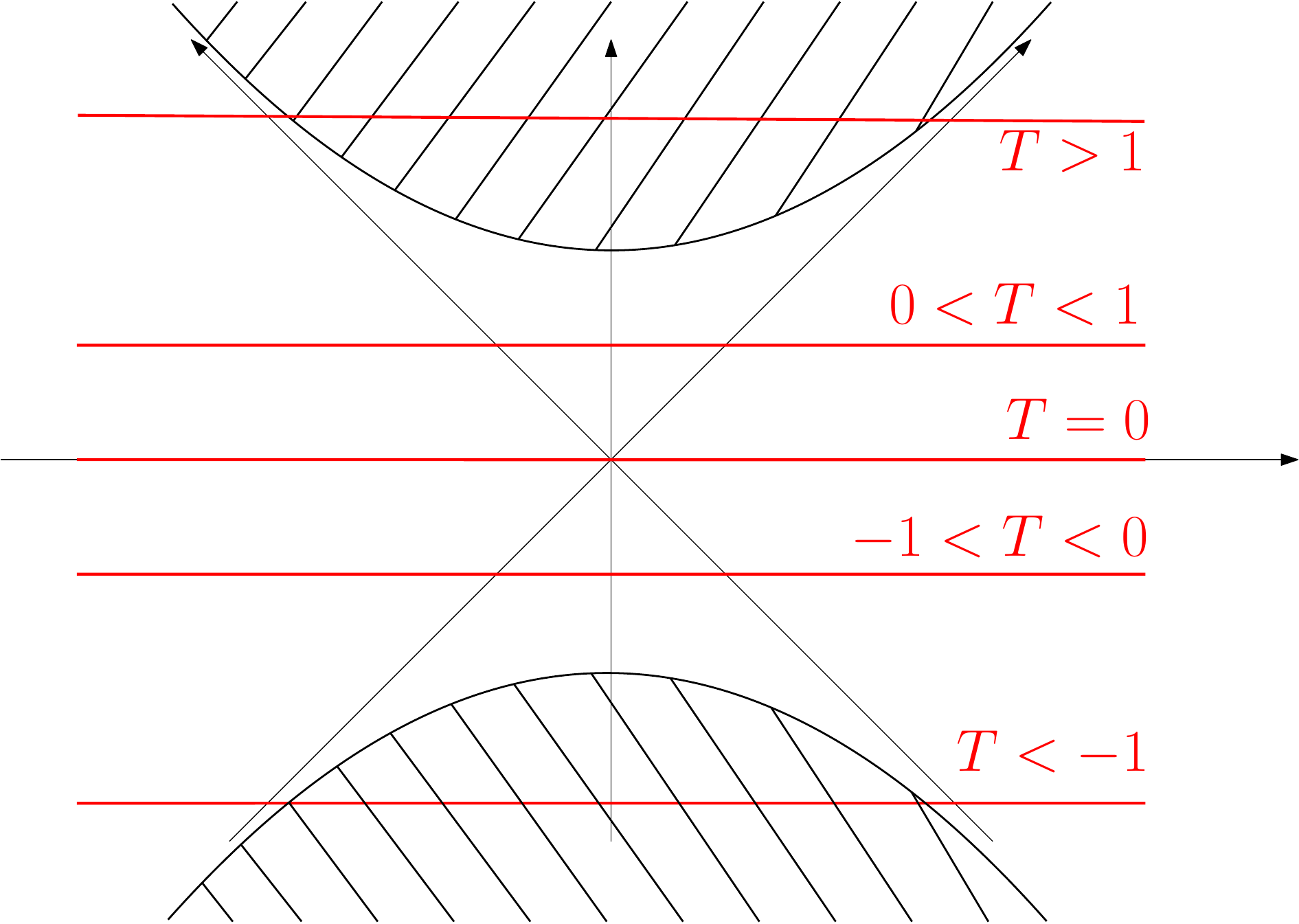}} & \includegraphics[width=0.3\columnwidth]{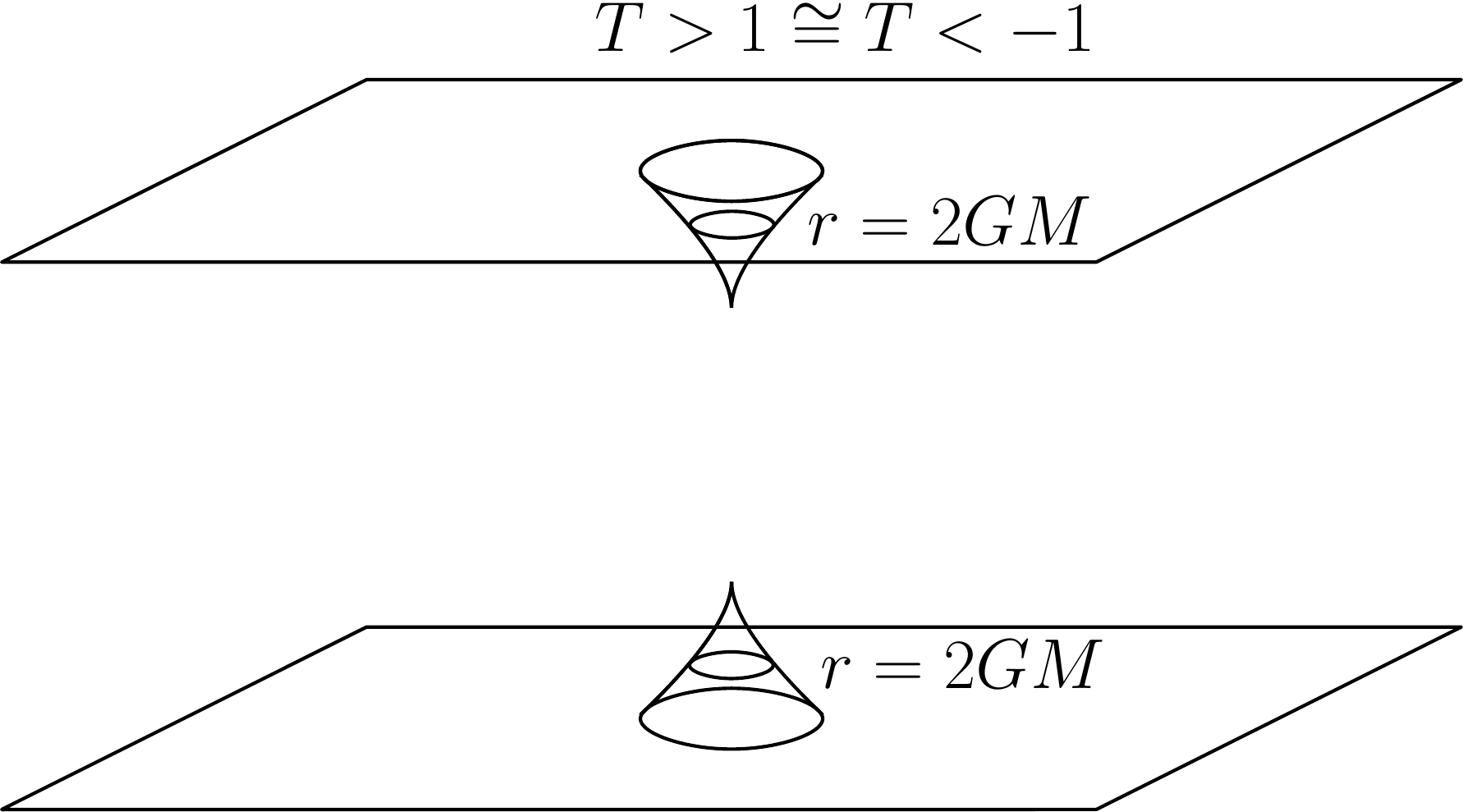}\tabularnewline
 & \tabularnewline
 & \includegraphics[width=0.3\columnwidth]{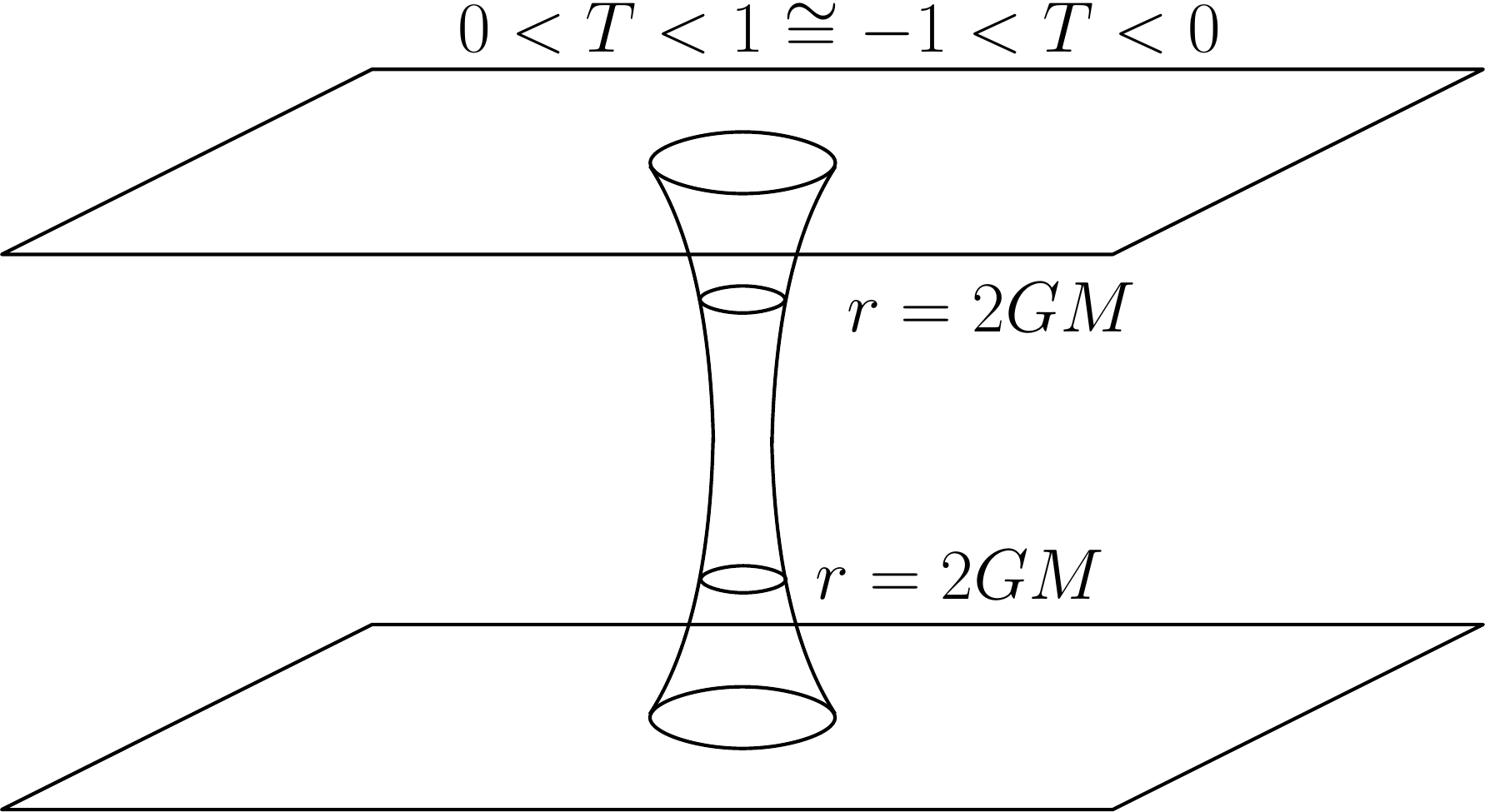}\tabularnewline
 & \tabularnewline
 & \includegraphics[width=0.3\columnwidth]{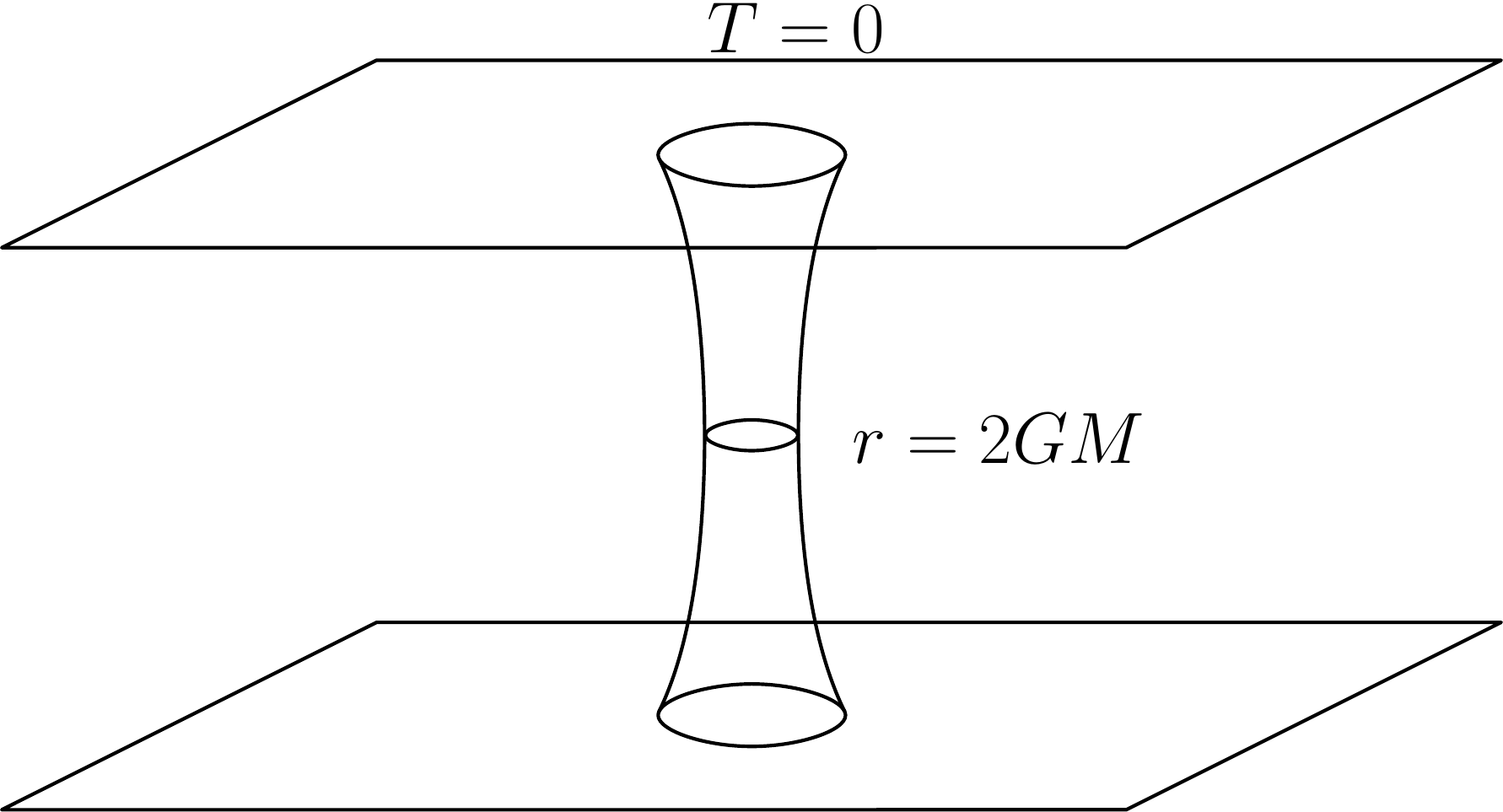}\tabularnewline
\end{tabular}
\par\end{centering}
\caption{\label{fig:Kruskal-Time-slices}Kruskal Time slices structure}
\end{figure}

Since 
\begin{align*}
dT^{2}-dR^{2}= & \frac{1}{4}\left[dU^{2}+dV^{2}+2dUdV-dU^{2}-dV^{2}+2dUdV\right]=dUdV
\end{align*}
\ifS
\else  we have \fi  the Kruskal metric
\begin{align*}
ds^{2}= & \frac{32\left(GM\right)^{3}}{r}e^{-\frac{r}{2GM}}\left(-dT^{2}+dR^{2}\right)+r^{2}d\Omega^{2},\ifS\else\fi
\end{align*}
defining $r$ implicitly from
\begin{align}
T^{2}-R^{2}=UV= & \left(1-\frac{r}{2GM}\right)e^{\frac{r}{2GM}}.\label{eq:ImplicitR}
\end{align}
\ifS
\else Radial null curves: $u=cst$ or $v=cst$ $\Leftrightarrow U=cst$
or $V=cst$ $\Leftrightarrow T=R+cst$ or $T=-R+cst$ 

but event horizon $r=2GM\Rightarrow T^{2}=R^{2}$ so is at $T=\pm R$
which is a null surface!

Constant $r$ surfaces are given by $T^{2}-R^{2}=cst$: hyperbolae
of $T-R$ plane

Constant $t$ surfaces in \encircle{I} are given by $\frac{T}{R}=\tanh\frac{t}{4GM}=cst$:
lines $T=\left(\tanh\frac{t}{4GM}\right)R$. Slope $\tanh\frac{t}{4GM}\underset{t\to\pm\infty}{\longrightarrow}\pm1$
so $t\to\pm\infty\Leftrightarrow r=2GM$

$T,R$ range avoids $r=0:T^{2}-R^{2}=1$

so $T^{2}-R^{2}-1<0\Leftrightarrow T^{2}<R^{2}+1,R\in\mathbb{R}$

We can now draw a spacetime diagram in the $T-R$ plane (Fig.~\ref{fig:Kruskal-spacetime-diagram})

Each point on the diagram is a 2-sphere:

Maximal extension of the Schwarzschild geometry, coords cover the
entire solution manifold

divide diagram in 4 regions:

\encircle{I}: $r>2GM$, original coords well defined: outer Schwarzschild

\encircle{II}: inside horizon: BH

Lightcones become spacelike and all end on $r=0$ singularity

For \encircle{II}: same procedure gives
\begin{align*}
ds^{2}= & -\frac{32\left(GM\right)^{3}}{r}e^{-\frac{r}{2GM}}\left(dT^{2}-dR^{2}\right)+r^{2}d\Omega^{2},\\
T^{2}-R^{2}= & \left(\frac{r}{2GM}-1\right)e^{\frac{r}{2GM}}.
\end{align*}
(exercise: show from tortoise procedure\exo )

\begin{align*}
\begin{array}{rl}
R: & \textrm{ spacelike }\\
T: & \textrm{ timelike }
\end{array}\ne\textrm{from Schw.:} & \begin{array}{rl}
t: & \textrm{ spacelike }\\
r: & \textrm{ timelike }
\end{array}
\end{align*}
$r=0$ singularity is spacelike

\encircle{III}: time reverse of \encircle{II}: white hole

\encircle{IV}: mirror image of \encircle{I}: other side of wormhole
(Einstein-Rosen bridge) instantaneously closed (no timelike curve
accross)

Spacetime slices: Fig.~\ref{fig:Kruskal-Time-slices}

Compactify Kruskal diagram for better analysis\fi 

\chapter{Conformal, Carter-Penrose, diagrams}

\ifS
\else Represent symmetric spacetime in a compact way
to explore global and causal structure $\Rightarrow$ keep lightcones
evidents.
\begin{itemize}
\item Symmetry $\longrightarrow$ ignore angular part: 1D time $T$, 1D
space $R$
\item Causal structure $\longrightarrow$ light cones keep at $dT=\pm dR\left\{ \begin{array}{rl}
r= & 2GM\\
t\to & +\infty
\end{array}\right.$
\end{itemize}
\fi 

\section{Conformal transformation}

\ifS
\else 
\begin{align*}
\tilde{g}_{ab}= & \omega^{2}\left(x\right)g_{ab} & \Leftrightarrow d\tilde{s}^{2}= & \omega^{2}\left(x\right)ds^{2}
\end{align*}
$x^{a}\left(\lambda\right)$ null curve $\Rightarrow g_{ab}\frac{dx^{a}}{d\lambda}\frac{dx^{b}}{d\lambda}=0$
tangent vector: null

$\Rightarrow\tilde{g}_{ab}\frac{dx^{a}}{d\lambda}\frac{dx^{b}}{d\lambda}=\omega^{2}g_{ab}\frac{dx^{a}}{d\lambda}\frac{dx^{b}}{d\lambda}=0$
: conformal transformation keeps light curves invariant\fi 

\ifS
\else 

\section{Minkowski metric}

In spherical coordinates
\begin{align*}
ds_{M}^{2}= & -dt^{2}+dr^{2}+r^{2}d\Omega^{2},
\end{align*}
 $d\Omega^{2}=d\theta^{2}+\sin^{2}\theta\,d\varphi^{2}$ metric on
unit 2-sphere

Radial light cones already at 45\textdegree : $dt^{2}=dr^{2}$

\fi 

\ifS
\else 

\subsection{Compactify $\left(t,r\right)$}

$\begin{array}{rl}
t\in & \mathbb{R}\\
r\in & \mathbb{R}^{+}
\end{array}$ Try $\begin{array}{rl}
\bar{t}= & \arctan t\\
\bar{r}= & \arctan r
\end{array}$ 

Since $d\tan x=\frac{dx}{\cos^{2}x}$,
\begin{align*}
d\bar{s}^{2}= & -\frac{1}{\cos^{2}\bar{t}}d\bar{t}^{2}+\frac{d\bar{r}^{2}}{\cos^{2}\bar{r}}+\tan^{2}\bar{r}d\Omega^{2}
\end{align*}
$\begin{array}{rl}
\bar{t}\in & \left]-\frac{\pi}{2};\frac{\pi}{2}\right[\\
\bar{r}\in & \left[0;\frac{\pi}{2}\right[
\end{array}$ but $\frac{d\bar{t}}{d\bar{r}}=\pm\frac{\cos^{2}\bar{t}}{\cos^{2}\bar{r}}\ne\pm1$
causality spacetime dependent.

\fi 
\begin{figure}
\begin{centering}
\includegraphics[width=0.5\columnwidth]{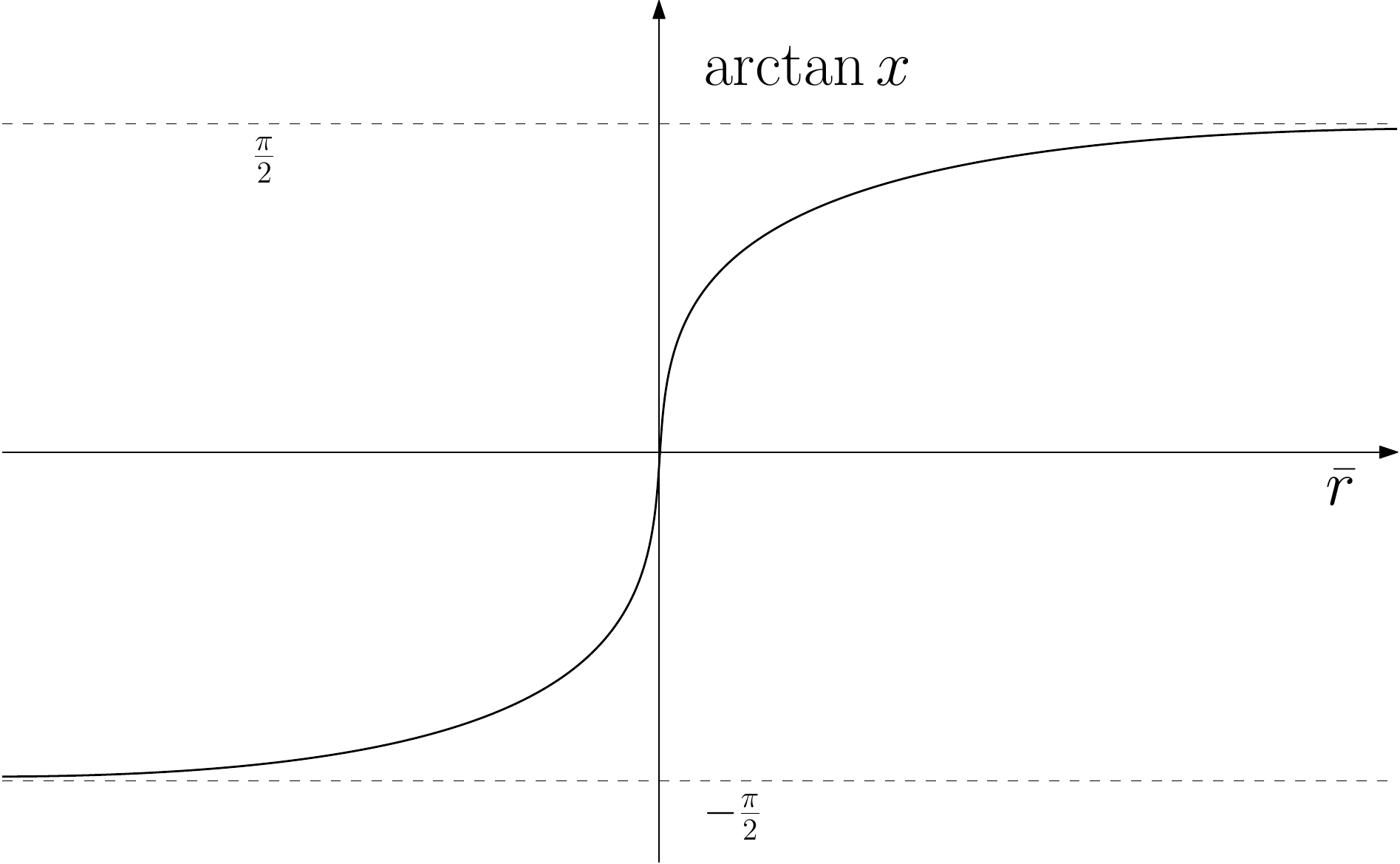}
\par\end{centering}
\caption{\label{fig:Example-of-compactification}Example of compactification
function: $\arctan x$}

\end{figure}

\ifS
\else 

\subsection{Compactify null coordinates}

Since we want to keep $dt^{2}=dr^{2}$ for null curves, use null coordinates,
constant on null curves:
\begin{align*}
u= & t-r & u\le & v & \Leftrightarrow r= & \frac{1}{2}\left(v-u\right)\ge0\\
v= & t+r & u,v\in & \mathbb{R} & t= & \frac{1}{2}\left(v+u\right)
\end{align*}
\fi 
\begin{figure}
\begin{centering}
\includegraphics[width=0.5\columnwidth]{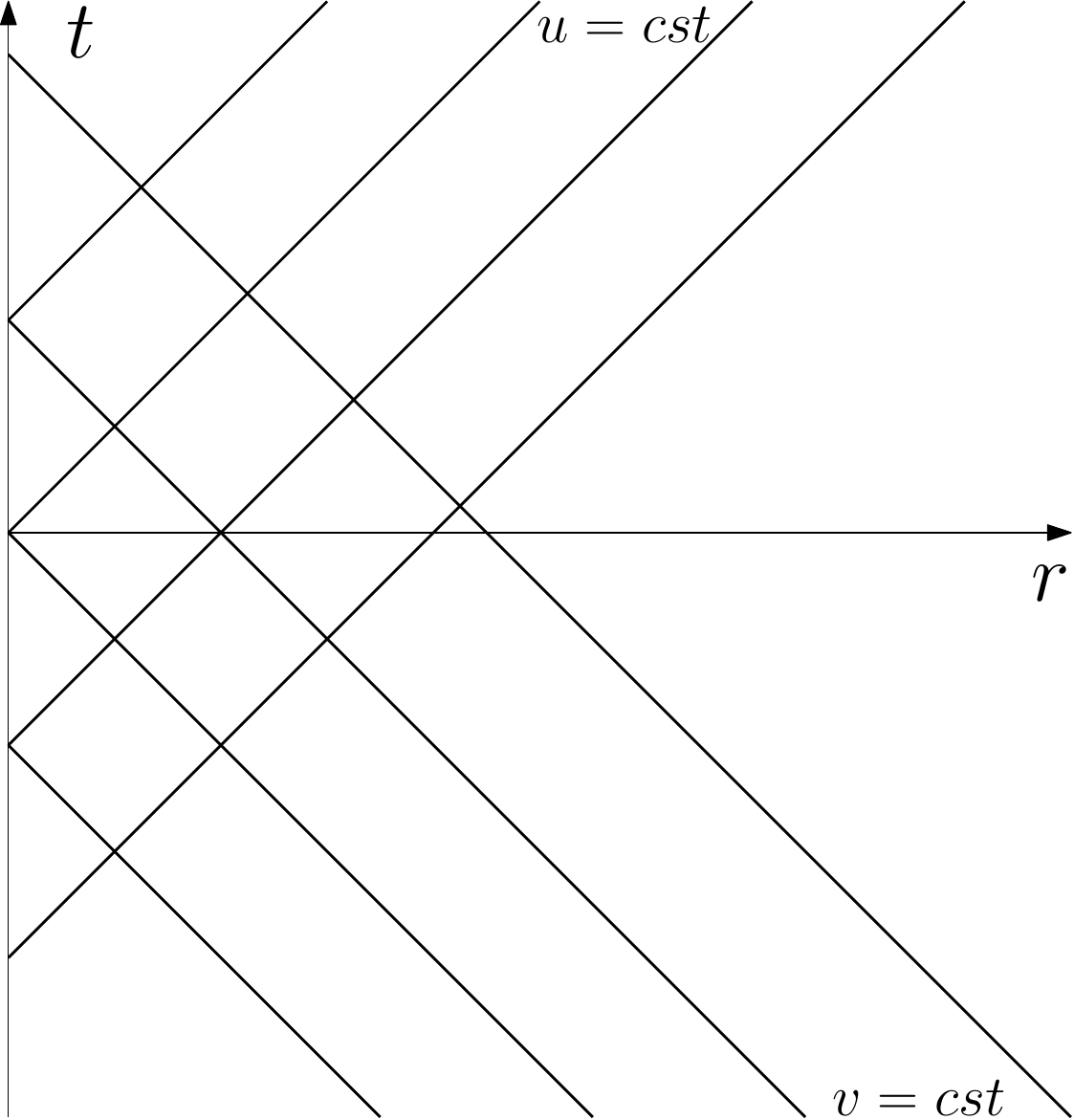}
\par\end{centering}
\caption{\label{fig:Spherical-Minkowski-null}Spherical Minkowski null curves}
\end{figure}
metric becomes
\begin{align*}
ds^{2}= & -dudv+\frac{1}{4}\left(v-u\right)^{2}d\Omega^{2},\ifS\else\fi
\end{align*}
(exercise: proof\exo )\ifS
\else then compactify with $\arctan x$
\begin{align*}
U= & \arctan u & U\le & V\\
V= & \arctan v & U,V\in & \left]-\frac{\pi}{2};\frac{\pi}{2}\right[\\
\Rightarrow dudv= & \frac{dUdV}{\cos^{2}U\cos^{2}V} & \textrm{ and }\left(v-u\right)^{2}= & \left(\tan V-\tan U\right)^{2}\\
 &  & = & \frac{1}{\cos^{2}U\cos^{2}V}\left(\sin V\cos U-\sin U\cos V\right)^{2}\\
 &  & = & \frac{\sin^{2}\left(V-U\right)}{\cos^{2}U\cos^{2}V}
\end{align*}
metric becomes now\fi 
\begin{align*}
ds^{2}= & \frac{1}{4\cos^{2}U\cos^{2}V}\left[-4dUdV+\sin^{2}\left(V-U\right)d\Omega^{2}\right]
\end{align*}
\ifS
\else Return to non-null coords.
\fi 
\begin{align*}
T= & V+U& \left|T\right|+R< & \pi\\
R= & V-U& R\in & \left[0;\pi\right[
\end{align*}
\ifS
\else so\footnote{since
\begin{align*}
\cos\left(a+b\right)= & \cos a\cos b-\sin a\sin b\\
\textrm{so }\cos\left(a-b\right)\cos\left(a-b\right)= & \cos^{2}a\cos^{2}b-\sin^{2}a\sin^{2}b\\
= & \left(\cos^{2}a-\sin^{2}a\right)\cos^{2}b+\left(\cos^{2}a-\sin^{2}a\right)\sin^{2}b\\
 & +\sin^{2}a\left(\cos^{2}b-\sin^{2}b\right)+\cos^{2}a\left(\cos^{2}b-\sin^{2}b\right)\\
 & +\sin^{2}a\sin^{2}b-\cos^{2}a\cos^{2}b\\
= & \cos2a\left(\cos^{2}b+\sin^{2}b\right)+\left(\cos^{2}a+\sin^{2}a\right)\cos2b\\
 & -\cos\left(a-b\right)\cos\left(a-b\right)\\
\Leftrightarrow2\cos\left(a-b\right)\cos\left(a-b\right)= & \cos2a+\cos2b.
\end{align*}
} \fi 
\begin{align*}
ds^{2}= & \omega^{-2}\left(T,R\right)\left[-dT^{2}+dR^{2}+\sin^{2}R\,d\Omega^{2}\right]\\
\omega= & 2\cos U\cos V\\
= & 2\cos\left[\frac{1}{2}\left(T-R\right)\right]\cos\left[\frac{1}{2}\left(T+R\right)\right]\\
= & \cos T+\cos R.\ifS\else\fi
\end{align*}
\ifS
\else Thus Minkowski is conformally transformed
into a compact metric\fi 
\begin{align*}
d\tilde{s}^{2}=\omega^{2}\left(T,R\right)ds^{2}= & -dT^{2}+dR^{2}+\sin^{2}R\,d\Omega^{2}
\end{align*}
\ifS
\else describing $\mathbb{R}\times S^{3}$ with curvature\fi 
\begin{align*}
\tilde{r}= & \sin R & \Leftrightarrow\hfill d\tilde{r}= & \cos RdR\\
 &  & \Leftrightarrow\frac{d\tilde{r}^{2}}{1-\tilde{r}^{2}}= & dR^{2},\ifS\else\fi
\end{align*}
\ifS\else so\fi 
\begin{align*}
d\tilde{s}^{2}= & -dT^{2}+\frac{d\tilde{r}^{2}}{1-\tilde{r}^{2}}+\tilde{r}^{2}d\Omega^{2}\ifS\else\fi
\end{align*}
\ifS
\else R.W. with positive curvature: Einstein static
universe (perfect fluid + cosmological constant)

Representing $\mathbb{R}\times S^{3}$ as a cylinder, Minkowski is
restricted by $\left|T\right|+R<\pi$ and $R>0\,$

Circles are constant $T$ slices 3-spheres

Unfolding Minkowski gives a triangle which is the conformal compactification
of Minkowski:

Minkowski + boundary: conformal infinity\fi 

~

\begin{tabular}{lcll}
$i^{+}$ & = & future timelike infinity ($T=\pi\Rightarrow R=0$) & \tabularnewline
$i^{0}$ & = & spatial infinity ($T=0\Rightarrow R=\pi$) & \tabularnewline
$i^{-}$ & = & past timelike infinity ($T=-\pi\Rightarrow R=0$) & \tabularnewline
$\mathscr{I}^{+}$ & = & future null infinity surface ($T=\pi-R,0<R<\pi$) & \tabularnewline
 &  & pronounced ``scri plus'',\hspace*{\fill} LaTe$\chi$ \textbackslash{}mathscr\{I\}\textasciicircum{}+ & :$\mathscr{I}^{+}$\tabularnewline
 &  & \hspace*{\fill}or \textbackslash{}mathcal\{I\}\textasciicircum{}+ & :$\mathcal{I}^{+}$\tabularnewline
$\mathscr{I}^{-}$ & = & past null infinity surface ($T=R-\pi,0<R<\pi$) & \tabularnewline
\end{tabular}

\begin{figure}
\begin{centering}
\noindent\begin{minipage}[t]{1.1\columnwidth}%
\begin{center}
\includegraphics[width=0.48\columnwidth]{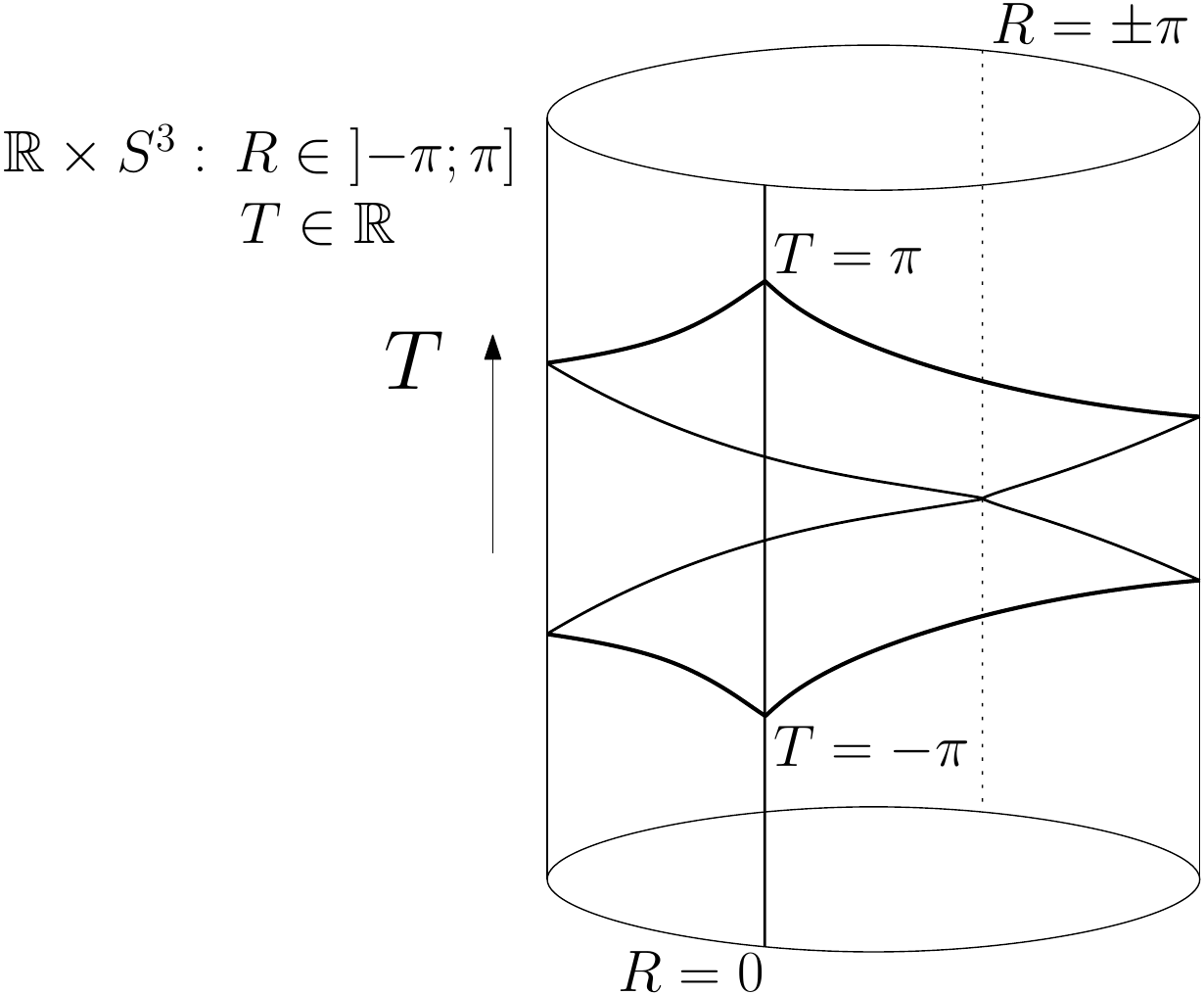}\quad{}\includegraphics[width=0.48\columnwidth]{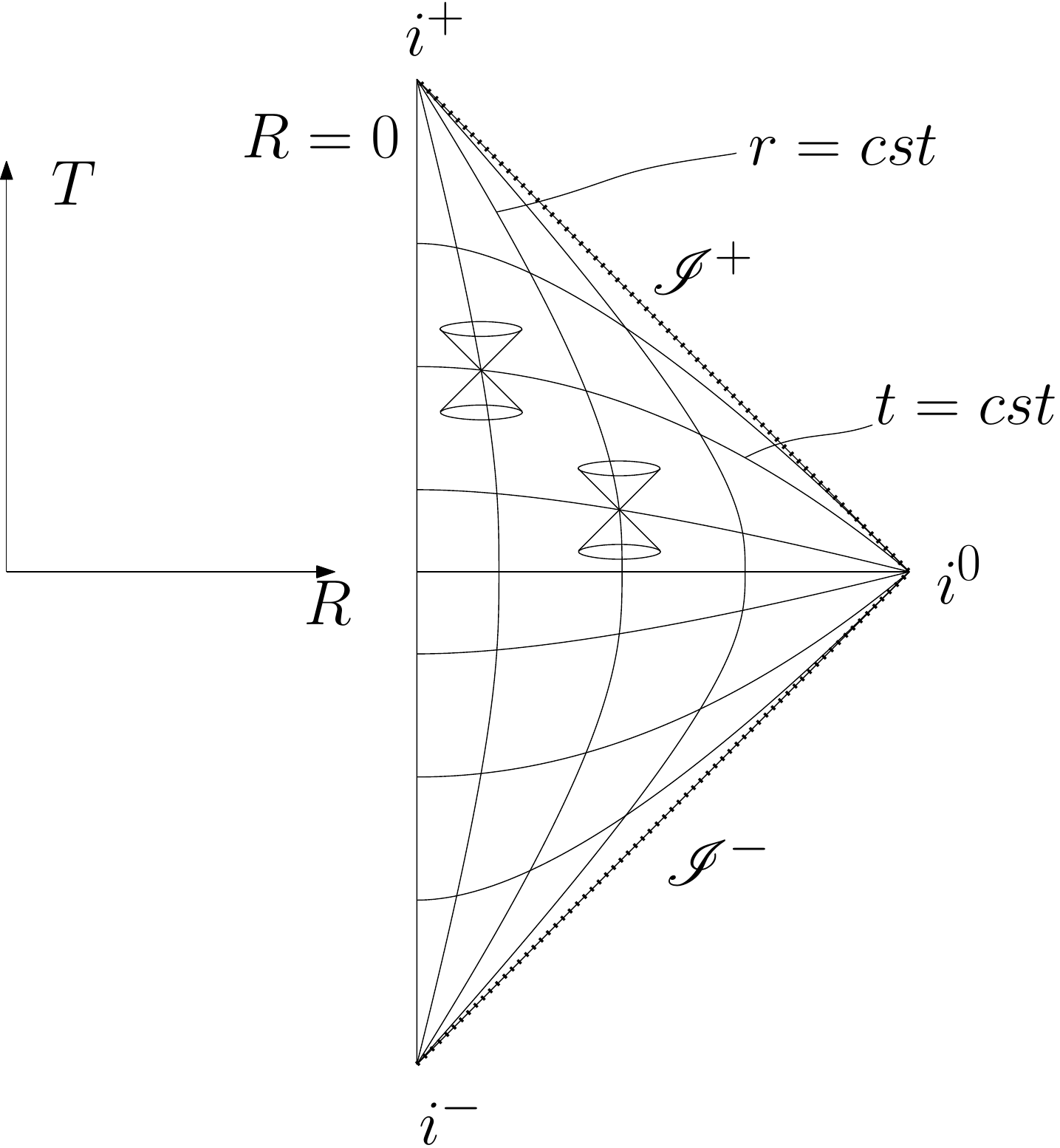}
\par\end{center}%
\end{minipage}
\par\end{centering}
\caption{\label{fig:4-cylinder-Minkowski-representat}Left: 4-cylinder Minkowski
representation; right: Penrose diagram for Minkowski space.}

\end{figure}
\ifS
\else Properties:
\begin{itemize}
\item Lightcones at 45\textdegree{} (radial null geodesics at $\pm45\text{\textdegree}$)
\item All timelike curves (geodesics) start at $i^{-}$ and end at $i^{+}$
\item All null geodesics begin on $\mathscr{I}^{-}$and end on $\mathscr{I}^{+}$
\\
(reflection accross $R=0$: $T=-R+c$ becomes $T=R+c$)
\item All spacelike geodesics begin and end at $i^{0}$ (reflection)
\end{itemize}
\fi Asymptotically flat spacetimes will conserve the $\mathscr{I}^{+}$,
$i^{0}$, $\mathscr{I}^{-}$ structure (e.g. Black Holes)

\section{Non-Trivial example: Power law, flat Robertson-Walker metric\label{sec:Non-Trivial-example:-Power}}

\begin{align*}
ds^{2}= & -dt^{2}+t^{2q}\left(dr^{2}+r^{2}d\Omega^{2}\right)\ifS\else\fi
\end{align*}
\ifS
\else where scale factor \fi  $a=t^{q}$, $0<q<1$. 

\ifS
\else Singular at $t=0$ so \fi 
\begin{align*}
t\in & \mathbb{R}^{*+}\\
r\in & \mathbb{R}^{+}
\end{align*}
\ifS
\else First
introduce conformal time so metric appears conformally flat\fi 
\begin{align*}
dt^{2}= & t^{2q}d\eta & \Rightarrow\frac{dt}{t^{q}}= & d\eta & \Rightarrow\eta= & \frac{t^{1-q}}{1-q} & \textrm{ or }t= & \left[\left(1-q\right)\eta\right]^{\frac{1}{1-q}}\ifS\else\fi
\end{align*}
\ifS\else then\fi 
\begin{align*}
ds^{2}= & t^{2q}\left(-d\eta^{2}+dr^{2}+r^{2}d\Omega^{2}\right)=\left[\left(1-q\right)\eta\right]^{\frac{2q}{1-q}}\left(-d\eta^{2}+dr^{2}+r^{2}d\Omega^{2}\right)\ifS\else\fi
\end{align*}
\ifS
\else Remark $\eta$ is time coord.: $\eta\in\mathbb{R}^{*+}$\fi 
\begin{align*}
g_{ab}\partial_{\eta}^{a}\partial_{\eta}^{b}< & 0.\ifS\else\fi
\end{align*}
\ifS
\else Comoving clock
$x^{a}\left(\lambda\right)=\left(\begin{array}{cc}
\eta\left(\lambda\right) & \vec{0}\end{array}\right)$ proper time: $\tau\propto t\propto\eta^{\frac{1}{1-q}}$ difference
between observables and coords!\fi 

\begin{figure}
\begin{centering}
\includegraphics[width=0.5\columnwidth]{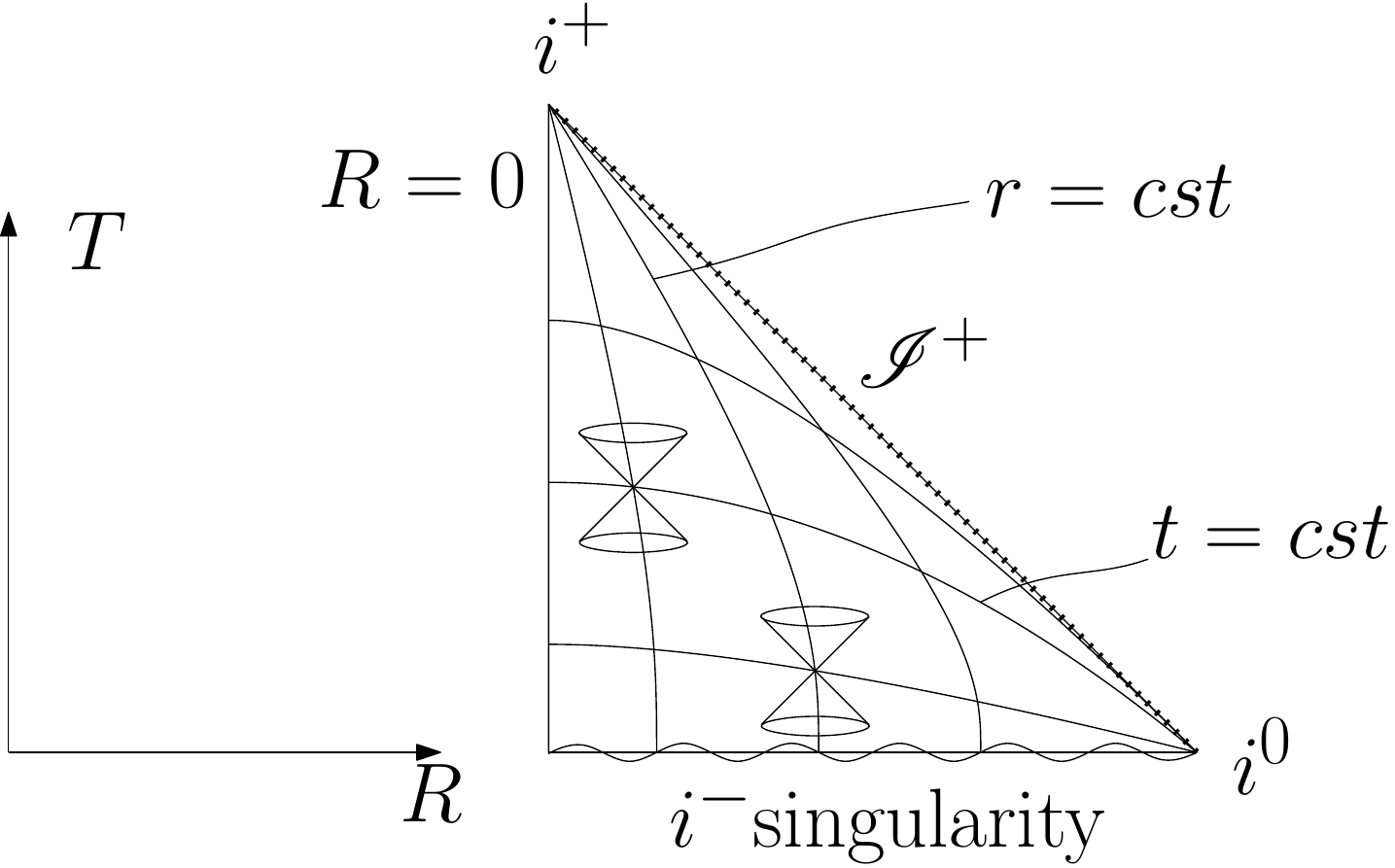}
\par\end{centering}
\caption{\label{fig:Penrose-diagram-for}Penrose diagram for power law, flat,
FLRW}

\end{figure}

\ifS
\else Use
the same procedure than Minkowski to compactify the lightlike directions
and keep light cones invariants
\begin{align*}
\begin{array}{rl}
u= & \eta-r\\
v= & \eta+r
\end{array} & \Leftrightarrow\begin{array}{rll}
r= & \frac{1}{2}\left(v-u\right)\ge0 & \Rightarrow u\le v,\,u,v\in\mathbb{R}\\
\eta= & \frac{1}{2}\left(v+u\right)>0 & \Rightarrow u>-v
\end{array}\\
\begin{array}{rl}
U= & \arctan u\\
V= & \arctan v
\end{array} & \begin{array}{rl}
U\le & V\\
U> & -V
\end{array}\:U,V\in\left]-\frac{\pi}{2};\frac{\pi}{2}\right[\\
\begin{array}{rl}
T= & V+U>0\\
R= & V-U\ge0
\end{array} & \begin{array}{r}
R\in\left[0;\pi\right[\\
T+R<\pi
\end{array}
\end{align*}
and the metric turns to (exercise: calculate $\omega$\exo )\fi 
\begin{align*}
ds^{2}= & \omega^{-2}\left(T,R\right)\left[-dT^{2}+dR^{2}+\sin^{2}R\,d\Omega^{2}\right]\ifS\else\fi
\end{align*}
\ifS
\else (conformal factor times Einstein
static universe) where trigonometry reveals\fi 
\begin{align*}
\omega\left(T,R\right)= & \left(\frac{\cos T+\cos R}{\left(1-q\right)\sin T}\right)^{\frac{q}{1-q}}\left(\cos T+\cos R\right)\ifS\else\fi
\end{align*}
\ifS
\else so the Penrose diagram reads (Fig~\ref{fig:Penrose-diagram-for})

Same as upper half of Minkowski but with initial past singularity.

It shows how past light cones may never intersect while future light
cones always do (horizon problem of cosmology)\fi 

\section{Case of Schwarzschild}

\ifS
\else To avoid the coordinate singularity of
the horizon we start from the null version of Kruskal coordinates
and compactify in the same manner as Minkowski \fi 
\begin{align*}
ds^{2}= & -\frac{32\left(GM\right)^{3}}{r}e^{-\frac{r}{2GM}}dUdV+r^{2}d\Omega^{2}, & U= & -\epsilon_{U}\sqrt{\left|\frac{r}{2GM}-1\right|}e^{\frac{r}{4GM}}e^{-\frac{t}{4GM}}=-\epsilon_{U}e^{-\frac{u}{4GM}}\\
\textrm{with }UV= & \left(\frac{r}{2GM}-1\right)e^{\frac{r}{2GM}}, & V= & \epsilon_{V}\sqrt{\left|\frac{r}{2GM}-1\right|}e^{\frac{r}{4GM}}e^{\frac{t}{4GM}}=\epsilon_{V}e^{\frac{v}{4GM}}
\end{align*}
with $\begin{array}{rl}
U_{1}= & \arctan U\\
V_{1}= & \arctan V
\end{array}$ and $\begin{array}{rl}
T_{1}= & V_{1}+U_{1}\\
R_{1}= & V_{1}-U_{1}
\end{array}$
\begin{align*}
ds^{2}= & -\frac{32\left(GM\right)^{3}}{r\cos^{2}U_{1}\cos^{2}V_{1}}e^{-\frac{r}{2GM}}dU_{1}dV_{1}+r^{2}d\Omega^{2}\\
= & -\frac{8\left(GM\right)^{3}e^{-\frac{r}{2GM}}}{r\cos^{2}U_{1}\cos^{2}V_{1}}\left(dT_{1}^{2}-dR_{1}^{2}\right)+r^{2}d\Omega^{2}.\ifS\else\fi
\end{align*}
\begin{figure}
\begin{centering}
\includegraphics[width=1\columnwidth]{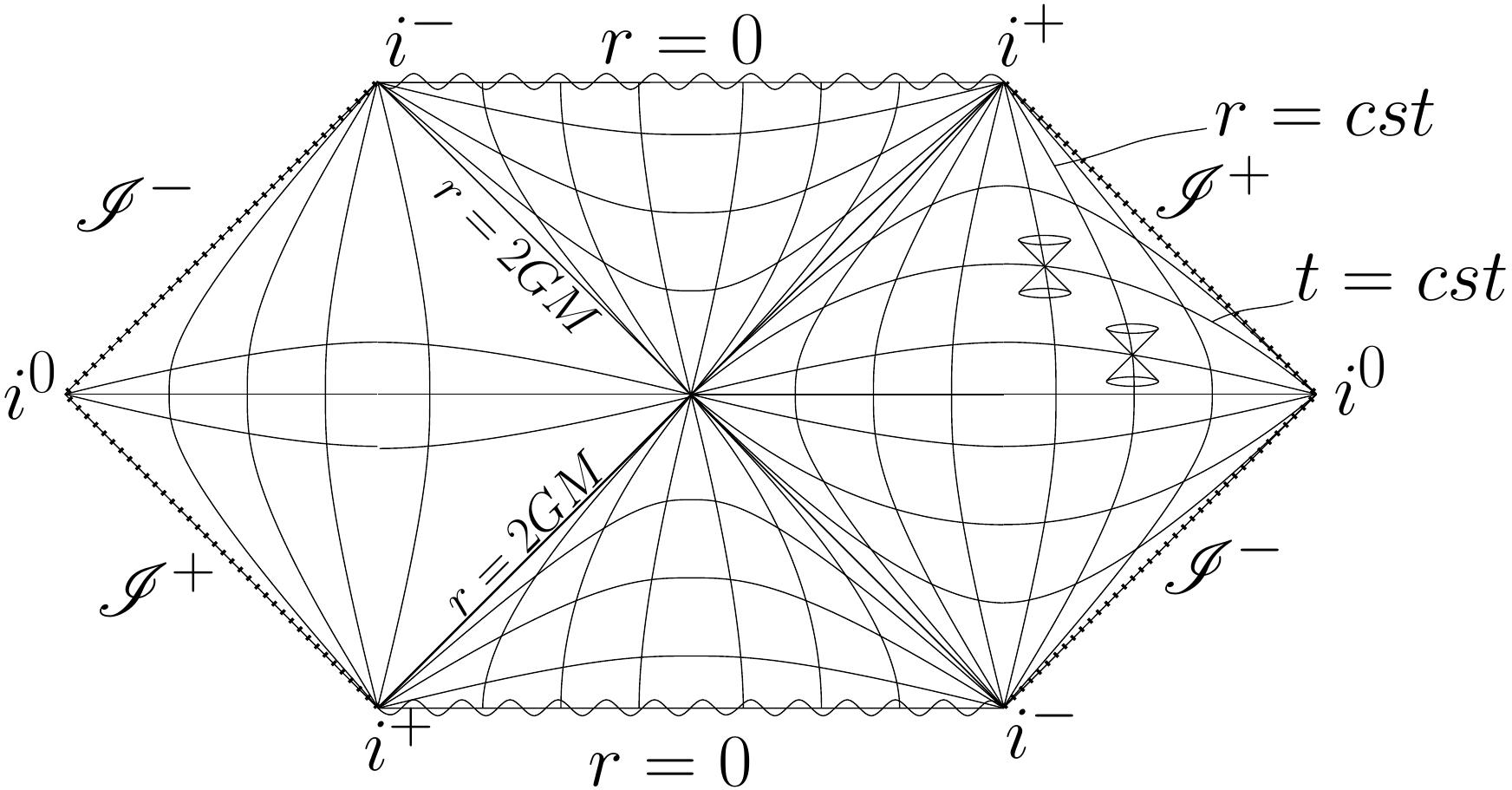}
\par\end{centering}
\caption{\label{fig:Penrose-Schwarzschild}Penrose diagram for the solution}
\end{figure}
Since $\tan\left(a+b\right)=\frac{\tan a+\tan b}{1-\tan a\tan b}$,
then $\begin{array}{rl}
T_{1}= & \arctan\left(\frac{U+V}{1-UV}\right)\\
R_{1}= & \arctan\left(\frac{V-U}{1+UV}\right)
\end{array}$

Like for Minkowski, spatial infinity resides at ($\epsilon_{U}=\epsilon_{V}$)
\begin{align*}
\left.\begin{array}{rl}
U_{1}\underset{r\to\infty}{\longrightarrow} & -\epsilon_{U}\frac{\pi}{2}\\
V_{1}\underset{r\to\infty}{\longrightarrow} & \epsilon_{V}\frac{\pi}{2}
\end{array}\right\} \Leftrightarrow & \left\{ \begin{array}{rl}
T_{1}\underset{r\to\infty}{\longrightarrow} & \epsilon_{V}\frac{\pi}{2}\left(1-1\right)=0\\
R_{1}\underset{r\to\infty}{\longrightarrow} & \epsilon_{V}\frac{\pi}{2}\left(1+1\right)=\epsilon_{V}\pi
\end{array}\right.
\end{align*}
time infinities stand at
\begin{align*}
\left.\begin{array}{rl}
U_{1}\underset{t\to\infty}{\longrightarrow} & 0\\
V_{1}\underset{t\to\infty}{\longrightarrow} & \epsilon_{V}\frac{\pi}{2}
\end{array}\right\} \Leftrightarrow & \left\{ \begin{array}{rl}
T_{1}\underset{t\to\infty}{\longrightarrow} & \epsilon_{V}\frac{\pi}{2}\\
R_{1}\underset{t\to\infty}{\longrightarrow} & \epsilon_{V}\frac{\pi}{2}
\end{array}\right.\\
\left.\begin{array}{rl}
U_{1}\underset{t\to-\infty}{\longrightarrow} & -\epsilon_{U}\frac{\pi}{2}\\
V_{1}\underset{t\to-\infty}{\longrightarrow} & 0
\end{array}\right\} \Leftrightarrow & \left\{ \begin{array}{rl}
T_{1}\underset{t\to-\infty}{\longrightarrow} & -\epsilon_{U}\frac{\pi}{2}\\
R_{1}\underset{t\to-\infty}{\longrightarrow} & \epsilon_{U}\frac{\pi}{2}
\end{array}\right.
\end{align*}
null infinities are
\begin{align*}
\begin{array}{rl}
U_{1}\underset{u\to\infty}{\longrightarrow} & 0,\\
U_{1}\underset{u\to-\infty}{\longrightarrow} & -\epsilon_{U}\frac{\pi}{2},\\
V_{1}\underset{v\to\infty}{\longrightarrow} & \epsilon_{V}\frac{\pi}{2},\\
V_{1}\underset{v\to\infty}{\longrightarrow} & 0,
\end{array} & \begin{array}{rl}
V_{1} & \textrm{ unchanged}\\
\\
U_{1} & \textrm{ unchanged}\\
\\
\end{array}\ifS\else\fi
\end{align*}
the central singularity yields
\begin{align*}
\begin{array}{rl}
UV\underset{r\to0}{\longrightarrow} & 1\\
(\epsilon_{U}= & \epsilon_{V})
\end{array}\; & \begin{array}{rl}
T_{1}\underset{r\to0}{\longrightarrow} & \epsilon_{V}\frac{\pi}{2}\\
R_{1}\left(r=0\right)= & \arctan\left[R\left(r=0\right)\right]
\end{array}
\end{align*}
We thus can build the Carter-Penrose diagram

\ifS
\else  Asymptotic flatness causes
the infinities to match Minkowski\fi 

\chapter{Birkhoff Theorem}

\cite{Birkhoff23}

Often misunderstood as relativistic version of Newton's ``iron spheres''
theorem, it is different in scope and hypotheses.

\section{Current general form}

\ifS
\else Most evolved current phrasing
by Bona \cite{Bona(1988)} for conformally reducible direct products
of spacetime:\fi 
\[
\left\{ \begin{array}{rl}
\hat{g}= & Y^{2}g\\
ds^{2}= & g_{ab}dx^{a}dx^{b}=\gamma_{AB}dx^{A}dx^{B}+h_{\alpha\beta}dy^{\alpha}dy^{\beta}
\end{array}\right.
\]
\ifS
\else where $h_{\alpha\beta}$ and $y^{\alpha}$ are metric
and coords. on 2D orbits $O_{2}$ of $G_{3}$, a group of motion,
and $\gamma_{AB}$, $x^{A}$ are metric and coords. on $V_{2}$, orthogonal
submanifold to $O_{2}$ from $g$\fi 
\begin{thm}
Metric with a group $G_{3}$ of motion on non-null orbits $O_{2}$
and with Ricci tensors of the type $\left[\left(\begin{array}{cc}
1 & 1\end{array}\right)\left(\begin{array}{cc}
1 & ,1\end{array}\right)\right]$ and $\left[\left(\begin{array}{cccc}
1 & 1 & 1 & ,1\end{array}\right)\right]$ admit a group $G_{4}$ provided that $d_{a}Y\ne0$.
\end{thm}

\section{Spherical case and classic form}

The original version states, more transparently, that
\begin{thm}
Any vacuum, spherically symmetric solution of \ifS
\else E.F.E.\fi  is also static, and Schwarzschild is
its unique form.
\end{thm}
Physically, it implies that radial pulsations of a spherical star
cannot propagate gravitational radiations. Conversely, \ifS
\else GW\fi  require at minimum quadrupolar variations.

The proof involves linking spherical symmetry with foliation of spacetime,
then the form of the metric to plug in \ifS
\else EFE\fi .

\subsection{Preliminary: Killing equation}
\begin{description}
\item [{\uline{Killing~vector}:}] \label{Def:Killingvector:-Any-symmetry}\ifS
\else Symmetry yields
invariant motion in direction $K^{a}$. Observer: $u^{a}$ timelike
vector\\
then $K^{a}u_{a}$ along $u^{a}$ is unchanged: $u^{a}$ motion along
$K^{a}$ is invariant:\fi  $\frac{du^{a}K_{a}}{d\tau}=0=\frac{du^{\left(K\right)}}{d\tau}\Leftrightarrow u^{a}\nabla_{a}\left(K^{b}u_{b}\right)=0=u^{a}u^{b}\nabla_{(a}K_{b)}$
\\
$\nabla_{(a}K_{b)}=0$\ifS
\else  Killing
equation: Definition\fi 
\end{description}

\subsection{Proof of the classic form}

\subsubsection{General form of the metric}
\begin{proof}
Spherical symmetry is the symmetry of the sphere $S_{2}$, that is
the rotation group of $\mathbb{R}^{3}:\;SO\left(3\right)$, which
generators follow the algebra (Killing vectors)
\begin{align*}
\left[R,S\right]= & T,\ifS\else\fi\\
\left[S,T\right]= & R,\ifS\else\fi\\
\left[T,R\right]= & S.\ifS\else\fi
\end{align*}

From Frobenius theorem, since the $SO\left(3\right)$ Killing vector
set closes, then their integral curves cover the submanifold $S_{2},$and
thus $S_{2}$ foliates spherically symmetric spacetimes.

The set of all spheres is 2 dimensional since $S_{2}$ is 2D, in 4D,
thus we can chose $\left(\begin{array}{cccc}
a, & b, & \theta, & \phi\end{array}\right)$ on $M$,$a,b$ labels each sphere.

\begin{figure}
\begin{centering}
\includegraphics[width=0.8\columnwidth]{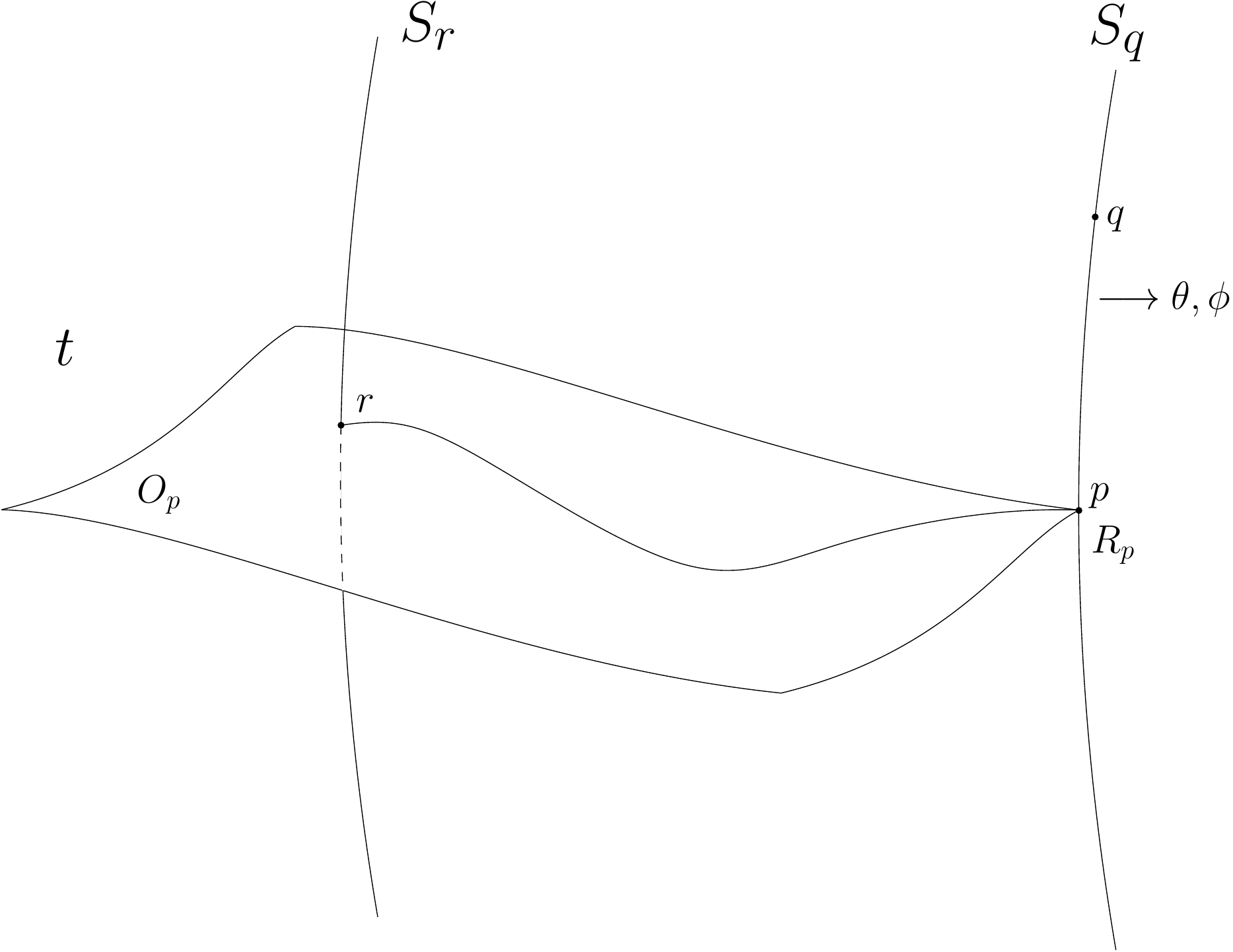}
\par\end{centering}
\caption{\label{fig:Foliation-of-spherically}Foliation of spherically symmetric
spacetime}
\end{figure}
\ifS
\else For
any $p$ on $S_{q}$

Invariant under $R_{p}$, $O_{p}$ its orthogonal space, any $r$
on $O_{p}$ is also invariant and define $S_{r}$ (as $R_{p}$ isometry
and $O_{p}\bot S_{q}$). $S_{r}\bot O_{p}$ as $R_{p}$ leaves also
$S_{r}$ invariant and tangent space to $S_{r}$ is orthogonal to
$O_{p}$.\fi 

The geodesic linking $p$ and $r$ maps $\theta,\phi$ onto neighbouring
spheres, thus they are global \ifS\else coords. \fi 
Its tangent vector's components along a chosen basis yields $\left(a,b\right)$. Since we decomposed
$M$ into $S_{q}$ and $O_{p}$ for any $q,p$, their \ifS
\else coords\fi  tangent vectors will be orthogonal between $\left(a,b\right)$ and $\left(\theta,\phi\right)$,
hence no cross terms in the metric
which should sport the form
\begin{align*}
ds^{2}= & g_{aa}\left(a,b\right)da^{2}+g_{ab}\left(a,b\right)\left(da\,db+db\,da\right)+g_{bb}\left(a,b\right)db^{2}+r^{2}\left(a,b\right)d\Omega^{2},\ifS\else\fi
\end{align*}
where $d\Omega^{2}=d\theta^{2}+\sin^{2}\theta d\phi^{2}$ is the metric
on $S_{2}$.

We can always choose, if $r\left(a,b\right)\to b\left(a,r\right)$
invertible (otherwise choose $r\left(a,b\right)\to a\left(r,b\right)$
and relabel $b\to a$), so 
\begin{align*}
ds^{2}= & g_{aa}\left(a,r\right)da^{2}+g_{ar}\left(a,r\right)\left(da\,dr+dr\,da\right)+g_{rr}\left(a,r\right)dr^{2}+r^{2}d\Omega^{2}.\ifS\else\fi
\end{align*}
\ifS\else Then we \fi choose $t\left(a,r\right)$ such
as to eliminate the cross terms: $dt=\partial_{a}t\,da+\partial_{r}t\,dr$
so 
\begin{align*}
dt^{2}= & \left(\partial_{a}t\right)^{2}da^{2}+\partial_{a}t\partial_{r}t\left(da\,dr+dr\,da\right)+\left(\partial_{r}t\right)^{2}dr^{2},\ifS\else\fi
\end{align*}
which yields that to obtain $m\,dt^{2}+n\,dr^{2}$ in $ds^{2}$, we
need
\begin{align*}
m\left(\partial_{a}t\right)^{2}= & g_{aa}, & n+m\left(\partial_{r}t\right)^{2}= & g_{rr}, & m\partial_{a}t\partial_{r}t= & g_{ar}
\end{align*}
\ifS\else which is a system of 3
eqs.\fi  for 3 unknowns $t\left(a,r\right),m\left(a,r\right)$ and
$n\left(a,r\right)$, completely determined up to initial conditions.

The metric of spherically symmetric $M$ in general writes
\begin{align*}
ds^{2}= & m\left(t,r\right)dt^{2}+n\left(t,r\right)dr^{2}+r^{2}d\Omega^{2}.\ifS\else\fi
\end{align*}
If we choose $t$ to be timelike and $r$ spacelike, given our Minkowskian
signature $\begin{array}{cccc}
- & + & + & +\end{array}$, yields
\begin{align*}
ds^{2}= & -e^{2\alpha\left(t,r\right)}dt^{2}+e^{2\beta\left(t,r\right)}dr^{2}+r^{2}d\Omega^{2}.\ifS\else\fi
\end{align*}
\end{proof}

\subsubsection{EFE and final proof of the theorem}
\begin{proof}
Next we need the EFE for vacuum. \ifS
\else Note
the metric here is time dependent!! but compare \fi with the static
form.

Christoffel symbol/connection
reads ($\partial_{t}=\dot{}$, $\partial_{r}=\,^{\prime}$)
\begin{gather*}
\begin{array}{rlrlrl}
\Gamma_{00}^{0}= & \dot{\alpha}, & \Gamma_{01}^{0}= & \alpha^{\prime}, & \Gamma_{11}^{0}= & e^{2\left(\alpha-\beta\right)}\dot{\beta},\\
\Gamma_{00}^{1}= & e^{2\left(\alpha-\beta\right)}\alpha^{\prime}, & \Gamma_{01}^{1}= & \dot{\beta}, & \Gamma_{11}^{1}= & \beta^{\prime},\\
\Gamma_{12}^{2}= & \frac{1}{r}, & \Gamma_{22}^{1}= & -re^{-2\beta}, & \Gamma_{13}^{3}= & \frac{1}{r},\\
\Gamma_{33}^{1}= & -re^{-2\beta}\sin^{2}\theta, & \Gamma_{33}^{2}= & -\sin\theta\cos\theta, & \Gamma_{23}^{3}= & \cot\theta,
\end{array}\ifS\else\fi
\end{gather*}
\ifS\else rest \fi are $0$ or symmetrics.

\ifS\else Riemann then
corresponds to\fi 
\begin{align*}
 & R_{\:101}^{0}=e^{2\left(\alpha-\beta\right)}\left[\ddot{\beta}+\dot{\beta}^{2}-\dot{\alpha}\dot{\beta}\right]+\left[\alpha^{\prime}\beta^{\prime}-\alpha^{\prime\prime}-\left(\alpha^{\prime}\right)^{2}\right],\ifS\else\fi\\
 & \begin{array}{rlrlcc}
R_{\:202}^{0}= & -re^{-2\beta}\alpha^{\prime}, & R_{\:303}^{0}= & -re^{-2\beta}\sin^{2}\theta\alpha^{\prime}, & R_{\:212}^{0}= & re^{-2\alpha}\dot{\beta},\\
R_{\:313}^{0}= & re^{-2\alpha}\sin^{2}\theta\dot{\beta}, & R_{\:212}^{1}= & re^{-2\beta}\beta^{\prime}, & R_{\:313}^{1}= & re^{-2\beta}\sin^{2}\theta\beta^{\prime},\\
R_{\:323}^{2}= & \left(1-e^{-2\beta}\right)\sin^{2}\theta,
\end{array}\ifS\else\fi
\end{align*}
\ifS\else rest \fi are $0$ or symmetric.

\ifS\else Ricci
tensor proceeds\fi 
\begin{align*}
R_{00}= & \left[\ddot{\beta}+\dot{\beta}^{2}-\dot{\alpha}\dot{\beta}\right]+e^{2\left(\alpha-\beta\right)}\left[\alpha^{\prime\prime}+\left(\alpha^{\prime}\right)^{2}-\alpha^{\prime}\beta^{\prime}+\frac{2}{r}\alpha^{\prime}\right],\ifS\else\fi\\
R_{11}= & -\left[\alpha^{\prime\prime}+\left(\alpha^{\prime}\right)^{2}-\alpha^{\prime}\beta^{\prime}-\frac{2}{r}\beta^{\prime}\right]+e^{2\left(\beta-\alpha\right)}\left[\ddot{\beta}+\dot{\beta}^{2}-\dot{\alpha}\dot{\beta}\right],\ifS\else\fi\\
R_{01}= & \frac{2}{r}\dot{\beta},\ifS\else\fi\\
R_{22}= & e^{-2\beta}\left[r\left(\beta^{\prime}-\alpha^{\prime}\right)-1\right]+1,\hfill R_{33}=R_{22}\sin^{2}\theta,\ifS\else\fi
\end{align*}
and the EFE yield $R_{ab}=0$ in vacuum. \ifS
\else 

From $R_{01}=0$: $\dot{\beta}=0$.\fi  

\ifS
\else Taking $\dot{R}_{22}=0$ with
$\dot{\beta}=0\Rightarrow\dot{\alpha}^{\prime}=0$ thus\fi  
\begin{align*}
\dot{\beta}= & 0 & \Rightarrow\beta= & \beta\left(r\right)\\
\dot{\alpha}^{\prime}= & 0 & \Rightarrow\alpha= & f\left(r\right)+g\left(t\right)
\end{align*}
\ifS
\else and time
can be redefined as $-e^{2\alpha}dt^{2}=-e^{2f\left(r\right)}e^{2g\left(t\right)}dt^{2}=-e^{2f\left(r\right)}d\bar{t}^{2}$
hence $\alpha\left(t,r\right)=f\left(r\right)$ and we obtain\fi 
\begin{align*}
ds^{2}= & -e^{2\alpha\left(r\right)}dt^{2}+e^{2\beta\left(r\right)}dr^{2}+r^{2}d\Omega^{2},\ifS\else\fi
\end{align*}
which is static.
\end{proof}

\chapter{Geodesics and congruences}

\ifS
\else``Straightest'' \fi lines followed by test particles and
light, in absence of other interactions (shortest distance). \ifS
\else It parallel transports its \fi own
tangent vector.

\section{Geodesic equation}

\subsection{Geodesic equation from parallel transport}

\subsubsection{Recall: parallel transport}
\begin{description}
\item [{~}] \ifS
\else Conservation of tensors \fi along a curve $x^{a}\left(\lambda\right)$
\begin{align*}
\left(\frac{D}{d\lambda}T\right)_{\qquad b_{1}\cdots b_{m}}^{a_{1}\cdots a_{n}}= & \frac{dx^{c}}{d\lambda}\nabla_{c}T_{\qquad b_{1}\cdots b_{m}}^{a_{1}\cdots a_{n}}=0
\end{align*}
\ifS
\else It preserves metric \fi $\frac{D}{d\lambda}g_{ab}=\frac{dx^{c}}{d\lambda}\nabla_{c}g_{ab}=0$\\
\ifS
\else and norm (type) of vectors
(so keeps time-/space-like and null)\fi 
\begin{align*}
\frac{D}{d\lambda}\left(V^{a}W_{a}\right)= & \frac{D}{d\lambda}\left(g_{ab}\right)V^{a}W^{b}+\frac{D}{d\lambda}\left(V^{a}\right)g_{ab}W^{b}+\frac{D}{d\lambda}\left(W^{b}\right)g_{ab}V^{a}\\
= & 0.\ifS\else\fi
\end{align*}
For vectors\ifS\else  it \fi reads $\frac{dV^{a}}{d\lambda}+\Gamma_{bc}^{a}\frac{dx^{b}}{d\lambda}V^{c}=0$
\end{description}

\subsubsection{Geodesic equation}

\ifS
\else For geodesics\fi 
\begin{align*}
\frac{D}{d\lambda}\frac{dx^{c}}{d\lambda}= & 0 & \Leftrightarrow\frac{d^{2}x^{a}}{d\lambda^{2}}+\Gamma_{bc}^{a}\frac{dx^{b}}{d\lambda}\frac{dx^{c}}{d\lambda}= & 0
\end{align*}

\begin{rem}
As \ifS
\else norm preserved in parallel transport, sign
is unchanged and \fi geodesics remain of same type
\end{rem}

\subsection{Geodesic equation from least action principle\label{subsec:Geodesic-equation-from}}

\ifS
\else Distance between events\fi :
\begin{align*}
d= & \int ds\textrm{ for spacelike paths}\ifS\else\fi\\
l= & \int ds=0\textrm{ for null paths}\ifS\else\fi\\
\tau= & \int d\tau=\int\sqrt{-ds^{2}}\textrm{ proper time, for timelike curves}\ifS\else\fi
\end{align*}
Curves changing type have undefined length

\ifS
\else Geodesic:
extremum of distance: \fi principle of Maupertuis (least action)

For curve $C\left(\lambda\right)$ with
tangent $T^{a}=\frac{dx^{a}}{d\lambda}$, as $ds^{2}=g_{ab}dx^{a}dx^{b}=g_{ab}T^{a}T^{b}d\lambda^{2}$,
\begin{align*}
d= & \int_{C}\left(g_{ab}T^{a}T^{b}\right)^{\frac{1}{2}}d\lambda.\ifS\else\fi
\end{align*}
For $T^{a}T_{a}=0$ (null)\ifS
\else 
\begin{align*}
l= & \int\left(g_{ab}T^{a}T^{b}\right)^{\frac{1}{2}}d\lambda=0.
\end{align*}
\fi For timelike $C$:
\begin{align*}
ds^{2}= & g_{ab}T^{a}T^{b}d\lambda^{2}<0\\
= & -d\tau^{2},\ifS\else\fi
\end{align*}
defines proper time
\begin{align*}
\Rightarrow\tau= & \int_{C}\left(-g_{ab}T^{a}T^{b}\right)^{\frac{1}{2}}d\lambda
\end{align*}
\ifS
\else choose $\lambda$ as $\tau$: $T^{a}=u^{a}$ 4-vel.\fi 
\begin{align*}
\Rightarrow\tau= & \int_{C}\left(-\left(-1\right)\right)^{\frac{1}{2}}d\tau=\int_{C}d\tau
\end{align*}
\ifS\else We can do that \fi as
length is independent of parameter $\lambda$
\begin{proof}
\ifS
\else New parameter $\mathfrak{s}\left(\lambda\right)\Rightarrow$
new tangent $S^{a}=\frac{d\lambda}{d\mathfrak{s}}T^{a}$ \fi 
\begin{align*}
l^{\prime}= & \int\sqrt{\left|g_{ab}S^{a}S^{b}\right|}d\mathfrak{s}=\int\sqrt{\left|g_{ab}T^{a}T^{b}\right|}\frac{d\lambda}{d\mathfrak{s}}d\mathfrak{s}=l\ifS\else\fi
\end{align*}
\end{proof}
\ifS\else Variation
of length for \fi timelike curves yields
\begin{align*}
\delta\tau= & \int-\frac{1}{2}\frac{\delta\left(g_{ab}u^{a}u^{b}\right)}{\sqrt{-g_{ab}u^{a}u^{b}}}d\tau=-\frac{1}{2}\int\delta\left(g_{ab}u^{a}u^{b}\right)d\tau=-\delta I\\
\textrm{with }I= & \frac{1}{2}\int g_{ab}u^{a}u^{b}d\tau=\int Ld\tau
\end{align*}
where $L$ is \ifS
\else Lagrangian\fi 

\ifS
\else Extremised
variation is then given by Euler-Lagrange eq. ($u^{a}=\frac{dx^{a}}{d\tau}=\dot{x}^{a}$):\fi 
\begin{align*}
\frac{\partial L}{\partial x^{a}}= & \frac{1}{2}g_{cb,a}u^{b}u^{c}, & \frac{\partial L}{\partial\dot{x}^{a}}= & g_{ab}u^{b} & \textrm{ so }\frac{d}{d\tau}\left(\frac{\partial L}{\partial\dot{x}^{a}}\right)= & g_{ab,c}u^{b}u^{c}+g_{ab}\dot{u}^{b}
\end{align*}
\begin{align*}
g_{ab}\dot{u}^{b}+\left[g_{ab,c}-\frac{1}{2}g_{cb,a}\right]u^{b}u^{c}= & 0\\
\Leftrightarrow\frac{d^{2}x^{a}}{d\lambda^{2}}+\frac{g^{ad}}{2}\left[g_{db,c}+g_{dc,b}-g_{cb,d}\right]= & 0,\ifS\else\fi
\end{align*}
where we recognise the Christoffel connection and \ifS
\else the geodesic eq.\fi 
\begin{thm}
\label{thm:As-the-Christoffel}As \ifS
\else Christoffel connection\fi  not a tensor, we can choose
\ifS\else coords\fi  for which locally $\Gamma=0$
\end{thm}
\begin{proof}
Since $\nabla_{a}V^{b}=\partial_{a}V^{b}+\Gamma_{ac}^{b}V^{c}$ \ifS
\else tensor,
so\fi 
\begin{align*}
\nabla_{a^{\prime}}V^{b^{\prime}}= & \frac{\partial x^{a}}{\partial x^{a^{\prime}}}\frac{\partial x^{b^{\prime}}}{\partial x^{b}}\nabla_{a}V^{b}=\partial_{a^{\prime}}x^{a}\partial_{b}x^{b^{\prime}}\nabla_{a}V^{b}
\end{align*}
\ifS
\else and $\frac{\partial}{\partial x^{a^{\prime}}}=\frac{\partial x^{a}}{\partial x^{a^{\prime}}}\frac{\partial}{\partial x^{a}}=\partial_{a^{\prime}}x^{a}\partial_{a}$,
$V^{b^{\prime}}=\partial_{b}x^{b^{\prime}}V^{b}$ but\fi 
\begin{align*}
\nabla_{a^{\prime}}V^{b^{\prime}}= & \partial_{a^{\prime}}V^{b^{\prime}}+\Gamma_{a^{\prime}c^{\prime}}^{b^{\prime}}V^{c^{\prime}}\\
= & \partial_{a^{\prime}}x^{a}\left(\partial_{b}x^{b^{\prime}}\underbrace{\partial_{a}V^{b}}+V^{b}\partial_{a}\partial_{b}x^{b^{\prime}}\right)+\Gamma_{a^{\prime}c^{\prime}}^{b^{\prime}}\partial_{c}x^{c^{\prime}}V^{c}\\
= & \partial_{a^{\prime}}x^{a}\partial_{b}x^{b^{\prime}}\left(\overbrace{\partial_{a}V^{b}}+\Gamma_{ac}^{b}V^{c}\right)
\end{align*}
\begin{align*}
\Leftrightarrow\Gamma_{a^{\prime}c^{\prime}}^{b^{\prime}}\partial_{c}x^{c^{\prime}}V^{c}+\partial_{a^{\prime}}x^{a}V^{c}\partial_{a}\partial_{c}x^{b^{\prime}}= & \partial_{a^{\prime}}x^{a}\partial_{b}x^{b^{\prime}}\Gamma_{ac}^{b}V^{c}
\end{align*}
\ifS
\else valid $\forall V^{c}$so
we isolate $\Gamma^{\prime}$ by $\times\partial_{d^{\prime}}x^{c}$\fi 
\begin{align*}
\Leftrightarrow\Gamma_{a^{\prime}d^{\prime}}^{b^{\prime}}= & \partial_{a^{\prime}}x^{a}\partial_{b}x^{b^{\prime}}\partial_{d^{\prime}}x^{c}\Gamma_{ac}^{b}-\partial_{a^{\prime}}x^{a}\partial_{d^{\prime}}x^{c}\partial_{a}\partial_{c}x^{b^{\prime}}
\end{align*}
which gives
\begin{align*}
\Gamma_{b^{\prime}c^{\prime}}^{a^{\prime}}= & \partial_{b^{\prime}}x^{b}\partial_{c^{\prime}}x^{c}\partial_{a}x^{a^{\prime}}\Gamma_{bc}^{a}-\partial_{b^{\prime}}x^{b}\partial_{c^{\prime}}x^{c}\partial_{b}\partial_{c}x^{a^{\prime}}
\end{align*}
\ifS
\else so it is always possible to
choose inertial coordinates at $p$, \fi which vanish $\Gamma$:
\begin{align*}
\left.\Gamma_{b^{\prime}c^{\prime}}^{a^{\prime}}\right|_{p}= & \partial_{b^{\prime}}x^{b}\partial_{c^{\prime}}x^{c}\left.\left(\partial_{a}x^{a^{\prime}}\Gamma_{bc}^{a}-\partial_{b}\partial_{c}x^{a^{\prime}}\right)\right|_{p}=0\\
\Leftrightarrow\frac{\partial_{b}\partial_{c}x^{a^{\prime}}}{\partial_{a}x^{a^{\prime}}}\left(p\right)= & \Gamma_{bc}^{a}\left(p\right)
\end{align*}
\ifS
\else defines
$x^{a^{\prime}}\left(x^{b}\right)$.\fi 
\end{proof}
Locally $\frac{d^{2}x^{a}}{d\lambda^{2}}=0$: $\frac{dx^{a}}{d\tau}=cst$
along \ifS
\else geodesic:
usual straight line 

$\Leftrightarrow$ inertial frame (local)

see Equivalence Principle

$\Rightarrow$ no local way to feel gravity: need to go to geodesic
deviation\fi 

\section{Geodesic deviation}

\ifS\else Measure \fi how
neighbouring geodesics drift away or converge.

\ifS\else Choose \fi a
one parameter geodesics family $\gamma_{s}\left(\tau\right)$ with
smooth parameter $s\in\mathbb{R}$ and proper time parameter $\tau\in\mathbb{R}$
\begin{align*}
\frac{\partial x^{a}}{\partial\tau}= & T^{a}\textrm{ tangent to geodesic }\gamma_{s}\\
\frac{\partial x^{a}}{\partial s}= & X^{a}\textrm{ infinitesimal displacement to next geodesic }\gamma_{s+ds}
\end{align*}
\ifS
\else Geodesic tangent\fi 
\begin{figure}
\begin{centering}
\includegraphics[width=0.5\columnwidth]{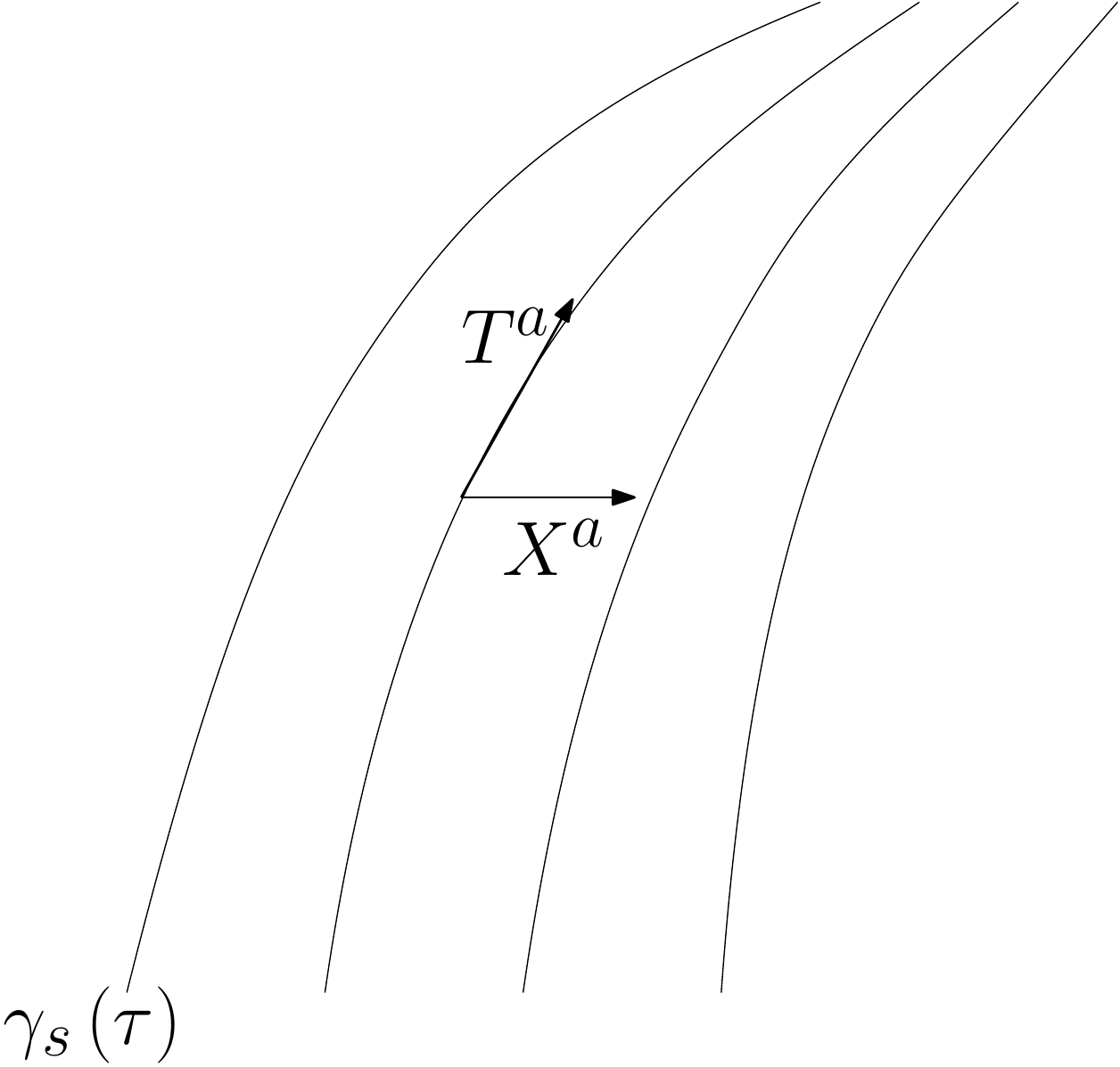}
\par\end{centering}
\caption{\label{fig:Geodesic-family-and}Geodesic family and its tangent vectors}
\end{figure}
\begin{align*}
T^{a}\nabla_{a}T^{b}= & 0
\end{align*}
\ifS
\else Affine parameterization $\tau$ of $\gamma_{s}\left(\tau\right)$
means the freedom $\tau^{\prime}=a\left(s\right)\tau+b\left(s\right)$,
however we can set it by choosing \fi $T^{a}X_{a}=0$ ($X^{a}$ orthogonal
to geodesics):
\begin{proof}
\begin{itemize}
\item $T^{a}T_{a}$ constant along geodesics:
$T^{b}\nabla_{b}\left(T^{a}T_{a}\right)=2\left(T^{a}T^{b}\nabla_{b}T_{a}\right)=0$
\item \ifS
\else $T^{a}$, $X^{a}$ are $\left(\partial_{\tau}\right)^{a},\left(\partial_{s}\right)^{a}$,
coordinate vector fields $\Rightarrow$ commute
\begin{align*}
\Leftrightarrow T^{a}\nabla_{a}X^{b}= & X^{a}\nabla_{a}T^{b}
\end{align*}
and geodesics don't change their type so we can always normalise $T^{a}$:
$n^{a}=\frac{T^{a}}{\sqrt{\left|T^{b}T_{b}\right|}}\Rightarrow n^{a}n_{a}=\pm1$
or $T^{a}T_{a}=0$ always along geodesics (reparameterise $\tau$
with norm; in fact as $\tau$ proper time, $T^{a}$ normalised)\\
thus\fi 
\begin{align*}
T^{b}\nabla_{b}\left(T^{a}X_{a}\right)= & \left(T^{b}\nabla_{b}T^{a}\right)X_{a}+T^{a}T^{b}\nabla_{b}X_{a}\\
= & T^{a}X^{b}\nabla_{b}T_{a}=\ifS\else\frac{1}{2}X^{b}\nabla_{b}\underset{=\pm1,0}{\underbrace{\left(T^{a}T_{a}\right)}}=0\fi
\end{align*}
\ifS\else so \fi $T^{a}X_{a}=cst$
along geodesics
\item at $\tau=0$ we can reparameterise \ifS
\else  $\tau^{\prime}=\tau+b\left(s\right)$ such
that $T^{a}X_{a}=0\Rightarrow$ everywhere (exercise: find $b\left(s\right)$
so $\tau^{\prime}=\tau+b\left(s\right)\Rightarrow T^{a}X_{a}=0$\exo )\fi 
\end{itemize}
\end{proof}
\ifS\else Relative \fi ``velocity'' of
infinitesimally nearby geodesics
\begin{align*}
V^{a}= & T^{b}\nabla_{b}X^{a}=\frac{dX^{a}}{d\tau}
\end{align*}

\ifS\else Relative \fi acceleration
\begin{align*}
a^{a}=\frac{d^{2}X^{a}}{d\tau^{2}}=T^{c}\nabla_{c}V^{a}= & T^{c}\nabla_{c}\left(T^{b}\nabla_{b}X^{a}\right)\\
= & T^{c}\nabla_{c}\left(X^{b}\nabla_{b}T^{a}\right)\qquad\qquad\qquad\qquad\textrm{ : commutation}\\
= & \left(T^{c}\nabla_{c}X^{b}\right)\nabla_{b}T^{a}+X^{b}T^{c}\nabla_{c}\nabla_{b}T^{a}\\
= & \underset{\textrm{comm.}}{\left(X^{c}\nabla_{c}T^{b}\right)}\nabla_{b}T^{a}+\underset{\textrm{Ricci ID}}{X^{b}T^{c}\nabla_{b}\nabla_{c}T^{a}-R_{cbd}^{\quad\,a}X^{b}T^{c}T^{d}}\\
= & X^{c}\nabla_{c}\left(T^{b}\nabla_{b}T^{a}\right)-R_{cbd}^{\quad\,a}X^{b}T^{c}T^{d}\\
= & R_{\,dcb}^{a}T^{d}T^{c}X^{b}\qquad\textrm{ T geodesics, Riemann symmetries}\\
= & R_{\,bcd}^{a}T^{b}T^{c}X^{d}.\qquad\qquad\qquad\qquad\textrm{ dummy renaming}
\end{align*}
\ifS\else So \fi geodesic
deviation equation:\ifS
\else 
\begin{align*}
\frac{d^{2}X^{a}}{d\tau^{2}}=T^{c}\nabla_{c}\left(T^{b}\nabla_{b}X^{a}\right)= & R_{\,bcd}^{a}T^{b}T^{c}X^{d}.
\end{align*}
Marks \fi the difference between accelerated frames and the effect
of gravity/curvature:
\begin{itemize}
\item accelerated frames accelerate neighbouring geodesics the same way\ifS
\else :\fi  $Rie=0$
\item initially parallel geodesics ($V^{a}=0$) will fail to remain parallel
$\Leftrightarrow Rie\ne0$
\item manifestation of gravitational tidal force\ifS
\else  (tidal equation)\fi 
\end{itemize}
This concerns 2D subspaces of geodesic
families\ifS
\else 

What happens for vector fields in 4D?\fi 

\section{Geodesic congruences}

\ifS
\else Congruence: in open
set of spacetime

each point lies in only one geodesic

$U^{a}=\frac{dx^{a}}{d\lambda}$ tangent to congruence $\gamma_{x}\left(\lambda\right)$
of geodesic $U^{b}\nabla_{b}U^{a}=0$

Separation vector $X^{a}$\fi 
\begin{align*}
\frac{dX^{a}}{d\lambda}=U^{b}\nabla_{b}X^{a}=X^{b}\nabla_{b}U^{a}= & \left(\nabla_{b}U^{a}\right)X^{b}=U_{\:;b}^{a}X^{b}
\end{align*}
\ifS
\else $\nabla_{b}U^{a}$
measure of failure of $X$ to be parallel transported\fi 
\begin{figure}
\begin{centering}
\includegraphics[width=0.5\columnwidth]{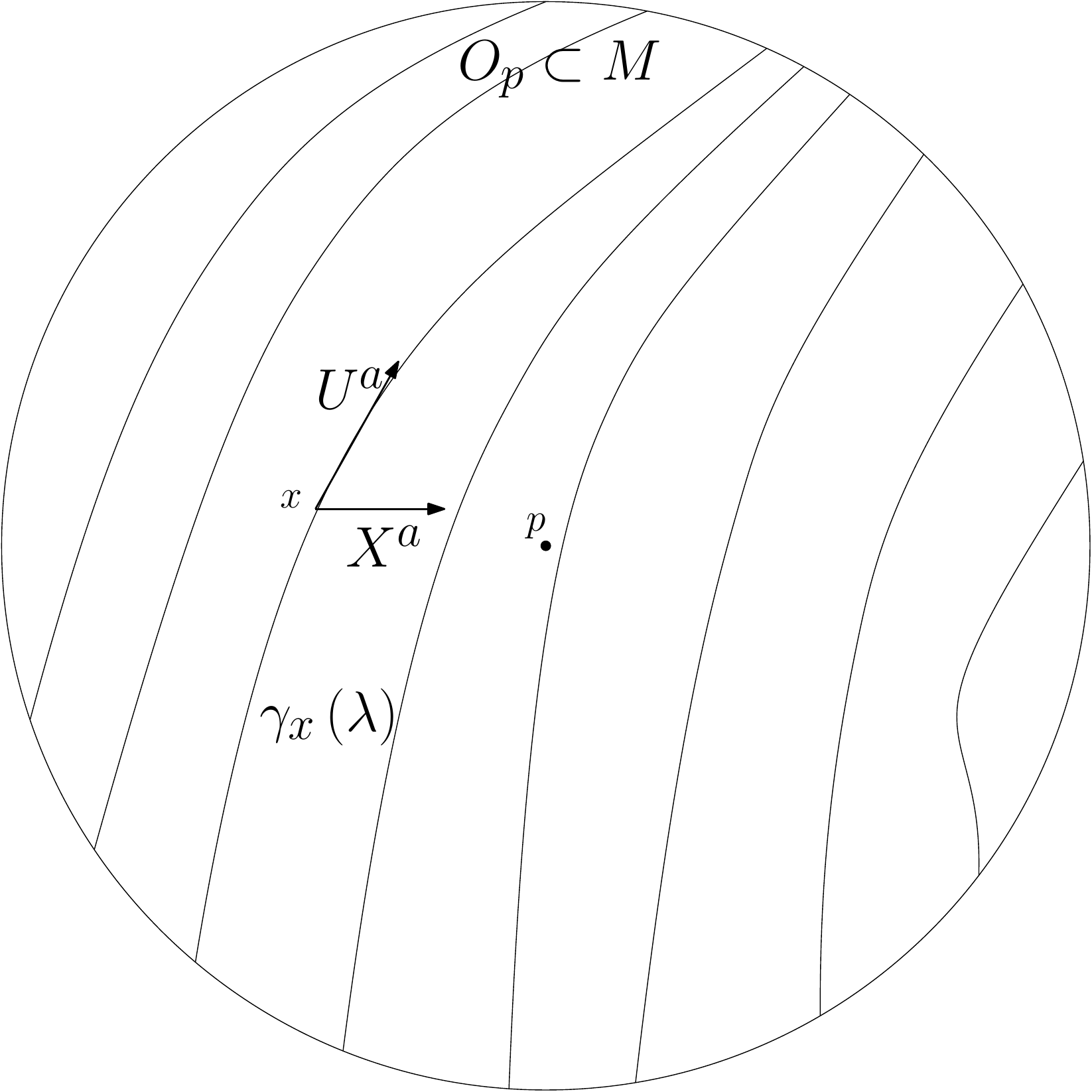}
\par\end{centering}
\caption{\label{fig:Congruence-in-an}Congruence in an open set. Note it fills
the whole spacetime open set $O_{p}\subset M$ and $X$ is 3D.}
\end{figure}

\subsection{Timelike geodesic congruences}

\ifS
\else timelike
$U\to u,\lambda\to\tau,u^{a}u_{a}=-1$ \fi 

\subsubsection{1+3 decomposition (Ellis) optical scalars/tensors\label{subsec:1+3-decomposition-(Ellis)}}

\ifS
\else Projector $\bot u^{a}$\fi 
\begin{figure}
\begin{centering}
\hspace*{-3cm}%
\begin{minipage}[t]{1.4\columnwidth}%
\begin{center}
\includegraphics[width=0.21\columnwidth]{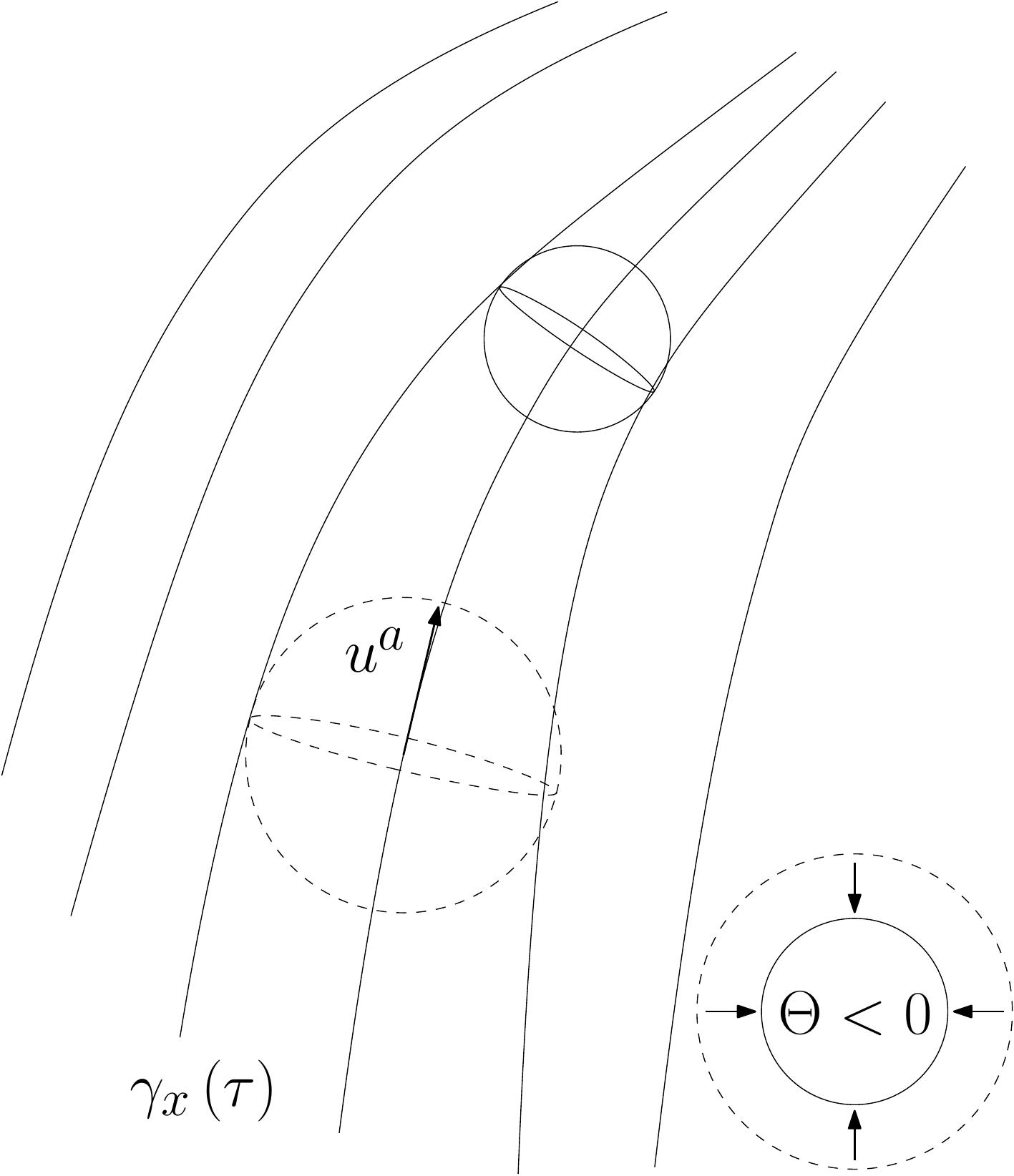}\qquad{}\includegraphics[width=0.21\columnwidth]{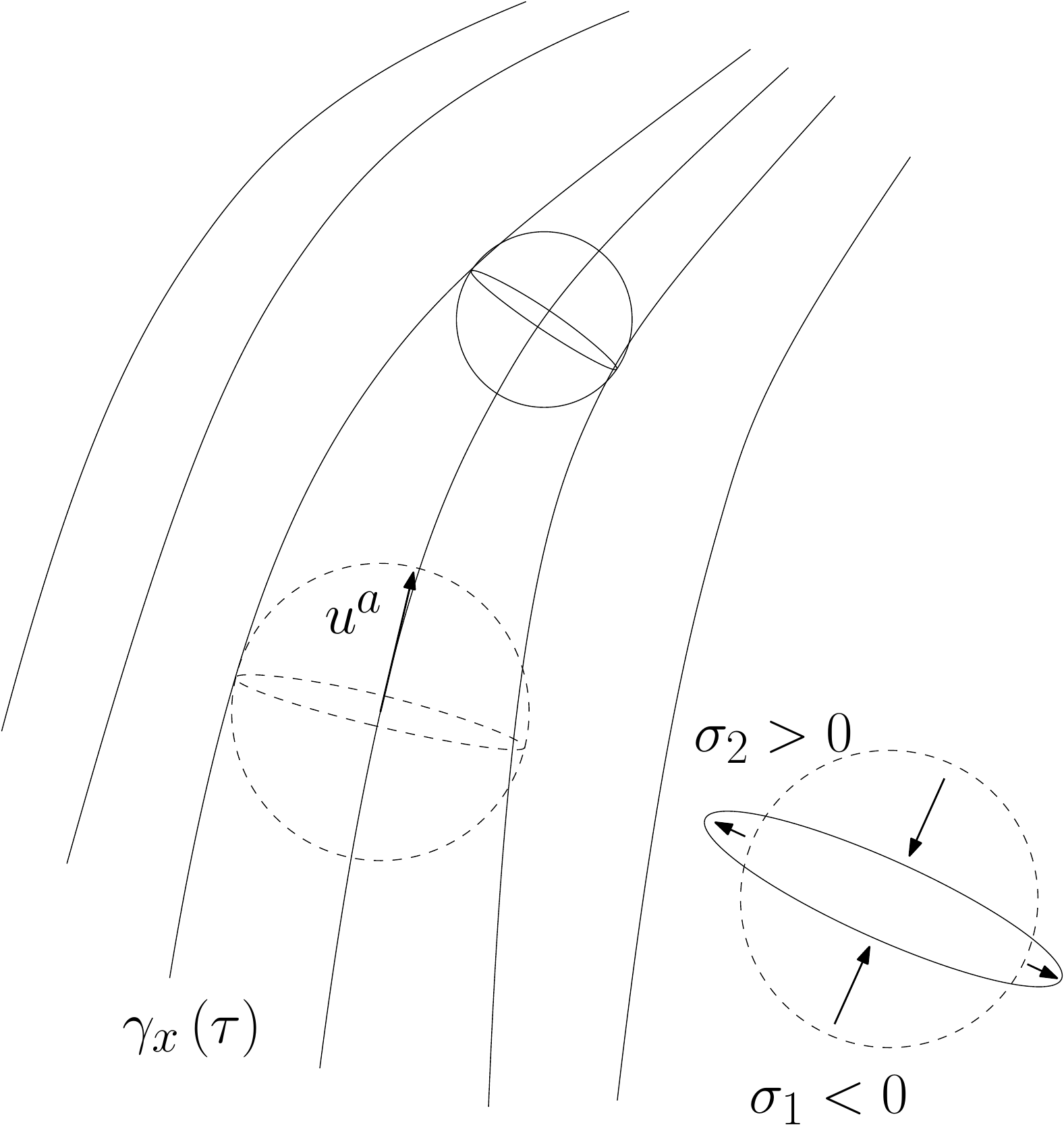}\qquad{}\includegraphics[width=0.21\columnwidth]{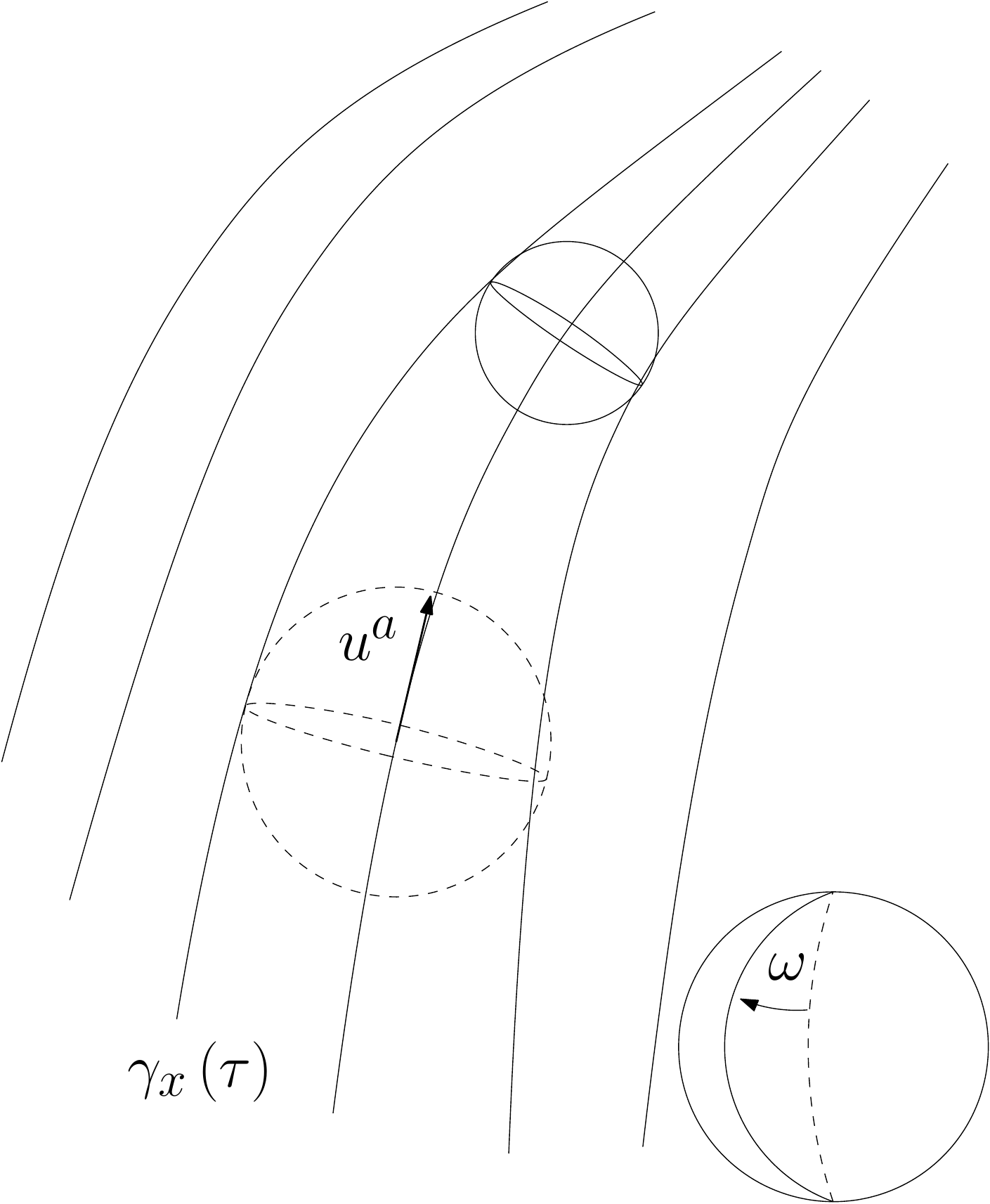}\qquad{}\includegraphics[width=0.21\columnwidth]{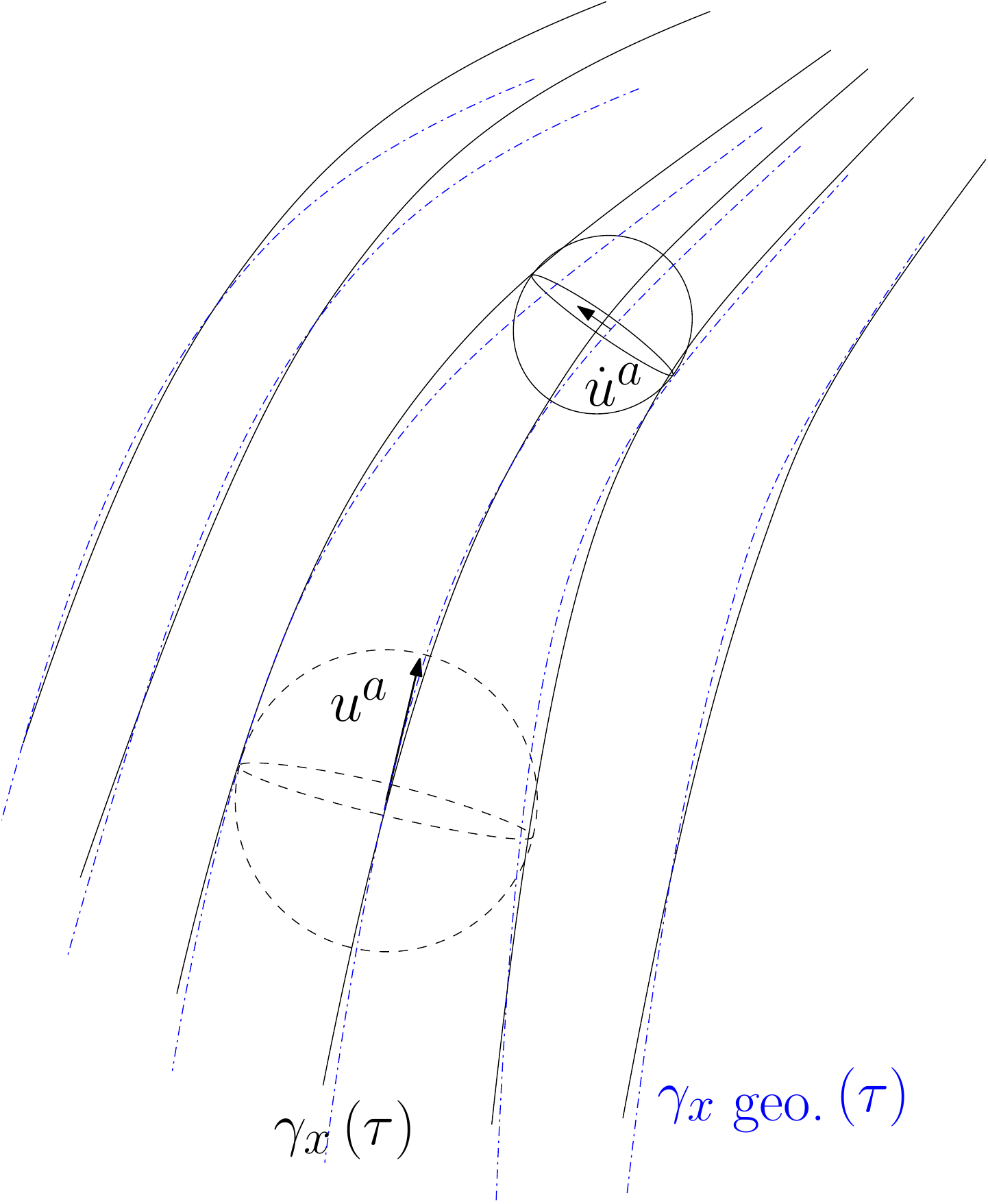}
\par\end{center}%
\end{minipage}
\par\end{centering}
\caption{\label{fig:Optical-scalars-physical}Optical scalars physical effects}
\end{figure}
\begin{align*}
h_{ab}= & g_{ab}+u_{a}u_{b}& h_{ab}u^{b}= & 0\\
 &  & h_{a}^{\:c}h_{cb}= & h_{ab}
\end{align*}
\begin{align*}
u^{a}u_{a;b}= & \frac{1}{2}\underset{-1}{\left(\underbrace{u^{a}u_{a}}\right)_{;b}}=0\\
u^{b}u_{a;b}= & \dot{u}_{a}=u^{b}\nabla_{b}u_{a}=0\textrm{ for geodesics}\Rightarrow u_{a;b}=h_{a}^{\:c}h_{b}^{\:d}u_{c;d}
\end{align*}
\begin{description}
\item [{Trace}] $h^{ab}u_{a;b}=u_{\:;a}^{a}=\Theta=\textrm{div }u$ \ifS
\else expansion\fi 
\item [{Symmetric~tracefree}] $h_{(a}^{\:c}h_{b)}^{\:d}u_{c;d}-h_{ab}\frac{\Theta}{h_{c}^{\:c}}=\sigma_{ab}$
\ifS
\else shear;
$\frac{1}{2}\sigma^{ab}\sigma_{ab}=\sigma^{2}$ shear scalar\fi 
\item [{Antisymmetric~part}] $u_{\left[a;b\right]}=\omega_{ab}$ \ifS
\else vorticity ; $\frac{1}{2}\omega^{ab}\omega_{ab}=\omega^{2}$
vorticity scalar\fi 

\ifS\else they are both \fi  spatial
tensors: $u.\sigma=u.\omega=0$
\item [{Projection~along~$u^{a}$}] $-u^{a}u^{b}u_{b}=u^{a}$,
\begin{align*}
-u^{a}u^{b}u_{b;c}= & 0\\
-u^{a}u^{c}u_{b;c}= & -u^{a}\dot{u}_{b}\textrm{ for non-geodesics}
\end{align*}
\end{description}
\ifS
\else In 4D: \fi 
\begin{description}
\item [{Volume~element}] 
\begin{align}
\eta_{abcd}= & \epsilon_{abcd}\sqrt{-g}\label{eq:Levi-CivitaVolumeTensor}\\
= & \eta_{\left[abcd\right]}\nonumber \\
\eta^{abcd}= & \frac{1}{\sqrt{-g}}\epsilon^{abcd}\nonumber 
\end{align}
\ifS
\else  $\epsilon_{abcd}$:  totally antisymmetric
tensor (+1 if permutation even; -1 if permutation odd; 0 if repeated
indices)\fi 
\item [{Vorticity~vector}] 
\begin{align*}
\omega^{a}= & \frac{1}{2}\eta^{abcd}u_{b}\omega_{cd}=-\frac{1}{2}\textrm{curl }u^{a}
\end{align*}
\ifS
\else inverted by (exercise: check\exo )\fi 
\begin{align*}
\eta_{abcd}\omega^{c}u^{d}= & \omega_{ab}
\end{align*}
\ifS\else orthogonalities\fi 
\begin{align*}
\omega_{ab}u^{b}= & 0\\
\omega_{a}u^{a}= & 0\\
\omega_{ab}\omega^{a}= & 0
\end{align*}
\ifS\else normalisation\fi 
\begin{align*}
\omega^{a}\omega_{a}= & \frac{1}{2}\omega^{ab}\omega_{ab}=\omega^{2}
\end{align*}
\end{description}

\subsubsection{\label{subsec:timelikeGeoCongrEvolution}Evolution}

\begin{align*}
\frac{d}{d\tau}\left(u_{a;b}\right)=u^{c}\nabla_{c}\left(u_{a;b}\right)= & u^{c}\nabla_{c}\nabla_{b}u_{a}\\
= & u^{c}\nabla_{b}\nabla_{c}u_{a}+u^{c}R_{\:abc}^{d}u_{d}\\
= & \nabla_{b}\left(u^{c}\nabla_{c}u_{a}\right)-\left(\nabla_{b}u^{c}\right)\left(\nabla_{c}u_{a}\right)-R_{adbc}u^{d}u^{c}\\
= & \underset{0\textrm{ for geodesics}}{\underbrace{\dot{u}_{a;b}}}-u_{a;c}u_{\:;b}^{c}-R_{acbd}u^{c}u^{d}
\end{align*}
\begin{description}
\item [{Trace:}] 
\begin{align}
\frac{d\Theta}{d\tau}= & -\frac{\Theta^{2}}{3}-2\left(\sigma^{2}-\omega^{2}\right)-R_{ab}u^{a}u^{b}\quad(+\dot{u}_{\:;a}^{a})\label{eq:Ray}
\end{align}
\ifS
\else Raychaudhuri
equation for geodesics\fi 
\item [{Symmetric~tracefree~part}] 
\begin{align}
\frac{D\sigma_{ab}}{d\tau}= & -\frac{2}{3}\Theta\sigma_{ab}-\sigma_{ac}\sigma_{\:b}^{c}-\omega_{ac}\omega_{\:b}^{c}+\frac{2}{3}h_{ab}\left(\sigma^{2}-\omega^{2}\right)+C_{acbd}u^{c}u^{d}+\frac{1}{2}\,^{3}R_{ab}\quad(+\,^{3}\dot{u}_{a;b})\label{eq:3shear}
\end{align}
where $\,^{3}R_{ab}/\,^{3}\dot{u}_{a;b}$ \ifS
\else is projected symmetric tracefree part (PSTF)
of ~ Ricci tensor/covariant derivative of acceleration\fi 
\begin{align*}
\,^{3}R_{ab}= & \left(h_{a}^{\:c}h_{b}^{\:d}-h_{ab}\frac{h^{cd}}{h_{c}^{\:c}}\right)R_{cd}\\
\,^{3}\dot{u}_{a;b}= & \left(h_{a}^{\:c}h_{b}^{\:d}-h_{ab}\frac{h^{cd}}{h_{c}^{\:c}}\right)\dot{u}_{c;d}
\end{align*}
\item [{Antisymmetric~part:}] 
\begin{align*}
\frac{D\omega_{ab}}{d\tau}= & -\frac{2}{3}\Theta\omega_{ab}-\sigma_{a}^{\:c}\omega_{cb}+\sigma_{b}^{\:c}\omega_{ca}\quad(+\dot{u}_{\left[a;b\right]})
\end{align*}
\end{description}

\paragraph{Example }

\ifS
\else In Raychaudhuri, the source term yields the
strong energy condition (SEC) from Einstein's Eqs.\fi 
\begin{align*}
R_{ab}u^{a}u^{b}= & \kappa^{2}\left(T_{ab}-\frac{T}{2}g_{ab}\right)u^{a}u^{b}
\end{align*}
\ifS
\else and the SEC requires\fi 
\begin{align*}
\left(T_{ab}-\frac{T}{2}g_{ab}\right)u^{a}u^{b}\ge & 0 & \textrm{so }R_{ab}u^{a}u^{b}\ge & 0
\end{align*}

\subsubsection{Aside: Energy conditions\label{subsec:Aside:-Energy-conditions}}

\ifS
\else Limits
to $T_{ab}$ \fi so as to get physically meaningful solutions

\paragraph{Energy momentum tensors and generic vector congruences\label{par:Energy-momentum-tensors}}

\ifS
\else 
\begin{description}
\item [{Archetypal~Examples:}]~
\end{description}
\fi 
\begin{description}
\item [{Perfect~fluid}] \ifS
\else  $(\rho,P,u_{a})$\fi 
\begin{align*}
T_{ab}= & \rho u_{a}u_{b}+P\,h_{ab}=\left(\rho+P\right)u_{a}u_{b}+P\,g_{ab}
\end{align*}
\item [{Vacuum~with~cosmological~constant}] 
\begin{align*}
T_{ab}= & \frac{\Lambda}{\kappa^{2}}u_{a}u_{b}-\frac{\Lambda}{\kappa^{2}}h_{ab}=-\frac{\Lambda}{\kappa^{2}}g_{ab}
\end{align*}
\item [{Imperfect~fluid}] with energy density $\rho$, isotropic pressure
$P$, heat flux $q_{a}$ and anisotropic stress $\Pi_{ab}$ (and cosmological
constant $\Lambda$) (with flow $u_{a}$, $u^{a}u_{a}=-1$)
\begin{align*}
T_{ab}= & \rho u_{a}u_{b}+P\,h_{ab}+2q_{(a}u_{b)}+\Pi_{ab}-\frac{\Lambda}{\kappa^{2}}g_{ab}
\end{align*}
\end{description}
We consider \ifS
\else  vector fields $t^{a}$,timelike,
and $l^{a}$, null,\fi 
\begin{align*}
t^{a}t_{a}< & 0 & l^{a}l_{a}= & 0
\end{align*}
For a fluid description, since any two timelike vectors \ifS
\else  of same light cone are related
by a given vector, we can thus decompose one in terms of the other
and\fi  a null vector:

$\forall t^{a}$ timelike $t^{a}-u^{a}=S^{a}$ 
\begin{align*}
t^{a}-u^{a}= & S^{0}u^{a}+\pi\negthinspace S\,e^{a}
\end{align*}
where \ifS
\else  $S^{0}=-S^{a}u_{a}$, $\pi\negthinspace S^{a}=h_{\:b}^{a}S^{b}=S^{a}-S^{0}u^{a}$\fi 
spacelike

\ifS\else so \fi  $\pi\negthinspace S=\sqrt{\pi\negthinspace S^{a}\pi\negthinspace S_{a}}>0$and we define $e^{a}=\frac{\pi\negthinspace S^{a}}{\pi\negthinspace S}$
such as $e^{a}e_{a}=1,u^{a}e_{a}=0$

Then
\begin{align*}
t^{a}t_{a}= & \left(1+S^{0}\right)^{2}u^{a}u_{a}+\pi\negthinspace S^{a}\pi\negthinspace S_{a}\\
= & -\left(1+S^{0}\right)^{2}+\left(\pi\negthinspace S\right)^{2}<0
\end{align*}
 as $t^{a}$ is timelike

\ifS\else So we have \fi $\left(\pi\negthinspace S\right)^{2}<\left(1+S^{0}\right)^{2}$

Since $t^{a}$ and $u^{a}$ \ifS
\else in same light cone so \fi $1+S^{0}>0$ and thus $1+S^{0}>\pi\negthinspace S$

\ifS
\else (In case $t^{a}$ in opposite
light cone from $u^{a}$, $1+S^{0}<0\Rightarrow1+S^{0}<-\pi\negthinspace S<\pi\negthinspace S$
so $1+S^{0}<\pi\negthinspace S$)\fi 

We can thus \ifS
\else write, defining \fi $l^{a}=u^{a}+e^{a}\Rightarrow l^{a}l_{a}=0$,
\begin{align*}
t^{a}= & \left(1+S^{0}\right)u^{a}+\pi\negthinspace S\,e^{a}\\
= & \left(1+S^{0}-\pi\negthinspace S\right)u^{a}+\pi\negthinspace S\left(u^{a}+e^{a}\right)\\
= & \left(1+S^{0}-\pi\negthinspace S\right)u^{a}+\pi\negthinspace S\,l^{a}
\end{align*}
\ifS
\else with \fi $\pi\negthinspace S,1+S^{0}-\pi\negthinspace S>0$

Then for any symmetric 2-tensor $A_{ab}$
\begin{align}
A_{ab}t^{a}t^{b}= & \left(1+S^{0}-\pi\negthinspace S\right)^{2}A_{ab}u^{a}u^{b}+\left(\pi\negthinspace S\right)^{2}A_{ab}l^{a}l^{b}+2\pi\negthinspace S\left(1+S^{0}-\pi\negthinspace S\right)A_{ab}u^{a}l^{b}\label{eq:AttDecomposition}
\end{align}
and we are left with 
\begin{align*}
 & A_{ab}u^{a}u^{b}\\
 & A_{ab}l^{a}l^{b}\\
 & A_{ab}u^{a}l^{b}
\end{align*}
\ifS
\else and for
any vector $V^{a}$, the perfect fluid \fi yields
\begin{align}
T_{ab}V^{b}= & \left(\rho+P\right)\left(V^{b}u_{b}\right)u_{a}+P\,V_{a}\nonumber \\
\Rightarrow T_{ac}T_{\:b}^{c}V^{a}V^{b}= & \left(\rho+P\right)^{2}\left(V^{b}u_{b}\right)^{2}u^{a}u_{a}+P^{2}V^{a}V_{a}+2P\left(\rho+P\right)\left(V^{b}u_{b}\right)^{2}\nonumber \\
= & \left(\rho+P\right)\left(V^{b}u_{b}\right)^{2}\left[-\left(\rho+P\right)+2P\right]+P^{2}V^{a}V_{a}\nonumber \\
= & -\left(\rho^{2}-P^{2}\right)\left(V^{b}u_{b}\right)^{2}+P^{2}V^{a}V_{a}\label{eq:ObservedMomentum}
\end{align}
\ifS
\else while
the Ricci tensor for a perfect fluid, given the EFE, gives ($T=-\rho+3P$)\fi 
\begin{align}
R_{ab}= & \kappa^{2}\left(T_{ab}-\frac{T}{2}g_{ab}\right)\nonumber \\
= & \kappa^{2}\left(\left(\rho+P\right)u_{a}u_{b}+\left(P+\frac{\rho-3P}{2}\right)g_{ab}\right)\nonumber \\
= & \kappa^{2}\left(\left(\rho+P\right)u_{a}u_{b}+\frac{\rho-P}{2}g_{ab}\right)\nonumber \\
= & \kappa^{2}\left(\frac{\rho+3P}{2}u_{a}u_{b}+\frac{\rho-P}{2}h_{ab}\right)\nonumber \\
= & \kappa^{2}\chi_{ab}\label{eq:reducedTidal}
\end{align}
Some of the most popular energy conditions are

\paragraph{The weak energy condition (WEC)}

requires for all timelike vectors $t^{a}$ 
\begin{align*}
\forall t^{a},t^{a}t_{a}< & 0 & T_{ab}t^{a}t^{b}\ge & 0
\end{align*}
\ifS
\else (average energy density as seen by an observer $t^{a}$must
be positive (see \cite{Rosa:2020hex}))\fi 

\subparagraph{For a perfect fluid:}

\ifS\else 
\begin{align*}
T_{ab}t^{a}t^{b}= & \left(1+S^{0}-\pi\negthinspace S\right)^{2}T_{ab}u^{a}u^{b}+\left(\pi\negthinspace S\right)^{2}T_{ab}l^{a}l^{b}+2\pi\negthinspace S\left(1+S^{0}-\pi\negthinspace S\right)T_{ab}u^{a}l^{b}
\end{align*}
but \fi 
\begin{align*}
T_{ab}u^{a}u^{b}= & \rho& T_{ab}l^{a}l^{b}= & \left(\rho+P\right)\left(u^{a}l_{a}\right)^{2}
\end{align*}
\begin{align*}
T_{ab}u^{a}l^{b}= & -\rho u^{a}l_{a}=-\rho u^{a}\left(u_{a}+e_{a}\right)=\rho
\end{align*}
Thus 
\begin{align*}
T_{ab}t^{a}t^{b}\ge & 0\Leftrightarrow & \boxed{\rho\ge0\textrm{ and }\left(\rho+P\right)\ge0} & 
\end{align*}

\paragraph{The null energy condition (NEC)}

requires for all null vector $L^{a}$ 
\begin{align*}
\forall L^{a},L^{a}L_{a}= & 0 & T_{ab}L^{a}L^{b}\ge & 0
\end{align*}

\subparagraph{For a perfect fluid:}

We have seen 
\begin{align*}
T_{ab}l^{a}l^{b}= & \left(\rho+P\right)\left(u^{a}l_{a}\right)^{2}\\
\Leftrightarrow & \boxed{\left(\rho+P\right)\ge0}
\end{align*}
\ifS\else note \fi $\rho$ and $P$ can be negative

\paragraph{The dominant energy condition (DEC)}

requires the WEC\ifS
\else  with $T^{ab}t_{b}$
non spacelike\fi , i.e.
\begin{align*}
T_{ac}T_{\:b}^{c}t^{a}t^{b}\le & 0
\end{align*}
\ifS \else (matter must move along timelike or null worldlines:
matter momentum density $-T^{ab}t_{b}$ must also be future directed
and non spacelike (see \cite{Rosa:2020hex}))\fi 

\subparagraph{For a perfect fluid:}

\begin{align*}
T_{ac}T_{\:b}^{c}t^{a}t^{b}= & -\left(\rho^{2}-P^{2}\right)\left(u^{a}t_{a}\right)^{2}+P^{2}t^{a}t_{a}\le0\\
\Rightarrow & \rho^{2}-P^{2}\ge0
\end{align*}
\ifS
\else Since WEC $\Leftrightarrow\rho\ge0\textrm{ and }\left(\rho+P\right)\ge0$

then $\rho-P\ge0$ so\fi 
\begin{align*}
\left.\begin{array}{rl}
\rho\ge & P\\
\rho\ge & -P
\end{array}\right\} \Leftrightarrow & \boxed{\rho\ge\left|P\right|}
\end{align*}

\paragraph{The null dominant energy condition (NDEC)}

requires the NEC\ifS
\else 
with $T^{ab}l_{b}$ non spacelike\fi , i.e.
\begin{align*}
T_{ac}T_{\:b}^{c}l^{a}l^{b}\le & 0
\end{align*}

\subparagraph{For a perfect fluid:}

\begin{align*}
T_{ac}T_{\:b}^{c}l^{a}l^{b}= & -\left(\rho^{2}-P^{2}\right)\left(u^{a}l_{a}\right)^{2}\le0\\
\Leftrightarrow & \rho^{2}-P^{2}\ge0
\end{align*}
\ifS
\else and $\rho+P\ge0\Rightarrow$ \fi either
\begin{align*}
\underline{\rho\ge0}, & \boxed{\rho\ge\left|P\right|}
\end{align*}
or $\underline{\rho<0}$ \ifS
\else so $\rho+P\ge0\Leftrightarrow P\ge-\rho>0$ but
$\rho^{2}-P^{2}\ge0\Leftrightarrow\rho-P\ge0\Rightarrow P\le\rho<0\textrm{ or }\rho+P\le0\Rightarrow$
\fi 
\begin{align*}
 & \boxed{P=-\rho}
\end{align*}

\paragraph{The strong energy condition (SEC)}

\ifS
\else requires\fi 
\begin{align*}
\forall t^{a},t^{a}t_{a}<0,T_{ab}t^{a}t^{b}\ge & \frac{1}{2}T_{\:c}^{c}t^{a}t_{a}\\
\Leftrightarrow\chi_{ab}t^{a}t^{b}= & \left(T_{ab}-\frac{T}{2}g_{ab}\right)t^{a}t^{b}\ge0\\
= & \left(1+S^{0}-\pi\negthinspace S\right)^{2}\chi_{ab}u^{a}u^{b}+\left(\pi\negthinspace S\right)^{2}\chi_{ab}l^{a}l^{b}+2\pi\negthinspace S\left(1+S^{0}-\pi\negthinspace S\right)\chi_{ab}u^{a}l^{b}
\end{align*}
\ifS \else (trace of tidal tensor, i.e. $R_{ab}$, must be non negative
as measured by $t^{a}$ (see \cite{Rosa:2020hex}))\fi 

\subparagraph{For a perfect fluid:}

\begin{align*}
\chi_{ab}u^{a}u^{b}= & \left(\frac{\rho+3P}{2}u_{a}u_{b}+\frac{\rho-P}{2}h_{ab}\right)u^{a}u^{b}=\frac{\rho+3P}{2}\\
\chi_{ab}l^{a}l^{b}= & \left(\left(\rho+P\right)u_{a}u_{b}+\frac{\rho-P}{2}g_{ab}\right)l^{a}l^{b}=\left(\rho+P\right)\left(u^{a}l_{a}\right)^{2}=\rho+P\\
\chi_{ab}u^{a}l^{b}= & -\frac{\rho+3P}{2}u^{a}l_{a}=\frac{\rho+3P}{2}
\end{align*}
Thus
\begin{align*}
\chi_{ab}t^{a}t^{b}\ge & 0\Leftrightarrow & \boxed{\rho+P\ge0\textrm{ and }\rho+3P\ge0} & 
\end{align*}
\ifS
\else Gravity attractive\fi 

\paragraph{Relations between energy conditions}

Note \ifS
\else that \fi 
\begin{align*}
\textrm{DEC}\Rightarrow & \textrm{WEC}\Rightarrow\textrm{NEC}\\
\textrm{NDEC}\Rightarrow & \textrm{NEC}\\
\textrm{SEC}\Rightarrow & \textrm{NEC}
\end{align*}
\ifS
\else Return
to Raychaudhuri source for geodesic congruences\fi 

\subsubsection{Energy condition and congruence evolution\label{subsec:Energy-condition-and}}

\ifS
\else Now by Frobenius theorem,
$u^{a}$ hypersurface orthogonal translates as\fi 
\begin{align*}
u_{[a}\nabla_{b}u_{c]}= & 0\\
= & \frac{1}{3}\left(u_{a}\nabla_{[b}u_{c]}+u_{b}\nabla_{[c}u_{a]}+u_{c}\nabla_{[a}u_{b]}\right)
\end{align*}
\ifS\else and by definition
of \fi  $\omega_{ab}=\nabla_{[b}u_{a]}$, this yields
\begin{align*}
\left(u_{a}\omega_{cb}+u_{b}\omega_{ac}+u_{c}\omega_{ba}\right)= & 0
\end{align*}
Since $u^{a}\omega_{ab}=0$, contracting with, say, $u^{a}$ gives
\begin{align*}
-\omega_{cb}= & 0
\end{align*}
and conversely if $\omega_{ab}=0$ then $u_{[a}\nabla_{b}u_{c]}=0$ 

\ifS\else Thus \fi  $\omega_{ab}=0\Leftrightarrow u^{a}$
hypersurface orthogonal.

\ifS\else Then: the Raychaudhuri
Eq. of \fi a hypersurface orthogonal geodesic congruence verifying
the SEC \ifS\else implies\fi 
\begin{align*}
\frac{d\Theta}{d\tau}+\frac{\Theta^{2}}{3}= & -2\sigma^{2}-R_{ab}u^{a}u^{b}\le0\\
\Leftrightarrow\frac{d\Theta}{d\tau}\le & -\frac{\Theta^{2}}{3}
\end{align*}
\ifS\else which \fi integrates into
\begin{align*}
\frac{d\tau}{3}\le & -\frac{d\Theta}{\Theta^{2}}=d\left(\frac{1}{\Theta}\right) & \Rightarrow\Theta^{-1}\ge & \Theta_{0}^{-1}+\frac{\tau}{3}
\end{align*}
Therefore starting from a converging ($\Theta_{0}<0$)
hypersurface orthogonal congruence obeying the SEC can never diverge
before hitting a caustic ($\Theta\to-\infty$, geodesic crossing\ifS
\else ) in \fi finite proper time ($\tau\le-3\Theta_{0}^{-1}$).

\subsection{Null geodesic congruences}

\ifS
\else Defined
\fi  as $k^{a}=\frac{dx^{a}}{d\lambda},k^{a}k_{a}=0$. \ifS
\else Problem to define \fi orthogonal
space, as self orthogonal 

\ifS
\else Take \fi  $l^{a}l_{a}=0$ null with $l^{a}k_{a}=-1$,
that is pointing in the opposite spatial direction \ifS
\else (frame dependent) and \fi parallel transported
along $k^{a}$: $k^{b}\nabla_{b}l^{a}=0$

\subsubsection{2+2 decomposition (Hayward)}

\ifS
\else Define \fi projector orthogonal to $\left(k,l\right)$,
onto
\begin{align*}
T_{\bot}= & \left\{ V^{a}|V^{a}k_{a}=0,V^{a}l_{a}=0\right\} 
\end{align*}

\subsubsection{A hint of 1+1+2 decomposition (Clarkson)\label{subsec:A-hint-of1+1+2}}

\ifS
\else From a timelike and spacelike
orthogonal pair \fi  $u^{a}u_{a}=-1$ $e^{a}e_{a}=1$ $u^{a}e_{a}=0$we can build 
\begin{align*}
\left\{ \begin{array}{rl}
k^{a}= & \frac{u^{a}+e^{a}}{\sqrt{2}}\\
l^{a}= & \frac{u^{a}-e^{a}}{\sqrt{2}}
\end{array}\right. & k^{a}l_{a}=-1
\end{align*}
\ifS
\else We saw in 1+3: $h_{ab}=g_{ab}+u_{a}u_{b}$
projector $\bot u^{a}$

Iterate in 3D space with metric $h_{ab}$

Define $N_{ab}=h_{ab}-e_{a}e_{b}$ \fi 
\begin{align*}
\begin{array}{rl}
N_{ab}e^{b}= & 0\\
N_{ab}u^{b}= & 0
\end{array} & \textrm{ so }\begin{array}{rl}
N_{ab}k^{b}= & 0\\
N_{ab}l^{b}= & 0
\end{array}\ifS\else\fi
\end{align*}
\ifS\else This looks right. Since\fi 
\begin{align*}
u^{a}= & \frac{k^{a}+l^{a}}{\sqrt{2}}\\
e^{a}= & \frac{k^{a}-l^{a}}{\sqrt{2}}
\end{align*}
\begin{align*}
N_{ab}= & g_{ab}+u_{a}u_{b}-e_{a}e_{b}\\
= & g_{ab}+\frac{\left(k_{a}+l_{a}\right)\left(k_{b}+l_{b}\right)}{2}-\frac{\left(k_{a}-l_{a}\right)\left(k_{b}-l_{b}\right)}{2}\\
= & g_{ab}+k_{(a}l_{b)}+k_{(a}l_{b)}\boxed{=g_{ab}+2k_{(a}l_{b)}}
\end{align*}
$\forall V^{a},W^{a}\in T_{\bot}$
\begin{align*}
N_{ab}V^{a}W^{b}= & \left(g_{ab}+2k_{(a}l_{b)}\right)V^{a}W^{b}\\
= & g_{ab}V^{a}W^{b}\\
N_{\:b}^{a}V^{b}= & g_{\:b}^{a}V^{b}=V^{a}\\
N_{\:b}^{a}k^{b}= & k^{a}+k^{a}\left(l_{b}k^{b}\right)=0\\
N_{\:b}^{a}l^{b}= & l^{a}+l^{a}\left(k_{b}l^{b}\right)=0\\
N_{\:c}^{a}N_{\:b}^{c}= & N_{\:c}^{a}\left(g_{\:b}^{c}+k^{c}l_{b}+l^{c}k_{b}\right)\\
= & N_{\:b}^{a}\hphantom{glk^{c}2\left(\nabla_{c}\left(k_{(d}\right)l_{b)}+k_{(d}\nabla_{c}l_{b)}\right)}
\end{align*}
\ifS
\else \vspace{-0.9cm}
\fi 
\begin{align*}
k^{c}\nabla_{c}N_{\:b}^{a}= & k^{c}\nabla_{c}\left(g^{ad}N_{db}\right)=g^{ad}k^{c}\nabla_{c}N_{db}\\
= & g^{ad}k^{c}\nabla_{c}\left(g_{db}+2k_{(d}l_{b)}\right)\\
= & g^{ad}k^{c}2\left(\nabla_{c}\left(k_{(d}\right)l_{b)}+k_{(d|}\nabla_{c}l_{|b)}\right)
\end{align*}
and since
\begin{align*}
k^{c}\nabla_{c}k_{d}= & 0\textrm{ (null geodesic)}\\
k^{c}\nabla_{c}l_{b}= & 0\textrm{ (parallel transport)}
\end{align*}
then $k^{c}\nabla_{c}N_{\:b}^{a}=0$

Again, the failure for a normal separation vector $X^{a}\in T_{\bot}$
to be parallel transported, as for timelike congruences, is measured
by $k_{\:;b}^{a}$ (see deviation and commuting vectors):
\begin{align*}
\frac{DX^{a}}{d\lambda}=k^{b}\nabla_{b}X^{a}=X^{b}\nabla_{b}k^{a}= & \boxed{k_{\:;b}^{a}}X^{b}
\end{align*}
\ifS
\else but here only its projection
is needed:\fi  
\begin{align*}
\frac{DX^{a}}{d\lambda}=k^{b}\nabla_{b}X^{a}= & k^{b}\nabla_{b}\left(N_{\:c}^{a}X^{c}\right)\\
= & N_{\:c}^{a}k^{b}\nabla_{b}X^{c}\\
= & N_{\:c}^{a}X^{b}\nabla_{b}k^{c}\\
= & N_{\:c}^{a}N_{\:d}^{b}k_{\:;b}^{c}X^{d}\\
= & \hat{k}_{\:;d}^{a}X^{d}
\end{align*}
where $\hat{k}_{\:;b}^{a}=N_{\:c}^{a}N_{\:b}^{d}k_{\:;d}^{c}$ 

Again \ifS
\else 2-trace \fi 
\begin{description}
\item [{2-Trace}] $N^{ab}k_{a;b}=\hat{k}_{\:;a}^{a}=\Theta_{\left(k\right)}$
\ifS
\else 2-expansion\fi 
\item [{Projected~symmetric~tracefree}] $\left(N_{(a}^{\:c}N_{b)}^{\:d}-N_{ab}\frac{N^{cd}}{N_{c}^{\:c}}\right)k_{c;d}=\hat{\sigma}_{ab}$
\ifS
\else 
2-shear\fi 
\item [{Antisymmetric~part}] $\hat{k}_{\left[a;b\right]}=\hat{\omega}_{ab}$
\ifS
\else 
2-vorticity\fi 
\end{description}
Note 
$N_{\:c}^{c}=h_{\:c}^{c}-e^{c}e_{c}=g_{\:c}^{c}+u^{c}u_{c}-e^{c}e_{c}=4-1-1=2$
dimensionality of $T_{\bot}$

As in \ifS
\else timelike case $k^{a}\hat{\omega}_{ab}=0$
and Frobenius lead to $\hat{\omega}_{ab}=0\Leftrightarrow k^{a}$
$T_{\bot}$ surface orthogonal\fi 

\subsubsection{Evolution}

\begin{align*}
\frac{D}{d\lambda}\left(\hat{k}_{a;b}\right)=k^{c}\nabla_{c}\left(\hat{k}_{a;b}\right)= & k^{e}\nabla_{e}\left(N_{a}^{\:c}N_{b}^{\:d}k_{c;d}\right)\\
= & N_{a}^{\:c}N_{b}^{\:d}k^{e}\nabla_{e}\nabla_{d}k_{c}\\
= & N_{a}^{\:c}N_{b}^{\:d}\left(k^{e}\nabla_{d}\nabla_{e}k_{c}+k^{e}R_{\:cde}^{f}k_{f}\right)\\
= & N_{a}^{\:c}N_{b}^{\:d}\left(\nabla_{d}\left(k^{e}\nabla_{e}k_{c}\right)-\left(\nabla_{d}k^{e}\right)\left(\nabla_{e}k_{c}\right)-R_{cfde}k^{f}k^{e}\right)\\
= & \underset{0\textrm{ for geodesics}}{\underbrace{N_{a}^{\:c}N_{b}^{\:d}\left(\frac{dk_{c}}{d\lambda}\right)_{;d}}}-\hat{k}_{a;c}\hat{k}_{\:;b}^{c}-N_{a}^{\:c}N_{b}^{\:d}R_{cfde}k^{f}k^{e}
\end{align*}
\ifS\else since\fi 
\begin{align*}
\nabla_{d}k^{e}\nabla_{e}k_{c}= & g^{ef}\nabla_{d}k_{f}\nabla_{e}k_{c}\\
= & \left(N^{ef}-2k^{(e}l^{f)}\right)\nabla_{d}k_{f}\nabla_{e}k_{c}
\end{align*}
and
\begin{align*}
k^{e}\nabla_{e}k_{c}= & 0\textrm{ (null geodesic)}\\
l^{e}\nabla_{e}k_{c}= & k^{e}\nabla_{e}l_{c}\textrm{ (commuting)}\\
= & 0\textrm{ (parallel transport)}
\end{align*}
\ifS\else so\fi 
\begin{align*}
\nabla_{d}k^{e}\nabla_{e}k_{c}= & \nabla_{d}k_{f}N^{fe}\nabla_{e}k_{c}
\end{align*}
Furthermore 
\begin{align*}
\mathrm{tr}NNRkk= & N^{ac}N_{a}^{\:d}R_{cfde}k^{f}k^{e}\\
= & N^{ac}R_{cfae}k^{f}k^{e}\\
= & \left(g^{ac}+2k^{((a|}l^{c)}\right)R_{[cf][ae]}k^{(f}k^{|e))}\\
= & g^{ac}R_{cfae}k^{f}k^{e}\\
= & \mathrm{tr}Rkk
\end{align*}
by anticommutation of Riemann's pairs\ifS
\else Following the same path than for the timelike congruences\fi :
\begin{description}
\item [{Trace:}] 
\begin{align}
\frac{d\Theta_{\left(k\right)}}{d\lambda}= & -\frac{\Theta_{\left(k\right)}^{2}}{2}-2\left(\hat{\sigma}^{2}-\hat{\omega}^{2}\right)-R_{ab}k^{a}k^{b}\label{eq:RayNull}
\end{align}
\ifS
\else Raychaudhuri
Eq. for null geodesics\fi \\
Note the ($k,2$)-expansion is naturally projected as $k^{e}\nabla_{e}k_{c}=0$
from \ifS\else geodesics and \fi $k^{a}\nabla_{b}k_{a}=\frac{1}{2}\nabla_{b}k^{a}k_{a}=0$
\ifS
\else from
norm, so\fi 
\begin{align*}
\Theta_{\left(k\right)}=N^{ab}\hat{k}_{a;b}= & N^{ab}N_{a}^{\:c}N_{b}^{\:d}k_{c;d}\\
= & N^{cd}k_{c;d}\\
= & \left(g^{ab}+2k^{(a}l^{b)}\right)k_{a;b}\\
= & g^{ab}k_{a;b}=k_{\:;a}^{a}
\end{align*}
Samely\ifS\else for
\fi $\hat{\sigma}^{2}$ and $\hat{\omega}^{2}$,
noting $\left\{ \right\} $ for either $\left(\right)$ or $\left[\right]$,
and for the same geodesic/ norm reasons
\begin{align*}
\hat{k}^{\left\{ a;b\right\} }\hat{k}_{\left\{ a;b\right\} }= & N^{ac}N^{ad}k_{\left\{ c;d\right\} }k_{\left\{ a;b\right\} }\\
= & g^{ac}g^{ad}k_{\left\{ c;d\right\} }k_{\left\{ a;b\right\} }\\
= & k^{\left\{ a;b\right\} }k_{\left\{ a;b\right\} }
\end{align*}
\ifS\else thus Raychaudhuri
\fi is independent of the arbitrary choice of $l^{a}$\\
The source term \ifS
\else is given through EFE \fi (recall $k^{a}k_{a}=0$)
\begin{align*}
R_{ab}k^{a}k^{b}= & \kappa^{2}\left(T_{ab}-\frac{T}{2}g_{ab}\right)k^{a}k^{b}=\kappa^{2}T_{ab}k^{a}k^{b}
\end{align*}
\ifS\else so this is \fi the base for the NEC:
\ifS
\else convergence to caustics more general\fi 
\item [{Projected~Symmetric~tracefree~part}] 
\begin{align}
\frac{D\hat{\sigma}_{ab}}{d\lambda}= & -\Theta_{\left(k\right)}\hat{\sigma}_{ab}-N_{a}^{\:c}N_{b}^{\:d}C_{cedf}k^{e}k^{f}\label{eq:2shearNull}
\end{align}
\ifS\else Shear evolution\fi 
\item [{Antisymmetric~part:}] 
\begin{align*}
\frac{D\hat{\omega}_{ab}}{d\lambda}= & -\Theta_{\left(k\right)}\hat{\omega}_{ab}
\end{align*}
\ifS\else vorticity
evolution\fi 
\end{description}
\ifS
\else If $l^{a}$
considered, by symmetry you get\fi 
\begin{align*}
\textrm{Raychaudhuri }\frac{d\Theta_{\left(l\right)}}{d\nu}= & -\frac{\Theta_{\left(l\right)}^{2}}{2}-2\left(\hat{\sigma}_{\left(l\right)}^{2}-\hat{\omega}_{\left(l\right)}^{2}\right)-R_{ab}l^{a}l^{b}\\
\textrm{Shear }\frac{D\hat{\sigma}_{\left(l\right)ab}}{d\nu}= & -\Theta_{\left(l\right)}\hat{\sigma}_{\left(l\right)ab}-N_{a}^{\:c}N_{b}^{\:d}C_{cedf}l^{a}l^{b}\\
\textrm{Vorticity }\frac{D\hat{\omega}_{\left(l\right)ab}}{d\nu}== & -\Theta_{\left(l\right)}\hat{\omega}_{\left(l\right)ab}
\end{align*}
\ifS
\else and
mixed Raychaudhuri can even be considered\fi 
\begin{align*}
\frac{1}{2}\left(\frac{d\Theta_{\left(l\right)}}{d\lambda}+\frac{d\Theta_{\left(k\right)}}{d\nu}-\frac{k_{a}}{2}\left[l,k\right]^{a}\Theta_{\left(l\right)}-\frac{l_{a}}{2}\left[k,l\right]^{a}\Theta_{\left(k\right)}\right)+\Theta_{\left(k\right)}\Theta_{\left(l\right)}= & \frac{g_{ab}k^{a}l^{b}}{r^{2}}+R_{ab}k^{a}l^{b}
\end{align*}

\chapter{Horizons and singularity theorems\label{chap:Horizons-and-singularity}}

\section{Horizons}

\subsection{Event horizons}

\begin{defn}
Event Horizon\label{def:Event-Horizon}

Hypersurface separating events connected to time infinity by timelike
curves from those not connected
\end{defn}
\warningsign Global
concept

\begin{figure}
\includegraphics{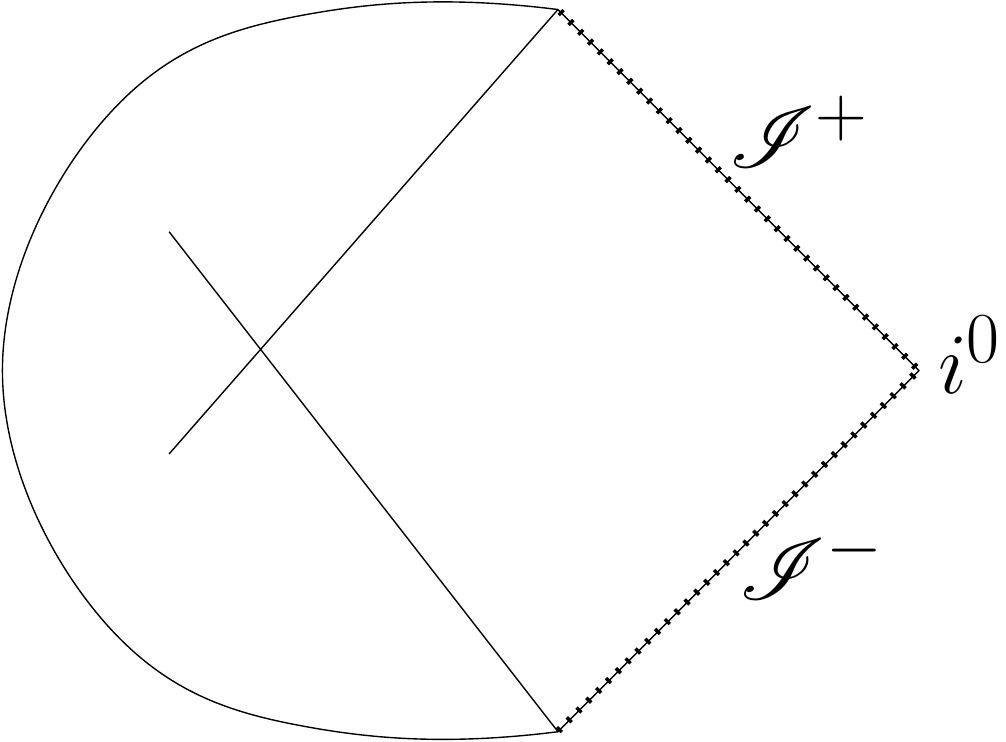}

\caption{\label{fig:Conformal-diagram-for}Conformal diagram for asymptotically
flat spacetimes}

\end{figure}

In asymptotically flat spacetimes with
event horizons (EH)\ifS
\else 

$\Rightarrow$conformal diagram with \fi Minkowski-like light and
space infinities $\left(i^{0},\mathscr{I}^{+},\mathscr{I}^{-}\right.$,
see Fig.~\ref{fig:Conformal-diagram-for}$\left.\vphantom{i^{0},\mathscr{I}^{+},}\right)$

\ifS\else Timelike
infinities aren't \fi necessary to define EH. 

\ifS\else Define \fi the causal past of an
open set $O$
\begin{align}
J^{-}\left(O\right)= & \left\{ x\in\mathcal{M}|\exists y\in O,\exists\gamma\left(\tau\right),\dot{\gamma}^{2}\le0,\gamma\left(\tau_{2}\right)=y,\exists\tau_{1}<\tau_{2},\gamma\left(\tau_{1}\right)=x\right\} \label{eq:J-DefCausal}
\end{align}
\ifS\else i.e. \fi set of events connected by \uline{past}
causal curves to events of $O$.

\ifS\else Causal
future of $O$\fi 
\begin{align}
J^{+}\left(O\right)= & \left\{ x\in\mathcal{M}|\exists y\in O,\exists\gamma\left(\tau\right),\dot{\gamma}^{2}\le0,\gamma\left(\tau_{1}\right)=y,\exists\tau_{2}>\tau_{1},\gamma\left(\tau_{2}\right)=x\right\} \label{eq:J+DefCausal}
\end{align}
\ifS\else this time, \fi events
connected by \uline{future} curves to $O$.

Their boundaries yield light cones of events of $O$
\begin{description}
\item [{Event~Horizons:}] \ifS
\else ~\fi \\
\ifS
\else Boundary of \fi $J^{-}\left(\mathscr{I}^{+}\right)$\ifS
\else :\\
\fi $\partial J^{-}\left(\mathscr{I}^{+}\right)$\ifS
\else :future horizon\fi \\
\ifS
\else Boundary of \fi $J^{+}\left(\mathscr{I}^{-}\right)$\ifS
\else :\\
\fi $\partial J^{+}\left(\mathscr{I}^{-}\right)$\ifS
\else :past horizon\fi 
\end{description}
\ifS\else Thus
\fi EH are null hypersurfaces
\begin{description}
\item [{$\Sigma$~Hypersurface:}] \ifS
\else Defined
by an equation \fi $f\left(x\right)=cst$\\
$\partial_{a}f$
normal vector to $\Sigma$
\item [{$\Sigma$~Null~Hypersurface:}] $\partial_{a}f\partial^{a}f=0$\ifS
\else  thus $\partial_{a}f$ also tangent
\fi to $\Sigma$
\end{description}
\begin{description}
\item [{Generators~of~Null~Hypersurface~$\Sigma$:}] \ifS
\else $\gamma\left(\lambda\right)$
\fi family of null geodesics\\
If $\Sigma=\cup\gamma\left(\lambda\right)$,
the $\gamma$'s are called generators of $\Sigma$\\
Their tangent vectors are proportional to the normal 
\begin{align*}
k^{a}=\frac{dx^{a}}{d\lambda}= & h\left(x\right)g^{ab}\partial_{b}f\Rightarrow k^{a}\textrm{ normal}
\end{align*}
A possible choice of $h\left(x\right)$ renders $\gamma$ affinely
parameterised
\begin{align*}
k_{a}k^{a}= & 0 & k^{b}\nabla_{b}k^{a}= & 0\textrm{ (geodesic)}
\end{align*}
\end{description}
$\partial\left[J^{-}\left(\mathscr{I}^{+}\right)\right]$
\ifS
\else future EH: $\gamma$ may end in
past, not in future\fi 

$\partial\left[J^{+}\left(\mathscr{I}^{-}\right)\right]$
\ifS
\else past EH: $\gamma$
may end in future, not in past\fi 
\begin{description}
\item [{EH~in~static~spacetimes~(Schwarzschild-type)}] ~\\
\ifS
\else Stationary metric $\Rightarrow\partial_{t}$
timelike \fi Killing vector at $r\to\infty$\\
\ifS\else  $\partial_{a}t$
\fi normal to $t=cst$ hypersurfaces\\
\ifS\else $\partial_{a}r$ \fi normal
to $r=cst$ hypersurfaceswhich are timelike at
$r\to\infty$ \ifS\else (Minkowski-like)\fi \\
\ifS\else with
norm\fi  $g^{ab}\partial_{a}r\partial_{b}r=g^{rr}$\\
$r=cst$
becomes lightlike at $r=r_{H}$: $g^{rr}\left(r_{H}\right)=0$\\
In Schwarzschild $g^{rr}=1-\frac{2GM}{r}\underset{r_{H}=2GM}{\longrightarrow}0$
\end{description}

\subsection{Importance of E.H.}

\begin{itemize}
\item They often appear to hide singularities (Cosmic Censorship Conjecture)
\item Singularities are inevitable (Singularity theorems)
\end{itemize}
\ifS\else $\Rightarrow$ \fi EH
are ubiquitous!

\subsubsection{Singularity theorems}

\begin{description}
\item [{Singularity~theorems:}] Generic physical initial conditions lead
to collapse into singularities, once reaching a certain point (Hawking-Penrose)
\end{description}
\begin{itemize}
\item Singularities are marked by geodesic incompleteness (they end at finite
affine parameter)
\item The point of no return is marked by the appearance of Trapped Surfaces
\end{itemize}

\paragraph{Trapped surfaces and Trapping Horizons}

\begin{description}
\item [{Trapped~Surfaces}] are $\boxed{\textrm{local}}$ conditions for
the appearance of Event Horizons
\end{description}
In spherical symmetry, foliated
in spheres \ifS\else with \fi dual
null 2+2 decomposition, $k^{a},l^{a}$
lightlike independent directions \ifS
\else with \fi $N_{ab}=g_{ab}+k_{(a}l_{b)}$\\
\ifS
\else 2-expansion along vector $v^{a}$:
$\Theta_{v}=\frac{2}{r}\mathcal{L}_{v}r$ , with $r$ areal radius\fi 
\begin{itemize}
\item If $\Theta_{k}\Theta_{l}<0$ on a sphere\ifS
\else : \fi normal or untrapped\\
\ifS\else Region \fi foliated with normal surfaces \ifS
\else = \fi normal region
\item If $\Theta_{k}\Theta_{l}>0$ on a sphere\ifS
\else : \fi $\boxed{\textrm{Trapped Surface}}$
\begin{itemize}
\item $\Theta_{k},\Theta_{l}<0$\ifS
\else : \fi future trapped or simply trapped
\item $\Theta_{k},\Theta_{l}>0$\ifS\else :
\fi past trapped or anti-trapped
\end{itemize}
\ifS\else Region \fi foliated with trapped (future/past)
surfaces \ifS\else = \fi $\boxed{\textrm{Trapped (future/past) Region}}$
\item If $\Theta_{k}\Theta_{l}=0$ on a sphere\ifS
\else : \fi $\boxed{\textrm{Marginal Surface}}$\\
Without loss of generality, assume $\Theta_{k}=0$
\begin{itemize}
\item $\mathcal{L}_{l}\Theta_{k}<0$\ifS
\else : \fi outer marginal surface
\item $\mathcal{L}_{l}\Theta_{k}>0$\ifS
\else : \fi inner marginal surface
\item $\mathcal{L}_{l}\Theta_{k}=0$\ifS
\else : \fi degenerated marginal surface
\end{itemize}
\end{itemize}
\begin{defn}
\label{def:Trapping-Horizon}Trapping Horizon

A trapping horizon (TH) is a Lorentzian 3-hypersurface foliated with
marginal surfaces
\end{defn}
\ifS
\else In static solutions EH=TH\fi 

\ifS
\else As TH is local while EH is global, \fi TH
is used to characterise Black Holes
\begin{description}
\item [{Examples}] In Minkowski \ifS\else with
\fi a 2-sphere thin shell emitting light outward and inward, governed
by $\Theta_{k},\Theta_{l}$.\\
\ifS\else Outwards \fi yield
$\Theta_{k}>0$\\
\ifS\else Inwards
\fi yield $\Theta_{l}<0$\\
\ifS
\else so Minkowski has $\Theta_{k}\Theta_{l}<0\Rightarrow$
normal region/spacetime\fi 
\item [{\phantom{Examples}}] \ifS
\else In Schwarzschild, inside \fi $r_{H}$,
we get\\
$\Theta_{k}<0$ and $\Theta_{l}<0\Rightarrow$
future trapped region\\
$r=r_{H}$\ifS
\else : $\Theta_{k}=0\Rightarrow$TH\fi 
\end{description}
We return to singularity theorems later. A possible escape to arisal
of singularities could rely on the discovery of quantum gravity\ifS
\else .

Or dark energy (violation of SEC)\fi 

\subsubsection{Cosmic censorship conjecture}

\begin{conjecture}
Naked singularities cannot form in gravitational collapse from generic,
initially nonsingular states in asymptotically flat spacetimes obeying
the DEC.
\end{conjecture}
\ifS\else Conjecture:
\fi yet unproven, for generic initial conditions

\paragraph{Naked singularity:}

\ifS
\else Singularity
that can send light to $\mathscr{I}^{+}$ unimpeeded by EH.\fi 

\begin{figure}
\begin{centering}
\includegraphics[scale=0.5]{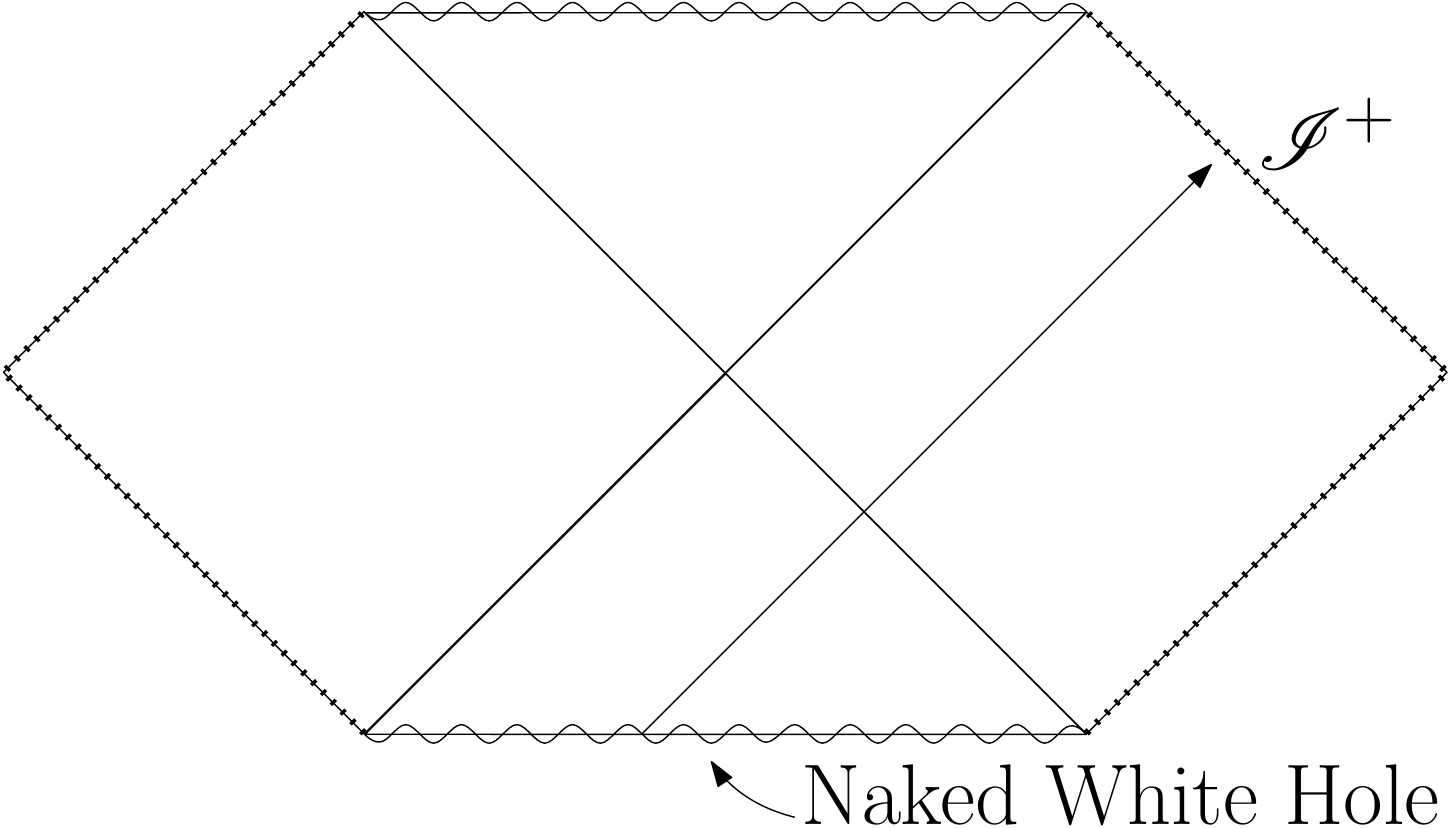}
\par\end{centering}
\caption{\label{fig:Schwarzschild-white-hole}Schwarzschild white hole Carter-Penrose
diagram}

\end{figure}

\begin{example}
~

- Schwarzschild White Hole

see Fig.~\ref{fig:Schwarzschild-white-hole}
\end{example}

\begin{example}
-Charged BH (Reissner-Nordstr\"om)

$Q$: electric charge

$M_{B}$: magnetic charge
\begin{align*}
ds^{2}= & -\Delta dt^{2}+\Delta^{-1}dr^{2}+r^{2}d\Omega^{2}\\
\Delta= & 1-\frac{2GM}{r}+\frac{G\left(Q^{2}+M_{B}^{2}\right)}{r^{2}}
\end{align*}
\ifS
\else Electric
and magnetic fields are radial (see Fig.~\ref{fig:Reissner-Nordstrm-Carter-Penrose-1})\fi 
\begin{align*}
\vec{E}= & E_{r}\vec{e_{r}}=F_{rt}\vec{e_{r}}=\frac{Q}{r^{2}}\vec{e_{r}}\\
\vec{B}= & B_{r}\vec{e_{r}}=\frac{F_{\theta\phi}}{r^{2}sin\theta}\vec{e_{r}}=\frac{M_{B}}{r^{2}}\vec{e_{r}}
\end{align*}
\end{example}
\ifS \else 
\begin{figure}
\hspace*{-3.5cm}\subfloat[case $Q^{2}+M_{B}^{2}>GM^{2}$]{\begin{centering}
\includegraphics[scale=0.4]{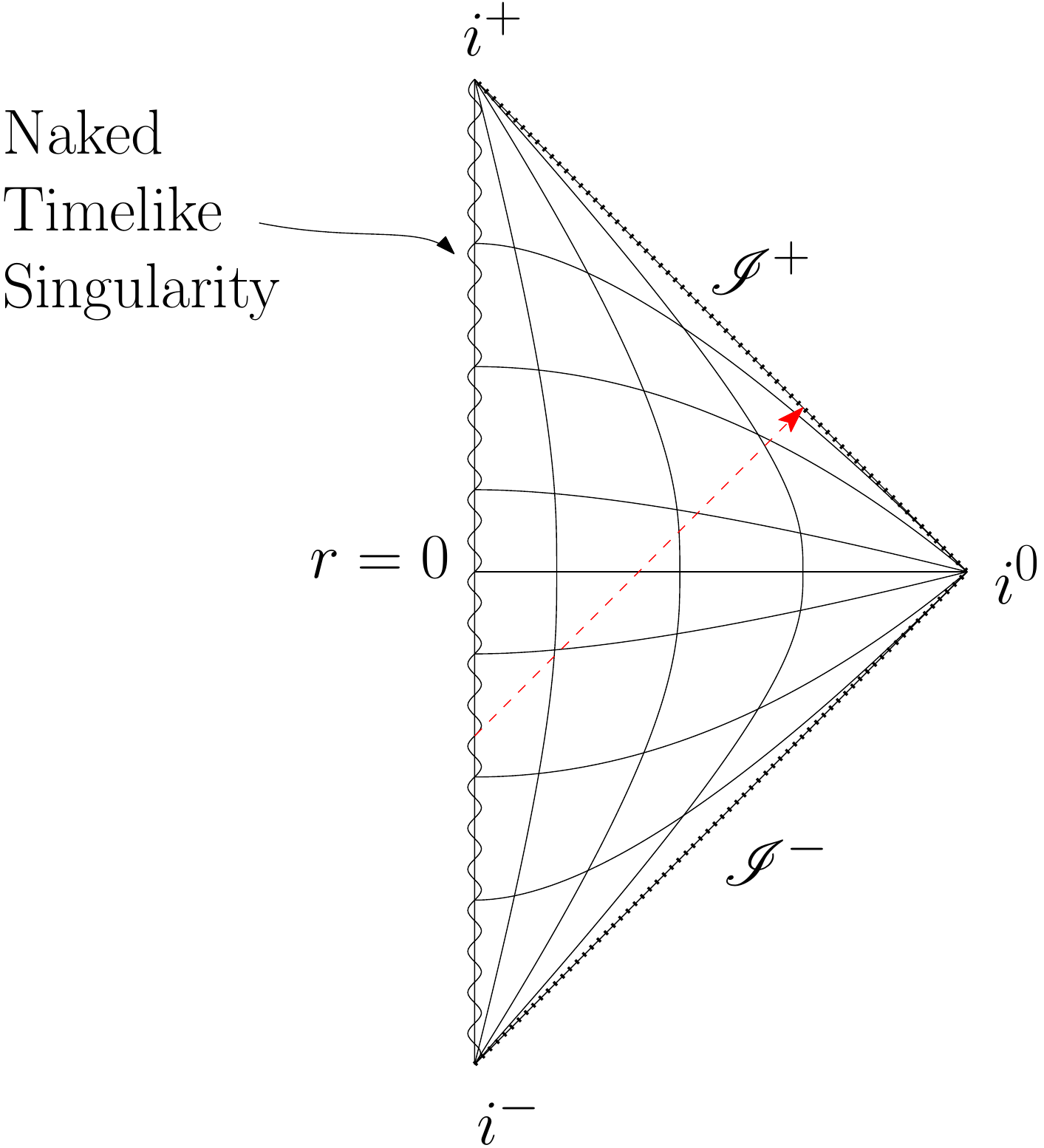}
\par\end{centering}
}\subfloat[case $Q^{2}+M_{B}^{2}=GM^{2}$]{\begin{centering}
\includegraphics[scale=0.4]{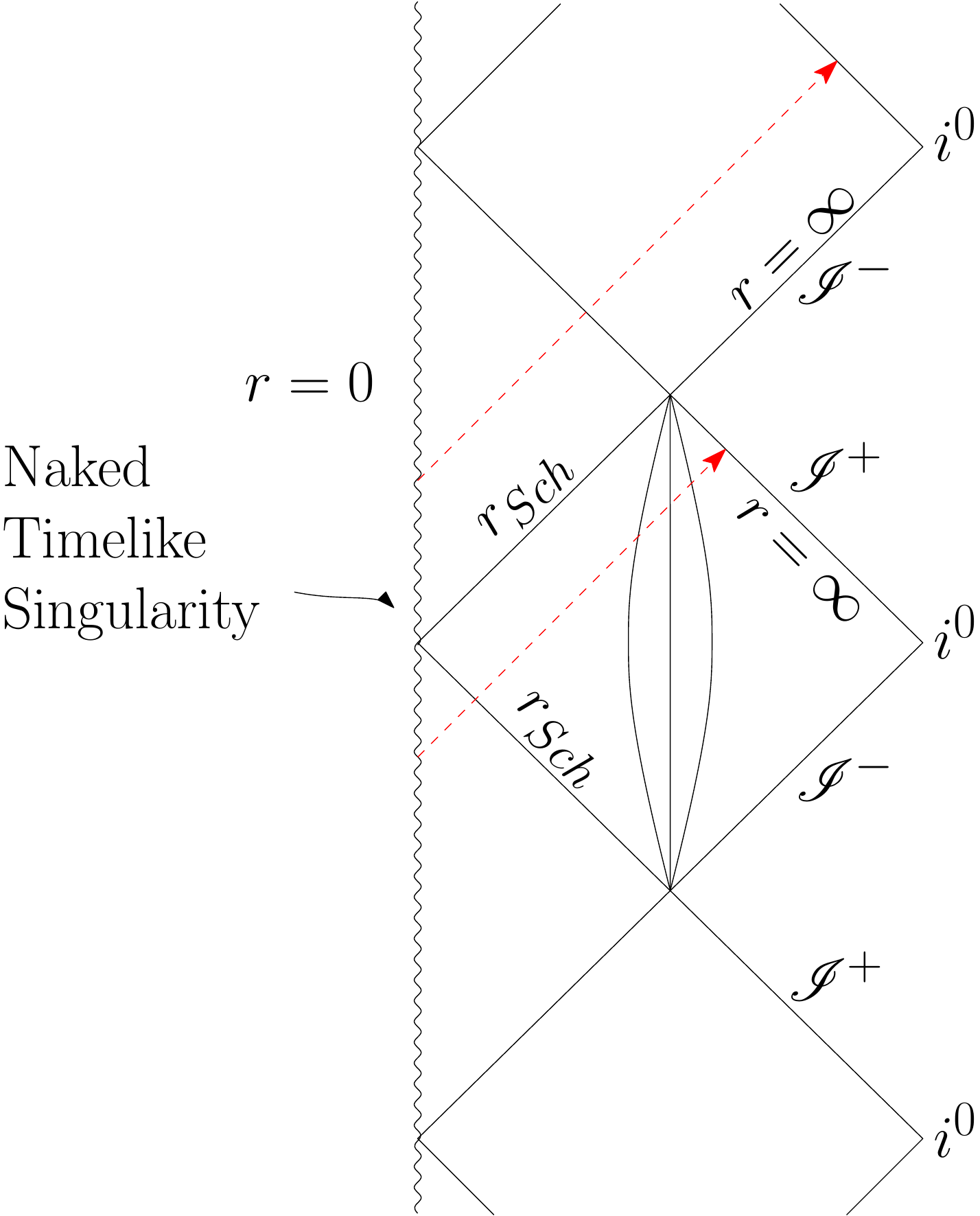}
\par\end{centering}
}\subfloat[case $Q^{2}+M_{B}^{2}<GM^{2}$]{\begin{centering}
\includegraphics[scale=0.4]{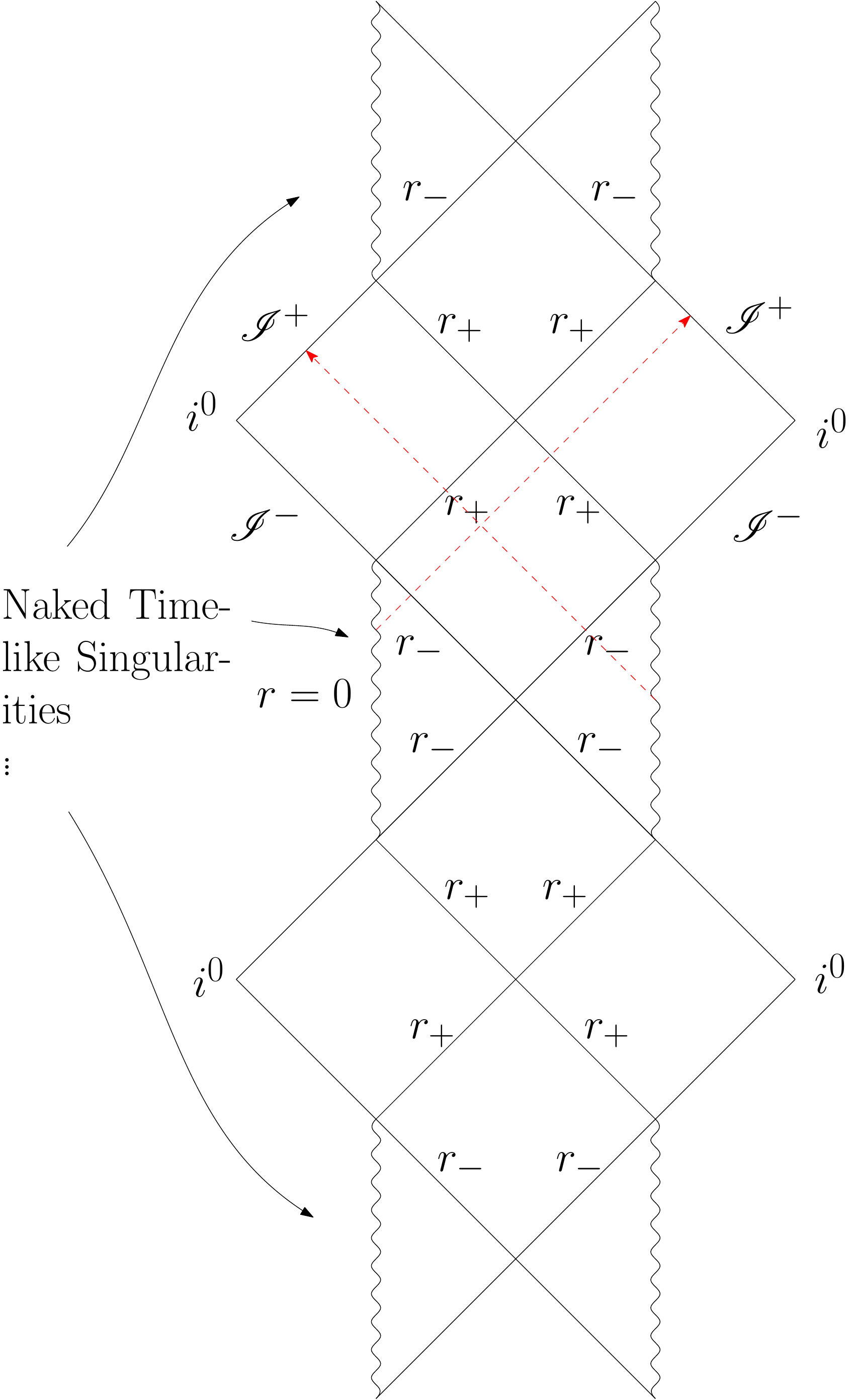}
\par\end{centering}
}

\caption{\label{fig:Reissner-Nordstrm-Carter-Penrose-1}Reissner-Nordstr\"om
Carter-Penrose diagrams}
\end{figure}
\fi 

\subsection{Killing horizons}

\begin{defn}
\label{def:A-Killing-Horizon}A Killing Horizon (K.H.) is a surface
$\Sigma$ where a Killing vector $\xi^{a}$ becomes null.
\end{defn}
\ifS
\else e.g. In Schwarzschild, $\xi^{a}=\partial_{t}^{a}$,
on K.H. $\xi^{a}$ is tangent and normal (null)\fi 

In general K.H., T.H. and E.H. are distinct.

In spacetimes with time-reversal symmetry, K.H. and E.H. are related.
\begin{itemize}
\item Every $\Sigma$ E.H., in stationary, asymptotically flat spacetime,
is a K.H. for some Killing field $\xi^{a}$
\item If spacetime is static, \ifS
\else $\xi^{a}=\partial_{t}^{a}=K^{a}$,
time translation at space infinity\fi 
\item If \ifS\else spacetime stationary,
\fi not static, then it is axisymmetric with rotation Killing field
$R^{a}=\partial_{\phi}^{a}$ and $\zeta^{a}=K^{a}+\Omega_{H}R^{a}$ for $\Omega_{H}$ constant
\end{itemize}
\begin{example}
In Minkowski $ds^{2}=-dt^{2}+dx^{2}+dy^{2}+dz^{2}$

Invariance under boost in $x$ direction\ifS
\else $\Rightarrow$\fi  Killing vector
$\xi=x\partial_{t}+t\partial_{x}$
\begin{align*}
\xi^{a}\xi_{a}= & -x^{2}+t^{2} & \Rightarrow\xi^{a}\textrm{ null on }t= & \pm x:\textrm{ K.H.}
\end{align*}
\ifS
\else  so \fi Minkowski is filled with K.H.s
\end{example}

\subsubsection{Surface gravity of K.H.}

\ifS\else  $\Sigma$ K.H. with \fi Killing
vector $\xi^{a}$ 

$\ifS\else\Leftrightarrow\xi^{a}\fi$ normal to $\Sigma$ and
$\xi^{a}\xi_{a}=0$ on $\Sigma$

Thus\ifS
\else ~$\nabla_{a}\xi^{b}\xi_{b}$
normal to $\Sigma\Rightarrow$ proportional to $\xi^{a}$, normal\fi 

$\exists\kappa,\nabla^{a}\xi^{b}\xi_{b}=-2\kappa\xi^{a}$ and\ifS
\else ~\fi Killing equation $\nabla^{a}\xi^{b}=-\nabla^{b}\xi^{a}$

\ifS \else so \fi $\xi_{b}\nabla^{b}\xi^{a}=\kappa\xi^{a}$\ifS
\else : geodesic eq.
\fi in non-affine parameterisation

\ifS
\else We have $\underset{\textrm{for scalar }\xi^{a}\xi_{a},\xi^{a}\textrm{ K.V. and }\nabla^{a}\textrm{ torsionless}}{\mathcal{L}_{\xi}\left(\nabla^{a}\xi^{b}\xi_{b}\right)=\nabla^{a}\mathcal{L}_{\xi}\left(\xi^{b}\xi_{b}\right)}\underset{\xi^{b}\xi_{b}=0\textrm{ along }\Sigma/\xi^{a}}{=0}=-2\left(\xi^{a}\mathcal{L}_{\xi}\kappa+\kappa\mathcal{L}_{\xi}\xi^{a}\right)$
and \fi $\mathcal{L}_{\xi}\xi^{a}=\xi^{b}\nabla_{b}\xi^{a}-\xi^{b}\nabla_{b}\xi^{a}=0$\ifS
\else 

$\Rightarrow\mathcal{L}_{\xi}\kappa=0$: $\kappa$ constant along
$\Sigma$ (exercise: show $\mathcal{L}_{\xi}\left(\nabla^{a}\xi^{b}\xi_{b}\right)=\nabla^{a}\mathcal{L}_{\xi}\left(\xi^{b}\xi_{b}\right)$\exo )\fi 

$\kappa$\ifS\else : \fi failure of Killing parameter $v$, defined by $\xi^{a}\nabla_{a}v=1$,
to agree with affine parameter $\lambda$ on
null generators of $\Sigma$
\begin{itemize}
\item \ifS\else Define \fi $k^{a}=\frac{dx^{a}}{d\lambda}$
on $\Sigma$ by $k^{a}=e^{-\kappa v}\xi^{a}$ \\
then 
\begin{align*}
k^{b}\nabla_{b}k^{a}= & e^{-2\kappa v}\left[\xi^{b}\nabla_{b}\xi^{a}-\xi^{a}\xi^{b}\nabla_{b}\left(\kappa v\right)\right]\\
= & e^{-2\kappa v}\left[\kappa\xi^{a}-\kappa\xi^{a}\underset{1}{\underbrace{\left(\xi^{b}\nabla_{b}v\right)}}\right]\textrm{ since }\mathcal{L}_{\xi}\kappa=\xi^{b}\nabla_{b}\kappa=0\\
= & 0
\end{align*}
so $k$ is
the affinely parameterised null geodesic tangent, generators of $\Sigma$,
and $k^{a}=\frac{dx^{a}}{d\lambda}$, $\xi^{a}=\frac{dx^{a}}{dv}=\frac{dx^{a}}{d\lambda}\frac{d\lambda}{dv}$\\
$\Rightarrow\frac{d\lambda}{dv}\propto e^{\kappa v}\Rightarrow\kappa\ne0,\lambda\propto e^{\kappa v}$
\end{itemize}
Since \ifS\else $\xi^{a}$ \fi hypersurface
$\bot$ at $\Sigma$, we can use Frobenius Theorem on $\Sigma$
\begin{align*}
\xi_{[a}\nabla_{b}\xi_{c]}= & 0\\
= & \frac{1}{3}\left(\xi_{a}\nabla_{[b}\xi_{c]}+\xi_{b}\nabla_{[c}\xi_{a]}+\xi_{c}\nabla_{[a}\xi_{b]}\right)
\end{align*}
\ifS
\else and Killing Eq.
\begin{align*}
\Rightarrow & \left\{ \begin{array}{rl}
\nabla_{[a}\xi_{b]}= & \nabla_{a}\xi_{b}\\
\nabla_{a}\xi_{b}= & -\nabla_{b}\xi_{a}
\end{array}\right.
\end{align*}
\fi 
\begin{align*}
\Leftrightarrow\xi_{c}\nabla_{a}\xi_{b}= & -\xi_{a}\nabla_{b}\xi_{c}-\xi_{b}\overset{-\nabla_{a}\xi_{c}}{\overbrace{\nabla_{c}\xi_{a}}}\\
= & \left(\xi_{b}\nabla_{a}-\xi_{a}\nabla_{b}\right)\xi_{c}\\
= & -2\xi_{[a}\nabla_{b]}\xi_{c}
\end{align*}
Contracting with $\nabla^{[a}\xi^{b]}=\nabla^{a}\xi^{b}$
\begin{align*}
\xi_{c}\nabla^{a}\xi^{b}\nabla_{a}\xi_{b}= & -2\left(\nabla^{a}\xi^{b}\right)\xi_{a}\nabla_{b}\xi_{c}\\
= & -2\underset{\textrm{geodesic}}{\underbrace{\left(\xi_{a}\nabla^{a}\xi^{b}\right)}}\left(\nabla_{b}\xi_{c}\right)\\
= & -2\kappa\overbrace{\xi^{b}\nabla_{b}\xi_{c}}\\
= & -2\kappa^{2}\xi_{c}\\
\Rightarrow\kappa^{2}= & -\frac{1}{2}\left(\nabla^{a}\xi^{b}\right)\left(\nabla_{a}\xi_{b}\right)\textrm{ Surface Gravity}
\end{align*}
Contracting the Frobenius tensor we
get
\begin{align*}
f=3\xi^{[a}\nabla^{b}\xi^{c]}\xi_{[a}\nabla_{b}\xi_{c]}= & \xi^{[a}\nabla^{b}\xi^{c]}\left(\xi_{a}\nabla_{[b}\xi_{c]}+\xi_{b}\nabla_{[c}\xi_{a]}+\xi_{c}\nabla_{[a}\xi_{b]}\right)\\
= & \frac{1}{3}\left[3\xi^{a}\xi_{a}\left(\nabla^{b}\xi^{c}\right)\left(\nabla_{b}\xi_{c}\right)+\left(\xi^{a}\nabla^{b}\xi^{c}\right)\xi_{b}\overset{-\nabla_{a}\xi_{c}}{\overbrace{\nabla_{c}\xi_{a}}}+\xi^{a}\overset{-\nabla^{c}\xi^{b}}{\overbrace{\nabla^{b}\xi^{c}}}\left(\xi_{c}\nabla_{a}\xi_{b}\right)\right.\\
 & +\xi^{b}\overset{-\nabla^{a}\xi^{c}}{\overbrace{\nabla^{c}\xi^{a}}}\left(\xi_{a}\nabla_{b}\xi_{c}\right)+\left(\xi^{b}\nabla^{c}\xi^{a}\right)\xi_{c}\overset{-\nabla_{b}\xi_{a}}{\overbrace{\nabla_{a}\xi_{b}}}\\
 & \left.\left(\xi^{c}\nabla^{a}\xi^{b}\right)\xi_{a}\overset{-\nabla_{c}\xi_{b}}{\overbrace{\nabla_{b}\xi_{c}}}+\xi^{c}\overset{-\nabla^{b}\xi^{a}}{\overbrace{\nabla^{a}\xi^{b}}}\left(\xi_{b}\nabla_{c}\xi_{a}\right)\right]\\
= & \xi^{a}\xi_{a}\left(\nabla^{b}\xi^{c}\right)\left(\nabla_{b}\xi_{c}\right)-2\left(\xi^{a}\nabla_{a}\xi_{c}\right)\left(\xi_{b}\nabla^{b}\xi^{c}\right)
\end{align*}
and since \ifS\else it is using
\fi Frobenius tensor, $f\underset{\Sigma}{\longrightarrow}0$

Since \ifS
\else $g=\xi^{a}\xi_{a}\underset{\Sigma}{\longrightarrow}0$
and $\nabla_{b}\left(\xi^{a}\xi_{a}\right)=-2\kappa\xi_{b}\ne0$ at
$\Sigma$ and \fi 
\begin{align*}
\nabla f= & 3\nabla\left(Frob^{abc}Frob_{abc}\right)\\
= & 6Frob^{abc}\nabla Frob_{abc}=0\textrm{ at }\Sigma
\end{align*}
we can apply l'Hospital rule to get
\begin{align*}
\lim_{\Sigma}\frac{f}{g}= & \lim_{\Sigma}\frac{\nabla f}{\nabla g}\\
= & 0\\
= & \lim_{\Sigma}\left[\left(\nabla^{b}\xi^{c}\right)\left(\nabla_{b}\xi_{c}\right)\right]-2\lim_{\Sigma}\left[\nicefrac{\left(\xi^{a}\nabla_{a}\xi_{c}\right)\left(\xi_{b}\nabla^{b}\xi^{c}\right)}{\xi^{d}\xi_{d}}\right]\\
= & -2\kappa^{2}-2\lim_{\Sigma}\left[\frac{\left(\xi^{a}\nabla_{a}\xi_{c}\right)\left(\xi_{b}\nabla^{b}\xi^{c}\right)}{\xi^{d}\xi_{d}}\right]\\
\Leftrightarrow\kappa^{2}= & \lim_{\Sigma}\left[-\frac{\left(\xi^{a}\nabla_{a}\xi_{c}\right)\left(\xi_{b}\nabla^{b}\xi^{c}\right)}{\xi^{d}\xi_{d}}\right]
\end{align*}
Away from $\Sigma$, \ifS\else $\xi^{a}$
\fi timelike, and we can define
\begin{align*}
\frac{\xi^{a}}{\sqrt{-\xi^{b}\xi_{b}}}= & u^{a}=\frac{dx^{a}}{d\tau}
\end{align*}
\ifS \else $\xi$ proportional to time orbit velocity\fi 

Thus\ifS
\else acceleration
\begin{align*}
a^{a}=\frac{d^{2}x^{a}}{d\tau^{2}}=\frac{du^{a}}{d\tau}= & u^{b}\nabla_{b}u^{a}\\
= & \frac{\xi^{b}\nabla_{b}\xi^{a}}{\left(-\xi^{c}\xi_{c}\right)}
\end{align*}
\fi and \ifS
\else thus\fi 
\begin{align*}
\kappa=\lim_{\Sigma}\sqrt{\frac{\left(\xi^{a}\nabla_{a}\xi_{c}\right)\left(\xi_{b}\nabla^{b}\xi^{c}\right)}{-\xi^{d}\xi_{d}}}= & \lim_{\Sigma}\sqrt{a_{c}a^{c}\left(-\xi^{c}\xi_{c}\right)}\\
= & \lim_{\Sigma}\left(aV\right)
\end{align*}
defining
\begin{align*}
a= & \sqrt{a_{c}a^{c}}\\
V= & \sqrt{-\xi^{c}\xi_{c}}
\end{align*}
Interpretation is clear in static, asymptotically flat spacetimes,
where \ifS
\else $\xi=\partial_{t}$
and normalised $\xi^{c}\xi_{c}\left(r\to\infty\right)=-1$

Static observer has $u^{a}u_{a}=-1$ so $\xi^{a}=V\left(x\right)u^{a}$
and \fi we regain $V=\sqrt{-\xi^{c}\xi_{c}}$~going
from 0 at $\Sigma$ to 1 at $r\to\infty$\ifS
\else 

Then acceleration, in general, is\fi 
\begin{align*}
a^{a}= & u^{b}\nabla_{b}u^{a}\\
= & u^{b}\nabla_{b}\left(V^{-1}\xi^{a}\right)\\
= & V^{-1}u^{b}\underset{-\nabla^{a}\xi_{b}\textrm{ Killing Eqs}}{\underbrace{\nabla_{b}\xi^{a}}}+u^{b}\xi^{a}\nabla_{b}V^{-1}\\
= & -V^{-2}\nabla^{a}\left(\frac{\xi^{b}\xi_{b}}{2}\right)-V^{-1}u^{b}u^{a}\nabla_{b}V\\
= & V^{-2}\nabla^{a}\left(\frac{V^{2}}{2}\right)-u^{b}u^{a}\nabla_{b}\ln V\\
= & V^{-1}\nabla^{a}V-u^{b}u^{a}\nabla_{b}\ln V\\
= & \nabla^{a}\ln V-u^{b}u^{a}\nabla_{b}\ln V
\end{align*}
\ifS
\else but\fi 
\begin{align*}
a^{a}u_{a}= & \frac{du^{a}}{d\tau}u_{a}=\frac{du^{a}u_{a}}{d\tau}=0\\
= & u_{a}\nabla^{a}\ln V+u^{b}\nabla_{b}\ln V\\
= & 2u_{a}\nabla^{a}\ln V\\
\Rightarrow a^{a}= & \nabla^{a}\ln V
\end{align*}
\ifS
\else Thus the force on a mass $m$ is \fi 
\begin{align*}
F= & ma=m\sqrt{a_{c}a^{c}}=V^{-1}m\sqrt{\nabla^{c}V\nabla_{c}V}
\end{align*}
In the case \ifS
\else $\xi=\partial_{t}$
(static spacetime), energy of a photon with \fi $p^{a}$ is
\begin{align*}
E= & -p^{a}\xi_{a}
\end{align*}
while its frequency, measured by an observer $u^{a}$
gives
\begin{align*}
\omega= & -p^{a}u_{a}\Rightarrow\omega=\frac{E}{V}
\end{align*}
so photons emitted \ifS
\else at $\lambda_{1}=\frac{2\pi}{\omega_{1}}$
are received at $\lambda_{2}=\frac{V_{2}}{V_{1}}\lambda_{1}$ thus
at $\infty$: $\lambda_{\infty}=\frac{\lambda_{1}}{V_{1}}$; $V$
is redshift seen \fi at $\infty$\ifS
\else 

So\fi 
\begin{align*}
VF= & m\sqrt{\nabla^{c}V\nabla_{c}V}=maV
\end{align*}
\ifS\else the force as measured at $\infty$

So \fi acceleration at $\Sigma$ as measured at $\infty$ is
\begin{align*}
\lim_{\Sigma}\left(aV\right)= & \kappa
\end{align*}
In the case \ifS
\else only stationary,
$K=\partial_{t}$ is null at ergosurface but $\xi=K+\Omega_{H}\partial_{\theta}$
(rotating BH) is the Killing vector and $\xi^{a}\xi_{a}=0$ on E.H.\fi 

If we use \ifS\else Schwarzschild\fi 
\begin{align*}
ds^{2}= & -\left(1-\frac{2GM}{r}\right)dt^{2}+\frac{dr^{2}}{1-\frac{2GM}{r}}+r^{2}d\Omega^{2}
\end{align*}
\ifS
\else $\xi^{a}=K^{a}=\left(1,\vec{0}\right)$
and static \fi velocity is defined by
\begin{align*}
g_{ab}u^{a}u^{b}= & -1=g_{ab}\frac{\xi^{a}\xi^{b}}{V^{2}}\Rightarrow-\frac{\left(1-\frac{2GM}{r}\right)}{V^{2}}=-1\\
\Leftrightarrow V= & \sqrt{1-\frac{2GM}{r}}\\
\Rightarrow u^{a}= & \left(\left(1-\frac{2GM}{r}\right)^{-\frac{1}{2}},\vec{0}\right)
\end{align*}
Thus the acceleration
reads
\begin{align*}
a^{a}=\nabla^{a}\ln V= & \frac{2}{2}\frac{GM}{r^{2}}\frac{\nabla^{a}r}{1-\frac{2GM}{r}}=\frac{GM\nabla^{a}r}{r^{2}\left(1-\frac{2GM}{r}\right)}
\end{align*}
\ifS
\else but
$\nabla^{a}r=g^{ab}\delta_{b}^{r}$\fi 
\begin{align*}
\Rightarrow a= & \frac{GM}{r^{2}\left(1-\frac{2GM}{r}\right)}\sqrt{g^{ab}\delta_{a}^{r}\delta_{b}^{r}}=\frac{GM}{r^{2}\left(1-\frac{2GM}{r}\right)}\sqrt{1-\frac{2GM}{r}}=\frac{GM}{r^{2}\sqrt{1-\frac{2GM}{r}}}=\frac{GM}{Vr^{2}}\ifS\else\fi\\
\Rightarrow\kappa= & \lim_{\Sigma}\left(aV\right)=\lim_{r\to2GM}\left(aV\right)=\frac{GM}{\left(2GM\right)^{2}}=\frac{1}{4GM}
\end{align*}
\ifS
\else $\kappa\searrow$ with $M$ as $r_{H}\nearrow$\fi 

\section{Singularity theorems}

Singularities, under generic conditions, are inevitable in General
Relativity, as given by the Big Bang types and Black Hole types theorems.

Here we will see 2 Big Bang types and 2 B.H. types theorems with various
strengthes of hypotheses. We will sketch some proofs and start with
definitions and lemmas.

\subsection{First Big Bang Singularity theorem}

\subsubsection{Definitions}

\begin{defn}
Causal curve

$\gamma\left(\tau\right)$ causal curve $\Leftrightarrow\forall\tau,\dot{\gamma}^{2}\le0$
\end{defn}
\noindent 
\begin{defn}
\label{def:Causal-future-(recallDef.,}Causal future\ifS
\else ~(recall)\fi 
\end{defn}
\noindent 
\begin{align*}
J^{+}\left(p\right)= & \left\{ q\in M|\exists\lambda,\forall\tau,\dot{\lambda}^{2}\le0,\lambda\left(0\right)=p,\lambda\left(1\right)=q\right\} 
\end{align*}
(related to set of future directed causal curves from
$p$)
\begin{defn}
\label{def:Chronological-future}Chronological future
\end{defn}
\begin{align*}
I^{+}\left(p\right)= & \left\{ q\in M|\exists\lambda,\forall\tau,\dot{\lambda}^{2}<0,\lambda\left(0\right)=p,\lambda\left(1\right)=q\right\} 
\end{align*}
(related to set of future directed timelike curves from
$p$\ifS
\else )

Thus $J^{+}\left(p\right)-I^{+}\left(p\right)=\left\{ q\in M|\exists\gamma,\textrm{ null geodesic },\lambda\left(0\right)=p,\lambda\left(1\right)=q\right\} $:
future light cone of $p$\fi 

\noindent \ifS
\else Samely for a set
\fi $S\subset M$
\begin{defn}
\label{def:Causal-or-chronological}Causal or chronological futures
of a set $S$
\begin{align*}
J^{+}\left(S\right)= & \underset{{\scriptscriptstyle q\in S}}{\cup}J^{+}\left(q\right)& I^{+}\left(S\right)= & \underset{{\scriptscriptstyle q\in S}}{\cup}I^{+}\left(q\right)
\end{align*}
\end{defn}
\begin{defn}
\label{def:Causal-or-chronological-1}Causal or chronological pasts
of an event $p$ or a set $S$
\end{defn}
\begin{align*}
J^{-}\left(p\right)= & \left\{ q\in M|\exists\lambda,\forall\tau,\dot{\lambda}^{2}\le0,\lambda\left(0\right)=q,\lambda\left(1\right)=p\right\} & I^{-}\left(p\right)= & \left\{ q\in M|\exists\lambda,\forall\tau,\dot{\lambda}^{2}<0,\lambda\left(0\right)=q,\lambda\left(1\right)=p\right\} \\
J^{-}\left(S\right)= & \underset{{\scriptscriptstyle q\in S}}{\cup}J^{-}\left(q\right)& I^{-}\left(S\right)= & \underset{{\scriptscriptstyle q\in S}}{\cup}I^{-}\left(q\right)
\end{align*}

\begin{figure}
\begin{centering}
\includegraphics[scale=0.25]{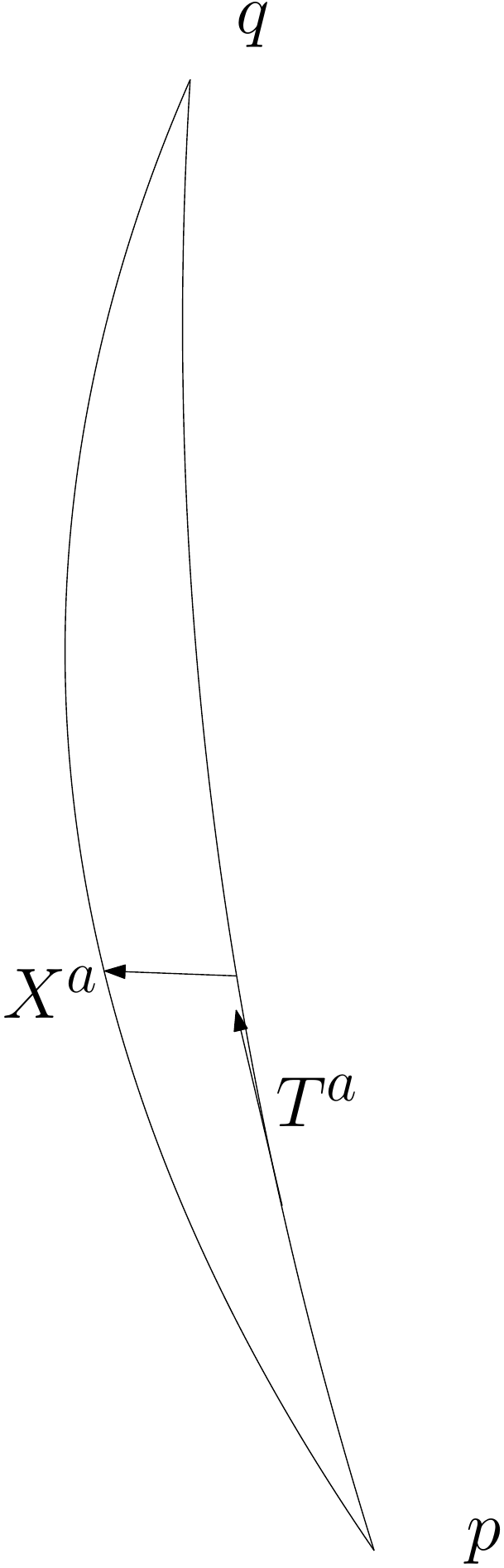}
\par\end{centering}
\caption{\label{fig:Conjugate-points-and}Conjugate points and Jacobi field}

\end{figure}

\begin{defn}
\label{def:Causally-or-chronologically}Causally or chronologically
connected parts of $M$ to $S$
\begin{align*}
J\left(S\right)= & J^{+}\left(S\right)\cup J^{-}\left(S\right)\\
I\left(S\right)= & I^{+}\left(S\right)\cup I^{-}\left(S\right)
\end{align*}
\end{defn}
\noindent 
\begin{defn}
Achronal Set

S achronal $\Leftrightarrow I^{+}\left(S\right)\cap S=\emptyset$:
no point in the future of another point
\end{defn}
\ifS
\else Generated \fi by spacelike or null curves
\begin{defn}
\label{def:Cauchy-surface}Cauchy surface

A closed achronal surface $\Sigma$ with $J^{+}\left(\Sigma\right)\cup J^{-}\left(\Sigma\right)=M$
is called a Cauchy surface
\end{defn}
All $M$ events can be causally connected
to $\Sigma$
\begin{defn}
\label{def:Globally-hyperbolic-spacetime}Globally hyperbolic spacetime

$\left(M,g_{ab}\right)$ globally hyperbolic $\Leftrightarrow\exists\Sigma\subset M$,
$\Sigma$ Cauchy
\end{defn}
\noindent \ifS
\else 
\begin{defn}
Conjugate points

$\gamma\subset M$, geodesic, $\dot{\gamma}=T$,

$X^{a}$ solution of geodesic deviation equation
\begin{align*}
T^{c}\nabla_{c}\left(T^{b}\nabla_{b}X^{a}\right)= & R_{\:bcd}^{a}T^{b}T^{c}X^{d}
\end{align*}
is called a Jacobi field
\begin{align*}
p,q\in M,\left(p,q\right)\textrm{ conjugates }\Leftrightarrow & \exists X^{a}\textrm{ Jacobi field},X^{a}\ne0,X^{a}\left(p\right)=X^{a}\left(q\right)=0
\end{align*}
\end{defn}
See Fig.~\ref{fig:Conjugate-points-and}\fi 

\begin{figure}
\begin{centering}
\includegraphics[scale=0.4]{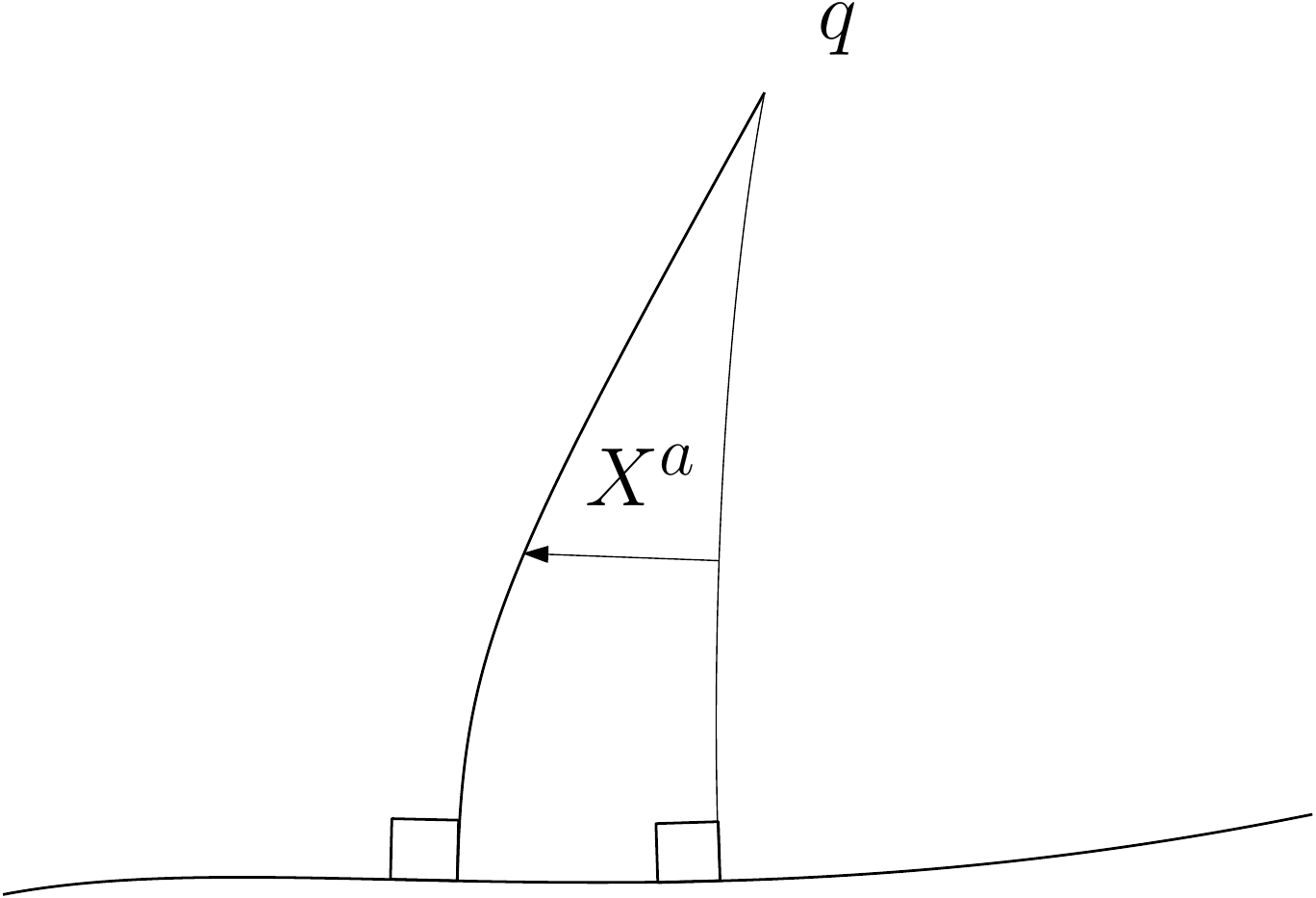}
\par\end{centering}
\caption{\label{fig:Conjugate-point-to}Conjugate point to a surface and Jacobi
field}
\end{figure}

\begin{defn}
\label{def:Point-conjugate-to-S}Point conjugate to a surface

$p\in M,\Sigma\subset M,p\in\gamma$ geodesic, $\gamma\bot\Sigma$
(i.e. $T^{a}\propto g^{ab}\partial_{b}\Sigma$\ifS
\else )\fi 

$p$ conjugate to $\Sigma\Leftrightarrow\exists X^{a}$ Jacobi field
of%
\begin{minipage}[t]{0.5\columnwidth}%
 $\gamma,X^{a}\left(\Sigma\right)\ne0,X^{a}\bot g^{ab}\partial_{b}\Sigma,X^{a}\left(p\right)=0$%
\end{minipage}
\end{defn}
\ifS
\else See Fig.~\ref{fig:Conjugate-point-to}\fi 

\begin{defn}
Normal neighbourhood

$U\subset M$, normal $\Leftrightarrow\forall q,r\in U,r\in J^{+}\left(q\right),\exists!\gamma$
geodesic, $\dot{\gamma}^{2}\le0,{\gamma\subset U,}{\gamma\left(0\right)=q,}\,{\gamma\left(1\right)=r}$
\end{defn}
\noindent 
\begin{defn}
Strongly causal spacetime

\begin{tabular}{c>{\raggedright}p{0.8\columnwidth}}
$\left(M,g_{ab}\right)$ strongly causal $\Leftrightarrow\forall p\in M,$ & $\forall O\left(p\right)\subset M$ (neighbourhood)\tabularnewline
 & $\exists V\left(p\right)\subset O\left(p\right),\forall\gamma$ causal
curve, $\exists!\gamma_{1},\gamma\cap V=\gamma_{1}$ \\
~\hspace*{\fill}($\gamma$ only intersects $V$ once)\tabularnewline
\end{tabular}
\end{defn}
\ifS
\else ($\Rightarrow$
foliation with causal curve families)\fi 
\begin{defn}
\label{def:Spaces-of-causal}Spaces of causal curves

$C\left(p,q\right)$, from $p$ to $q$, or $C\left(\Sigma,q\right)$
from $\Sigma$ to $q$, are defined as the sets of continuous, future
directed causal curves from $\left|\begin{array}{c}
p\\
\Sigma
\end{array}\right.$to $q$.
\end{defn}
\noindent 
\begin{description}
\item [{Property:}] $M$ globally hyperbolic $\Rightarrow C\left(\;,q\right)$
compact\ifS\else ~(``closed'' and ``bounded'')\fi 
\end{description}
\noindent 
\begin{defn}
Length of causal curves

$\lambda\left(t\right)\,C^{0}$, causal curve, $\dot{\lambda}=T^{a}\partial_{a}=\partial_{t}^{a}\partial_{a},\tau\left[\lambda\right]=\int_{p}^{q}\left(-T^{a}T_{a}\right)^{\frac{1}{2}}dt$
\end{defn}
\noindent 
\begin{defn}
\label{def:Upper-semi-continuity-of}Upper semi-continuity of $\tau$

$\tau$ upper semi-continuous on $C\left(p,q\right)$

\begin{tabular}{cl}
$\Leftrightarrow\forall\lambda\in C\left(p,q\right),\forall\epsilon>0$ & $\exists O\left(\lambda\right)\subset C\left(p,q\right)$ (open neighbourhood)\tabularnewline
 & $\forall\lambda^{\prime}\in O\left(\lambda\right),\tau\left[\lambda^{\prime}\right]\le\tau\left[\lambda\right]+\epsilon$\tabularnewline
\end{tabular}
\end{defn}
\noindent 
\begin{description}
\item [{Illustration}] $\lambda^{\prime}$
null by parts (segments): 
\begin{align*}
\tau\left[\lambda^{\prime}\right]= & 0\\
\tau\left[\lambda\right]= & \tau>0
\end{align*}
\ifS\else See
Fig.~\ref{fig:Illustration-of-upper}\fi 
\begin{figure}
\begin{centering}
\includegraphics[scale=0.4]{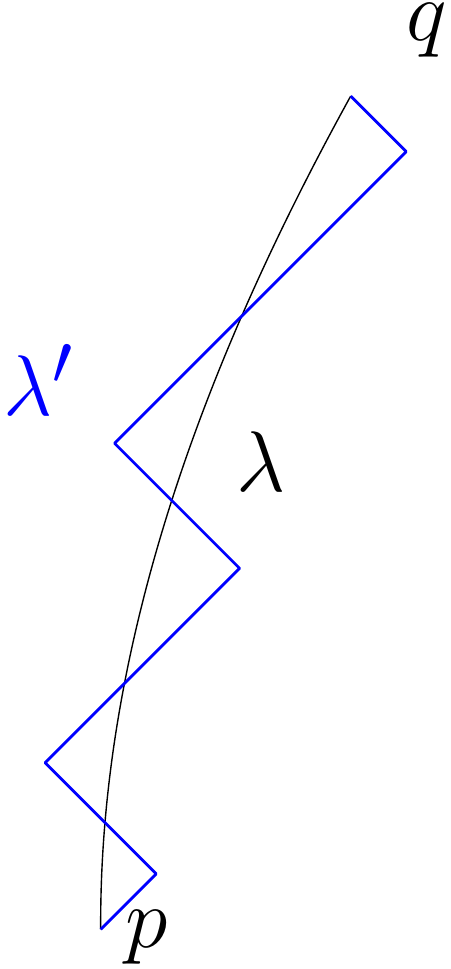}
\par\end{centering}
\caption{\label{fig:Illustration-of-upper}Illustration of upper semicontinuity}
\end{figure}
\end{description}

\subsubsection{Lemmas}

\begin{lem}
\label{lem:1Upper-semicontinuity}Upper semicontinuity

$\left(M,g_{ab}\right)$ strongly causal, $p,q\in M,q\in J^{+}\left(p\right)\Rightarrow\tau$
upper semi-continuous on $C\left(p,q\right)$
\end{lem}
\begin{lem}
\label{lem:2Property-geodesic-and}Property geodesic and $\tau$

$\gamma$ geodesic $\Leftrightarrow\gamma$ extremises $\tau$
\end{lem}
\noindent 
\begin{lem}
\label{lem:3Existence-of-conjugate}Existence of conjugate to $\Sigma$
in finite $\tau$

$\left(M,g_{ab}\right)$ satisfies S.E.C., $\Sigma$ spacelike hypersurface
with $K=\Theta<0$ at $q\in\Sigma$ ($K$ trace of extrinsic curvature
for future normal geodesics to $\Sigma$

$\Rightarrow\exists p$ conjugate to $\Sigma$ along $\gamma\bot\Sigma,q\in\gamma$
within $\tau\left[\gamma_{qp}\right]\le\frac{3}{\left|K\right|}$
\end{lem}
\begin{proof}
We saw from Raychaudhuri equation for S.E.C.,
hypersurface $\bot$, geodesic congruences (see Sec.~\ref{subsec:Energy-condition-and})
that they converge in finite proper time $\tau\le\frac{3}{\left(-\Theta_{0}\right)}=\frac{3}{\left|K\right|}$
\ifS
\else and here it is the condition
on $q\in\Sigma$, note \fi that $X^{a}\left(p\right)=0\Leftrightarrow\Theta\left(p\right)=-\infty$

Hence the condition being valid for $q\in\Sigma$ ($,K=\Theta<0$
with $\gamma\bot\Sigma$)

$\Rightarrow p$ exists.
\end{proof}
\noindent 
\begin{lem}
\label{lem:4Geodesics-with-no}Geodesics with no inner conjugates
maximise $\tau$

$\gamma\in C\left(p,q\right),\gamma$ smooth timelike

$\left\{ \lambda_{\alpha}\left(t\right)\right\} \subset C\left(p,q\right),$
smooth, one parameter family of smooth timelike curves
\begin{align*}
\lambda_{\alpha}\left(a\right)= & p, & \lambda_{\alpha}\left(b\right)= & q, & \lambda_{0}= & \gamma
\end{align*}
$\gamma$ maximises $\tau\left[\lambda_{\alpha}\right]\Leftrightarrow\gamma$
geodesic with no conjugate point between $p$ and $q$
\end{lem}
\begin{proof}
Sketch of proof

\ifS
\else $\lambda_{0}=\gamma,\left(\partial_{t}\right)^{a}=T^{a},\left(\partial_{\alpha}\right)^{a}=X^{a},\mathcal{L}_{T}X^{a}=T^{b}\nabla_{b}X^{a}-X^{b}\nabla_{b}T^{a}=0$
by commutation\fi 

$\forall\alpha,\left.\begin{array}{rl}
q= & \lambda_{\alpha}\left(b\right)\\
p= & \lambda_{\alpha}\left(a\right)
\end{array}\right\} \Rightarrow X^{a}\left(a\right)=X^{a}\left(b\right)=0$; \ifS
\else define \fi $\tau\left(\alpha\right)=\tau\left[\lambda_{\alpha}\right]=\int_{a}^{b}f\left(\alpha,t\right)dt=\int_{a}^{b}\left(-T^{a}T_{a}\right)^{\frac{1}{2}}dt$

\ifS
\else Calculate\fi 
\begin{align*}
\frac{d\tau}{d\alpha}=\int_{a}^{b}\partial_{\alpha}fdt= & \int_{a}^{b}X^{a}\nabla_{a}\left(-T^{b}T_{b}\right)^{\frac{1}{2}}dt\\
= & -\int_{a}^{b}\frac{X^{a}T^{b}}{f}\nabla_{a}T_{b}dt\\
= & -\int_{a}^{b}\frac{T^{a}T^{b}}{f}\nabla_{a}X_{b}dt\textrm{ by commutation}\\
= & -\int_{a}^{b}T^{a}\nabla_{a}\left(\frac{T^{b}X_{b}}{f}\right)dt+\int_{a}^{b}T^{a}X_{b}\nabla_{a}\left(\frac{T^{b}}{f}\right)dt\\
= & -\int_{a}^{b}\partial_{t}\left(\frac{T^{b}X_{b}}{f}\right)dt+\int_{a}^{b}T^{a}X_{b}\nabla_{a}\left(\frac{T^{b}}{f}\right)dt\\
= & \left[-\frac{T^{b}X_{b}}{f}\right]_{a}^{b}+\int_{a}^{b}T^{a}X^{b}\nabla_{a}\left(\frac{T_{b}}{f}\right)dt
\end{align*}
The boundary terms vanish as $X^{a}\left(a\right)=X^{a}\left(b\right)=0$
\begin{align*}
\Rightarrow\frac{d\tau}{d\alpha}= & \int_{a}^{b}X^{b}T^{a}\nabla_{a}\left(\frac{T_{b}}{f}\right)dt
\end{align*}
but \ifS
\else $T^{a}\nabla_{a}\left(\frac{T_{b}}{\sqrt{-T^{c}T_{c}}}\right)$
is geodesic equation for arbitrary param.\fi 

So $\left.\frac{d\tau}{d\alpha}\right|_{\begin{array}{c}
\alpha=0\\
(\lambda_{0}=\gamma)
\end{array}}=0\Leftrightarrow\gamma$ geodesic ($\gamma$ geodesic locally maximises $\tau$)\ifS
\else 

Now compute\fi 
\begin{align*}
\frac{d^{2}\tau}{d\alpha^{2}}= & \int_{a}^{b}X^{c}\nabla_{c}\left[X^{b}T^{a}\nabla_{a}\left(\frac{T_{b}}{f}\right)\right]dt\\
= & \int_{a}^{b}\left(X^{c}\nabla_{c}X^{b}\right)T^{a}\nabla_{a}\left(\frac{T_{b}}{f}\right)dt+\int_{a}^{b}X^{b}\left(X^{c}\nabla_{c}T^{a}\right)\nabla_{a}\left(\frac{T_{b}}{f}\right)dt\\
 & +\int_{a}^{b}X^{b}T^{a}X^{c}\nabla_{c}\nabla_{a}\left(\frac{T_{b}}{f}\right)dt
\end{align*}
\ifS
\else At $\alpha=0$, $\lambda_{0}$
\fi geodesic $\left(T^{a}\nabla_{a}\left(\frac{T_{b}}{f}\right)=0\right)$
and using Ricci identity, commutation of $X$ and $T$ and relabeling
dummy indices
\begin{align*}
\left.\frac{d^{2}\tau}{d\alpha^{2}}\right|_{\alpha=0}= & \int_{a}^{b}X^{b}\left(T^{c}\nabla_{c}X^{a}\right)\nabla_{a}\left(\frac{T_{b}}{f}\right)dt+\int_{a}^{b}X^{b}T^{a}X^{c}\nabla_{a}\nabla_{c}\left(\frac{T_{b}}{f}\right)dt\\
 & +\int_{a}^{b}X^{b}T^{a}X^{c}R_{cab}^{\quad\,d}\frac{T_{d}}{f}dt\quad\textrm{Ricci}\\
= & \int_{a}^{b}X^{b}T^{c}\nabla_{c}\left[X^{a}\nabla_{a}\left(\frac{T_{b}}{f}\right)\right]dt+\int_{a}^{b}X^{b}T^{a}X^{c}R_{cab}^{\quad\,d}\frac{T_{d}}{f}dt\quad\textrm{relabeling}
\end{align*}
We can reexpress
\begin{align*}
\left[X^{a}\nabla_{a}\left(\frac{T_{b}}{f}\right)\right]= & \frac{X^{a}}{f}\nabla_{a}T_{b}-\frac{T_{b}}{f^{2}}X^{a}\nabla_{a}f\\
= & \frac{T^{a}}{f}\nabla_{a}X_{b}-\frac{T_{b}}{f^{2}}X^{a}\nabla_{a}\left(-T^{c}T_{c}\right)^{\frac{1}{2}}\quad\textrm{commuting}\\
= & \frac{T^{a}}{f}\nabla_{a}X_{b}+\frac{T_{b}}{f^{2}}X^{a}\frac{T^{c}}{f}\nabla_{a}T_{c}\\
= & \frac{T^{a}}{f}\nabla_{a}X_{b}+\frac{T_{b}}{f^{2}}\frac{T^{c}}{f}\left(T^{a}\nabla_{a}X_{c}\right)\quad\textrm{commuting}\\
= & \frac{T^{a}}{f}\nabla_{a}X_{b}+\frac{T_{b}}{f^{2}}\left[T^{a}\nabla_{a}\left(\frac{T^{c}X_{c}}{f}\right)-X_{c}\underset{\textrm{geodesic at }\alpha=0}{\underbrace{T^{a}\nabla_{a}\left(\frac{T^{c}}{f}\right)}}\right]\\
= & \frac{T^{a}}{f}\nabla_{a}X_{b}+\frac{T_{b}}{f^{2}}T^{a}\nabla_{a}\left(\frac{T^{c}X_{c}}{f}\right)
\end{align*}
choosing in addition $f=1$ along $\lambda_{0}$, geodesic by affine
parameterisation, and $T\bot X$: $T^{c}X_{c}=0$ along $\lambda_{0}$,
we get, relabeling dummy indices
\begin{align*}
\left.\frac{d^{2}\tau}{d\alpha^{2}}\right|_{\alpha=0}= & \int_{a}^{b}X^{a}\left\{ T^{c}\nabla_{c}\left[T^{b}\nabla_{b}X_{a}\right]+R_{cbad}T^{b}T^{d}X^{c}\right\} dt
\end{align*}
The Riemann term can be rewritten
\begin{align*}
R_{cbad}T^{b}T^{d}X^{c}= & R_{adcb}T^{b}T^{d}X^{c}\quad\textrm{Riemann symmetry}\\
\rightarrow & R_{\:bcd}^{a}T^{b}T^{d}X^{c}\quad\textrm{relabeling }b\leftrightarrow d\\
\rightarrow & -R_{\:bdc}^{a}T^{b}T^{d}X^{c}\quad\textrm{Riemann antisymmetry}
\end{align*}
We recognise the operator that
gives $0$ on Jacobi fields: $\mathlarger{\mathlarger{\mathlarger{\mathscr{O}}}}X=0$
\begin{align*}
\Rightarrow\left.\frac{d^{2}\tau}{d\alpha^{2}}\right|_{\alpha=0}= & \int_{a}^{b}X_{a}\mathlarger{\mathlarger{\mathlarger{\mathscr{O}}}}X^{a}dt
\end{align*}
\ifS
\else If $\lambda_{0}=\gamma$
geodesic \fi with conjugate point $r$ between $p$ and $q$\ifS
\else 

$\Rightarrow\exists X_{0}^{a}\ne0,\mathlarger{\mathlarger{\mathlarger{\mathscr{O}}}}X_{0}^{a}=0,X_{0}^{a}\left(p\right)=X_{0}^{a}\left(r\right)=0$\fi ,
taking $X_{0}^{a}=0$ between $r$ and $q$\ifS
\else 

then smoothing of the $r$ point into $\gamma^{\prime}$ yields \fi $\gamma^{\prime}-\gamma:X=X_{0}+\epsilon$
\begin{figure}
\begin{centering}
\includegraphics[scale=0.5]{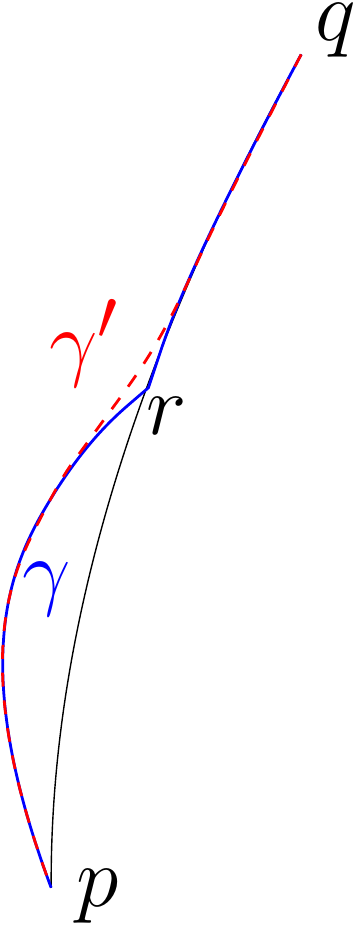}
\par\end{centering}
\caption{\label{fig:Smoothing-from-geodesic}Smoothing from geodesic with inner
conjugate}
\end{figure}
(see Fig.~\ref{fig:Smoothing-from-geodesic})since $\left.\begin{array}{rl}
\dot{\epsilon}^{a}\epsilon_{a}= & 0\\
\ddot{\epsilon}^{a}\dot{\epsilon}_{a}= & 0
\end{array}\right\} $ we take $\ddot{\epsilon}^{a}=\kappa\epsilon^{a},\kappa>0$ when $\epsilon$
grows (before $r$)
\begin{align*}
RTT\epsilon\epsilon\ll1\Rightarrow R_{\:bcd}^{a}T^{b}T^{c}\epsilon^{d}\epsilon_{a}= & 0\Rightarrow R_{\:bcc}^{a}T^{b}T^{c}\epsilon^{d}\propto\dot{\epsilon}^{a}
\end{align*}
since smoothing is infinitesimal, $\epsilon^{a}\epsilon_{a}\gg\dot{\epsilon}^{a}\dot{\epsilon}_{a}$
\ifS
\else for most of the part where $\epsilon\ne0$, and $\mathlarger{\mathlarger{\mathlarger{\mathscr{O}}}}X^{a}=\mathlarger{\mathlarger{\mathlarger{\mathscr{O}}}}X_{0}^{a}+\mathlarger{\mathlarger{\mathlarger{\mathscr{O}}}}\epsilon^{a}\simeq\kappa\epsilon^{a}$\fi 

Thus~to first order $X_{a}\mathlarger{\mathlarger{\mathlarger{\mathscr{O}}}}X^{a}\simeq X_{a}\kappa\epsilon^{a}\simeq\kappa X_{0}^{a}\epsilon_{a}\ne0$
before $r$, \ifS\else thus \fi when $\kappa>0$

Since this \ifS
\else is growing phase for $\epsilon$ then $X_{0},\epsilon$
in same direction\fi , then $X_{0}^{a}\epsilon_{a}>0$ and thus $\left.\frac{d^{2}\tau}{d\alpha^{2}}\right|_{\alpha=0}>0$\ifS
\else ~and so $\tau\left[\gamma^{\prime}\right]>\tau\left[\gamma\right]=\max\tau$:
\fi contradiction!

Conversely, $\gamma$ geodesic without
conjugate point between $p$ and $q$ with $T^{a}$ tangent. Introduce
orthonormal
basis $\bot T^{a}:\left(e_{\mu}^{\:a}\right)$~parallely
propagated along $\gamma$ $\left(T^{b}\nabla_{b}e_{\mu}^{\:a}=0=\frac{de_{\mu}^{\:a}}{d\tau}\right)$,
choosing $X^{a}T_{a}=0$\ifS
\else 

$X^{a}=X^{\mu}e_{\mu}^{\:a}$ verify geodesic deviation\fi 
\begin{align*}
\frac{dX^{a}}{d\tau}= & \frac{dX^{\mu}}{d\tau}e_{\mu}^{\:a}+X^{\mu}\frac{de_{\mu}^{\:a}}{d\tau}=\frac{dX^{\mu}}{d\tau}e_{\mu}^{\:a}\quad\textrm{parallel transport}\\
\Rightarrow\frac{d^{2}X^{a}}{d\tau^{2}}=\frac{d^{2}X^{\mu}}{d\tau^{2}}e_{\mu}^{\:a}= & R_{\:bcc}^{a}T^{b}T^{c}X^{d}\\
= & e_{\mu}^{\:a}R_{\:bc\nu}^{\mu}e_{\:d}^{\nu}T^{b}T^{c}X^{d}\\
= & R_{\:bc\nu}^{\mu}T^{b}T^{c}X^{\nu}e_{\mu}^{\:a}\\
\Leftrightarrow\frac{d^{2}X^{\mu}}{d\tau^{2}}= & R_{\:bc\nu}^{\mu}T^{b}T^{c}X^{\nu}
\end{align*}
\ifS\else since\fi  $X^{\mu}\left(\tau\right)$ depends
linearly on initial data $X^{\mu}\left(0\right)$ and $\frac{dX^{\mu}}{d\tau}\left(0\right)$\ifS
\else ~where at $p$,
\fi  $X^{\mu}\left(0\right)=X^{\mu}\left(p\right)=0$

\ifS\else then
\fi $X^{\mu}\left(\tau\right)=A_{\:\nu}^{\mu}\left(\tau\right)\frac{dX^{\mu}}{d\tau}\left(0\right)$
with $A_{\:\nu}^{\mu}\left(0\right)=0,\frac{dA_{\:\nu}^{\mu}}{d\tau}\left(0\right)=\delta_{\nu}^{\mu}$\ifS
\else 

$q$ conjugate $\Leftrightarrow\exists\frac{dX^{\mu}}{d\tau}\left(0\right)\ne0$
(non-trivial initial data), $X^{\mu}\left(q\right)=0$\fi 

\ifS
\else From \fi $X^{\mu}=A_{\:\nu}^{\mu}\frac{dX^{\mu}}{d\tau}\left(0\right)\Leftrightarrow\det A\left(q\right)=0$\ifS
\else Thus \fi $\nexists r$ conjugate of $p$ before $q\Leftrightarrow\forall r\in\left]p,q\right[,\det A\left(r\right)\ne0$

\ifS
\else Define $Y^{\mu}=\left(A^{-1}\right)_{\,\nu}^{\mu}X^{\nu}\Leftrightarrow X^{\mu}=A_{\:\nu}^{\mu}Y^{\nu}$
(all possible initial conditions considered)

Thus
\begin{align*}
\left.\frac{d^{2}\tau}{d\alpha^{2}}\right|_{\alpha=0}= & \int_{a}^{b}e_{\mu}^{\:a}X^{\mu}\left(\mathlarger{\mathlarger{\mathlarger{\mathscr{O}}}}X\right)_{\nu}e_{\:a}^{\nu}d\tau=\int_{a}^{b}\delta_{\nu}^{\mu}X^{\mu}\left(\mathlarger{\mathlarger{\mathlarger{\mathscr{O}}}}X\right)_{\nu}d\tau=\int_{a}^{b}X^{\mu}\left(\mathlarger{\mathlarger{\mathlarger{\mathscr{O}}}}X\right)_{\mu}d\tau
\end{align*}
\fi and introducing $X^{\mu}=A_{\:\nu}^{\mu}\frac{dX^{\mu}}{d\tau}\left(0\right)$
in the geodesic deviation equation, we get
\begin{align*}
\frac{d^{2}A_{\:\nu}^{\mu}}{d\tau^{2}}\frac{dX^{\mu}}{d\tau}\left(0\right)= & R_{\:bc\beta}^{\mu}T^{b}T^{c}A_{\:\nu}^{\beta}\frac{dX^{\mu}}{d\tau}\left(0\right) & \Leftrightarrow\frac{d^{2}A_{\:\nu}^{\mu}}{d\tau^{2}}= & R_{\:bc\beta}^{\mu}T^{b}T^{c}A_{\:\nu}^{\beta}
\end{align*}
\ifS
\else and\fi 
\begin{align*}
\left(\mathlarger{\mathlarger{\mathlarger{\mathscr{O}}}}X\right)_{\mu}= & \frac{d^{2}X_{\mu}}{d\tau^{2}}-R_{\mu bc\nu}T^{b}T^{c}X^{\nu}\\
= & \left(\frac{d^{2}A_{\mu\nu}}{d\tau^{2}}-R_{\mu bc\beta}A_{\:\nu}^{\beta}\right)Y^{\nu}+2\frac{dA_{\mu\nu}}{d\tau}\frac{dY^{\nu}}{d\tau}+A_{\mu\nu}\frac{d^{2}Y^{\nu}}{d\tau^{2}}
\end{align*}
so
\begin{align*}
\left.\frac{d^{2}\tau}{d\alpha^{2}}\right|_{\alpha=0}= & \int_{a}^{b}A^{\mu\alpha}Y_{\alpha}\left(2\frac{dA_{\mu\beta}}{d\tau}\frac{dY^{\beta}}{d\tau}+A_{\mu\beta}\frac{d^{2}Y^{\beta}}{d\tau^{2}}\right)d\tau
\end{align*}
From \cite{hawking} Proposition 4.5.8, this can be shown negative
definite $\Rightarrow$ any curve away from $\gamma$ has $\tau<\tau\left[\gamma\right]$
\end{proof}
\noindent 
\begin{lem}
\label{lem:5Geodesics-without-inner}Geodesics without inner conjugate
to $\Sigma$ maximise $\tau$

$p\in M,\forall q\in\Sigma$, smooth, spacelike hypersurface, $\gamma\in C\left(q,p\right),\gamma$
smooth, timelike

$\lambda_{\alpha}\left(t\right)$ smooth one parameter family of smooth
timelike curves, $\left\{ \lambda_{\alpha}\left(t\right)\right\} \subset C\left(q,p\right),\lambda_{\alpha}\left(a\right)=q,\lambda_{\alpha}\left(b\right)=p,\lambda_{0}=\gamma$

$\gamma$ maximises $\tau\left[\lambda_{\alpha}\right]\Leftrightarrow\gamma$
geodesic $\bot\Sigma$ with no conjugate points between $\Sigma$
and $p$
\end{lem}
\begin{proof}
\ifS\else Similar \fi to Lemma~\ref{lem:4Geodesics-with-no}
(Geodesics with no inner conjugates maximise $\tau$)
\end{proof}
\noindent 
\begin{lem}
\label{lem:6Maximum-of-}Maximum of $\tau$ on $C\left(\Sigma,p\right)$
is geodesics without inner conjugate to $\Sigma$ 

$\left(M,g_{ab}\right)$ strongly causal, $p\in M,\Sigma$ achronal,
smooth, spacelike hypersurface

$\tau$ defined on $C\left(\Sigma,p\right)$

$\tau$ maximum on $\gamma\in C\left(\Sigma,p\right)\Rightarrow\gamma$
geodesic, $\bot\Sigma$ with no conjugate points between $\Sigma$
and $p$
\end{lem}
\begin{proof}
Sketch of proof

$\gamma\in C\left(\Sigma,p\right)\Rightarrow\gamma$ continuous
\begin{itemize}
\item If $\gamma$ smooth, Lemma \ref{lem:5Geodesics-without-inner}
(Geodesics without inner conjugate to $\Sigma$ maximise $\tau$)
gives $\tau$ maximum on $\gamma\Leftrightarrow\gamma$ geodesic $\bot\Sigma$
without conjugate
\item If $\gamma$ not smooth: we are going to prove
that any piecewise curve cannot have larger $\tau$ than a single
smooth geodesic.
\end{itemize}
$\forall U$ convex normal neighbourhood

$r,s,t\in U,s\in J^{+}\left(r\right),t\in J^{+}\left(r\right)\cap J^{+}\left(s\right),\gamma_{rt},\gamma_{rs},\gamma_{st}$
geodesics with $\gamma\cap C\left(r,t\right)=\gamma_{rt}$ 
\begin{figure}
\begin{centering}
\includegraphics[scale=0.5]{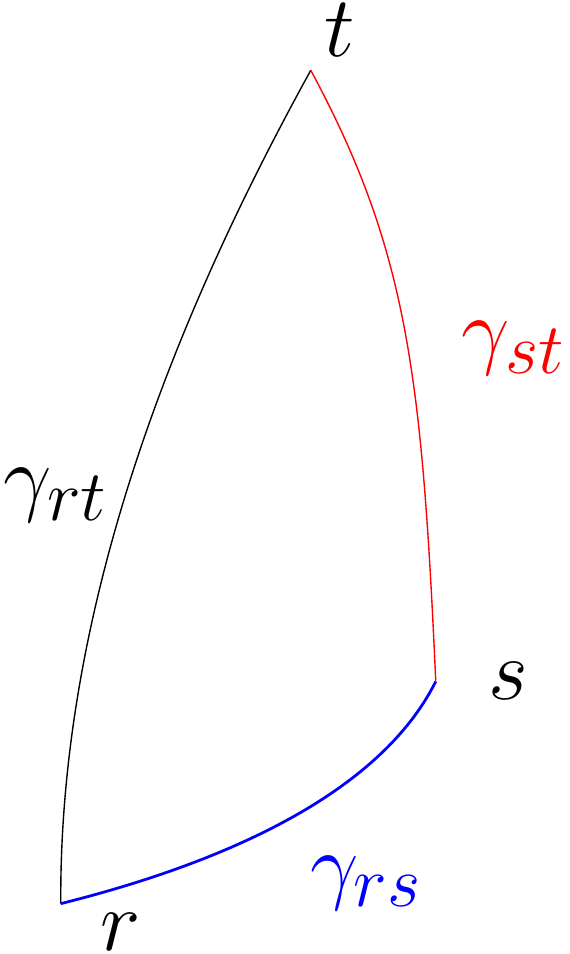}
\par\end{centering}
\caption{\label{fig:Piecewise-vs-Geodesic}Piecewise vs Geodesic}
\end{figure}
(see Fig.~\ref{fig:Piecewise-vs-Geodesic})

\begin{align*}
\tau\left[\gamma_{rt}\right]> & \tau\left[\gamma_{rs}\right]+\tau\left[\gamma_{st}\right]
\end{align*}
\begin{proof}
that $\tau\left[\gamma_{0}\right]>\tau\left[\gamma_{1}\right]+\tau\left[\gamma_{2}\right]$
\begin{figure}
\begin{centering}
\subfloat[\label{fig:Tangent-vectors-and}Tangent vectors and induced parameterisation]{\begin{centering}
\includegraphics[scale=0.5]{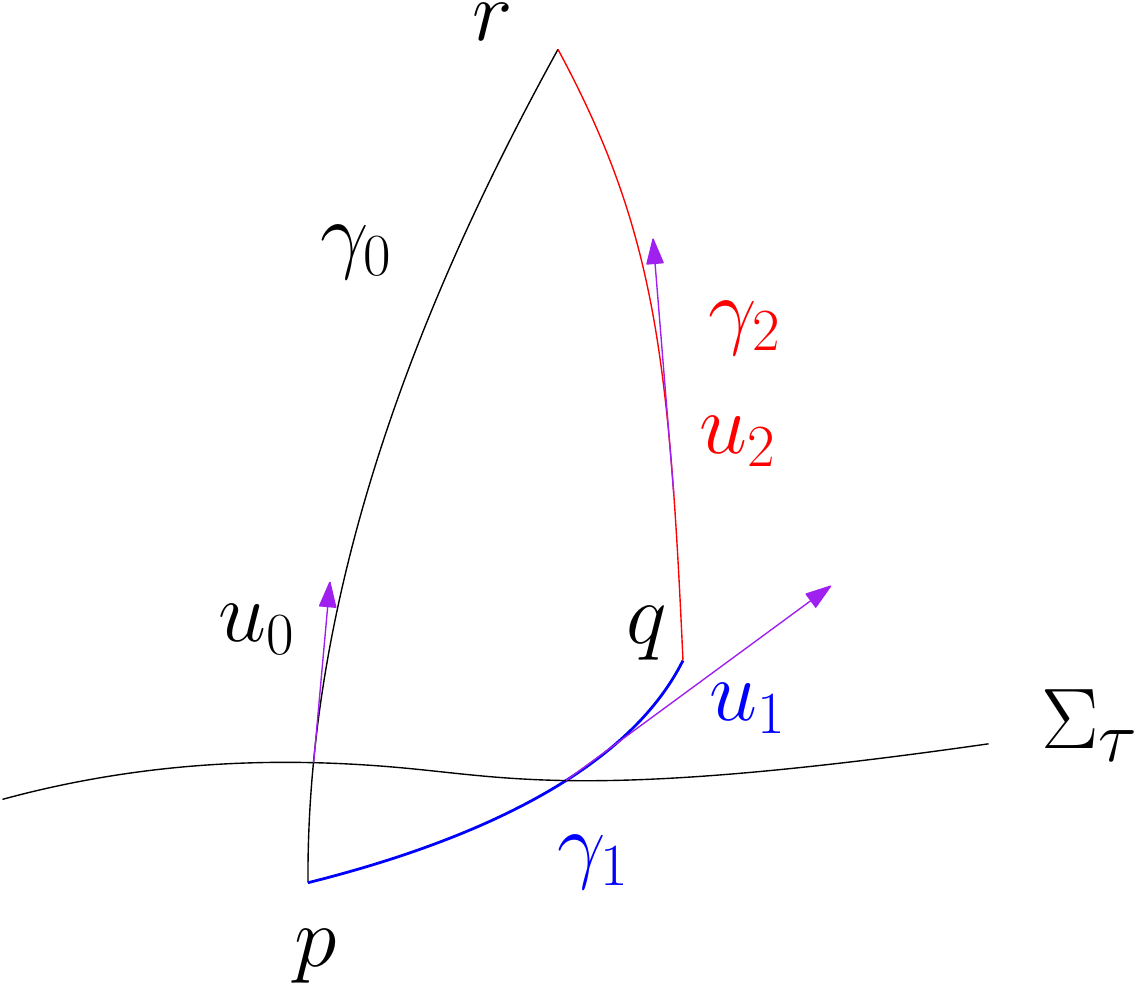}
\par\end{centering}
}\quad{}\subfloat[\label{fig:Induced-parameterisation-and}Induced parameterisation
and scalar product]{\begin{centering}
\includegraphics[scale=0.5]{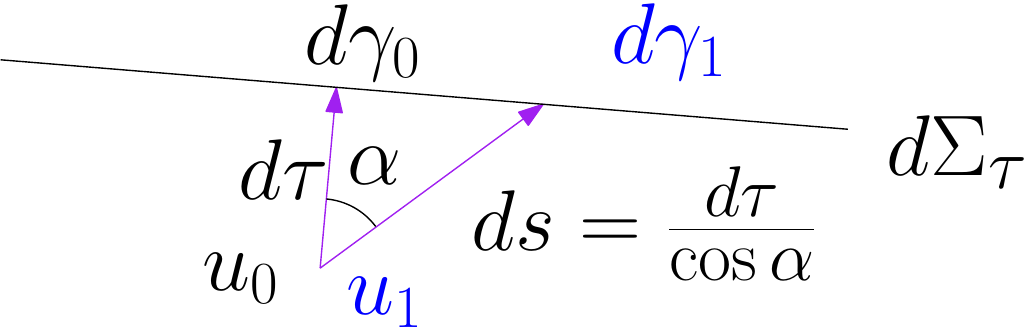}
\par\end{centering}
}
\par\end{centering}
\caption{\label{fig:Proof-of-Piecewise}Proof of Piecewise vs Geodesic }
\end{figure}
(see Fig.~\ref{fig:Proof-of-Piecewise})

\ifS\else Use
proper time on $\gamma_{0}$ parameterisation\fi 

Since $U$ convex normal neighbourhood$\exists\Sigma_{\tau}$ a family of hypersurfaces normal to $u_{0}=\left.\frac{dx}{d\tau}\right|_{\gamma_{0}}$
inducing parameterisation on $\gamma_{1}$ and $\gamma_{2}$ (see
Fig.~\ref{fig:Tangent-vectors-and}):
\begin{align*}
u_{1}= & \left.\frac{dx}{d\tau_{1}}\right|_{\gamma_{1}} & u_{2}= & \left.\frac{dx}{d\tau_{2}}\right|_{\gamma_{2}}
\end{align*}
From decomposition of any timelike vector into
\ifS\else another \fi normalised timelike vector and a spacelike
vector, as in Sec.~\ref{par:Energy-momentum-tensors}
\begin{align*}
u_{1}= & \left(u_{0}.u_{1}\right)u_{0}+\pi\negthinspace S_{1}e\ifS\else\fi\\
\textrm{with }\pi\negthinspace S_{1}e= & u_{1}-\left(u_{0}.u_{1}\right)u_{0},\ifS\else\fi\\
e.e= & 1,\ifS\else\fi\\
\pi\negthinspace S_{1}= & \sqrt{\pi\negthinspace S_{1}e.\pi\negthinspace S_{1}e}>0
\end{align*}
\ifS
\else as $u_{0}$ and $u_{1}$
in same light cone with same parameter $\tau$\fi 
\begin{align*}
\left|u_{0}.u_{1}\right|= & -u_{0}.u_{1}<1
\end{align*}
\ifS
\else as (following Fig.~\ref{fig:Induced-parameterisation-and})\fi 
\begin{align*}
u_{0}^{a}= & \left.\frac{dx^{a}}{d\tau}\right|_{\gamma_{0}}\\
u_{1}^{a}= & \left.\frac{dx^{a}}{ds}\right|_{\gamma_{1}}=\left.\frac{dx^{a}}{d\tau}\frac{d\tau}{ds}\right|_{\gamma_{1}}\\
\Rightarrow\left|u_{0}.u_{1}\right|= & \left|\left|u_{0}\right|.\left(\left|u_{0}\right|\cos\alpha\right)\cos\alpha\right|\\
= & \cos^{2}\alpha<1
\end{align*}
\ifS\else Thus\fi 
\begin{align*}
\tau_{1}=\int_{p}^{q}\sqrt{-u_{1}.u_{1}}d\tau= & \int_{p}^{q}\sqrt{\left(u_{0}.u_{1}\right)^{2}-\left(\pi\negthinspace S_{1}\right)^{2}}d\tau\\
\le & \int_{p}^{q}\left|u_{0}.u_{1}\right|d\tau\\
< & \int_{p}^{q}d\tau
\end{align*}
and samely
\begin{align*}
\tau_{2}=\int_{q}^{r}\sqrt{-u_{2}.u_{2}}d\tau< & \int_{q}^{r}d\tau
\end{align*}
\ifS
\else so\fi 
\begin{align*}
\tau_{1}+\tau_{2}< & \int_{p}^{q}d\tau+\int_{q}^{r}d\tau=\tau_{0}
\end{align*}
\end{proof}
\ifS
\else By upper semicontinuity, \fi $\forall\mu\in C\left(r,t\right),\tau\left[\mu\right]\le\tau\left[\gamma_{rt}\right]$\ifS
\else 

If $\tau\left[\mu\right]=\tau\left[\gamma_{rt}\right]$ with $\mu\ne\gamma_{rt},\exists q\in\mu,q\notin\gamma_{rt},\gamma_{1},\gamma_{2}$
geodesics, $\gamma_{1}=\gamma_{rs},\gamma_{2}=\gamma_{st}$\fi 
\begin{align*}
\left.\begin{array}{c}
\gamma_{1}\textrm{ maximises }\tau\textrm{ between }r\textrm{ and }s\\
\gamma_{2}\textrm{ maximises }\tau\textrm{ between }s\textrm{ and }t
\end{array}\right|\Rightarrow & \tau\left[\gamma_{1}\right]+\tau\left[\gamma_{2}\right]\ge\tau\left[\mu\right]=\tau\left[\gamma_{rt}\right]
\end{align*}
\ifS
\else Contradiction!

$\Rightarrow\gamma$ unique curve\fi , hence the result is proven
\end{proof}
\begin{lem}
\label{lem:7Existence-of-maximum}Existence of maximum $\tau$ causal
curves

$\left(M,g_{ab}\right)$ globally hyperbolic, $p\in M$

$\Sigma\subset M$ Cauchy surface $\Rightarrow\exists\gamma\in C\left(\Sigma,p\right),\tau\left[\gamma\right]$
maximum on $C\left(\Sigma,p\right)$
\end{lem}
\begin{proof}
Sketch of proof

\ifS
\else $C\left(\Sigma,p\right)$ compact and from Lemma~\ref{lem:1Upper-semicontinuity}
(Upper semicontinuity)

$\tau$ is upper semi-continuous $\Rightarrow$ maximum is attained
inside $C\left(\Sigma,p\right)$\fi 
\end{proof}
\noindent We are now ready to prove the first Big Bang Singularity
theorem

\subsubsection{First Big Bang Singularity theorem}
\begin{thm}
\label{thm:BBST1Big-Bang-Singularity}Big Bang Singularity theorem
1

$\left(M,g_{ab}\right)$ globally hyperbolic with S.E.C.

Suppose $\exists\Sigma,C^{2}$(smooth) Cauchy (initial) surface with
$-\Theta\le C<0,C$ constant, everywhere on $\Sigma$ (trace of extinsic
curvature for past directed normal geodesic $K=-\Theta$)

Then $\forall\gamma\subset J^{-}\left(\Sigma\right),\dot{\gamma}^{2}\le0,\tau=\int_{-\infty}^{\Sigma}\sqrt{-\dot{\gamma}^{2}}dt\le\frac{3}{\left|C\right|}$

In particular: past directed timelike geodesics are incomplete
\end{thm}
\begin{proof}
\ifS
\else Suppose
$\exists\lambda\subset J^{-}\left(\Sigma\right),\dot{\lambda}^{2}\le0,\tau=\int_{-\infty}^{\Sigma}\sqrt{-\dot{\lambda}^{2}}dt>\frac{3}{\left|C\right|}$

$p=\lambda\left(t_{1}\right),\tau\left(p\right)\equiv\int_{p}^{\Sigma}\sqrt{-\dot{\lambda}^{2}}dt>\frac{3}{\left|C\right|}\left(\ge\frac{3}{\Theta}\right)$\fi 

From Lemma~\ref{lem:7Existence-of-maximum} (Existence of maximum
$\tau$ causal curve) $\exists\gamma$ of maximum $\tau\ge\frac{3}{\left|C\right|}$

From Lemma~\ref{lem:6Maximum-of-} (Maximum of $\tau$ on $C\left(\Sigma,p\right)$
is geodesics without inner conjugate to $\Sigma$) $\gamma$ geodesic
$\bot\Sigma$ with no conjugate point

\ifS\else But \fi from Lemma~\ref{lem:3Existence-of-conjugate}
(Existence of conjugate to $\Sigma$ in finite $\tau$) $\exists p,\tau\left(p\right)\le\frac{3}{\Theta}$,
conjugate to $\Sigma$resulting in a contradiction, which prooves the theorem
\end{proof}
However the globally hyperbolic condition is very strong! Hawking
(1967) found a weaker condition

\subsection{More general Big Bang Singularity theorem}

\subsubsection{Definitions}

\begin{defn}
Future endpoint of $\gamma$

$p\in M$ future endpoint of $\gamma\Leftrightarrow\forall O\left(p\right)\subset M$
open, $\exists t_{0},\forall t>t_{0},\gamma\left(t\right)\in O\left(p\right)$
\end{defn}
\noindent 
\begin{defn}
Future inextendible curve $\gamma$

$\gamma$ future inextendible curve $\Leftrightarrow\nexists$ future
endpoint of $\gamma$
\end{defn}
\noindent 
\begin{defn}
\label{def:Domain-of-dependence}Domain of dependence

Future
\begin{align*}
D^{+}\left(S\right)= & \left\{ p\in M|\forall\gamma,\textrm{ past inextendible causal, }\exists\tau_{p},\gamma\left(\tau_{p}\right)=p,\gamma\cap S\ne\emptyset\right\} 
\end{align*}
\ifS
\else (every past past
inextendible causal curve through $p$ intersects $S$, see Fig.~\ref{fig:Future-domain-of})\fi 
\begin{figure}
\begin{centering}
\subfloat[\label{fig:Future-domain-of}Future domain of dependence]{\begin{centering}
\includegraphics[width=0.45\columnwidth]{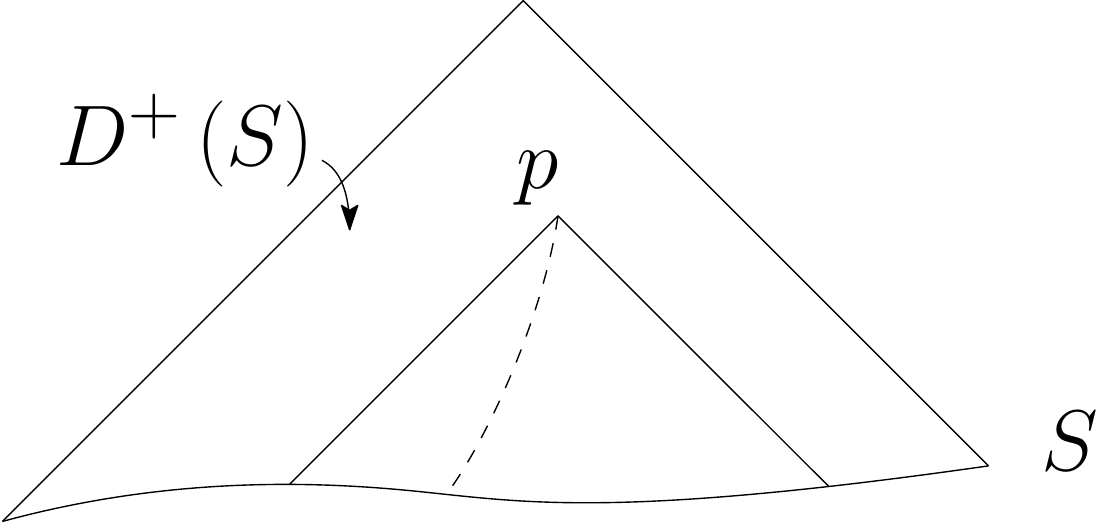}
\par\end{centering}

}\quad{}\subfloat[\label{fig:Past-domain-of}Past domain of dependence]{\begin{centering}
\includegraphics[width=0.45\columnwidth]{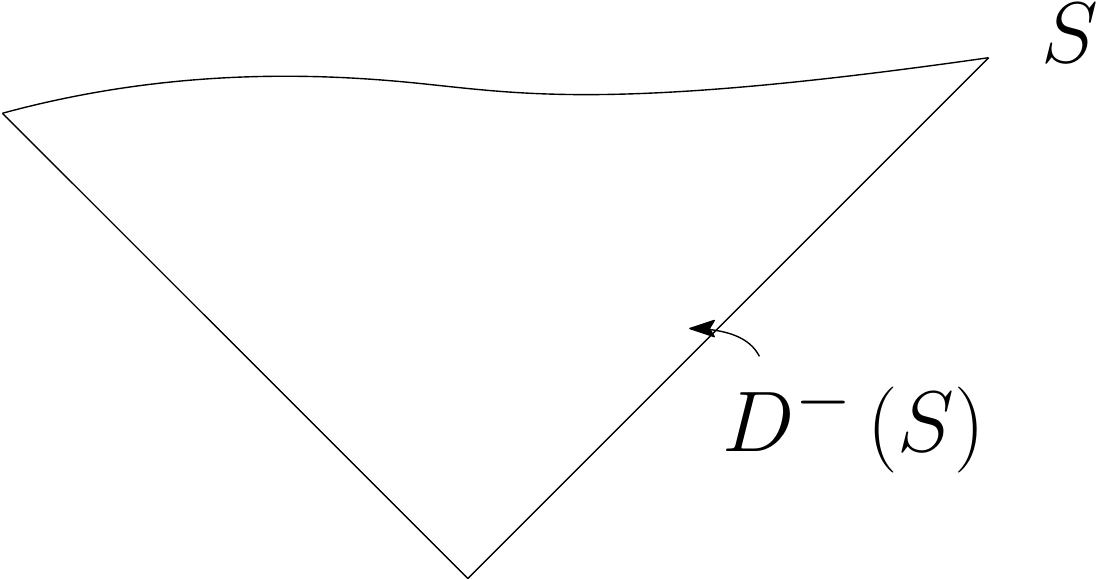}
\par\end{centering}
}
\par\end{centering}
\caption{Domains of dependence}

\end{figure}

Past
\begin{align*}
D^{-}\left(S\right)= & \left\{ p\in M|\forall\gamma,\textrm{ future inextendible causal, }\exists\tau_{p},\gamma\left(\tau_{p}\right)=p,\gamma\cap S\ne\emptyset\right\} 
\end{align*}
\ifS
\else (see
Fig.~\ref{fig:Past-domain-of})\fi 

Full
\begin{align*}
D\left(S\right)= & D^{+}\left(S\right)\cup D^{-}\left(S\right)
\end{align*}
\end{defn}
\begin{defn}
Interior of $A$
\begin{align*}
\mathrm{Int}\left[A\right]= & \underset{{\scriptscriptstyle \forall O\subset A}}{\cup}O
\end{align*}
\ifS
\else Union of all open sets contained in $A$\fi 
\end{defn}
~
\begin{defn}
Closure of $A$
\begin{align*}
\overline{A}= & A+\lim A=\underset{{\scriptscriptstyle \forall C\supset A}}{\cap}C
\end{align*}
\ifS
\else Intersection of all closed sets containing $A$\fi 
\end{defn}
\begin{defn}
\label{def:Boundary-ofA}Boundary of $A$
\begin{align*}
\dot{A}= & \partial A=\overline{A}-\mathrm{Int}\left[A\right]
\end{align*}
\end{defn}
\begin{figure}
\begin{centering}
\includegraphics[scale=0.5]{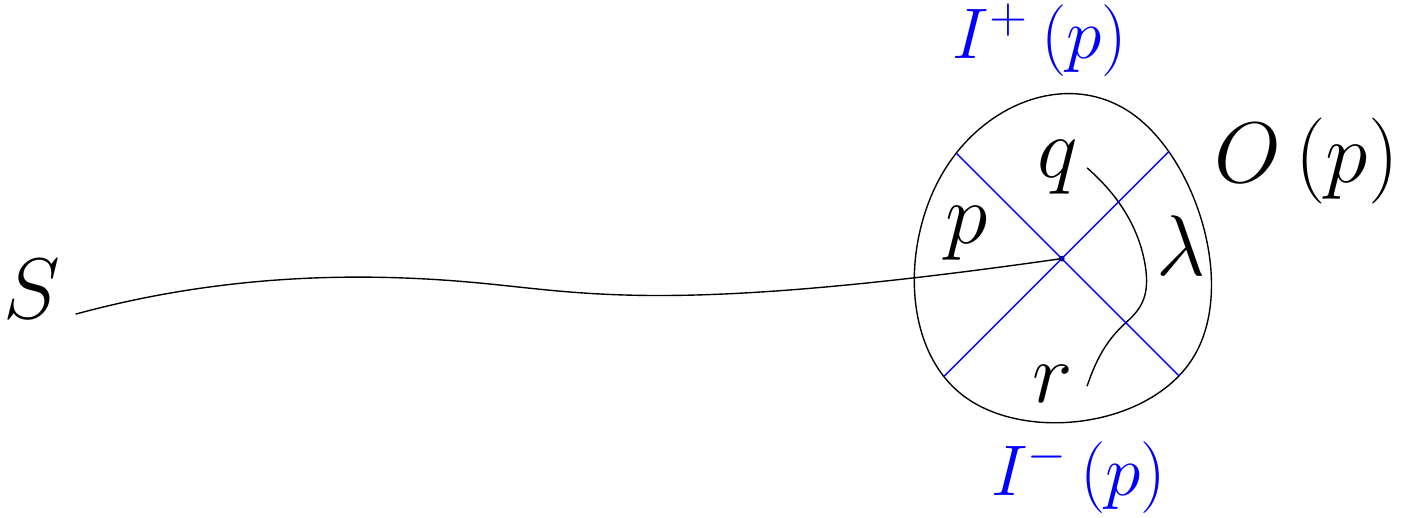}
\par\end{centering}
\caption{\label{fig:Edge-of}Edge of $S$}
\end{figure}

~
\begin{defn}
Edge of $S$\ifS
\else 

\begin{align*}
\hspace{-2cm}p\in\mathrm{Edge}\left(S\right)\Leftrightarrow\forall O\left(p\right),\exists q\in I^{+}\left(p\right)\cap O\left(p\right),\exists r\in I^{-}\left(p\right)\cap O\left(p\right), & \exists\lambda,\dot{\lambda}^{2}<0,\lambda\subset O\left(p\right),\\
 & \lambda\left(0\right)=r,\lambda\left(1\right)=q,\lambda\cap\left(S\cup\left\{ p\right\} \right)=\emptyset
\end{align*}
\fi 
\end{defn}
\ifS
\else See Fig.~\ref{fig:Edge-of}\fi 
\begin{defn}
\label{def:Cauchy-Horizon(C.H.)}Cauchy Horizon

\ifS
\else $S$ \fi closed, achronal

Future:
\begin{align*}
H^{+}\left(S\right)= & \overline{D^{+}\left(S\right)}-I^{-}\left[D^{+}\left(S\right)\right]
\end{align*}
\ifS\else (see
Fig.~\ref{fig:Future-Cauchy-Horizon})\fi 
\begin{figure}
\begin{centering}
\includegraphics[scale=0.5]{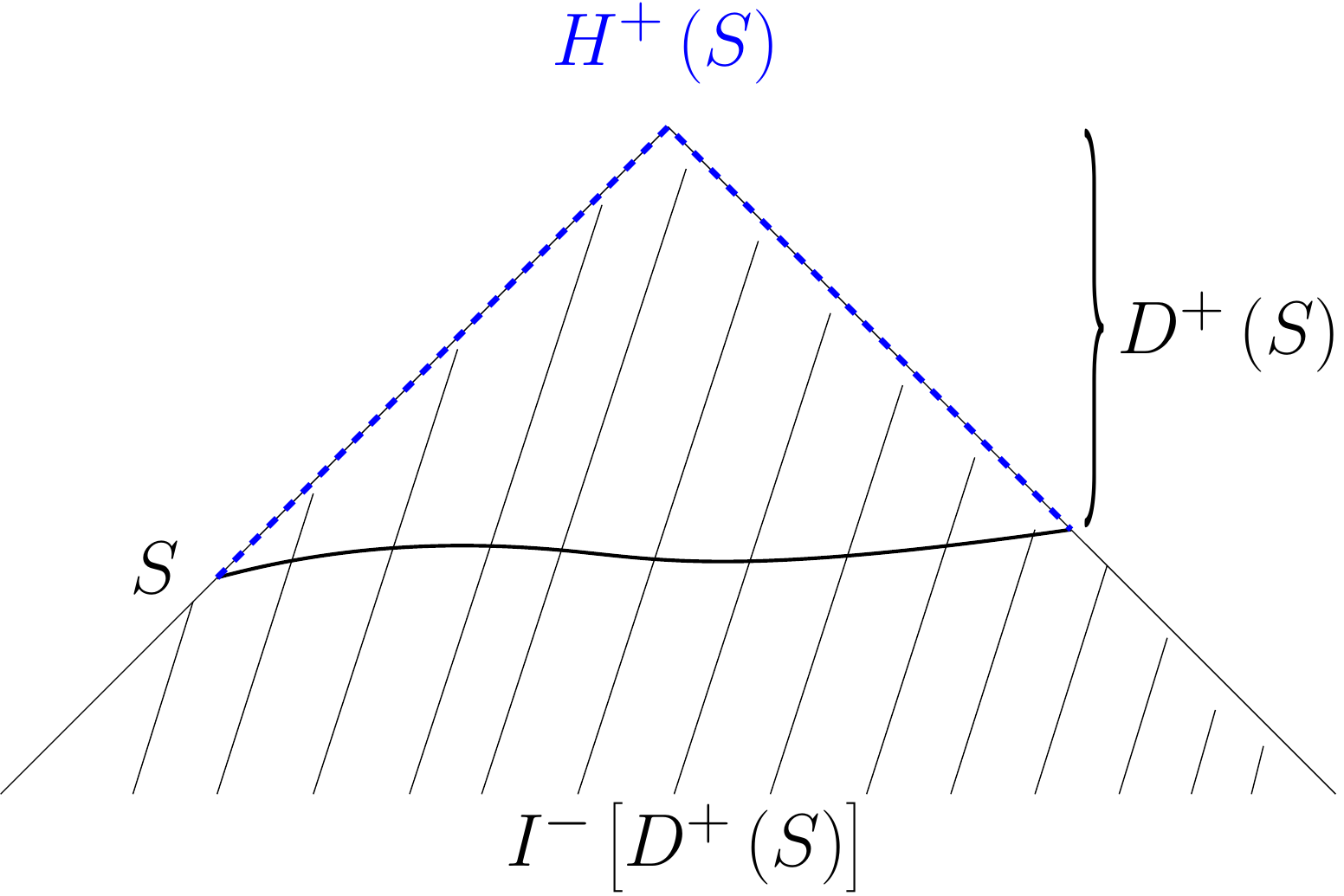}
\par\end{centering}
\caption{\label{fig:Future-Cauchy-Horizon}Future Cauchy Horizon}

\end{figure}

Past:
\begin{align*}
H^{-}\left(S\right)= & \overline{D^{-}\left(S\right)}-I^{+}\left[D^{-}\left(S\right)\right]
\end{align*}

Full:
\begin{align*}
H\left(S\right)= & H^{+}\left(S\right)\cup H^{-}\left(S\right)
\end{align*}
\end{defn}
\noindent 
\begin{description}
\item [{Property}] Cauchy horizons and geodesics\\
\uline{Future Cauchy Horizon}
\begin{align*}
\hspace{-3cm}H^{+}\left(S\right)= & \left\{ p\in M|\exists\lambda,\dot{\lambda}^{2}=0\textrm{ geodesic, }p\in\lambda,\lambda\subset H^{+}\left(S\right)\Leftrightarrow\lambda\left\{ \begin{array}{l}
\textrm{past inextendible}\\
\textrm{or}\\
\textrm{with past endpoint on }\mathrm{Edge}\left(S\right)
\end{array}\right\} \right\} 
\end{align*}
\uline{Past Cauchy Horizon}
\begin{align*}
\hspace{-3cm}H^{-}\left(S\right)= & \left\{ p\in M|\exists\lambda,\dot{\lambda}^{2}=0\textrm{ geodesic, }p\in\lambda,\lambda\subset H^{-}\left(S\right)\Leftrightarrow\lambda\left\{ \begin{array}{l}
\textrm{future inextendible}\\
\textrm{or}\\
\textrm{with future endpoint on }\mathrm{Edge}\left(S\right)
\end{array}\right\} \right\} 
\end{align*}
\end{description}

\subsubsection{Lemmas}

\begin{lem}
\label{lem:8Cauchy-horizon-and}Cauchy horizon and domain of dependence

$H\left(S\right)=\dot{D}\left(S\right)$: full
Cauchy horizon is boundary of domain
of dependence
\end{lem}
\begin{lem}
\label{lem:9Cauchy-horizon-decomposition}Cauchy horizon decomposition

$\forall p\in H^{\pm}\left(S\right),p\in\lambda\subset H^{\pm}\left(S\right)\ifS\else\,\fi\lambda$
either %
\begin{tabular}[t]{l}
-past/future inextendible, or\tabularnewline
-past/future endpoint $\in\mathrm{Edge}\left(S\right)$\tabularnewline
\end{tabular}
\end{lem}
\noindent 
\begin{lem}
\label{lem:10Causal-curves-in}Causal curves in compact submanifolds

$\left(M,g_{ab}\right)$ strongly causal, $K\subset M$ compact

$\Rightarrow\left(\forall\mathscr{C}\subset K,\dot{\mathscr{C}}^{2}\le0\textrm{ (causal)}\Rightarrow\textrm{past and future endpoints of }\mathscr{C}\in K\right)$
\end{lem}
\noindent 

\subsubsection{More general Big Bang Singularity theorem}

\begin{thm}
\label{thm:Big-Bang-Singularity2}Big Bang Singularity theorem 2 (Hawking
1967, more general)

$\left(M,g_{ab}\right)$ strongly causal with S.E.C.

Suppose $\exists S$ compact, edgeless, achronal, smooth spacelike
hypersurface with $-\Theta\le C<0$ everywhere on $S$ (trace of extinsic
curvature for past directed normal geodesic $K=-\Theta$)

Then $\exists\gamma\subset J^{-}\left(S\right),\dot{\gamma}^{2}\le0,\gamma$
inextendible geodesic and 
\begin{align*}
\tau= & \int_{-\infty}^{S}\sqrt{-\dot{\gamma}^{2}}dt\le\frac{3}{\left|C\right|}
\end{align*}
\end{thm}
\noindent 
\begin{proof}
Sketch of proof

\ifS
\else Suppose
$\forall\gamma,\dot{\gamma}^{2}\le0$, past directed inextendible
geodesic $\bot S$ with $\tau\left[\gamma\right]>\frac{3}{\left|C\right|}$\fi 

\ifS
\else $\left(\mathrm{Int}\left[D\left(S\right)\right],g_{ab}\right)$
globally hyperbolic ($S$: Cauchy surface of $\mathrm{Int}\left[D\left(S\right)\right]$
by construction of $D\left(S\right)$)\fi 

\ifS
\else Then
it satisfies Theorem~\ref{thm:BBST1Big-Bang-Singularity} $\Rightarrow\forall\gamma$
inextendible past directed geodesic, with $\dot{\gamma}^{2}\le0$,
it must leave $\mathrm{Int}\left[D\left(S\right)\right]\left(\textrm{as }\tau\left[\left.\gamma\right|_{\mathrm{Int}\left[D\left(S\right)\right]}\right]\le\frac{3}{\left|C\right|}\right)$\fi 

As\ifS
\else ~$H\left(S\right)$ is boundary of $D\left(S\right)$
from Lemma~\ref{lem:8Cauchy-horizon-and} (Cauchy horizon and domain
of dependence), then they must intersect $H^{-}\left(S\right)$ before
$\tau\left[\gamma_{Sq}\right]>\frac{3}{\left|C\right|}\Rightarrow H^{-}\left(S\right)\ne\emptyset$
as $H^{-}\left(S\right)=\dot{D}^{-}\left(S\right)$\fi 
\begin{align}
\forall p\in H^{-}\left(S\right),\forall l\in C\left(S,p\right),\tau\left[l\right]\le & \frac{3}{\left|C\right|}\label{eq:limitTauCauchyH}
\end{align}
\ifS\else so $\exists\tau_{0}$,
\fi least upper bound of $\tau\left[l\right]$
\begin{figure}
\begin{centering}
\includegraphics[scale=0.5]{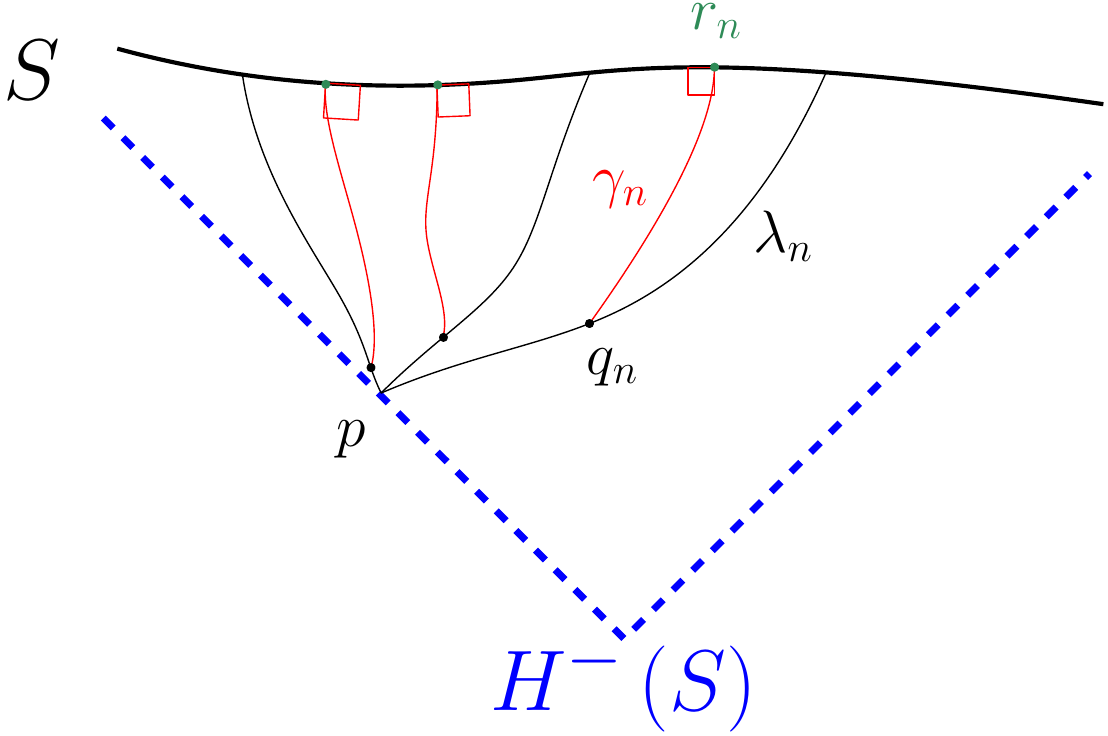}
\par\end{centering}
\caption{\label{fig:Sequence-of-timelike}Sequence of timelike curves from
$S$ to $p$ supporting geodesics converging to $p$.}
\end{figure}

\ifS
\else $\left\{ \lambda_{n}\right\} $ sequence
of timelike curves from $S$ to $p,\underset{n\to\infty}{\lim}\tau\left[\lambda_{n}\right]=\tau_{0}$\fi 

\ifS
\else $q_{n}\in\lambda_{n}$,
$q_{n}\ne p$, $\underset{n\to\infty}{\lim}q_{n}=p,$ $q_{n}\in I^{+}\left(p\right)\Rightarrow q_{n}\in\mathrm{Int}\left[D^{-}\left(S\right)\right]$
(as $q_{n}\leftrightarrow p$ by $\dot{\lambda}_{n}^{2}<0$)\fi 

From Lemma~\ref{lem:7Existence-of-maximum} (Existence of maximum
$\tau$ causal curves)

\ifS
\else $\exists\gamma_{n}\in C\left(S,q_{n}\right),\tau\left[\gamma_{n}\right]$
maximum on $C\left(S,q_{n}\right)$, and Lemma~\ref{lem:6Maximum-of-}
(existence of geodesics without inner conjugate to $\Sigma$) gives
$\gamma_{n}$ geodesic $\bot S$\fi 

\ifS
\else Thus $\underset{n\to\infty}{\lim}\tau\left[\gamma_{n}\right]=\tau_{0}$
\fi (see Fig.~\ref{fig:Sequence-of-timelike})

Naming \ifS
\else $r_{n}=\gamma_{n}\cap S$;
since $S$ compact $\underset{n\to\infty}{\lim}r_{n}=r\in S$\fi 

\ifS
\else $\exists\gamma$ geodesic
$\bot S,r\in S\Rightarrow$ by continuity of $r_{n}$ and of the $\bot$
tangent vectors to $\gamma_{n}$: \fi $\gamma\cap H^{-}\left(S\right)=p$
and $\tau\left[\gamma\right]=\underset{n\to\infty}{\lim}\tau\left[\gamma_{n}\right]=\tau_{0}$

\noindent\fbox{\begin{minipage}[t]{1\columnwidth - 2\fboxsep - 2\fboxrule}%
\ifS \else Thus \fi $\forall p\in H^{-}\left(S\right),\exists\gamma$
geodesic $\bot S,\tau\left[\gamma\right]=\tau_{0}$ least upper bound
of $\tau$ over $C\left(S,p\right)$%
\end{minipage}}

\ifS
\else $\left\{ \widetilde{\gamma}_{n}\right\} $
any sequence of geodesics $\bot S$, maximum length, to $p_{n}\in H^{-}\left(S\right)$\fi 

\ifS
\else $\widetilde{r}_{n}=S\cap\widetilde{\gamma}_{n}$;
since $S$ compact $\underset{n\to\infty}{\lim}\widetilde{r}_{n}=\widetilde{r}\in S$\fi 

\ifS
\else $\exists\widetilde{\gamma}$ geodesic $\bot S,\widetilde{r}\in\widetilde{\gamma},\widetilde{\gamma}\cap H^{-}\left(S\right)=p$
(\fi see Fig.~\ref{fig:Sequence-of-geodesics})
\begin{figure}
\begin{centering}
\includegraphics[scale=0.5]{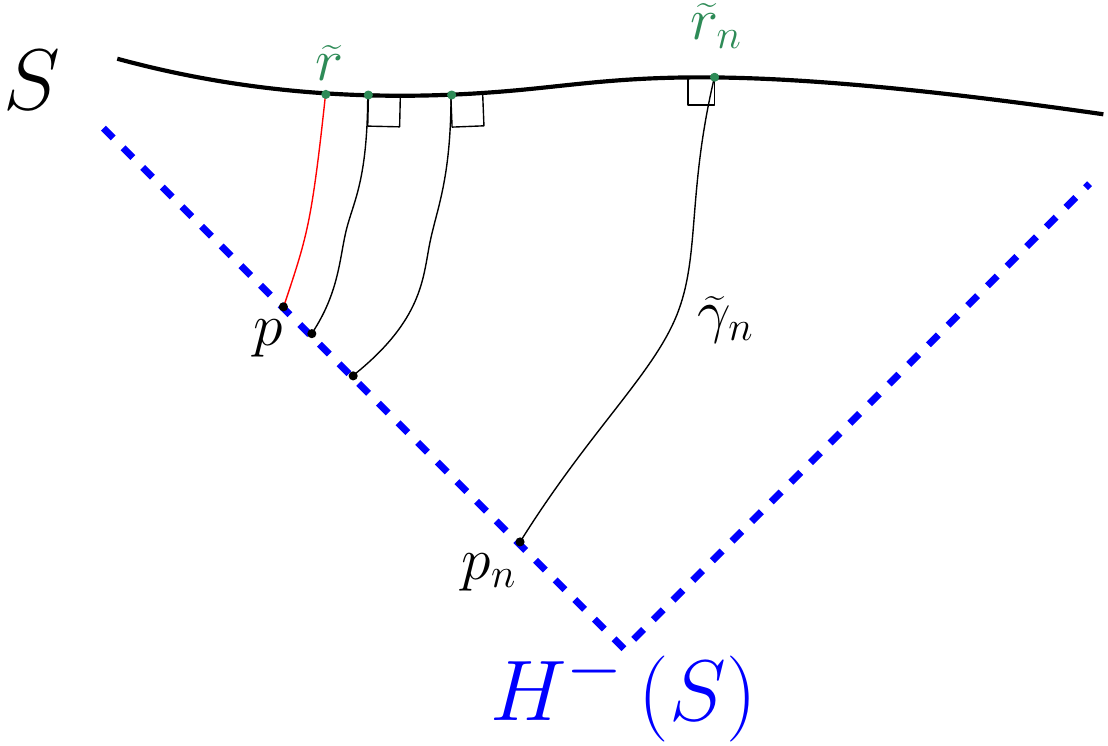}
\par\end{centering}
\caption{\label{fig:Sequence-of-geodesics}Sequence of geodesics from $S$
to $p_{n}$ converging to a geodesic from $S$ to $p$.}
\end{figure}

\ifS
\else By
continuity of $\widetilde{r}_{n}$ and $\widetilde{\gamma}_{n}$,
$p$ accumulation point of $\left\{ p_{n}\right\} \Rightarrow\boxed{H^{-}\left(S\right)\textrm{ compact}}$\fi 

Since $\mathrm{Edge}\left(S\right)=\emptyset$,
by Lemma~\ref{lem:9Cauchy-horizon-decomposition} (Cauchy horizon
decomposition) $\exists\gamma,\dot{\gamma}^{2}=0$
inextendible future geodesic, $\gamma\subset H^{-}\left(S\right)$

Since $\left(M,g_{ab}\right)$ strongly
causal and $H^{-}\left(S\right)\subset M$, compact 

from Lemma~\ref{lem:10Causal-curves-in} (Causal curves in compact
submanifolds) \ifS
\else $\forall\lambda$ causal, $\lambda\in H^{-}\left(S\right)$
\fi must have past and future endpoints in $H^{-}\left(S\right)$,
thus contradicting that $\gamma$ is inextendible!

\fbox{\begin{minipage}[t]{0.56\textwidth}%
$\exists\gamma$ past inextendible, $\dot{\gamma}^{2}\le0,\bot S,\tau\left[\gamma\right]<\frac{3}{\left|C\right|}$%
\end{minipage}}

\ifS
\else At least $\gamma$, past directed timelike geodesic
is incomplete\fi 
\end{proof}
\ifS
\else Those theorems predict the existence of Big Bang
(in a cosmological context).\fi 

\noindent Next we deal with gravitational collapsewith the first singularity
theorem discovered by Penrose (1965)

\subsection{First Black Hole Singularity theorem}

\subsubsection{Definitions}

\begin{defn}
Length of a null curve

$\gamma\left(\lambda\right),\dot{\gamma}^{2}=0,\lambda\left[\gamma\right]=\int_{\gamma}d\lambda$
\end{defn}
\begin{defn}
\label{def:Time-orientable-manifolds}Time orientable manifolds

$\left(M,g_{ab}\right)$ time orientable $\Leftrightarrow\exists t^{a}$
on $M$, smooth, non-vanishing, timelike vector field
\end{defn}
\noindent 
\begin{defn}
Metric ``opener''

In $\left(M,g_{ab}\right),\exists t^{a},t^{a}t_{a}<0$,
define $\widetilde{g}_{ab}=g_{ab}-t_{a}t_{b}$
\end{defn}
$\widetilde{g}_{ab}$
has Lorentz signature and light cones strictly larger than $g_{ab}$:

$\forall l^{a},l^{a}l_{a}=0\Rightarrow\widetilde{g}_{ab}l^{a}l^{b}=l^{a}l_{a}-\left(t^{a}l_{a}\right)^{2}=-\left(t^{a}l_{a}\right)^{2}<0$

Thus if $\left(M,g_{ab}\right)$ \ifS\else on \fi verge
of having timelike closed curves then $\left(M,\widetilde{g}_{ab}\right)$
contains timelike closed curves
\begin{defn}
Stably causal spacetime

$\left(M,g_{ab}\right)$ stably causal $\Leftrightarrow\exists t^{a},t^{a}t_{a}<0$,
continuous non-vanishing vector field, $\left(M,\widetilde{g}_{ab}\right)$
has no closed timelike curves
\end{defn}

\subsubsection{Lemmas}

\begin{figure}
\begin{centering}
\includegraphics[scale=0.5]{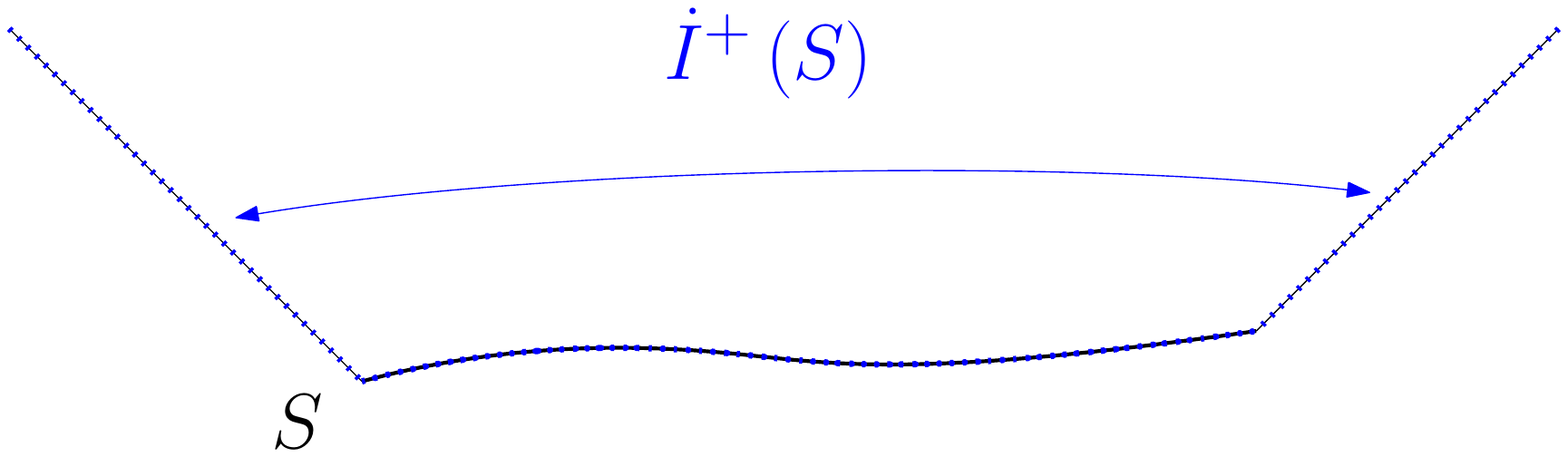}
\par\end{centering}
\caption{\label{fig:Boundary-of-time}Boundary of time orientable spacetime
submanifold chronological future.}

\end{figure}

\begin{lem}
\label{lem:11Existence-of-conjugate}Existence of conjugate to $S$
in finite $\lambda$

$\left(M,g_{ab}\right)$ with N.E.C., $S$ smooth 2D spacelike submanifold
with $\Theta_{+}\left(q\right)=\Theta_{0}<0,q\in S$ (outgoing null
geodesic)

$\Rightarrow\exists p$ conjugate to $S$ along $\mu\left(\lambda\right),\dot{\mu}^{2}=0,q\in\mu$
for $\lambda\le\frac{2}{\left|\Theta_{0}\right|}$
\end{lem}
\begin{proof}
\ifS\else (by analogy to \fi Lemma~\ref{lem:3Existence-of-conjugate}:
Existence of conjugate to $\Sigma$ in finite $\tau$\ifS
\else )\fi 
\end{proof}
\begin{lem}
\label{lem:12Null-geodesics-with}Null geodesics with no inner conjugates
to $K$

$\left(M,g_{ab}\right)$ globally hyperbolic, $K$ compact, orientable,
2D, spacelike, $K\subset M$

$p\in\dot{I}^{+}\left(K\right)\Rightarrow\exists\mu,p\in\mu,\dot{\mu}^{2}=0$
future geodesic, $\mu\cap K\ne\emptyset,\mu\bot K$ with no conjugate
between $K$ and $p$
\end{lem}
\begin{proof}
\ifS\else (by analogy to \fi Lemma~\ref{lem:4Geodesics-with-no}:
Geodesics with no inner conjugates maximise $\tau$\ifS
\else )\fi 
\end{proof}
\noindent 
\begin{lem}
\label{lem:13Globally-hyperbolic-time}Globally hyperbolic time orientability

Globally hyperbolic spacetimes are time orientable
\end{lem}
\noindent 
\begin{lem}
\label{lem:14Boundary-of-time}Boundary of time orientable spacetime
submanifold

$\left(M,g_{ab}\right)$ time orientable, $S\subset M\Rightarrow\dot{I}^{+}\left(S\right)$
achronal, 3D, $C^{0}$ submanifold of $M$
\end{lem}
\ifS\else See \fi Fig.~\ref{fig:Boundary-of-time}

\begin{lem}
\label{lem:15Topology-of-globally}Topology of globally hyperbolic
spacetimes

$\left(M,g_{ab}\right)$ globally hyperbolic $\Rightarrow\left(M,g_{ab}\right)$
stably causal

Furthermore, $\exists f$ global time function, $f=cst$: Cauchy surface
$\Sigma\Rightarrow M$ foliated by $\Sigma$ with topology of $\mathbb{R}\times\Sigma$
\end{lem}

\subsubsection{First Black Hole Singularity theorem}

\begin{thm}
\label{thm:Black-Hole-Singularity}Black Hole Singularity theorem
1 (Penrose 1965)

$\left(M,g_{ab}\right)$ connected, globally hyperbolic, with $\Sigma$
non-compact Cauchy surface and N.E.C.

$\exists T\subset M$ trapped surface, $\Theta_{0}<0$ maximum of
$\Theta_{+},\Theta_{-}$, outgoing and ingoing null geodesic expansions,
each geodesic $\bot T$

$\Rightarrow\exists\gamma,\dot{\gamma}^{2}=0$ future inextendible
$\bot T$ with affine length
\begin{align*}
\lambda\left[\gamma\right]\le & \frac{2}{\left|\Theta_{0}\right|}
\end{align*}
\end{thm}
\noindent 
\begin{proof}
\ifS
\else Suppose
$\forall\gamma,\dot{\gamma}^{2}=0$ future geodesic $\bot T$, and
$\lambda\left[\gamma\right]\ge\frac{2}{\left|\Theta_{0}\right|}$\fi 

\ifS
\else Then define the maps\fi 
\begin{align*}
f_{+}:T\times\left[0;\frac{2}{\left|\Theta_{0}\right|}\right]\longrightarrow & M\\
q,a\longrightarrow & f_{+}\left(q,a\right)=\gamma_{\mathrm{out}}\left(a\right),\\
 & \dot{\gamma}_{\mathrm{out}}^{2}=0,\gamma_{\mathrm{out}}\bot T,\gamma_{\mathrm{out}}\cap T=q
\end{align*}
\begin{align*}
f_{-}:T\times\left[0;\frac{2}{\left|\Theta_{0}\right|}\right]\longrightarrow & M\\
q,a\longrightarrow & f_{-}\left(q,a\right)=\gamma_{\mathrm{in}}\left(a\right),\\
 & \dot{\gamma}_{\mathrm{in}}^{2}=0,\gamma_{\mathrm{in}}\bot T,\gamma_{\mathrm{in}}\cap T=q
\end{align*}
Since $\left[0;\frac{2}{\left|\Theta_{0}\right|}\right]$
\ifS
\else compact and $f_{+},f_{-}$ continuous, then\fi 
\begin{align*}
A= & f_{+}\left\{ T\times\left[0;\frac{2}{\left|\Theta_{0}\right|}\right]\right\} \cup f_{-}\left\{ T\times\left[0;\frac{2}{\left|\Theta_{0}\right|}\right]\right\} 
\end{align*}
compact.

By Lemma~\ref{lem:11Existence-of-conjugate} (Existence of conjugate
to $T$ in finite $\lambda$) and Lemma~\ref{lem:12Null-geodesics-with}
(Null geodesics with no inner conjugates to $K$), $\dot{I}^{+}\left(T\right)$,
the boundary of future of $T$,
is made of future null geodesic, \ifS\else hence
in \fi $A$

\ifS
\else So
$\dot{I}^{+}\left(T\right)\subset A$ and since $\dot{I}^{+}\left(T\right)$
closed $\Rightarrow\boxed{\dot{I}^{+}\left(T\right)\textrm{ compact}}$\fi 

Using Lemma~\ref{lem:13Globally-hyperbolic-time} (Globally hyperbolic
time orientability) and \ifS
\else definition of time orientability, \fi we define $t^{a}$ smooth,
$t^{a}t_{a}<0$, on $M$

\ifS
\else Since $\dot{I}^{+}\left(T\right)$
achronal (Lemma~\ref{lem:14Boundary-of-time}: Boundary of time orientable
spacetime submanifold) i.e. the boundary of the future of $T$ is
spacelike, or null

$\left[\left(\textrm{then }I^{+}\left(\dot{I}^{+}\left(T\right)\right)\ne\emptyset\Rightarrow I^{+}\left(\dot{I}^{+}\left(T\right)\right)\cap\dot{I}^{+}\left(T\right)=\emptyset\right)!\right]$

$\Rightarrow\forall\gamma$ integral curve of $t^{a}$ ($\dot{\gamma}=t^{a}\partial_{a}$),
$\gamma\cap\dot{I}^{+}\left(T\right)=\left\{ \begin{array}{l}
q\\
\textrm{or}\\
\emptyset
\end{array}\right.$ (at most once)

while $\gamma\cap\Sigma=p$ by definition of $\Sigma$ Cauchy\fi 

\ifS
\else Define the map\fi 
\begin{align*}
\Psi:\dot{I}^{+}\left(T\right)\longrightarrow & \Sigma\\
\gamma\left(\tau_{1}\right)=q\longrightarrow & p=\gamma\left(\tau_{0}\right)
\end{align*}

\ifS
\else $S\subset\Sigma,S=\Psi\left[\dot{I}^{+}\left(T\right)\right]$
with topology induced by $\Sigma$\fi 

Then $\overline{\Psi}=\Psi:\dot{I}^{+}\left(T\right)\longrightarrow S$
homeomorphism

\ifS
\else $\dot{I}^{+}\left(T\right)$ compact $\Rightarrow S$
compact, with $S\subset\Sigma\Rightarrow S$ closed\fi 

\ifS
\else $\dot{I}^{+}\left(T\right)$ is $C^{0}$ from Lemma~\ref{lem:14Boundary-of-time},
so $\forall q\in\dot{I}^{+}\left(S\right),\exists\mathcal{O}\left(q\right)$
\fi homeomorphic to an open ball of $\mathbb{R}^{3}$\ifS
\else 

$\overline{\Psi}$ homeomorphism $\Rightarrow$ same holds for $S$,
hence $S$ open\fi 

However since\ifS
\else ~$M$
connected, from Lemma~\ref{lem:15Topology-of-globally} (Topology
of globally hyperbolic spacetimes) $\Sigma$ also connected\fi 

Therefore\ifS
\else ~$\dot{I}^{+}\left(T\right)\ne\emptyset\Rightarrow S=\Sigma$:
contradiction! as $S$ compact and $\Sigma$ non-compact\fi 

$\Rightarrow\exists\gamma,\dot{\gamma}^{2}=0$ future geodesic $\bot T$,
$\lambda\left[\gamma\right]\le\frac{2}{\left|\Theta_{0}\right|}$

i.e. there exists at least one future null geodesic that is \uline{incomplete}
\end{proof}
Again, the global hyperbolicity hypothesis is very strongHawking and Penrose found a weaker condition \ifS
\else that
we just give here (for proof, see \cite{hawking})\fi 

\subsection{More general Big Bang and Black Hole Singularity theorem}

\subsubsection{Definitions}

\begin{defn}
Timelike generic condition (T.G.C.)

$\left(M,g_{ab}\right)$ satisfies timelike generic condition \\
~\hspace*{\fill}$\Leftrightarrow\forall\gamma\subset M,\xi^{2}=\dot{\gamma}^{2}<0$
geodesic, $\exists p\in\gamma,R_{abcd}\xi^{a}\xi^{d}\left(p\right)\ne0$
\end{defn}
\begin{defn}
Null generic condition (N.G.C.)

$\left(M,g_{ab}\right)$ satisfies null generic condition \\
~\hspace*{\fill}$\Leftrightarrow\forall\mu\subset M,k^{2}=\dot{\mu}^{2}=0$,
$\exists p\in\mu,k_{[e}R_{a]bc[d}k_{f]}k^{b}k^{c}\left(p\right)\ne0$
\end{defn}
\begin{gather*}
\left(\Leftrightarrow\left\{ \begin{array}{l}
R_{ab}k^{a}k^{b}\left(p\right)\ne0\\
\textrm{or}\\
k_{[e}C_{a]bc[d}k_{f]}k^{b}k^{c}\left(p\right)\ne0
\end{array}\right.\right)
\end{gather*}

\subsubsection{Lemmas}
\begin{lem}
\label{lem:16Existence-of-conjugates}Existence of conjugates for
complete geodesics with S.E.C.

\hspace*{-3cm}$\left(M,g_{ab}\right)$ verifies $\left\{ \begin{array}{l}
\textrm{S.E.C.}\\
\textrm{T.G.C.}
\end{array}\right.\Rightarrow\left(\forall\gamma,\dot{\gamma}^{2}<0\textrm{ complete geodesic }\Rightarrow\exists p,q\in\gamma,p\ne q,X^{a}\left(p\right)=X^{a}\left(q\right)=0\textrm{ Jacobi field}\right)$
\end{lem}
\noindent 
\begin{lem}
\label{lem:17Existence-of-conjugates}Existence of conjugates for
complete null geodesics with N.E.C.

\hspace*{-3cm}$\left(M,g_{ab}\right)$ verifies $\left\{ \begin{array}{l}
\textrm{N.E.C.}\\
\textrm{N.G.C.}
\end{array}\right.\Rightarrow\left(\forall\mu,\dot{\mu}^{2}=0\textrm{ complete geodesic }\Rightarrow\exists p,q\in\mu,p\ne q,X^{a}\left(p\right)=X^{a}\left(q\right)=0\textrm{ Jacobi field}\right)$
\end{lem}
\noindent \ifS
\else Now we have the tools
to write the more general theorem\fi 

\subsubsection{More general Big Bang and Black Hole Singularity theorem}
\begin{thm}
\label{thm:Black-Hole-BB}Black Hole Singularity
theorem 2 (Hawking and Penrose 1970) \cite[Theorem 2 p266]{hawking}

$\left(M,g_{ab}\right)$ verifies 
\begin{enumerate}
\item \label{enu:S.E.C.}S.E.C.
\item \label{enu:Timelike-and-Null}Timelike and Null generic conditions
(and Lemmas~\ref{lem:16Existence-of-conjugates} and \ref{lem:17Existence-of-conjugates})
\item \label{enu:-closed-(no}$\nexists\gamma,\dot{\gamma}^{2}\le0,\gamma$
closed (no closed causal curves)
\item At least one of
\begin{enumerate}
\item $\exists S\subset M,S$ closed, achronal, $\mathrm{Edge}\left(S\right)=\emptyset$
($M$ closed universe)
\item $\exists S\subset M,S$ trapped surface
\item \label{enu:-future-(past)}$\exists p\in M,\forall\lambda_{\alpha}\subset M,p\in\lambda_{\alpha},\dot{\lambda}_{\alpha}^{2}=0$
future (past) directed $\Theta_{\lambda_{\alpha}}<0$
\end{enumerate}
\end{enumerate}
$\Rightarrow\exists\gamma,\dot{\gamma}^{2}<0$ or $\dot{\gamma}^{2}=0$
geodesic, $\gamma$ incomplete
\end{thm}
This theorem actually can also be applied to B.B. singularities: our
universe looks very much like an FLRW, almost flat ($k\simeq0$) in
our causal past, and seems to obey theorem \ref{thm:Black-Hole-BB}
conditions \ref{enu:S.E.C.}-\ref{enu:-closed-(no} as well as \ref{enu:-future-(past)}
for past events $p$ even after matter-radiation decoupling (past
directed $\Theta<0$). Normal matter is certainly obeying the S.E.C.

\ifS
\else 

Maybe escaped with %
\begin{tabular}[t]{l}
exotic matter?\tabularnewline
quantum gravity?\tabularnewline
bounce?\tabularnewline
\end{tabular}\fi 

\chapter[Kerr BH, Penrose Process \& BH Thermodyn.]{Kerr Black Hole, Penrose Process and Black Hole Thermodynamics}

\section{Rotating Black Hole: Kerr-Newman}

\subsection{Metric of charged, rotating BH}

\cite{Kerr:1963ud}, \cite{Newman:1965my}Axially symmetric stationary solutions to E.F.E. in vacuum
result in
\begin{align}
ds^{2}= & -\left(1-\frac{C}{\rho^{2}}\right)dt^{2}-\frac{Ca\sin^{2}\theta}{\rho^{2}}\left(dtd\phi+d\phi dt\right)+\frac{\rho^{2}}{\Delta}dr^{2}+\rho^{2}d\theta^{2}+\frac{\sin^{2}\theta}{\rho^{2}}\left[\left(r^{2}+a^{2}\right)^{2}-\Delta a^{2}\sin^{2}\theta\right]d\phi^{2}\label{eq:KerrNewmanMetric}\\
 & \begin{array}{rrl}
\textrm{with} & \rho^{2}\left(r,\theta\right)= & r^{2}+a^{2}\cos^{2}\theta\\
 & C\left(r\right)= & 2GMr-G\left(e^{2}+m^{2}\right)\textrm{ the BH charge decomposed into}\begin{array}[t]{rl}
e: & \textrm{electric charge}\\
m: & \textrm{magnetic charge}\\
M: & \textrm{mass charge}
\end{array}\\
 & \Delta\left(r\right)= & r^{2}+a^{2}-C
\end{array}\nonumber 
\end{align}

\subsection{Definitions of Charges}

\ifS
\else They are defined
with\fi 
\begin{thm}
Stokestheorem

\ifS\else $M$ $n$ \fi dimensional manifold,
$\partial M$ its boundary\ifS
\else 

$\omega$ $\left(n-1\right)$-form\fi 
\begin{align*}
\int_{M}d\omega= & \int_{\partial M}\omega
\end{align*}
\end{thm}
\ifS
\else There\fi 
exists a 1-form $V,\omega=\star V$ using the Hodge dual, i.e. the
 volume $n$-form. $V=\left(-1\right)^{s+n-1}\star\omega$
\ifS
\else $s$: signature\fi 

\begin{figure}
\begin{centering}
\includegraphics[scale=0.5]{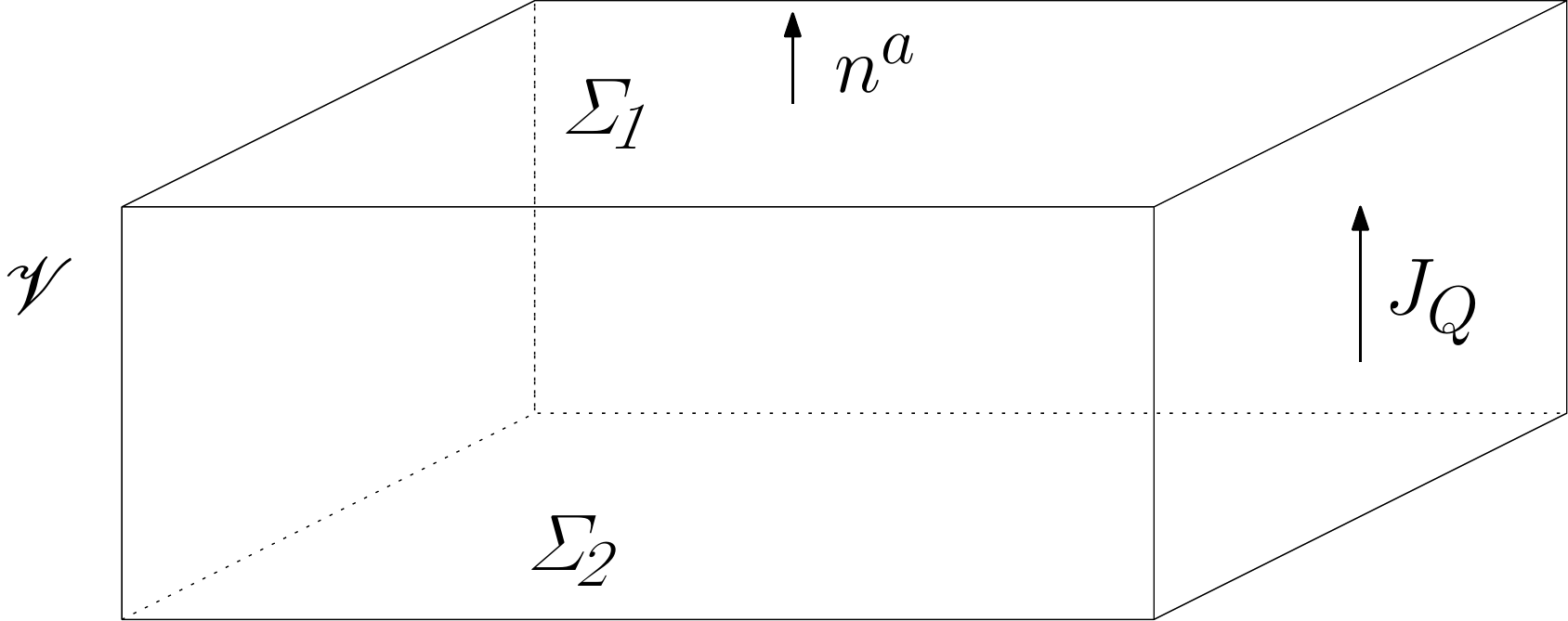}
\par\end{centering}
\caption{\label{fig:Volume-to-apply}Volume to apply Stokes Theorem}

\end{figure}

\begin{figure}
\begin{centering}
\includegraphics[scale=0.5]{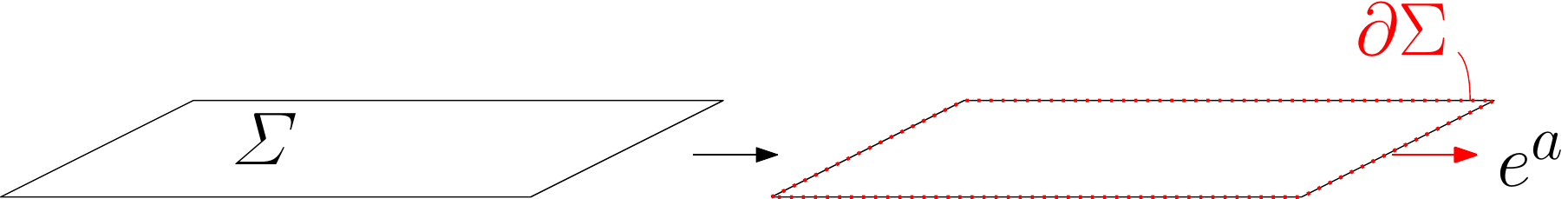}
\par\end{centering}
\caption{\label{fig:Iteration-on-a}Iteration on a surface of Stokes theorem}

\end{figure}

$\omega_{a_{1}\cdots a_{n-1}}dx^{a_{1}}\cdots dx^{a_{n-1}}=\eta_{ba_{1}\cdots a_{n-1}}V^{b}dx^{a_{1}}\cdots dx^{a_{n-1}}$
and $d\omega=d\star V\Leftrightarrow$
\begin{cor}
to Stokes theorem

\begin{align*}
\int_{M}d^{n}x\sqrt{\left|g\right|}\nabla_{a}V^{a}= & \int_{\partial M}d^{n-1}y\sqrt{\left|\gamma\right|}n_{a}V^{a}
\end{align*}
where %
\begin{tabular}[t]{l}
$g$ metric on $M$\tabularnewline
$\gamma$ metric on $\partial M$\tabularnewline
$n^{a}$ normal to $\partial M$\tabularnewline
\end{tabular}
\end{cor}
\ifS
\else so we define the charge on a hypersurface $\Sigma$ of $M$
as
\begin{align*}
Q=-\int_{\Sigma}d^{n-1}y\sqrt{\left|\gamma\right|}n_{a}J_{Q}^{a}= & -\int_{\Sigma}\star J_{Q}\\
= & -\int_{\partial\Sigma}d^{n-2}z\sqrt{\left|\gamma^{\left(\partial\Sigma\right)}\right|}n_{a}e_{b}S_{Q}^{ab}\\
\textrm{where }J_{Q}^{a}= & \nabla_{b}S_{Q}^{ab}
\end{align*}
\fi \ifS
\else using again Stokes for $\mathscr{V}$
interior of $\Sigma_{1},\Sigma_{2}$ and joining current lines (see
Fig.~\ref{fig:Volume-to-apply}; $J_{Q}$ current vector conserved:
$\nabla_{a}J_{Q}^{a}=0\Leftrightarrow d\star J_{Q}=0$)\fi 
\begin{align*}
0=\int_{\mathscr{V}}d\star J_{Q}=\int_{\partial\mathscr{V}}\star J_{Q}=\int_{\Sigma_{1}}\star J_{Q}-\int_{\Sigma_{2}}\star J_{Q}= & Q_{2}-Q_{1}
\end{align*}
\ifS
\else then (note $\mathscr{V}\ne M$) we apply
Stokes again to get the result (see Fig.~\ref{fig:Iteration-on-a})\fi 

\subsubsection{From conserved currents}

For electric and magnetic charges, the conserved currents are
\begin{align*}
J_{e}^{a}= & \nabla_{b}F^{ab}\textrm{, conserved as } & \nabla_{(a}\nabla_{b)}F^{\left[ab\right]}= & 0=\nabla_{a}J_{e}^{a}\\
J_{m}^{a}= & \nabla_{b}\star F^{ab}\textrm{, conserved as } & \nabla_{(a}\nabla_{b)}\star F^{\left[ab\right]}= & 0=\nabla_{a}J_{m}^{a}\\
\textrm{with }\star F^{ab}= & \frac{1}{2}\eta^{abcd}F_{cd}
\end{align*}
that define the electric and magnetic charges
\begin{align*}
e= & -\int_{\Sigma}d^{n-1}y\sqrt{\left|\gamma\right|}n_{a}\nabla_{b}F^{ab}=-\int_{\partial\Sigma}d^{n-2}z\sqrt{\left|\gamma^{\left(\partial\Sigma\right)}\right|}n_{a}e_{b}F^{ab}
\end{align*}
\ifS
\else (sending
$\partial\Sigma$ at infinity $ds^{2}$ is Minkowski-like and we recover
the correct sign)\fi 
\begin{align*}
m= & -\int_{\partial\Sigma}d^{n-2}z\sqrt{\left|\gamma^{\left(\partial\Sigma\right)}\right|}n_{a}e_{b}\frac{1}{2}\eta^{abcd}F_{cd}=-\int_{\partial\Sigma}d^{n-2}z\sqrt{\left|\gamma^{\left(\partial\Sigma\right)}\right|}\frac{\eta^{ab}F_{ab}}{2}.
\end{align*}
For the rest,
unless specified, we set $e=m=0$ but keep general charge $C$ when its expression is not needed.

\subsubsection{Killing vectors, conserved mass and spin}

\ifS
\else As the metric \fi is independent
of $t,\phi$, it admits $K=\partial_{t}$ and $L=\partial_{\phi}$
as Killing vectors (recall definition in Sec.~\ref{Def:Killingvector:-Any-symmetry}).

This is used to define
\begin{itemize}
\item the mass as Komar energy
\item the spin $a=\frac{J}{GM}$ from $J$, Komar
angular momentum
\end{itemize}
The conserved currents are based on the Ricci tensor

\ifS
\else $J_{KR}^{a}=K_{b}R^{ab}$
for $K$ the timelike Killing vector: Komar energy\fi , valid for
asymptotically flat spacetimes

\ifS
\else $J_{KR}^{a}$ conserved as\fi 
\begin{align*}
\nabla_{a}J_{KR}^{a}= & \underset{0\text{ from Killing Eq.}}{\underbrace{\nabla_{[a}K_{b]}R^{\left(ab\right)}}}+\underset{0\text{ from Bianchi Id.}}{\underbrace{K_{b}\nabla_{a}R^{ab}}}\\
= & 0
\end{align*}

\begin{proof}
From \ifS
\else Bianchi
Id. \fi $\nabla_{a}R^{ab}=\frac{1}{2}\nabla^{b}R$\ifS
\else ~so we need \fi $K_{b}\nabla^{b}R\overset{?}{=}0$

\ifS
\else From the contracted Bianchi Id., we have $\nabla_{[a}R_{bc]}^{\quad bc}=0$
\begin{align*}
\Leftrightarrow K^{a}\nabla_{a}R+K^{a}\nabla_{b}R_{ca}^{\quad bc}+K^{a}\nabla_{c}R_{ab}^{\quad bc}= & 0
\end{align*}
\fi 
\begin{align*}
\Leftrightarrow K^{a}\nabla_{a}R= & K^{a}\nabla_{b}R_{ac}^{\quad bc}+K^{a}\nabla_{c}R_{ab}^{\quad cb}\textrm{ (antisym. of Riemann)}\\
= & 2K^{a}\nabla_{b}R_{a}^{\:b}\textrm{ (Riemann}\to\textrm{ Ricci)}\\
= & 2\nabla_{b}\left(K^{a}R_{a}^{\:b}\right)-2\underset{0\text{ from Killing Eq.}}{\underbrace{R^{\left(ab\right)}\nabla_{[b}K_{a]}}}
\end{align*}
and since Ricci Id. gives
\begin{eqnarray*}
(\nabla_{a}\nabla_{b}K_{c}-\underset{0\text{ from Killing Eq.}}{\underbrace{\nabla_{b}\nabla_{a}K_{c}}} & = & (R_{abcd}\\
+\underset{0\text{ from Killing Eq.}}{\underbrace{\nabla_{c}\nabla_{b}K_{a}}}-\overbrace{\nabla_{b}\nabla_{c}K_{a}} &  & +R_{cbad}\\
+\overbrace{\nabla_{c}\nabla_{a}K_{b}}\underset{=+\nabla_{a}\nabla_{b}K_{c}\text{ from Killing Eq.}}{\underbrace{-\nabla_{a}\nabla_{c}K_{b}}}= &  & +R_{cabd})K^{d}\\
\Leftrightarrow2\nabla_{a}\nabla_{b}K_{c} & = & \left(R_{cbad}+\underset{\begin{array}[t]{c}
=-R_{cbda}\\
=R_{cbad}
\end{array}\textrm{ from }R_{c\left[dab\right]}}{\underbrace{R_{cdab}+R_{cabd}}}\right)K^{d}\textrm{ from Riemann symmetries}\\
 & = & 2R_{cbad}K^{d}\\
\Rightarrow\nabla_{b}\nabla_{a}K^{b} & = & R_{\:abd}^{b}K^{d}=R_{ab}K^{b}
\end{eqnarray*}
Therefore
\begin{align*}
K^{a}\nabla_{a}R=2\nabla_{b}\left(K^{a}R_{a}^{\:b}\right)= & 2\nabla_{(b}\nabla_{c)}\nabla^{[b}K^{c]}\\
= & 0\text{ from Killing Eq.}
\end{align*}
Thus we get
\begin{align*}
\nabla_{b}\left(K^{a}R_{a}^{\:b}\right)= & \frac{1}{2}K^{a}\nabla_{a}R=0=\nabla_{a}J_{KR}^{a}
\end{align*}
\end{proof}
For the spacelike (rotation) Killing vector

$J_{LR}^{a}=L_{b}R^{ab}$ for $K$ yields Komar angular momentum,
conserved as
\begin{align*}
\nabla_{a}J_{LR}^{a}= & \underset{0\text{ from Killing Eq.}}{\underbrace{\nabla_{[a}L_{b]}R^{\left(ab\right)}}}+L_{b}\nabla_{a}R^{ab}\\
= & \frac{L^{a}}{2}\nabla_{a}R=0\text{ as proof only uses Killing Eq.}
\end{align*}
They define the conserved mass and spin as
\begin{align*}
J_{KR}^{a}= & \nabla_{b}\nabla^{a}K^{b}\Rightarrow S_{M}^{ab}=\nabla^{a}K^{b}\textrm{ and } & J_{LR}^{a}= & \nabla_{b}\nabla^{a}L^{b}\Rightarrow S_{J}^{ab}=\nabla^{a}L^{b}
\end{align*}
\begin{gather*}
\left\{ \begin{array}{l}
\boxed{M=-\frac{1}{4\pi G}\int_{\partial\Sigma}d^{2}z\sqrt{\left|\gamma^{\left(\partial\Sigma\right)}\right|}n_{a}e_{b}\nabla^{a}K^{b}}\\
\boxed{J=-\frac{1}{8\pi G}\int_{\partial\Sigma}d^{2}z\sqrt{\left|\gamma^{\left(\partial\Sigma\right)}\right|}n_{a}e_{b}\nabla^{a}L^{b}}\textrm{ with }\boxed{a=\frac{GJ}{GM}=\frac{J}{M}}
\end{array}\right.
\end{gather*}
For $\begin{array}[t]{l}
e\ne0\\
m\ne0
\end{array}$, defining $F_{ab}=2\partial_{[a}A_{b]}$ one can get
\begin{align*}
A_{a}= & \frac{er-ma\cos\theta}{\rho^{2}}dt_{a}+\frac{-ear\sin^{2}\theta+m\left(r^{2}+a^{2}\right)\cos^{2}\theta}{\rho^{2}}d\phi_{a}
\end{align*}
We keep $e=m=0$ for the rest.

\subsection{Boyer-Lindquist coordinates}

$\left(t,r,\theta,\phi\right)$ are the Boyer-Lindquist coordinates.

For $a\to0$ the metric reduces to Schwarzschild \textcolor{green}{\uline{(verify)}}\exo 

\subsubsection{Analysis of Minkowski space in Boyer-Lindquist coordinates}

\begin{figure}
\begin{centering}
\includegraphics[scale=0.5]{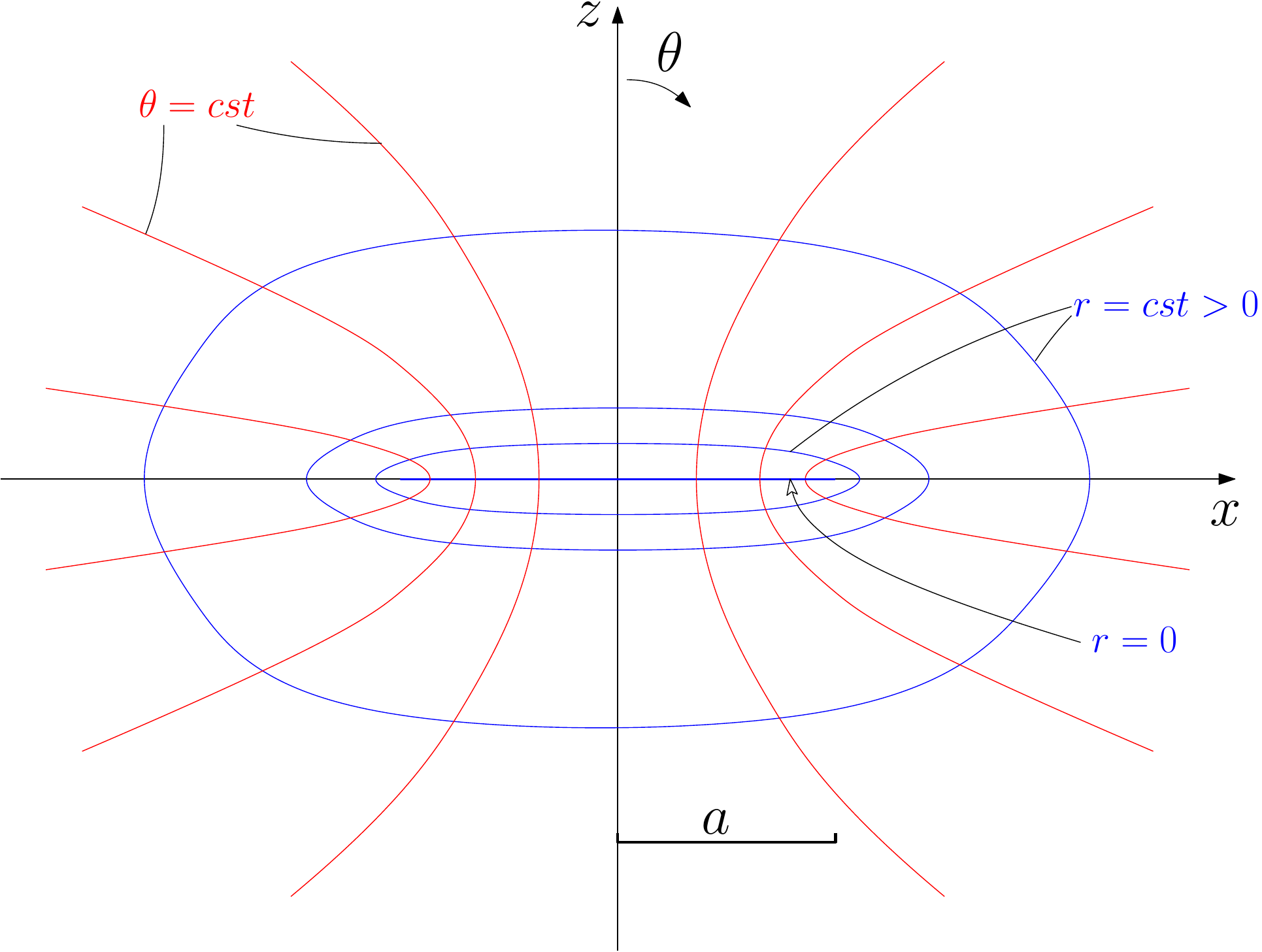}
\par\end{centering}
\caption{\label{fig:-xzplane-of}$x-z$ plane of Boyer-Lindquist coordinates}

\end{figure}

For $\begin{array}[t]{l}
M\\
\left(C\right)
\end{array}\to0$ they reduce to flat spacetime but
\begin{align*}
ds^{2}= & -dt^{2}+\frac{r^{2}+a^{2}\cos^{2}\theta}{r^{2}+a^{2}}dr^{2}+\left(r^{2}+a^{2}\cos^{2}\theta\right)d\theta^{2}+\left(r^{2}+a^{2}\right)\sin^{2}\theta d\phi^{2}
\end{align*}
is not simply Minkowski: it is so using ellipsoidal coordinates (in
spacelike hyperplanes)
\begin{align*}
\left\{ \begin{array}{rl}
x= & \sqrt{r^{2}+a^{2}}\sin\theta\cos\phi\\
y= & \sqrt{r^{2}+a^{2}}\sin\theta\sin\phi\\
z= & r\cos\theta
\end{array}\right.\Rightarrow ds^{2}= & -dt^{2}+dx^{2}+dy^{2}+dz^{2}
\end{align*}
They represent:
\begin{itemize}
\item revolution ellipsoids for $r=cst$
\begin{align*}
\left(\frac{x}{\sqrt{r^{2}+a^{2}}}\right)^{2}+\left(\frac{y}{\sqrt{r^{2}+a^{2}}}\right)^{2}+\left(\frac{z}{r}\right)^{2}= & 1
\end{align*}
\item revolution hyperboloids for $\theta=cst$
\begin{align*}
\left(\frac{x}{a\sin\theta}\right)^{2}+\left(\frac{y}{a\sin\theta}\right)^{2}-\left(\frac{z}{a\cos\theta}\right)^{2}= & \left(\frac{r}{a}\right)^{2}+1-\left(\frac{r}{a}\right)^{2}=1
\end{align*}
\item circles for $\theta,r=cst$
\begin{align*}
\frac{x^{2}}{\left(r^{2}+a^{2}\right)\sin^{2}\theta}+\frac{y^{2}}{\left(r^{2}+a^{2}\right)\sin^{2}\theta}= & 1
\end{align*}
\end{itemize}
\uline{so in the plane \mbox{$\phi=0\Rightarrow y=0:\left(x,z\right)$}
plane}

we restrict ellipsoids to ellipses at $r=cst:\left(\frac{x}{\sqrt{r^{2}+a^{2}}}\right)^{2}+\left(\frac{z}{r}\right)^{2}=1$
of semi axes $r,\sqrt{r^{2}+a^{2}}$ and thus semifocus $\sqrt{\left(\sqrt{r^{2}+a^{2}}\right)^{2}-r^{2}}=a$

while hyperboloids become hyperbolae at $\theta=cst$: $\begin{array}[t]{rl}
\theta= & 0\Rightarrow x=0:\,z\textrm{ axis}\\
\theta= & \frac{\pi}{2}\Rightarrow z=0:\,x\textrm{ axis}
\end{array}$ $\left(\frac{x}{a\sin\theta}\right)^{2}-\left(\frac{z}{a\cos\theta}\right)^{2}=1$
with linear excentricity (origin to focus distance) $\sqrt{\left(a\sin\theta\right)^{2}+\left(a\cos\theta\right)^{2}}=a$,
asymptote slope $\frac{a\cos\theta}{a\sin\theta}=\cot\theta$ and
the $r=0$ locus defined as
\begin{align*}
z= & 0\\
x\in & \left[-a;a\right] & x^{2}+y^{2}= & a^{2}\sin^{2}\theta\le a^{2}:\textrm{ Disk}\\
y\in & \left[-a;a\right]
\end{align*}
(see Fig.~\ref{fig:-xzplane-of})

\begin{figure}
\begin{centering}
\includegraphics[scale=0.5]{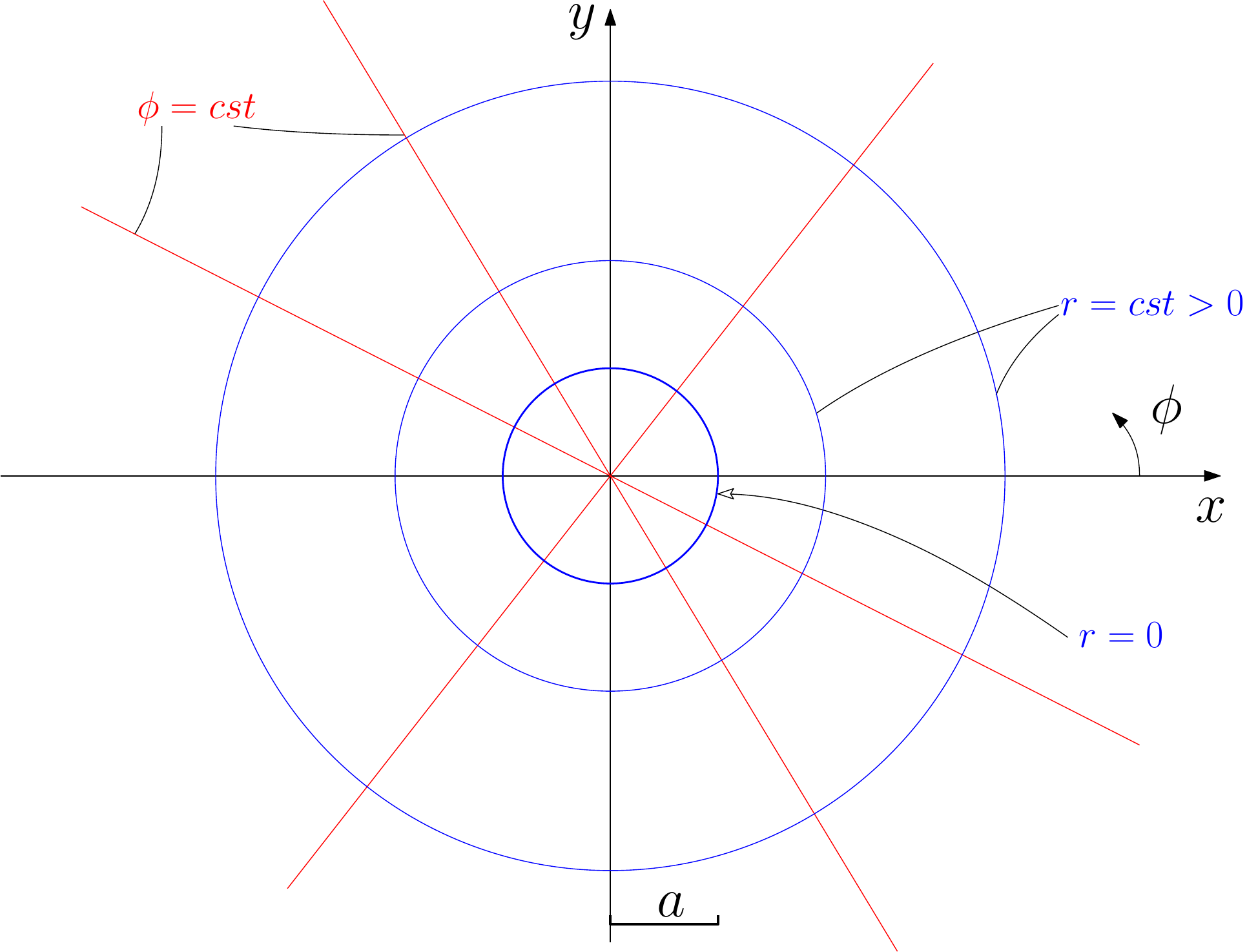}
\par\end{centering}
\caption{\label{fig:-xyplane-of}$x-y$ plane of Boyer-Lindquist coordinates}
\end{figure}

\uline{In \mbox{$\theta=\frac{\pi}{2}$} plane \mbox{$\Rightarrow r=0:\left(x,y\right)$}
plane}

we restrict the ellipsoids to circles at $\theta,r=cst:\frac{x^{2}}{r^{2}+a^{2}}+\frac{y^{2}}{r^{2}+a^{2}}=1$
of radius $R_{c}=\sqrt{r^{2}+a^{2}}\ge a$

and for $\phi=cst$, we have straight lines of slope $\frac{y}{x}=\tan\phi$
(see Fig.~\ref{fig:-xyplane-of})

\subsubsection{Killing vectors/tensor of Boyer-Lindquist coordinates}

We have seen that $K=\partial_{t}$ and $L=\partial_{\phi}$ are Killing
vectors

Defining the repeated principal null vectors
\begin{align*}
k^{a}= & \frac{1}{\Delta}\left(\underset{\begin{array}{c}
\shortparallel\\
R_{c}^{2}
\end{array}}{\left(r^{2}+a^{2}\right)}K^{a}+\Delta\partial_{r}+aL^{a}\right)\\
l^{a}= & \frac{1}{2\rho^{2}}\left(R_{c}^{2}K^{a}-\Delta\partial_{r}+aL^{a}\right)
\end{align*}
one can verify $k^{a}k_{a}=0=l^{a}l_{a}$ and $k^{a}l_{a}=-1$ ($k,l$
null) and $\sigma_{ab}=2\rho^{2}k_{(a}l_{b)}+r^{2}g_{ab}$ verifies
$\sigma_{\left(ab;c\right)}=0$ so $\sigma_{ab}$ is a Killing tensor

\subsection{Horizons, stationary limit and singularity}

\subsubsection{Horizons}

We have event horizons for $g^{rr}=0$

(recall Schwarzschild: $g^{rr}=1-\frac{2GM}{r}$)

Here $g^{rr}=\frac{\Delta}{\rho^{2}}=0\Rightarrow\Delta=0$
\begin{align*}
\Leftrightarrow r^{2}-2GMr+a^{2}= & 0=\left(r-GM\right)^{2}+a^{2}-\left(GM\right)^{2}\\
\Leftrightarrow\left(r-GM\right)^{2}= & \left(GM\right)^{2}-a^{2}\begin{array}[t]{ll}
<0 & :\textrm{ Naked singularity}\\
=0 & :\textrm{ unstable, so we restrict to}\\
>0 & :\textrm{ two horizons}
\end{array}\\
\Rightarrow r_{\pm}= & GM\pm\sqrt{\left(GM\right)^{2}-a^{2}}\textrm{ : null surfaces}
\end{align*}
We have 2 event horizons

Since metric is stationary but not static, $r_{\pm}$ not Killing
Horizons for $K^{a}$

In fact, since 
\begin{align*}
g_{tt}= & -\left(1-\frac{C}{\rho^{2}}\right)=-\frac{1}{\rho^{2}}\left(\rho^{2}-C\right)=-\frac{1}{\rho^{2}}\left(r^{2}+a^{2}\cos^{2}\theta-C\right)\\
= & -\frac{\left(r^{2}+a^{2}-C-a^{2}\sin^{2}\theta\right)}{\rho^{2}}=-\frac{1}{\rho^{2}}\left(\Delta-a^{2}\sin^{2}\theta\right)
\end{align*}
we have at horizons
\begin{align*}
K^{a}K_{a}= & g_{tt}=-\frac{1}{\rho^{2}}\left(\Delta-a^{2}\sin^{2}\theta\right)=\left\Vert K\right\Vert ^{2}\ne0\textrm{ (at }\Delta=0)
\end{align*}
and in fact already
\begin{align*}
\left\Vert K\right\Vert ^{2}= & \frac{a^{2}}{\rho^{2}}\sin^{2}\theta\ge0\textrm{ : }K\textrm{ is }\begin{array}[t]{l}
\textrm{spacelike at }r_{\pm}\\
\textrm{null at }r_{\pm},\theta=\left\{ \begin{array}[t]{l}
0\\
\pi
\end{array}\right.\textrm{: poles of }z\textrm{ axis}
\end{array}
\end{align*}

\subsubsection{Stationary limit}

$\left\Vert K\right\Vert ^{2}=0$ defines the stationary limit surface
\begin{align*}
\Rightarrow\Delta= & a^{2}\sin^{2}\theta\Leftrightarrow\left(r-GM\right)^{2}=\left(GM\right)^{2}-a^{2}\cos^{2}\theta
\end{align*}
\fbox{\begin{minipage}[t]{5cm}%
Take + solution: $r_{e}\ge r_{+}$%
\end{minipage}}

Defines circles at $\theta=cst$ in $\left(x,y\right)$ planes (for
$z=r\cos\theta$) with maximum radius at $\theta=\frac{\pi}{2}$ ($\Rightarrow r=0$)
\begin{align*}
r_{e,max}= & \left.r_{e}\right|_{\theta=\frac{\pi}{2}}=2GM=r_{Sch}
\end{align*}

\paragraph{In the $\phi=0$ plane ($y=0$)}

$r_{e}$ goes from $x$ maximum to $x=0$ ($\Rightarrow\theta=0$)
and so %
\fbox{\begin{minipage}[t]{2cm}%
$z=r_{+}=r_{e}$%
\end{minipage}} touches the outer Horizon

\paragraph{In general: }

the stationary limit
\begin{align*}
r_{e}= & GM+\sqrt{\left(GM\right)^{2}-a^{2}\cos^{2}\theta}\\
 & \left\{ \begin{array}{rl}
x= & r_{e}\left(\theta\right)\sin\theta\\
z= & r_{e}\left(\theta\right)\cos\theta
\end{array}\right.\theta\in\left[0;2\pi\right]
\end{align*}
This surface contains the Ergosphere, where pure radial motion is
not possible, only in the direction of rotation $a$, but $r<r_{+}$
still can escape (see Fig.~\ref{fig:Horizons,-stationary-limit})
\begin{figure}
\begin{centering}
\begin{tabular}{>{\centering}m{0.46\columnwidth}}
\hspace*{-1.9cm}\includegraphics[width=0.9\columnwidth]{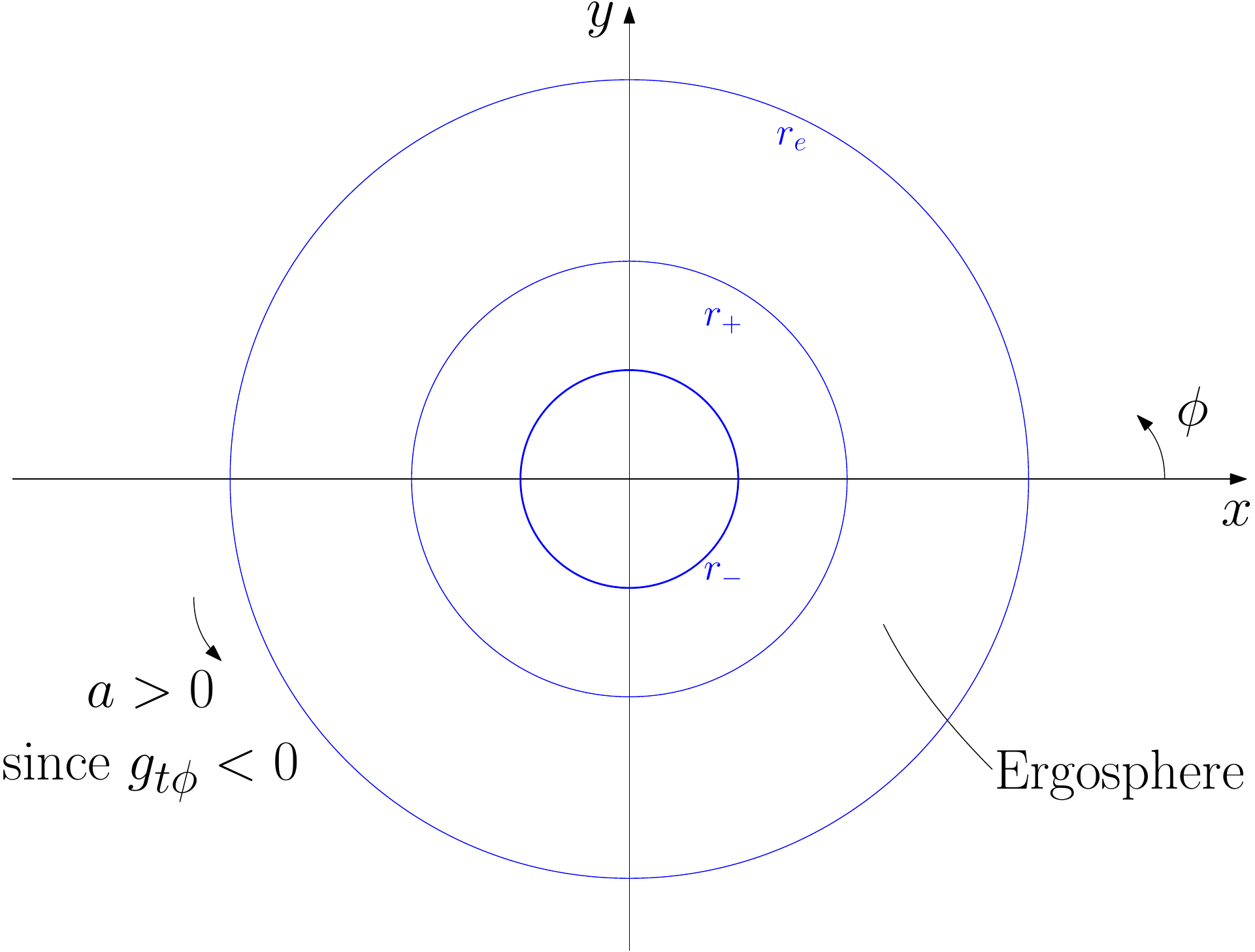}\tabularnewline
\includegraphics[width=0.6\columnwidth]{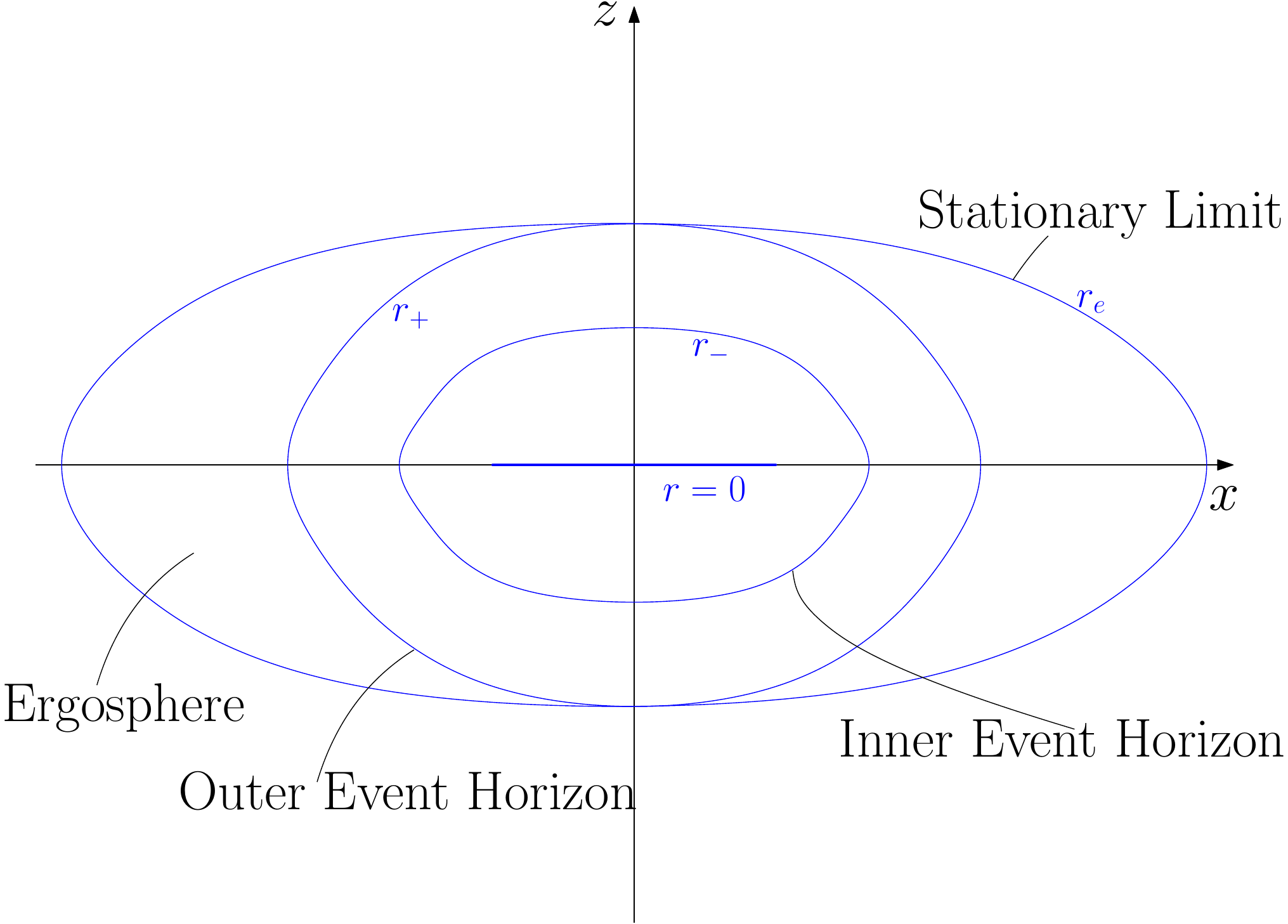}\tabularnewline
\end{tabular}
\par\end{centering}
\caption{\label{fig:Horizons,-stationary-limit}Horizons, stationary limit
and Ergosphere}

\end{figure}

\subsubsection{True singularity}

$\rho^{2}=0=r^{2}+a^{2}\cos^{2}\theta$ is a true singularity: $R^{abcd}R_{abcd}\underset{\rho\to0}{\longrightarrow}\infty$
\begin{align*}
\rho^{2}=0\Leftrightarrow & \left\{ \begin{array}{rl}
r= & 0\\
\textrm{and}\\
\theta= & \frac{\pi}{2}\left(\pi\right)
\end{array}\right.
\end{align*}
$r=0$:
\begin{align*}
\left\{ \begin{array}{rl}
z= & 0\\
x= & a\sin\theta\cos\phi\\
y= & a\sin\theta\sin\phi
\end{array}\right.: & x^{2}+y^{2}=a^{2}\sin^{2}\theta\;\theta\in\left[0;\pi\right]\;\boxed{\textrm{Disk}}
\end{align*}
$\theta=\frac{\pi}{2}\left(\pi\right)$:
\begin{align*}
\left\{ \begin{array}{rlcc}
z= & 0\\
x= & \sqrt{r^{2}+a^{2}}\cos\phi\; & r^{2}\in & \mathbb{R}^{+}\\
y= & \sqrt{r^{2}+a^{2}}\sin\phi & \phi\in & \left[0;2\pi\right]
\end{array}\right.: & x^{2}+y^{2}=r^{2}+a^{2}=R_{c}^{2},\textrm{ circles of}\:R_{c}\ge a\;\boxed{\textrm{Plane }z=0-\textrm{Disk}}
\end{align*}
$\left.\begin{array}{rl}
r= & 0\\
\theta= & \frac{\pi}{2}\left(\pi\right)
\end{array}\right\} \Rightarrow$
\begin{align*}
\left\{ \begin{array}{rl}
z= & 0\\
x= & a\cos\phi\\
y= & a\sin\phi
\end{array}\right.: & \;\boxed{\textrm{Ring}}
\end{align*}
so $\rho=0$: %
\fbox{\begin{minipage}[t]{2.5cm}%
Ring Singularity%
\end{minipage}}

Only seen from $\theta=\frac{\pi}{2}\left(\pi\right)$?

\paragraph{In $y=0$ plane ($\phi=0$): }

If we go to the ring at fixed $x$:
\begin{align*}
x=a=\sqrt{r^{2}+a^{2}}\sin\theta\Leftrightarrow & r^{2}+a^{2}=\frac{a^{2}}{\sin^{2}\theta}=a^{2}\left(1+\frac{\cos^{2}\theta}{\sin^{2}\theta}\right)\\
\Leftrightarrow & r^{2}=a^{2}\frac{\cos^{2}\theta}{\sin^{2}\theta}\\
\Rightarrow & \cos\theta=\pm\frac{r}{a}\left|\sin\theta\right|\\
\Rightarrow z=r\cos\theta= & a\cos\theta\frac{\left|\cos\theta\right|}{\left|\sin\theta\right|}\underset{\theta\to\frac{\pi}{2}}{\longrightarrow}0\\
= & \pm\frac{r^{2}}{a}\left|\sin\theta\right|\underset{r\to0}{\longrightarrow}0:\;\textrm{hit singularity Ring}
\end{align*}

\subsubsection{Analytic continuation and timelike loops}

Analytic continuation allows to go through the disk to $r<0$ for
$\theta\ne\frac{\pi}{2}\left(\pi\right)$ like in Schwarzschild.

However $r<0\Rightarrow\Delta>0$: no horizons!

Moreover for $r<0$ and small, fixed $\theta$ and $t$ gives $dt=dr=d\theta=0$
and $\theta$ close to $\frac{\pi}{2}$ ($z=0$ plane) we get
\begin{align*}
ds^{2}\simeq & \frac{\sin^{2}\theta}{r^{2}+a^{2}\cos^{2}\theta}\left[\left(r^{2}+a^{2}\right)^{2}-\left(r^{2}+a^{2}-2GMr\right)a^{2}\sin^{2}\theta\right]d\phi^{2}\\
= & \frac{\sin^{2}\theta}{r^{2}+a^{2}\cos^{2}\theta}\left[\left(r^{2}+a^{2}\right)\left(r^{2}+a^{2}-a^{2}\sin^{2}\theta\right)+2GMra^{2}\sin^{2}\theta\right]d\phi^{2}\\
= & \sin^{2}\theta\left[r^{2}+a^{2}+\frac{2GMra^{2}\sin^{2}\theta}{r^{2}+a^{2}\cos^{2}\theta}\right]d\phi^{2}
\end{align*}
with the restriction
\begin{align*}
\theta\sim & \frac{\pi}{2}\Rightarrow\left\{ \begin{array}{rl}
\cos^{2}\theta\ll & 1\\
\sin^{2}\theta\sim & 1
\end{array}\right.\\
0<-r & \ll1
\end{align*}
\begin{align*}
ds^{2}= & a^{2}\left(1-\cos^{2}\theta\right)\left[1+\left(\frac{r}{a}\right)^{2}+\frac{2GM}{r\left(\frac{1}{1-\cos^{2}\theta}+\left(\frac{a}{r}\right)^{2}\frac{\cos^{2}\theta}{1-\cos^{2}\theta}\right)}\right]d\phi^{2}\quad\textrm{take }\frac{a^{2}\cos^{2}\theta}{r^{2}}\ll1\\
\simeq & a^{2}\left[1+\frac{2GM}{r}\right]d\phi^{2}<0
\end{align*}
so $L^{a}$ timelike for a region of $r<0$ around $z=0$. Since $L^{a}$
goes in circles; it creates the possibility of timelike loops!

\subsubsection{Ergosphere, framedragging and outer horizon angular velocity}

\paragraph{Another look at the Ergosphere}

There
\begin{align*}
K^{a}K_{a}=g_{tt}=\frac{1}{\rho^{2}}\left(a^{2}\sin^{2}\theta-\Delta\right) & \ge0:\;r_{+}<r\le r_{e}\\
 & <\frac{a^{2}\sin^{2}\theta}{\rho^{2}}
\end{align*}
Thus all metric terms are positive except the cross terms.

For a timelike observer $t^{a}=\frac{dx^{a}}{d\tau}$
\begin{align*}
g_{ab}t^{a}t^{b}<0\Rightarrow2g_{t\phi}t^{t}t^{\phi}= & 2g_{t\phi}\frac{dt}{d\tau}\frac{d\phi}{d\tau}\\
= & 2g_{t\phi}\left(t^{a}\nabla_{a}t\right)\left(t^{b}\nabla_{b}\phi\right)<0
\end{align*}
but $\nabla_{a}t=\partial_{a}t=\left(\begin{array}{c}
1\\
\vec{0}
\end{array}\right)\Rightarrow t^{a}\nabla_{a}t=t^{t}\ge0$ since $t^{a}$ future timelike and $g_{t\phi}<0\Rightarrow\boxed{\frac{d\phi}{d\tau}>0}$

\paragraph{In the $\theta=\frac{\pi}{2}$ plane ($z=0$)}

Take null rotating curves: photon emitted in $L^{a}$ direction
\begin{align*}
ds^{2}= & g_{tt}dt^{2}+g_{t\phi}\left(dt\,d\phi+d\phi\,dt\right)+g_{\phi\phi}d\phi^{2}=0\\
\Leftrightarrow & \left(\frac{d\phi}{dt}\right)^{2}+2\frac{g_{t\phi}}{g_{\phi\phi}}\frac{d\phi}{dt}+\frac{g_{tt}}{g_{\phi\phi}}=0\\
\Rightarrow\frac{d\phi}{dt}= & -\frac{g_{t\phi}}{g_{\phi\phi}}\pm\sqrt{\left(\frac{g_{t\phi}}{g_{\phi\phi}}\right)^{2}-\frac{g_{tt}}{g_{\phi\phi}}}>0\;\textrm{for }g_{tt}>0\textrm{ and }g_{t\phi}<0
\end{align*}
Photons emitted in opposite directions rotate in the same direction

Furthermore, on the stationary limit ($g_{tt}=\left\Vert K\right\Vert ^{2}=0$)
$\Delta=a^{2}\sin^{2}\theta$ so
\begin{align*}
\frac{d\phi}{dt}= & \left\{ \begin{array}{c}
0\\
\textrm{or}\\
-2\frac{g_{t\phi}}{g_{\phi\phi}}=\frac{2Ca}{\left(r^{2}+a^{2}\right)^{2}-\Delta a^{2}\sin^{2}\theta}
\end{array}\right.
\end{align*}
Since $\Delta=r^{2}+a^{2}-C$ and $\theta=\frac{\pi}{2}$
\begin{align*}
\Rightarrow\left.\frac{d\phi}{dt}\right|_{e}= & \frac{2Ca}{\left(\Delta+C\right)^{2}-\Delta a^{2}\sin^{2}\theta}\underset{{\scriptscriptstyle \Delta=a^{2}\sin^{2}\theta}}{=}\frac{2Ca}{C\left(C+2a^{2}\sin^{2}\theta\right)}\\
\underset{{\scriptscriptstyle \theta=\frac{\pi}{2}}}{=} & \frac{2a}{C+2a^{2}}
\end{align*}
we have
\begin{align*}
C= & 2GMr_{e}\\
= & 2GM\left(GM+\sqrt{\left(GM\right)^{2}-a^{2}\cos^{2}\theta}\right)\\
\underset{{\scriptscriptstyle \theta=\frac{\pi}{2}}}{=} & \left(2GM\right)^{2}
\end{align*}
so $\frac{d\phi}{dt}=\frac{a}{2\left(GM\right)^{2}+a^{2}}$ with the
same sign as $a$: photons rotate in the same direction as B.H. rotation

\fbox{\begin{minipage}[t]{8cm}%
This illustrates the ``dragging of the inertial frame''%
\end{minipage}}

Locally nonrotating future pointing observer far from the B.H. can
be chosen as $v^{a}=-\frac{\nabla^{a}t}{\sqrt{-\nabla^{b}t\nabla_{b}t}}$
since we have seen $t^{b}\nabla_{b}t>0$ so $\left(\nabla^{b}t\right)\delta_{a}^{t}<0$
\begin{align*}
v^{a}= & -\frac{g^{at}}{\sqrt{-g^{tt}}}=\left(\begin{array}{c}
\frac{dt}{d\tau}\\
0\\
0\\
\frac{d\phi}{d\tau}
\end{array}\right)=\left(\begin{array}{c}
\sqrt{-g^{tt}}\\
0\\
0\\
-\frac{g^{t\phi}}{\sqrt{-g^{tt}}}
\end{array}\right)\underset{{\scriptscriptstyle r=\infty}}{\longrightarrow}\left(\begin{array}{c}
1\\
\vec{0}
\end{array}\right)
\end{align*}
Angular velocity is
\begin{align*}
\frac{d\phi}{dt}= & \frac{\frac{d\phi}{d\tau}}{\frac{dt}{d\tau}}=\frac{g^{t\phi}}{g^{tt}}
\end{align*}
so we need to invert the metric from
\begin{align*}
g_{ab}= & \left(\begin{array}{cccc}
g_{tt} & g_{t\phi} & 0 & 0\\
g_{t\phi} & g_{\phi\phi} & 0 & 0\\
0 & 0 & g_{rr} & 0\\
0 & 0 & 0 & g_{\theta\theta}
\end{array}\right)
\end{align*}
into
\begin{align*}
g^{ab}= & \left(\begin{array}{cccc}
g^{tt} & g^{t\phi} & 0 & 0\\
g^{t\phi} & g^{\phi\phi} & 0 & 0\\
0 & 0 & \frac{1}{g_{rr}} & 0\\
0 & 0 & 0 & \frac{1}{g_{\theta\theta}}
\end{array}\right)\Rightarrow\left\{ \begin{array}{rl}
g^{tt}g_{tt}+g_{t\phi}g^{t\phi}= & 1\\
g_{\phi\phi}g^{t\phi}+g_{t\phi}g^{tt}= & 0\\
g_{tt}g^{t\phi}+g_{t\phi}g^{\phi\phi}= & 0\\
g_{t\phi}g^{t\phi}+g_{\phi\phi}g^{\phi\phi}= & 1
\end{array}\right.\\
\Rightarrow-g^{tt}\frac{g_{t\phi}}{g_{\phi\phi}}= & g^{t\phi}=-g^{\phi\phi}\frac{g_{t\phi}}{g_{tt}}
\end{align*}
and so $g^{tt}\left(g_{tt}-\frac{g_{t\phi}^{2}}{g_{\phi\phi}}\right)=1=g^{\phi\phi}\left(g_{\phi\phi}-\frac{g_{t\phi}^{2}}{g_{tt}}\right)$
\begin{align*}
\Rightarrow g^{ab}= & \left(\begin{array}{cccc}
\frac{g_{\phi\phi}}{g_{tt}g_{\phi\phi}-g_{t\phi}^{2}} & \frac{-g_{t\phi}}{g_{tt}g_{\phi\phi}-g_{t\phi}^{2}} & 0 & 0\\
\frac{-g_{t\phi}}{g_{tt}g_{\phi\phi}-g_{t\phi}^{2}} & \frac{g_{tt}}{g_{tt}g_{\phi\phi}-g_{t\phi}^{2}} & 0 & 0\\
0 & 0 & \frac{1}{g_{rr}} & 0\\
0 & 0 & 0 & \frac{1}{g_{\theta\theta}}
\end{array}\right)
\end{align*}
We therefore get
\begin{align*}
\Omega=\frac{d\phi}{dt}=\frac{-g_{t\phi}}{g_{\phi\phi}}= & \frac{Ca}{\left(r^{2}+a^{2}\right)^{2}-\Delta a^{2}\sin^{2}\theta}\;\textrm{and }\Delta=r^{2}+a^{2}-C\\
= & \frac{a\left(r^{2}+a^{2}-C\right)}{\left(r^{2}+a^{2}\right)^{2}-\Delta a^{2}\sin^{2}\theta}
\end{align*}
on outer horizon, $r=r_{+}$, $\Delta=0$
\begin{gather*}
\Rightarrow\boxed{\Omega_{H}=\frac{a}{r_{+}^{2}+a^{2}}}
\end{gather*}

\subsection{Carter-Penrose diagram}

Kerr has no spherical symmetry: Carter-Penrose $\left(T,R\right)$
diagram cannot capture differences in geometry in $\left(\theta,\phi\right)$!

\begin{figure}
\hspace*{-4.5cm}%
\begin{tabular}{>{\centering}m{0.6\columnwidth}>{\centering}m{0.6\columnwidth}}
\includegraphics[height=0.8\textheight]{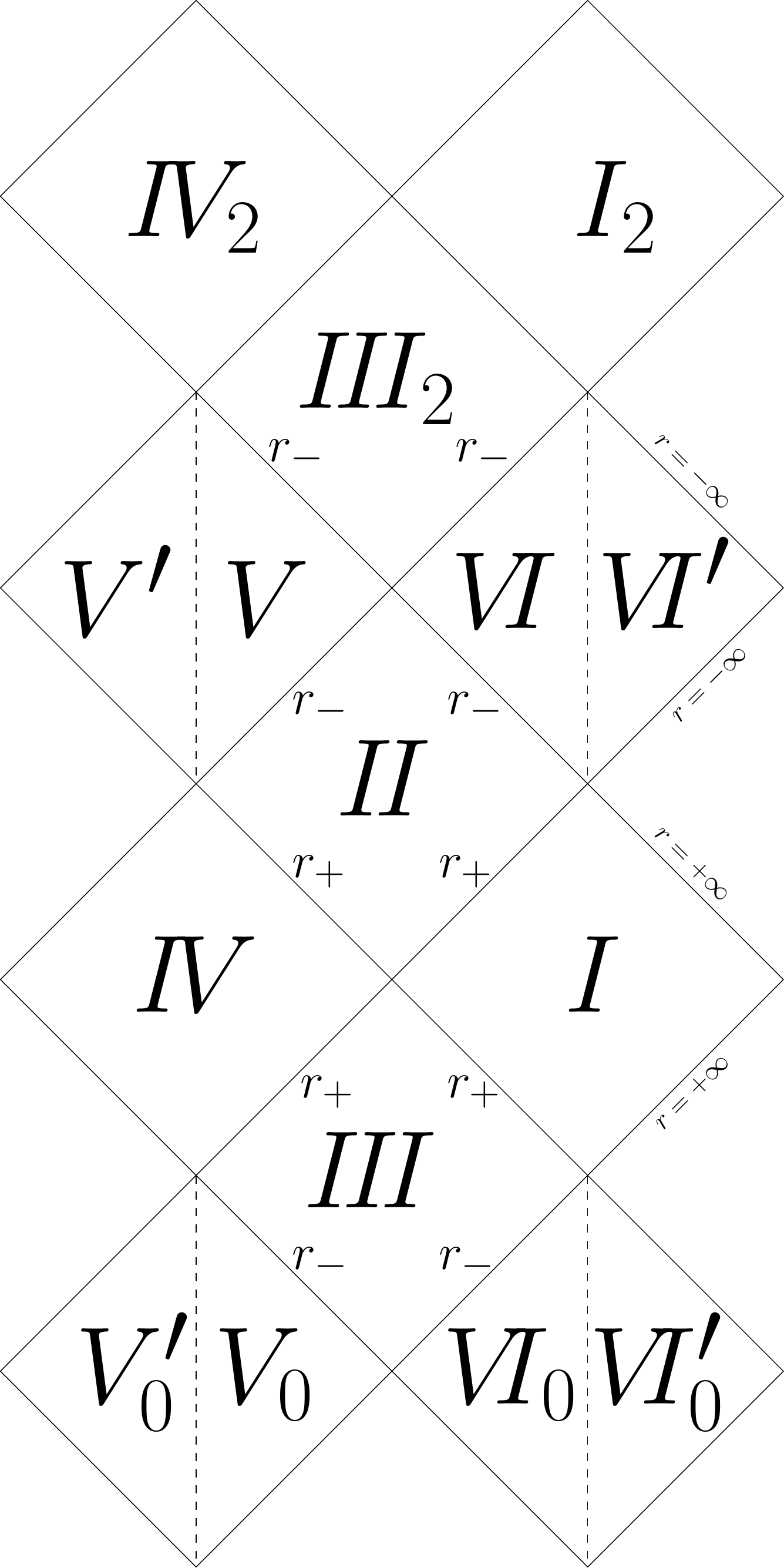} & \includegraphics[height=1\textheight]{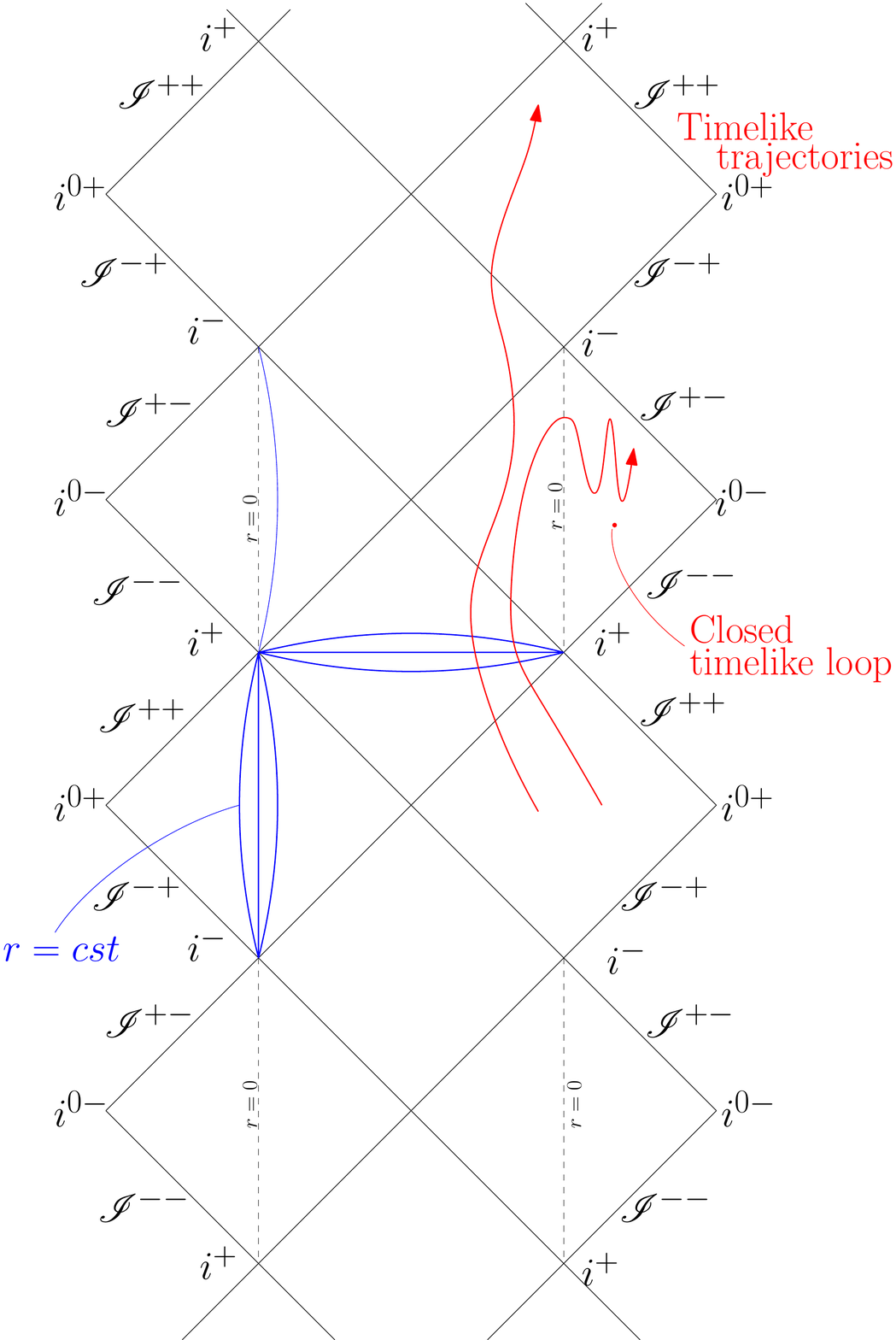}\tabularnewline
\end{tabular}

\caption{\label{fig:Kerr-solution-Carter-Penrose}Kerr solution Carter-Penrose
diagram}
\end{figure}

We can still compactify $\left(t,r\right)$ (see Fig.~ \ref{fig:Kerr-solution-Carter-Penrose})

\subsubsection{Similar to Schwarzschild}

Minkowski-like asymptotic region \encircle{I} limited by $r_{+}$
horizon with \encircle{II} inside $r_{+}$ being the B.H., region
\encircle{III} also giving a W.H., by extension through the coordinate
singularity at $\Delta=0$.

As in the Schwarzschild case, extending from \encircle{II} and \encircle{III}
we get a symmetric asymptotic region \encircle{IV}

\subsubsection{Different from Schwarzschild}

\encircle{II} and \encircle{III} are not barred by spacelike singularity/
next horizon/coordinate singularity $r_{-}$ can be reached (null
surface).

Extending further, we find regions \encircle{V} and \encircle{VI}
where $r<r_{-}$.

They contain the $\rho=0$ (at $r=0$) true singularity, however $r=0$
is only singular at $\theta=\frac{\pi}{2}$, so for $\theta\ne\frac{\pi}{2}$
it can be traversed to $r<0$ where no horizon exists.

Regions \encircle{V} and \encircle{VI} for $r<0$ (noted \encircle{V$^{\prime}$}
and \encircle{VI$^{\prime}$}) are also asymptotically flat and
the ring singularity is naked for its observers, with a mass $-M$.

Regions \encircle{V} and \encircle{VI} can be further extended
to another region \encircle{III$_{2}$} and corresponding regions
\encircle{I$_{2}$}, \encircle{IV$_{2}$},..., infinitely up to
future and down to past.

\subsubsection{Infinities}

Regions \encircle{I} and \encircle{IV}, similarly to Schwarzschild,
get a space infinity $i^{0}$, a future time $i^{+}$ and past time
$i^{-}$ infinities and future and past null infinity surfaces.

However asymptotically flat regions \encircle{V$^{\prime}$} and
\encircle{VI$^{\prime}$} have similar structure with space, future
and past null infinities.

We differentiate them by adding an exponent for the corresponding
space $r-$infinity:
\begin{align*}
i^{0}\left(r=\pm\infty\right)= & i^{0\pm}; & \mathscr{I}^{\pm}\left(r=\pm\infty\right)= & \mathscr{I}^{\pm\pm}.
\end{align*}

\subsubsection{Timelike curves}

\encircle{I}, \encircle{IV}, \encircle{V$^{\prime}$}, \encircle{VI$^{\prime}$}
have the same $r=cst$ curves as in Schwarzschild. Between $r_{+}$
and $r_{-}$ (\encircle{II}, \encircle{III}), as in Schwarzschild,
$r=cst$ becomes spacelike.

As in Reisner-Nordstr\" om charged B.H., timelike trajectories can
travel through to a W.H..

However, if they cross $r=0$ to $r<0$, timelike loops become possible
($\dot{\phi}$ loops) and paradoxes are appearing.

\subsection{Geodesic motion}

Any geodesic trajectory follows
\begin{align*}
u^{a}= & \dot{t}\partial_{t}^{a}+\dot{\phi}\partial_{\phi}^{a}+\dot{r}\partial_{r}^{a}+\dot{\theta}\partial_{\theta}^{a}=\frac{dx^{a}}{d\lambda}
\end{align*}
For timelike geodesics, $\lambda=\tau,u^{a}u_{a}=-1$

As $u^{a}$ and $K^{a}$, timelike Killing vector, are future pointing,
$K_{a}u^{a}<0$ at radial infinity and can there define $u^{a}$'s
energy per unit mass, conserved along $u^{a}$:
\begin{align*}
E= & -u^{a}K_{a}=g_{ab}u^{a}K^{b}=\left(1-\frac{C}{\rho^{2}}\right)\dot{t}+\frac{Ca\sin^{2}\theta}{\rho^{2}}\dot{\phi}
\end{align*}
The spacelike Killing vector $L^{a}$ define $u^{a}$'s angular momentum
per unit mass
\begin{align*}
J= & u^{a}L_{a}=g_{ab}u^{a}L^{b}=-\frac{Ca\sin^{2}\theta}{\rho^{2}}\dot{t}+\frac{\sin^{2}\theta}{\rho^{2}}\left[\left(r^{2}+a^{2}\right)^{2}-\Delta a^{2}\sin^{2}\theta\right]\dot{\phi}
\end{align*}
so we have the system
\begin{gather*}
\left\{ \begin{array}{rl}
E= & \frac{Ca\sin^{2}\theta}{\rho^{2}}\dot{\phi}+\left(1-\frac{C}{\rho^{2}}\right)\dot{t}\\
J= & -\frac{Ca\sin^{2}\theta}{\rho^{2}}\dot{t}+\frac{\sin^{2}\theta}{\rho^{2}}\left[\left(r^{2}+a^{2}\right)^{2}-\Delta a^{2}\sin^{2}\theta\right]\dot{\phi}
\end{array}\right.
\end{gather*}

\subsubsection{Equatorial motion}

We restrict to the $\theta=\frac{\pi}{2}$ ($\Rightarrow z=0$) plane
$\Rightarrow\rho^{2}=r^{2}$

$\Delta=r^{2}+a^{2}-C$ with $C=2GMr$ (no charge from $F^{ab}$)
$\Rightarrow\Delta=r^{2}+a^{2}-2GMr$ 

We have seen $g_{\phi\phi}=\sin^{2}\theta\left[r^{2}+a^{2}+\frac{2GMra^{2}\sin^{2}\theta}{r^{2}+a^{2}\cos^{2}\theta}\right]$
so the system reads
\begin{gather*}
\left\{ \begin{array}{rl}
E= & \frac{2GMa}{r}\dot{\phi}+\left(1-\frac{2GM}{r}\right)\dot{t}=A\dot{t}+B\dot{\phi}\\
J= & \left[r^{2}+a^{2}+\frac{2GMa^{2}}{r}\right]\dot{\phi}-\frac{2GMa}{r}\dot{t}=-B\dot{t}+C\dot{\phi}
\end{array}\right.
\end{gather*}
and the metric restricts to
\begin{align*}
g_{ab}= & \left(\begin{array}{cccc}
-A & -B & 0 & 0\\
-B & C & 0 & 0\\
0 & 0 & \frac{r^{2}}{\Delta} & 0\\
0 & 0 & 0 & r^{2}
\end{array}\right)
\end{align*}
The determinant of the system is
\begin{align*}
AC+B^{2}= & \left(1-\frac{2GM}{r}\right)\left[r^{2}+a^{2}+\frac{2GMa^{2}}{r}\right]+\left(\frac{2GMa}{r}\right)^{2}\\
= & r^{2}+a^{2}+\frac{2GMa^{2}}{r}-\left[r^{2}+a^{2}\right]\frac{2GM}{r}\\
= & r^{2}+a^{2}-2GMr=\Delta
\end{align*}
so we have
\begin{gather*}
\left\{ \begin{array}{rl}
CE-BJ= & \left[AC+B^{2}\right]\dot{t}=\Delta\dot{t}\\
AJ+BE= & \left[AC+B^{2}\right]\dot{\phi}=\Delta\dot{\phi}
\end{array}\right.\\
\Leftrightarrow\left\{ \begin{array}{rl}
\dot{t}= & \frac{1}{\Delta}\left[\left(r^{2}+a^{2}+\frac{2GMa^{2}}{r}\right)E-\frac{2GMa}{r}J\right]\\
\dot{\phi}= & \frac{1}{\Delta}\left[\left(1-\frac{2GM}{r}\right)J+\frac{2GMa}{r}E\right]
\end{array}\right.
\end{gather*}
\begin{tabular}{rlcc}
Allowing for  & timelike &  & \multirow{3}{*}{$u^{a}u_{a}=-s=\left\{ \begin{array}{c}
-1\\
1\\
0
\end{array}\right.$}\tabularnewline
 & spacelike & geodesics, & \tabularnewline
and  & null &  & \tabularnewline
\end{tabular}

we can write (recall $\theta=\frac{\pi}{2}\Rightarrow\dot{\theta}=0$)
\begin{align*}
-s= & g_{ab}u^{a}u^{b}\\
= & -A\dot{t}^{2}-2B\dot{t}\dot{\phi}+C\dot{\phi}^{2}+\frac{r^{2}}{\Delta}\dot{r}^{2}\;\textrm{and using the system of invariants}\\
= & -\underset{E}{\underbrace{\left[A\dot{t}+B\dot{\phi}\right]}}\dot{t}+\underset{J}{\underbrace{\left[C\dot{\phi}-B\dot{t}\right]}}\dot{\phi}+\frac{r^{2}}{\Delta}\dot{r}^{2}\\
= & -E\dot{t}+J\dot{\phi}+\frac{r^{2}}{\Delta}\dot{r}^{2}
\end{align*}
Now using its solution
\begin{align*}
\Leftrightarrow\dot{r}^{2}= & \frac{\Delta}{r^{2}}\left(E\dot{t}-J\dot{\phi}-s\right)\\
= & \frac{1}{r^{2}}\left(E\left[CE-BJ\right]-J\left[AJ+BE\right]-\Delta s\right)\\
= & \frac{1}{r^{2}}\left(CE^{2}-AJ^{2}-2BEJ-\Delta s\right)
\end{align*}
Defining now from their expressions $A=1-Y$, $B=aY$, $C=X+a^{2}Y$
so $\Delta=X-r^{2}Y$ $\Leftrightarrow Y=\frac{2GM}{r},\,X=r^{2}+a^{2}$
\begin{align*}
\Rightarrow\dot{r}^{2}= & \frac{1}{r^{2}}\left[\left(X+a^{2}Y\right)E^{2}-\left(1-Y\right)J^{2}-2aYEJ-\left(X-r^{2}Y\right)s\right]\\
= & \frac{1}{r^{2}}\left[X\left(E^{2}-s\right)+Y\left\{ \underset{\left(J-aE\right)^{2}}{\underbrace{\left(a^{2}E^{2}+J^{2}-2aEJ\right)}}+r^{2}s\right\} -J^{2}\right]\\
\Leftrightarrow & \frac{\dot{r}^{2}}{2}+V\left(E,J,r\right)=0
\end{align*}
where
\begin{align*}
V= & \frac{1}{r^{2}}\left[\frac{J^{2}}{2}-sr^{2}\frac{Y}{2}+\frac{1}{2}\left(s-E^{2}\right)X-\frac{Y}{2}\left(J-aE\right)^{2}\right]\\
= & \frac{J^{2}}{2r^{2}}-s\frac{GM}{r}+\frac{1}{2}\left(s-E^{2}\right)\left(1+\left(\frac{a}{r}\right)^{2}\right)-\frac{GM}{r}\left(J-aE\right)^{2}
\end{align*}
As in Schwarzschild: motion in an effective 1D potential.

\subsubsection{Non equatorial motion}

Add also conserved quantity from Killing tensor $\sigma$: since $\sigma_{\left(ab;c\right)}=0$,
along geodesics with tangent $u^{a}$
\begin{align*}
u^{c}\nabla_{c}\left(\sigma_{ab}u^{a}u^{b}\right)\equiv & u^{c}\nabla_{c}\left(\mathcal{\mathscr{C}}\right)\\
= & u^{(a}u^{b}u^{c)}\underset{\textrm{Killing Eq.}}{\underbrace{\nabla_{c}\sigma_{ab}}}+u^{a}\sigma_{ab}\underset{\textrm{Geo. Eq.}}{\underbrace{u^{c}\nabla_{c}u^{b}}}+u^{b}\sigma_{ab}\underset{\textrm{Geo. Eq.}}{\underbrace{u^{c}\nabla_{c}u^{a}}}\\
= & 0\\
\Rightarrow\mathscr{C}= & \sigma_{ab}u^{a}u^{b}\textrm{ conserved}
\end{align*}

\section{Penrose process and Energy extraction from Black Holes}

From geodesic motion of test particle of mass $m$, its 4-momentum
writes $p^{a}=mu^{a}$ and we similarly define its energy
\begin{align*}
\mathscr{E}= & -p^{a}K_{a}
\end{align*}
and angular momentrum
\begin{align*}
\mathscr{J}= & p^{a}L_{a}
\end{align*}
We thus have the system
\begin{gather*}
\left\{ \begin{array}{rl}
\mathscr{E}= & m\frac{Ca\sin^{2}\theta}{\rho^{2}}\dot{\phi}+m\left(1-\frac{C}{\rho^{2}}\right)\dot{t}\\
\mathscr{J}= & -m\frac{Ca\sin^{2}\theta}{\rho^{2}}\dot{t}+m\frac{\sin^{2}\theta}{\rho^{2}}\left[\left(r^{2}+a^{2}\right)^{2}-\Delta a^{2}\sin^{2}\theta\right]\dot{\phi}
\end{array}\right.
\end{gather*}
Since $K^{a}$ becomes spacelike in the Ergosphere, we can have $\mathscr{E}<0$
inside of it.

If the B.H. absorb an $\mathscr{E}<0$ particle, we can extract energy
from it.

Suppose we throw a particle with $p_{0}^{a}$ into the Ergosphere.

Its energy $\mathscr{E}_{0}=-p_{0}^{a}K_{a}>0$ since outside the
Ergosphere $K^{a}$ timelike

Now we split it in 2 parts. Conservation of momentum yields
\begin{align*}
p_{0}^{a}= & p_{1}^{a}+p_{2}^{a}
\end{align*}
contracting with $K^{a}$
\begin{align*}
\mathscr{E}_{0}= & \mathscr{E}_{1}+\mathscr{E}_{2}
\end{align*}
If we set the first fragment to have $\mathscr{E}_{1}<0$ then $\mathscr{E}_{2}=\mathscr{E}_{0}+\left|\mathscr{E}_{1}\right|>\mathscr{E}_{0}$
and we have $M_{BH}=M_{INITIAL}-\left|\mathscr{E}_{1}\right|:\left|\mathscr{E}_{1}\right|$
has been extracted extracted from the B.H.

This is called the Penrose process

\subsection{Killing Horizon}

$r_{+}$ is a Killing Horizon

$\chi_{a}=K_{a}+\Omega_{H}L_{a}$ is Killing vector and null at $r_{+}\Leftrightarrow r_{+}$
horizon is a Killing horizon
\begin{proof}
:
\begin{itemize}
\item $\chi_{\left(a;b\right)}=K_{\left(a;b\right)}+\Omega_{H}L_{\left(a;b\right)}$
since $\Omega_{H}=cst$. So $\chi_{a}$ Killing
\item Since
\begin{align*}
K^{a}L_{a}= & g_{t\phi}=-\frac{Ca\sin^{2}\theta}{\rho^{2}}\\
K^{a}K_{a}= & g_{tt}=-\frac{\Delta-a^{2}\sin^{2}\theta}{\rho^{2}}\qquad\left(\textrm{with }\Delta=r^{2}+a^{2}-C\right)\\
L^{a}L_{a}= & g_{\phi\phi}=\sin^{2}\theta\left[r^{2}+a^{2}+\frac{Ca^{2}\sin^{2}\theta}{\rho^{2}}\right]=\sin^{2}\theta\left[\Delta+\frac{C}{\rho^{2}}\left(\rho^{2}+a^{2}\sin^{2}\theta\right)\right]
\end{align*}
and recall $\Omega_{H}=-\left.\frac{g_{t\phi}}{g_{\phi\phi}}\right|_{r_{+}}$
so we can write
\begin{align*}
\chi^{a}\chi_{a}= & K^{a}K_{a}+\Omega_{H}^{2}L^{a}L_{a}+2\Omega_{H}K^{a}L_{a}\\
= & g_{tt}+\left(\frac{g_{t\phi}}{g_{\phi\phi}}\right)^{2}\left(r_{+}\right)g_{\phi\phi}-2\left(\frac{g_{t\phi}}{g_{\phi\phi}}\right)\left(r_{+}\right)g_{t\phi}
\end{align*}
so at $r_{+}$(where $\Delta=0$)
\begin{align*}
\chi^{a}\chi_{a}=\left.g_{tt}\right|_{r_{+}}-\left.\frac{g_{t\phi}^{2}}{g_{\phi\phi}}\right|_{r_{+}}= & \frac{a^{2}\sin^{2}\theta}{\rho^{2}}-\frac{\left(Ca\sin^{2}\theta\right)^{2}}{\rho^{2}\sin^{2}\theta C\left(\rho^{2}+a^{2}\sin^{2}\theta\right)}\\
= & \frac{a^{2}\sin^{2}\theta}{\rho^{2}}\left(1-\frac{C}{r_{+}^{2}+a^{2}}\right)\\
= & \frac{a^{2}\sin^{2}\theta}{\rho^{2}}\left.\left(1-\frac{C}{\Delta+C}\right)\right|_{r_{+}}\\
= & 0
\end{align*}
\end{itemize}
Moreover, $\chi$ is future directed (as $K^{a}$) so any particle
enterring the B.H. at $r_{+}$ have
\begin{align*}
0\ge p_{1}^{a}\chi_{a}=p_{1}^{a}\left(K_{a}+\Omega_{H}L_{a}\right)= & -\mathscr{E}_{1}+\Omega_{H}\mathscr{J}_{1}\\
\Rightarrow\mathscr{J}_{1}\le & \frac{\mathscr{E}_{1}}{\Omega_{H}}
\end{align*}
Negative $\mathscr{E}_{1}$ particles carry $\mathscr{J}_{1}<0$ (moving
against B.H. rotation)

After the B.H. swallows $p_{1}$, it settles to a new Kerr solution
with $\delta M=\mathscr{E}_{1}$ and $\delta J=\mathscr{J}_{1}$
\begin{gather*}
\Rightarrow\boxed{\delta J\le\frac{\delta M}{\Omega_{H}}}
\end{gather*}
\end{proof}

\subsection{BH horizon area changes}

At $r_{+}$ the induced metric is given for $dt=0=dr$
\begin{align*}
\left.ds^{2}\right|_{\begin{array}{rl}
dt= & dr=0\\
r= & r_{+}
\end{array}}= & \gamma_{ij}dx^{i}dx^{j}=\rho_{+}^{2}d\theta^{2}+\sin^{2}\theta\frac{\left(r_{+}^{2}+a^{2}\right)^{2}}{\rho_{+}^{2}}d\phi^{2}\\
\Rightarrow\sqrt{\left|\gamma\right|}= & \sqrt{\left|g_{\theta\theta}^{+}g_{\phi\phi}^{+}\right|}=\sqrt{\sin^{2}\theta\left(r^{2}+a^{2}\right)^{2}}=\left(r_{+}^{2}+a^{2}\right)\left|\sin\theta\right|
\end{align*}
so the horizon area calculates
\begin{align*}
A_{+}= & \int\sqrt{\left|\gamma\right|}d\theta d\phi=4\pi\left(r_{+}^{2}+a^{2}\right)
\end{align*}
Define the irreducible mass of the B.H., since $r_{+}=GM+\sqrt{\left(GM\right)^{2}-a^{2}}$,
\begin{align*}
M_{irr}^{2}=\frac{A_{+}}{16\pi G^{2}}= & \frac{r_{+}^{2}+a^{2}}{4G^{2}}\\
= & \frac{1}{4G^{2}}\left(\left(GM\right)^{2}+\left(GM\right)^{2}-a^{2}+a^{2}+2GM\sqrt{\left(GM\right)^{2}-a^{2}}\right)\\
= & \frac{1}{2}\left(M^{2}+M\sqrt{M^{2}-\left(\frac{a}{G}\right)^{2}}\right)\\
\Rightarrow M_{irr}^{2}= & \frac{1}{2}\left(M^{2}+\sqrt{M^{4}-\left(\frac{Ma}{G}\right)^{2}}\right)
\end{align*}
recall $Ma=J$ and since $\Omega_{H}=\frac{a}{r_{+}^{2}+a^{2}}$,
\begin{align*}
M_{irr}^{2}=\frac{r_{+}^{2}+a^{2}}{4G^{2}}=\frac{a\Omega_{H}^{-1}}{4G^{2}}\Leftrightarrow & \frac{a\Omega_{H}^{-1}}{2G^{2}}=M^{2}+\sqrt{M^{4}-\left(\frac{J}{G}\right)^{2}}
\end{align*}
Differentiating $M_{irr}^{2}$
\begin{align*}
\Rightarrow2M_{irr}\delta M_{irr}= & M\delta M+\frac{1}{4}\frac{4M^{3}\delta M-\frac{2J\delta J}{G^{2}}}{\sqrt{M^{4}-\left(\frac{J}{G}\right)^{2}}}\\
= & \frac{\left(M^{3}+M\sqrt{M^{4}-\left(\frac{J}{G}\right)^{2}}\right)\delta M-\frac{J}{2G^{2}}\delta J}{\sqrt{M^{4}-\left(\frac{J}{G}\right)^{2}}}\\
= & \frac{M\frac{a\Omega_{H}^{-1}}{2G^{2}}\delta M-\frac{Ma}{2G^{2}}\delta J}{M\sqrt{M^{2}-\left(\frac{a}{G}\right)^{2}}}\\
= & \frac{a}{2G\sqrt{\left(GM\right)^{2}-a^{2}}}\left(\Omega_{H}^{-1}\delta M-\delta J\right)\;\textrm{and since }\delta J\le\frac{\delta M}{\Omega_{H}}\\
\Leftrightarrow\delta M_{irr}= & \frac{a}{4GM_{irr}\sqrt{\left(GM\right)^{2}-a^{2}}}\left(\Omega_{H}^{-1}\delta M-\delta J\right)\ge0
\end{align*}
The irreducible mass can never be reduced

This is a corrolary to the area theorem (Hawking 1971)

Inverting $M_{irr}$ definition,
\begin{align*}
 & \left(2M_{irr}^{2}-M^{2}\right)^{2}=M^{4}-\left(\frac{J}{G}\right)^{2}\\
\Leftrightarrow & 4M_{irr}^{4}-4M_{irr}^{2}M^{2}=-\left(\frac{J}{G}\right)^{2}\\
\Leftrightarrow & M^{2}=M_{irr}^{2}+\left(\frac{J}{2GM_{irr}}\right)^{2}\ge M_{irr}^{2}
\end{align*}
so the mass of the B.H. cannot reduce below $M_{irr}$ from Penrose
process.

At most we can get $J=0$, extracting from it
\begin{align*}
\Delta M_{max}=M-M_{irr}= & M-\frac{1}{\sqrt{2}}\sqrt{M^{2}+\sqrt{M^{4}-\left(\frac{J}{G}\right)^{2}}}
\end{align*}
leaving a Schwarzschild B.H.

From a maximally rotating B.H.
\begin{align*}
\left(GM_{0}\right)^{2}=a_{0}^{2}\Leftrightarrow & GM_{0}^{2}=J_{0}
\end{align*}
Maximal extraction of mass-energy can reach
\begin{align*}
\frac{\Delta M_{max}}{M_{0}}= & 1-\frac{1}{\sqrt{2}}\simeq29\%.
\end{align*}

\subsection{Superradiance}

Equivalent of Penrose process for radiation exists. If absorbed radiation
is of negative energy, reflected part can gain intensity.

\subsection{Gravitational waves in binary coalescence}

The area theorem can be used to get upper limit of GW energy emitted
from 2 B.H. merger. Start from 2 Schwarzschild B.H. that merge and
settle down to another Schwarzschild merger B.H.
\begin{align*}
E_{rad}= & M_{1}+M_{2}-M
\end{align*}
Initial area is
\begin{align*}
A_{+i}= & A_{+1}+A_{+2}=16\pi G^{2}\left(M_{1}^{2}+M_{2}^{2}\right)
\end{align*}
Final area
\begin{align*}
A_{+f}= & 16\pi G^{2}M^{2}
\end{align*}
By the area theorem, $A_{+f}\ge A_{+i}\Rightarrow M\ge\sqrt{M_{1}^{2}+M_{2}^{2}}$
so $E_{rad}=M_{1}+M_{2}-M\le M_{1}+M_{2}-\sqrt{M_{1}^{2}+M_{2}^{2}}$

For $M_{1}=M_{2}$, at most $\frac{E_{rad}}{M_{1}+M_{2}}\le1-\frac{1}{\sqrt{2}}\simeq29\%$
can be radiated away

\section{Black Hole Thermodynamics}

\subsection{Kerr-Newman BH surface gravity}

From the Chapter \ref{chap:Horizons-and-singularity} on horizons,
we can deduce that, since $r_{+}is$a Killing horizon ($\chi^{a}$
null at $r_{+}$ and Killing) we can define its surface gravity as
$\kappa$:

$\nabla^{a}\left(\chi^{b}\chi_{b}\right)=-2\kappa\chi^{a}$ such as
$\mathcal{L}_{\chi}\kappa=0$ on $r_{+}$: $\kappa$ constant at the
horizon

Recall
\begin{itemize}
\item $\Omega_{H}=-\left.\frac{g_{t\phi}}{g_{\phi\phi}}\right|_{r_{+}}$
but $\Omega=-\frac{g_{t\phi}}{g_{\phi\phi}}=\frac{Ca\sin^{2}\theta}{\rho^{2}g_{\phi\phi}}$
in general for the rotation around the B.H.
\item $g_{\phi\phi}=\frac{\sin^{2}\theta}{\rho^{2}}\left[\left(r^{2}+a^{2}\right)^{2}-\Delta a^{2}\sin^{2}\theta\right]$
Define $X=\left(r^{2}+a^{2}\right)^{2}-\Delta a^{2}\sin^{2}\theta=\left(r^{2}+a^{2}\right)\left(r^{2}+a^{2}-a^{2}\sin^{2}\theta\right)+Ca^{2}\sin^{2}\theta$
so $g_{\phi\phi}=\frac{\sin^{2}\theta}{\rho^{2}}X$ and $\Omega=\frac{Ca}{X}$\\
\end{itemize}
We also have $g_{t\phi}=-\Omega g_{\phi\phi}$ and $\chi^{a}=K^{a}+\Omega_{H}L^{a}$
so since $K=\partial_{t},L=\partial_{\phi}$ %
\begin{tabular}{ccc}
we have  & $K^{a}K_{a}=g_{tt}$ & and so\tabularnewline
 & $L^{a}K_{a}=g_{t\phi}$ & \tabularnewline
 & $L^{a}L_{a}=g_{\phi\phi}$ & \tabularnewline
\end{tabular}
\begin{align*}
\chi^{a}\chi_{a}= & g_{tt}+2\Omega_{H}g_{t\phi}+\Omega_{H}^{2}g_{\phi\phi}\textrm{ and using }g_{t\phi}=-\Omega g_{\phi\phi}\\
= & g_{tt}-2\Omega_{H}\Omega g_{\phi\phi}+\Omega_{H}^{2}g_{\phi\phi}\textrm{ so we can write }\\
\chi^{2}= & \left(\Omega-\Omega_{H}\right)^{2}g_{\phi\phi}+g_{tt}-\Omega^{2}g_{\phi\phi}
\end{align*}
From the definitions above we have $\Omega^{2}g_{\phi\phi}=\frac{\left(Ca\right)^{2}}{X}\frac{\sin^{2}\theta}{\rho^{2}}$
so we can calculate
\begin{align*}
g_{tt}-\Omega^{2}g_{\phi\phi}= & -\left(1-\frac{C}{\rho^{2}}\right)-\frac{\left(Ca\right)^{2}}{X}\frac{\sin^{2}\theta}{\rho^{2}}\\
= & \frac{1}{\rho^{2}X}\left[\left(C-\rho^{2}\right)X-\left(Ca\right)^{2}\sin^{2}\theta\right]
\end{align*}
We proceed, noting $R^{2}=r^{2}+a^{2}$
\begin{gather*}
\Rightarrow\left\{ \begin{array}{rl}
C= & R^{2}-\Delta\\
\rho^{2}= & R^{2}-a^{2}\sin^{2}\theta\\
X= & R^{2}\rho^{2}+Ca^{2}\sin^{2}\theta=R^{4}-\Delta a^{2}\sin^{2}\theta
\end{array}\right.
\end{gather*}
so
\begin{align*}
g_{tt}-\Omega^{2}g_{\phi\phi}= & \frac{1}{\rho^{2}X}\left[\left(C-\rho^{2}\right)\left(R^{2}\rho^{2}+Ca^{2}\sin^{2}\theta\right)-\left(Ca\right)^{2}\sin^{2}\theta\right]\\
= & \frac{1}{\rho^{2}X}\left[CR^{2}\rho^{2}-\rho^{2}\left(R^{2}\rho^{2}+Ca^{2}\sin^{2}\theta\right)\right]\\
= & \frac{1}{X}\left[R^{2}\left(C-\rho^{2}\right)-Ca^{2}\sin^{2}\theta\right]\\
= & \frac{1}{X}\left[R^{2}\left(a^{2}\sin^{2}\theta-\Delta\right)-Ca^{2}\sin^{2}\theta\right]\\
= & \frac{1}{X}\left[\Delta a^{2}\sin^{2}\theta-R^{2}\Delta\right]=-\frac{\rho^{2}\Delta}{X}\\
\Rightarrow\chi^{a}\chi_{a}= & \left(\Omega-\Omega_{H}\right)^{2}\frac{\sin^{2}\theta}{\rho^{2}}X-\frac{\rho^{2}\Delta}{X}\underset{{\scriptscriptstyle \begin{array}{c}
r\to r_{+}\\
\left(\Leftrightarrow\begin{array}{c}
\Delta\to0\\
\Omega\to\Omega_{H}
\end{array}\right)
\end{array}}}{\longrightarrow}0
\end{align*}
as expected

Now since $\kappa$ is constant over the horizon, we can restrict
to the $\theta=\frac{\pi}{2}$ plane to compute
\begin{align*}
\nabla_{a}\chi^{2}= & e_{r}\partial_{r}\chi^{2}\\
= & \left[\left(\Omega-\Omega_{H}\right)\left\{ 2\partial_{r}\Omega\frac{X}{\rho^{2}}+\left(\Omega-\Omega_{H}\right)\partial_{r}\left(\frac{X}{\rho^{2}}\right)\right\} -\partial_{r}\left(\frac{\rho^{2}}{X}\right)\Delta-\frac{\rho^{2}}{X}\partial_{r}\Delta\right]e_{r}
\end{align*}
since it only depends on $r$ now

Furthermore, since at $r_{+},\Omega=\Omega_{H}$ and $\Delta=0\Rightarrow\nabla_{a}\chi^{2}=-\frac{\rho^{2}}{X}\partial_{r}\Delta e_{r}$
and since $\chi$ is null at $r_{+}\Rightarrow\chi$ normal to itself
so $\chi\propto$normal to horizon $r_{+}=cst:e_{r}$

This can be written as $\chi^{a}=Ne_{r}^{a}$ and $\left|\chi^{2}\right|=N^{2}\left|g^{rr}\right|=\frac{N^{2}}{\left|g_{rr}\right|}=\frac{N^{2}\Delta}{\rho^{2}}$.
Since $\Delta>0$ and $\chi^{2}<0$ for $r>r_{+}$,
\begin{proof}
As $\chi^{2}=\left(\Omega-\Omega_{H}\right)^{2}\frac{\sin^{2}\theta}{\rho^{2}}X-\frac{\rho^{2}\Delta}{X}$,
with $X=\left(r^{2}+a^{2}\right)^{2}-\Delta a^{2}\sin^{2}\theta$,
$\rho^{2}=r^{2}+a^{2}\cos^{2}\theta$, $\Delta=r^{2}+a^{2}-C$, $C=2GMr$
and $\Omega=\frac{Ca}{X}$,

at $r=r_{+},\Delta=0\Leftrightarrow C_{+}=r_{+}^{2}+a^{2}=2GMr_{+}$
with $r_{+}=GM+\sqrt{\left(GM\right)^{2}-a^{2}}$
\begin{align*}
\Delta= & 0\Leftrightarrow C_{+}=r_{+}^{2}+a^{2}=2GMr_{+}\textrm{ }\textrm{ with }r_{+}=GM+\sqrt{\left(GM\right)^{2}-a^{2}}\\
X_{+}= & \left(r_{+}^{2}+a^{2}\right)^{2}=C_{+}^{2}\\
\textrm{and }\Omega= & \Omega_{H}=\frac{C_{+}a}{C_{+}^{2}}=\frac{a}{r_{+}^{2}+a^{2}}
\end{align*}
then at $r=r_{+}+\epsilon$ to first order in $\epsilon$
\begin{align*}
\Delta= & \Delta_{+}+\left(2r_{+}-2GM\right)\epsilon=2\left(r_{+}-GM\right)\epsilon=2\sqrt{\left(GM\right)^{2}-a^{2}}\epsilon\\
X= & \left(r_{+}^{2}+a^{2}+2r_{+}\epsilon\right)^{2}-2\sqrt{\left(GM\right)^{2}-a^{2}}a^{2}\sin^{2}\theta\epsilon\\
= & C_{+}^{2}+\left[\frac{4r_{+}}{C_{+}}-2\sqrt{\left(GM\right)^{2}-a^{2}}a^{2}\sin^{2}\theta\right]\epsilon\\
= & C_{+}^{2}\left(1+2\frac{\left[2r_{+}-C_{+}\sqrt{\left(GM\right)^{2}-a^{2}}a^{2}\sin^{2}\theta\right]}{C_{+}^{3}}\epsilon\right)\\
\Rightarrow\Omega= & \frac{2GMr_{+}a\left(1+\frac{\epsilon}{r_{+}}\right)}{C_{+}^{2}\left(1+2\frac{\left[2r_{+}-C_{+}\sqrt{\left(GM\right)^{2}-a^{2}}a^{2}\sin^{2}\theta\right]}{C_{+}^{3}}\epsilon\right)}\\
= & \Omega_{H}\left(1+\left\{ \frac{1}{r_{+}}-2\frac{\left[2r_{+}-C_{+}\sqrt{\left(GM\right)^{2}-a^{2}}a^{2}\sin^{2}\theta\right]}{C_{+}^{3}}\right\} \epsilon\right)\Rightarrow\Omega-\Omega_{H}\propto\epsilon
\end{align*}
so $\left(\Omega-\Omega_{H}\right)^{2}\frac{\sin^{2}\theta}{\rho^{2}}X\propto\epsilon^{2}$
\begin{align*}
\Rightarrow\chi^{2}= & -\frac{\rho_{+}^{2}}{X_{+}}\left(\Delta_{+}+2\sqrt{\left(GM\right)^{2}-a^{2}}\epsilon\right)=-\frac{\rho_{+}^{2}}{C_{+}^{2}}2\sqrt{\left(GM\right)^{2}-a^{2}}\epsilon<0\textrm{ for }\epsilon>0.
\end{align*}
\end{proof}
\begin{align*}
\Rightarrow-\frac{N^{2}\Delta}{\rho^{2}}= & \left(\Omega-\Omega_{H}\right)^{2}g_{\phi\phi}-\frac{\rho^{2}\Delta}{X}\\
\Rightarrow N^{2}= & -\rho^{2}\frac{\left(\Omega-\Omega_{H}\right)^{2}}{\Delta}g_{\phi\phi}+\frac{\rho^{4}}{X}\xrightarrow[\begin{array}{c}
\Delta\to0\\
\Omega\to\Omega_{H}
\end{array}]{}\frac{\rho^{4}}{X}
\end{align*}
so if we define $\kappa$ from $\nabla_{a}\chi^{2}=-2\kappa\chi_{a}$,
$\kappa$ surface gravity, we get $\nabla_{a}\chi^{2}=-\frac{\rho^{2}}{X}\partial_{r}\Delta\left(e_{r}\right)_{a}=-2\kappa\frac{\rho^{2}}{\sqrt{X}}\left(e_{r}\right)_{a}$

$\Rightarrow\kappa=\frac{\partial_{r}\Delta}{2\sqrt{X}}$ with $\partial_{r}\Delta=\partial_{r}\left(r^{2}+a^{2}-C\right)$
while $C=2GMr-G\left(e^{2}+m^{2}\right)$ and $\Delta=0\Rightarrow r=r_{+}=GM+\sqrt{\left(GM\right)^{2}-a^{2}-G\left(e^{2}+m^{2}\right)}$

so $\partial_{r}\Delta_{+}=2r_{+}-2GM=2\sqrt{\left(GM\right)^{2}-a^{2}-G\left(e^{2}+m^{2}\right)}$
and $X_{+}=R_{+}^{4}-\Delta a^{2}\sin^{2}\theta=R_{+}^{4}\Rightarrow\sqrt{X_{+}}=R_{+}^{2}=r_{+}^{2}+a^{2}$

then 
\begin{align*}
\sqrt{X_{+}}= & \left(GM\right)^{2}+\left(GM\right)^{2}-a^{2}-G\left(e^{2}+m^{2}\right)+2GM\sqrt{\left(GM\right)^{2}-a^{2}-G\left(e^{2}+m^{2}\right)}+a^{2}\\
= & 2GM\left(GM+\sqrt{\left(GM\right)^{2}-a^{2}-G\left(e^{2}+m^{2}\right)}\right)-G\left(e^{2}+m^{2}\right)
\end{align*}
and finally
\begin{gather*}
\boxed{\kappa=\frac{\sqrt{\left(GM\right)^{2}-a^{2}-G\left(e^{2}+m^{2}\right)}}{2GM\left(GM+\sqrt{\left(GM\right)^{2}-a^{2}-G\left(e^{2}+m^{2}\right)}\right)-G\left(e^{2}+m^{2}\right)}}
\end{gather*}

\subsection{Parallel with thermodynamics}

We have seen $\Omega_{H}=\left.\frac{Ca}{X}\right|_{+}=\frac{R_{+}^{2}a}{R_{+}^{4}}=\frac{a}{\sqrt{X_{+}}}$\\
so $\kappa=\frac{\sqrt{\left(GM\right)^{2}-a^{2}-G\left(e^{2}+m^{2}\right)}}{\sqrt{X_{+}}}=\sqrt{\left(GM\right)^{2}-a^{2}-G\left(e^{2}+m^{2}\right)}\frac{\Omega_{H}}{a}$

Moreover, setting $e=m=0$, since $2M_{irr}\delta M_{irr}=\frac{\delta A}{16\pi G^{2}}=\frac{a}{2G\sqrt{\left(GM\right)^{2}-a^{2}}\Omega_{H}}\left(\delta M-\Omega_{H}\delta J\right)$

we then have
\begin{align*}
\frac{\delta A}{16\pi G^{2}}= & \frac{1}{2G\kappa}\left(\delta M-\Omega_{H}\delta J\right)\\
\Leftrightarrow\delta M= & \frac{\kappa}{8\pi G}\delta A+\Omega_{H}\delta J
\end{align*}
Recall First law of thermodynamics
\begin{align*}
dE= & TdS-PdV
\end{align*}
where $E$ energy, $S$ entropy, $T$ temperature, $P$ pressure and
$V$ volume of the system. 
\begin{itemize}
\item $PdV$ is elementary work done on system
\item $\Omega_{H}\delta J$ can be seen as work done on B.H. by sending
$\delta J<0$ particles in it.
\item $\delta M$ is variation of mass-energy of B.H.
\end{itemize}
Then we can propose the analogy
\begin{gather*}
\begin{array}{rcl}
E & \longleftrightarrow & M\\
S & \longleftrightarrow & \frac{A}{4G}\\
T & \longleftrightarrow & \frac{\kappa}{2\pi}
\end{array}
\end{gather*}
Normalisation is fixed from the Black Body Temperature of Hawking
radiation being $T=\frac{k\kappa}{2\pi cV}$ where $V=\sqrt{-K^{a}K_{a}}\xrightarrow[r\to\infty]{}1$

We can then extend the analogy (see Table~\ref{tab:Black-Hole-Thermodynamics})
\begin{table}
\hspace*{-2cm}%
\begin{tabular}{c|c|c}
Law & Thermodynamics & B.H.\tabularnewline
\hline 
$0^{th}$ & $T$ constant in equilibrium system & $\kappa=cst$ on horizon of stationary B.H.\tabularnewline
\hline 
$1^{st}$ & $dE=TdS+$work done & $dM=\frac{\kappa}{8\pi G}dA+\Omega_{H}dJ$\tabularnewline
\hline 
$2^{nd}$ & $\delta S\ge0$ in any process & $\delta A\ge0$ in any process\tabularnewline
\hline 
$3^{rd}$ & impossible to achieve $T=0$ by physical process & impossible to achieve $\kappa=0$ by physical process\tabularnewline
\end{tabular}

\caption{\label{tab:Black-Hole-Thermodynamics}Black Hole Thermodynamics analogy}

\end{table}

\begin{description}
\item [{$0^{th}$~law}] we have shown $\mathcal{L}_{\chi}\kappa=0$ for
stationary B.H.
\item [{$1^{st}$~law}] we just shown it
\item [{$2^{nd}$~law}] consequence of B.H. area theorem, we have shown
\begin{align*}
\frac{\delta A}{16\pi G^{2}}= & \delta M_{irr}^{2}>0
\end{align*}
in any process of Penrose
\item [{$3^{rd}$~law}] only extremum B.H., that is $\left(GM\right)^{2}=a^{2}+G\left(e^{2}+m^{2}\right)$,
yields $\kappa=0$.\\
Wald \cite{wald-1974,wald-book} shown it gets harder and harder from
non-extremal B.H. to go near extremal
\end{description}
However, at $\kappa=0$, $A\ne0$ contrary to $S=0$ for $T=0$ in
thermodynamics!

Finally Bekenstein \cite{bekenstein-1973} showed a generalised entropy
combines B.H. and matter
\begin{align*}
S^{\prime}= & S+\frac{1}{4}k\frac{c^{3}A}{G\hbar}
\end{align*}
such as $\delta S^{\prime}\ge0$

\chapter{Relativistic Thermodynamics}

John M. Stewart \cite{Stewart1971} \textquotedbl{}Non-Equilibrium
Relativistic Kinetic Theory\textquotedbl{}

Two main approaches to extending thermodynamics to relativistic fluids
and gravitation.

This is distinct from Black Hole thermodynamics: non-singular fluids,
starting from perfect fluid

Connected to Fourier heat transport, fluid entropy and conservation
or increase of entropy.

Assume particlenumber conservation (decays and particle physics not
treated)

\begin{tabular}{ll}
Recall & $a,b\in\left[0;3\right]$ spacetime indices\tabularnewline
 & $\mu,\nu\in\left[1;3\right]$ space indices\tabularnewline
\end{tabular}

\begin{tabular}{ll}
$N^{a}=nu^{a}$ particle flow, & $u^{a}$: flow 4-velocity\tabularnewline
 & $n$ particle number density\tabularnewline
\end{tabular}

$u^{a}$ timelike: $u^{a}u_{a}=-1$

\begin{tabular}{ll}
At rest, $N^{\mu}=0$\quad{}conservation: & $N_{\,,0}^{0}=0$ (Newtonian)\tabularnewline
 & $N_{\,;a}^{a}=0$ (Relativistic)\tabularnewline
\end{tabular}

\section{Eckart's model}

\subsection{Energy-momentum tensor}

\begin{tabular}{ll}
Decomposed into:  & -perfect fluid\tabularnewline
 & -rest\tabularnewline
\end{tabular}
\begin{align*}
T^{ab}= & T_{PF}^{ab}+\Delta T^{ab}\\
N^{a}= & nu^{a}+\Delta N^{a} & \to\Delta N^{a}= & 0\textrm{ assumed for conserved particle number}\\
 &  & \Rightarrow N^{a}\sslash n^{a}
\end{align*}
\begin{align*}
\Rightarrow T^{ab}= & \left(\rho+P\right)u^{a}u^{b}+Pg^{ab}+\Delta T^{ab}=\rho u^{a}u^{b}+Ph^{ab}+\Delta T^{ab}
\end{align*}
\begin{tabular}{lll}
for & $\rho$: energy density & $h^{ab}=g^{ab}+u^{a}u^{b}$: projector $\bot u^{a}$\tabularnewline
 & $P$: isotropic pressure & \tabularnewline
\end{tabular}

\begin{tabular}{ll}
We have the  & Bianchi identity for P.F.: $T_{PF\,;b}^{ab}=P^{;a}+\left[\left(\rho+P\right)u^{a}u^{b}\right]_{;b}$\tabularnewline
 & particle number conservation: $\begin{array}[t]{rl}
N_{\,;a}^{a}= & nu_{\,;a}^{a}+u^{a}n_{;a}\\
= & \dot{n}+nu_{\,;a}^{a}\\
= & 0
\end{array}$\tabularnewline
\end{tabular}

Projected along the flow
\begin{align*}
u_{a}T_{PF\,;b}^{ab}= & u^{a}P_{;a}-\left[\left(\rho+P\right)u^{b}\right]_{;b}+\left(\rho+P\right)u^{b}\underset{0}{\underbrace{u_{\,;b}^{a}u_{a}}}\\
= & \dot{P}-\left[\left(\rho+P\right)u^{b}\right]_{;b}\\
= & u^{b}P_{;b}-\left[\frac{\rho+P}{n}nu^{b}\right]_{;b}=u^{b}P_{;b}-\left[\frac{\rho+P}{n}\right]_{;b}N^{b}\textrm{ (}N\textrm{ conserved)}\\
= & -N^{b}\left(\frac{\rho}{n}\right)_{;b}+u^{b}P_{;b}-\frac{P_{;b}}{n}nu^{b}-P\left(\frac{1}{n}\right)_{;b}N^{b}\\
= & -N^{a}\left[\left(\frac{\rho}{n}\right)_{;a}+P\left(\frac{1}{n}\right)_{;a}\right]
\end{align*}
Now Gibb's equation applied to densities reads
\begin{align*}
du= & Tds-Pdv & u= & \frac{\rho}{n}:\textrm{ energy per particle}\\
 &  & s= & k\sigma:\textrm{ entropy per particle (}k\textrm{ Planck constant)}\\
 &  & v= & \frac{1}{n}:\textrm{ volume per particle}
\end{align*}
\begin{gather*}
\Rightarrow d\left(\frac{\rho}{n}\right)=kTd\sigma-Pd\left(\frac{1}{n}\right)\\
\boxed{d\left(\frac{\rho}{n}\right)+Pd\left(\frac{1}{n}\right)=kTd\sigma}
\end{gather*}
We recognise the term in the energy-momentum conservation along the
flow
\begin{align*}
u_{a}T_{PF\,;b}^{ab}= & -nu^{a}\left[\left(\frac{\rho}{n}\right)_{;a}+P\left(\frac{1}{n}\right)_{;a}\right]=-nu^{a}kT\sigma_{;a}\\
\overset{\textrm{P. cons.}}{=} & -kT\left(nu^{a}\sigma\right)_{;a}
\end{align*}

\subsection{Full energy-momentum tensor}

\begin{align*}
u_{a}T_{\quad;b}^{ab}= & 0\\
\Leftrightarrow\left(nu^{a}\sigma\right)_{;a}= & \frac{1}{kT}u_{a}\Delta T_{\quad;b}^{ab}
\end{align*}
This can be rewritten to define Eckart's entropy flow/current
\begin{align*}
\left(nku^{a}\sigma-\frac{1}{T}u_{b}\Delta T^{ab}\right)_{;a}= & \frac{1}{T}u_{a}\Delta T_{\quad;b}^{ab}-\frac{1}{T}u_{b}\Delta T_{\quad;a}^{ab}-\frac{1}{T}u_{b;a}\Delta T^{ab}+\frac{1}{T^{2}}u_{b}\Delta T^{ab}T_{;a}\\
= & -\frac{1}{T}u_{a;b}\Delta T^{ab}+\frac{1}{T^{2}}u_{a}\Delta T^{ab}T_{;b}\\
= & S_{\:;a}^{a}
\end{align*}
Defining $R^{a}$, that depends up to 1st order in $\Delta T^{ab}$,
\begin{align*}
S^{a}= & S_{PF}^{a}+\frac{1}{T}R^{a} & (R^{a}= & -u_{b}\Delta T^{ab})
\end{align*}
The 2nd law of thermodynamics is then expressed as
\begin{gather*}
\boxed{S_{\:;a}^{a}\ge0}
\end{gather*}

\begin{description}
\item [{Recall}] General 2-tensor decomposition along $u^{a}$\\
$V_{ab}=Au_{a}u_{b}+B_{a}u_{b}+C_{b}u_{a}+F_{ab}$ always possible
with $B_{a}u^{a}=C_{a}u^{a}=0$ and $F_{ab}=h_{a}^{\:c}h_{b}^{\:d}V_{cd}\Rightarrow F_{ab}u^{a}=F_{ab}u^{b}=0$\\
We can decompose further into symmetric and antisymmetric parts\\
$F_{ab}=F_{\left(ab\right)}+F_{\left[ab\right]}$ and the symmetric
part yields the trace and projected symmetric tracefree (PSTF):\\
$F_{\left(ab\right)}=\frac{1}{3}h_{ab}F_{\,c}^{c}+F_{\left\langle ab\right\rangle }$
with $F_{\left\langle ab\right\rangle }=h_{a}^{\:c}h_{b}^{\:d}\left[F_{\left(cd\right)}-\frac{1}{3}h_{cd}F^{ef}h_{ef}\right]$
so\\
$V_{ab}=Au_{a}u_{b}+B_{a}u_{b}+C_{b}u_{a}+\frac{h_{ab}}{3}h^{cd}V_{cd}+V_{\left\langle ab\right\rangle }+\underset{0\textrm{ for }V\textrm{ sym.}}{\underbrace{h_{a}^{\:c}h_{b}^{\:d}V_{\left[cd\right]}}}$\\
\uline{Restrict to \mbox{$V$} symmetric}\\
$V_{ab}=Au_{a}u_{b}+2B_{(a}u_{b)}+\frac{h_{ab}}{3}h^{cd}V_{cd}+V_{\left\langle ab\right\rangle }$
\end{description}
Apply to $T^{ab}$

$T^{ab}=T_{PF}^{ab}+\Delta T^{ab}$ yields
\begin{align*}
\Delta T_{ab}=\pi h_{ab}+2q_{(a}u_{b)}+ & \pi_{ab}\,\left(+Au_{a}u_{b}\right)\\
 & \textrm{with }\pi_{ab}=\pi_{\left\langle ab\right\rangle }
\end{align*}

If we choose $u^{a}$such that

\hspace*{\fill}%
\begin{tabular}{ccc}
out of equilibrium quantities & match & quantities in equilibrium\tabularnewline
$\widetilde{\rho}$ & $\longrightarrow$ & $\rho$\tabularnewline
$\widetilde{n}$ & $\longrightarrow$ & $n$\tabularnewline
$\widetilde{P}$ & $\longrightarrow$ & $P+\pi$\tabularnewline
\end{tabular}\hspace*{\fill}

then $A=0$ and we have
\begin{align*}
T_{ab}= & \rho u_{a}u_{b}+\underset{\underset{\begin{array}[t]{c}
\textrm{bulk viscous }\\
\textrm{pressure}
\end{array}}{\downarrow}}{\left(P+\pi\right)h_{ab}}+\underset{\underset{\begin{array}[t]{c}
\textrm{heat}\\
\textrm{flow}
\end{array}}{\downarrow}}{2q_{(a}u_{b)}}+\underset{\underset{\begin{array}[t]{c}
\textrm{anisotropic}\\
\textrm{stress}
\end{array}}{\downarrow}}{\pi_{ab}}
\end{align*}
and the entropy flow $\boxed{S^{a}=\underset{S}{\underbrace{k\sigma}}nu^{a}+\frac{R^{a}}{T}}$
with $S$ specific entropy scalar

Following Eckart's postulate
\begin{align*}
R^{a}= & -u_{b}\Delta T^{ab}\\
= & -u_{b}\left(\pi h^{ab}+2q^{(a}u^{b)}+\pi^{ab}\right)\\
= & -u_{b}q^{a}u^{b}-u_{b}q^{b}u^{a}=q^{a}
\end{align*}
so $\boxed{S^{a}=SN^{a}+\frac{q^{a}}{T}}$ \uline{Eckart Entropy}

From the second law of thermodynamics application:
\begin{align*}
S_{\,;a}^{a}= & \left(-\frac{1}{T}u_{a;b}+\frac{1}{T^{2}}u_{a}T_{;b}\right)\Delta T^{ab}=-\Delta T^{ab}\left[\frac{1}{T}u_{a}\right]_{;b}\ge0\\
\Rightarrow TS_{\,;a}^{a}= & -\left[\pi h^{ab}+2q^{(a}u^{b)}+\pi^{ab}\right]\left(u_{a;b}-u_{a}\frac{T_{;b}}{T}\right)
\end{align*}
Here we further decompose the flow derivative
\begin{align*}
u_{a;b}= & -\dot{u}_{a}u_{b}+\frac{\Theta}{3}h_{ab}+\sigma_{\left\langle ab\right\rangle }+\omega_{\left[ab\right]}
\end{align*}

\begin{tabular}{cll}
with & $\dot{u}_{a}=u^{b}u_{a;b}$ & : acceleration\tabularnewline
 & $\Theta=u_{\,;a}^{a}$ & : expansion\tabularnewline
 & $\sigma_{\left\langle ab\right\rangle }=u_{\left\langle a;b\right\rangle }$ & : shear\tabularnewline
 & $\omega_{\left[ab\right]}=h_{a}^{\:c}h_{b}^{\:d}u_{\left[c;d\right]}$ & : vorticity\tabularnewline
\end{tabular}

So
\begin{align*}
TS_{\,;a}^{a}= & -\left[\pi\Theta+q^{a}\dot{u}_{a}+\pi^{ab}\sigma_{ab}\right]+\frac{1}{T}\left[\pi h^{ab}+2q^{(a}u^{b)}+\pi^{ab}\right]\left(u_{a}T_{;b}\right)\\
= & -\left[\pi\Theta+q^{a}\left(\dot{u}_{a}+\frac{T_{;a}}{T}\right)+\pi^{ab}\sigma_{ab}\right]\ge0
\end{align*}
This constraint leads to Eckact's assumptions (with each $u_{a;b}$
terms independent)
\begin{gather*}
\left.\begin{array}{rl}
\pi= & -\zeta\Theta\\
q_{a}= & -\chi\left(T_{;a}+T\dot{u}_{a}\right)\\
\pi_{ab}= & -2\eta\sigma_{ab}
\end{array}\right\} \underline{\textrm{Eckart constitutive equations}}\textrm{ with}\left.\begin{array}{rl}
\zeta & \textrm{: bulk viscosity}\\
\chi & \textrm{: heat conductivity}\\
\eta & \textrm{: shear viscosity}
\end{array}\right\} >0
\end{gather*}
where we recognise the equations for

\begin{tabular}{ll}
$q_{a}$: & Fourier transport equation\tabularnewline
$\pi_{ab}$: & Newton shear viscosity\tabularnewline
$\pi$: & Stokes law of fluid mechanics\tabularnewline
\end{tabular}

so the entropy derivative is quadratic in $\pi,q^{a},\pi^{ab}$
\begin{gather*}
\boxed{S_{\,;a}^{a}=\frac{\pi^{2}}{\zeta T}+\frac{q^{a}q_{a}}{\chi T}+\frac{\pi^{ab}\pi_{ab}}{2\eta T}}\left(=0\textrm{ for perfect fluid}\right)
\end{gather*}
($\Rightarrow$P.F. always evolve adiabatically) while the entropy
flow is 1st order in $\pi,q^{a},\pi^{ab}$ through $R^{a}$
\begin{gather*}
\boxed{S^{a}=SN^{a}+\frac{q^{a}}{T}}
\end{gather*}

\subsection{Problem with Eckact's model}

Algebraic relations between flow and fluid quantities

$\Rightarrow$ infinite propagation speed

$\Rightarrow$unstable perturbations from equilibrium of the flow:
$\begin{array}[t]{rl}
\dot{u}= & 0\\
\Theta= & 0\\
\sigma= & \text{0}
\end{array}$

$\rightarrow$need for a better model: Israel-Stewart 

\section{Israel-Stewart model}

Find $R^{a}$ up to 2nd order. Add to Eckart 2nd order terms: postulate
the form
\begin{align*}
R^{a}=q^{a} & -\frac{u^{a}}{2}\left(\beta_{0}\pi^{2}+\beta_{1}q_{b}q^{b}+\beta_{2}\pi^{bc}\pi_{bc}\right) & \textrm{(quadratic terms)}\\
 & +\alpha_{0}\pi q^{a}+\alpha_{1}\pi^{ab}q_{b}(+\alpha_{2}\underset{\begin{array}{c}
0\\
{\scriptscriptstyle \textrm{orthogonal}}
\end{array}}{\pi\underbrace{u_{b}\pi^{ab}}}) & \textrm{(cross terms)}
\end{align*}
which yields the entropy flow
\begin{align*}
S^{a}=SN^{a}+\frac{1}{T}\left[q^{a}\vphantom{\frac{u^{a}}{2}}\right. & -\frac{u^{a}}{2}\left(\beta_{0}\pi^{2}+\beta_{1}q_{b}q^{b}+\beta_{2}\pi^{bc}\pi_{bc}\right)\\
 & \left.\vphantom{\frac{u^{a}}{2}}+\alpha_{0}\pi q^{a}+\alpha_{1}\pi^{ab}q_{b}\right]\qquad\textrm{\ensuremath{\rightarrow}heat/viscosity coupling}
\end{align*}
and along the flow (comoving observer's entropy)
\begin{align*}
u_{a}S^{a}= & -nS+\frac{1}{2T}\left(\beta_{0}\pi^{2}+\beta_{1}q_{b}q^{b}+\beta_{2}\pi^{bc}\pi_{bc}\right)
\end{align*}
Since it doesn't depend on the $\alpha_{i}$ terms, we neglect them
and \uline{assume} for the rest that $\boxed{\alpha_{i}=0}$

Thus
\begin{align*}
S^{a}= & S_{E}^{a}-\frac{u^{a}}{2T}\left(\beta_{0}\pi^{2}+\beta_{1}q_{b}q^{b}+\beta_{2}\pi^{bc}\pi_{bc}\right)
\end{align*}
and
\begin{align*}
TS_{\,;a}^{a}= & TS_{E\,;a}^{a}-\frac{T}{2}\left[\frac{u^{a}}{T}\left(\beta_{0}\pi^{2}+\beta_{1}q_{b}q^{b}+\beta_{2}\pi^{bc}\pi_{bc}\right)\right]_{;a}\\
= & TS_{E\,;a}^{a}-\Delta\left(TS_{\,;a}^{a}\right)
\end{align*}
We have seen previously that
\begin{align*}
TS_{E\,;a}^{a}= & -\left[\pi\Theta+q^{a}\left(\dot{u}_{a}+\frac{T_{;a}}{T}\right)+\pi^{ab}\sigma_{ab}\right]
\end{align*}
and
\begin{align*}
\Delta\left(TS_{\,;a}^{a}\right)= & \frac{T}{2}\left(\frac{u^{a}}{T}\right)_{;a}\left(\beta_{0}\pi^{2}+\beta_{1}q_{b}q^{b}+\beta_{2}\pi^{bc}\pi_{bc}\right)+u^{a}\left(\beta_{0}\pi\pi_{;a}+\beta_{1}q^{b}q_{b;a}+\beta_{2}\pi^{bc}\pi_{bc;a}\right)\\
= & \beta_{0}\pi\left(\dot{\pi}+\frac{T}{2}\pi\left(\frac{u^{a}}{T}\right)_{;a}\right)+\beta_{1}q^{b}\left(\dot{q}_{b}+\frac{T}{2}q_{b}\left(\frac{u^{a}}{T}\right)_{;a}\right)+\beta_{2}\pi^{bc}\left(\dot{\pi}_{bc}+\frac{T}{2}\pi_{bc}\left(\frac{u^{a}}{T}\right)_{;a}\right)
\end{align*}
Therefore
\begin{align*}
TS_{\,;a}^{a}= & -\pi\left[\Theta+\beta_{0}\dot{\pi}+\frac{T}{2}\left(\frac{\beta_{0}u^{a}}{T}\right)_{;a}\pi\right]\\
 & -q^{a}\left[\frac{T_{;a}}{T}+\dot{u}_{a}+\beta_{1}\dot{q}_{a}+\frac{T}{2}\left(\frac{\beta_{1}u^{b}}{T}\right)_{;b}q_{a}\right]\\
 & -\pi^{ab}\left[\sigma_{ab}+\beta_{2}\dot{\pi}_{ab}+\frac{T}{2}\left(\frac{\beta_{2}u^{c}}{T}\right)_{;c}\pi_{ab}\right]\ge0
\end{align*}
The last condition imposed by the 2nd law of thermodynamics 

In order to keep $TS_{\,;a}^{a}$ quadratic and thus the 2nd law,
we can choose each square bracket proportional to their respective
factors $\pi,q^{a},\pi^{ab}$ (recall $q^{a}=h_{\:b}^{a}q^{b},\pi^{ab}=h_{\:c}^{a}h_{\:d}^{b}\pi^{cd}$)
\begin{gather*}
\left\{ \begin{array}{rlr}
\zeta\beta_{0}\dot{\pi}+\pi= & -\zeta\Theta-\zeta\left[\frac{T}{2}\left(\frac{\beta_{0}u^{a}}{T}\right)_{;a}\pi\right] & \left(\textrm{from }\pi=-\zeta\left[\right]_{\pi}\right)\\
\chi T\beta_{1}h_{a}^{\:b}\dot{q}_{b}+q_{a}= & -\chi\left(h_{a}^{\:b}T_{;b}+T\dot{u}_{a}\right)-\chi\left[\frac{T^{2}}{2}\left(\frac{\beta_{1}u^{b}}{T}\right)_{;b}q_{a}\right] & \left(\textrm{from }q_{a}=-\chi T\left[\right]_{q}\right)\\
2\eta\beta_{2}h_{a}^{\:c}h_{b}^{\:d}\dot{\pi}_{cd}+\pi_{ab}= & -2\eta\sigma_{ab}-\eta\left[T\left(\frac{\beta_{2}u^{c}}{T}\right)_{;c}\pi_{ab}\right] & \left(\textrm{from }\pi_{ab}=-2\eta\left[\right]_{\pi_{ab}}\right)
\end{array}\right.
\end{gather*}
\begin{tabular}[t]{cl}
\fbox{\begin{minipage}[t]{3.8cm}%
Israel-Stewart equations%
\end{minipage}} with the same & $\zeta$: bulk viscosity\tabularnewline
 & $\chi$: heat conductivity\tabularnewline
 & $\eta$: shear viscosity\tabularnewline
\end{tabular}

which ensure
\begin{align*}
TS_{\,;a}^{a}= & \frac{\pi^{2}}{\zeta}+\frac{q^{a}q_{a}}{\chi}+\frac{\pi^{ab}\pi_{ab}}{2\eta}\ge0
\end{align*}
for $\zeta,\chi,\eta>0$ and define the relaxation times
\begin{gather*}
\left\{ \begin{array}{rlr}
\tau_{0}= & \zeta\beta_{0}\\
\tau_{1}= & \chi T\beta_{1} & \textrm{ and recall }\tau\sim\frac{1}{n\sigma v}\\
\tau_{2}= & 2\eta\beta_{2}
\end{array}\right.
\end{gather*}
If we neglect the 2nd R.H.S. terms (collisional time small) we obtain
the truncated I-S equations
\begin{gather*}
\left\{ \begin{array}{rl}
\tau_{0}\dot{\pi}+\pi= & -\zeta\Theta\\
\tau_{1}h_{a}^{\:b}\dot{q}_{b}+q_{a}= & -\chi\left(h_{a}^{\:b}T_{;b}+T\dot{u}_{a}\right)\\
\tau_{2}h_{a}^{\:c}h_{b}^{\:d}\dot{\pi}_{cd}+\pi^{ab}= & -2\eta\sigma_{ab}
\end{array}\right.
\end{gather*}
which introduce dynamic corrections to the Eckart equations.

We assume to be close to equilibrium
\begin{align*}
\left|\pi\right|\ll & P\\
\left(\pi^{ab}\pi_{ab}\right)^{\frac{1}{2}}\ll & P\\
\left(q^{a}q_{a}\right)^{\frac{1}{2}}\ll & \rho
\end{align*}
and thus
\begin{align*}
S_{\,;a}^{a}=\left(SN^{a}+\frac{R^{a}}{T}\right)_{;a}= & n\dot{S}+\left(\frac{R^{a}}{T}\right)_{;a}\textrm{(part. nb con.)}\\
= & n\dot{S}+\left(\frac{q^{a}}{T}\right)_{;a}+\left(\frac{\Delta R^{a}}{T}\right)_{;a}\simeq n\dot{S}+\left(\frac{q^{a}}{T}\right)_{;a}
\end{align*}
and
\begin{align*}
TS_{\,;a}^{a}= & TS_{E\,;a}^{a}-\Delta\left(TS_{\,;a}^{a}\right)\simeq TS_{E\,;a}^{a}
\end{align*}
we have in Eckart and Israel-Stewart models in that case
\begin{gather*}
\boxed{Tn\dot{S}=-\pi\Theta-q^{a}\dot{u}_{a}-q_{\,;a}^{a}-\pi^{ab}\sigma_{ab}}
\end{gather*}
Specific entropy evolution: measures evolution of entropy of the fluid
along the flow
\begin{example}
Entropy contained in comoving FLRW volume
\begin{align*}
\Sigma=\int\left(-S^{a}u_{a}\right)\frac{dV}{v}= & \left[Sn-\frac{1}{2}\left(\beta_{0}\pi^{2}+\beta_{1}q_{b}q^{b}+\beta_{2}\pi^{bc}\pi_{bc}\right)\right]a^{3}\\
= & a^{3}nS:\\
 & \boxed{\Sigma=a^{3}nS}\textrm{ with the previous assumptions}
\end{align*}
\end{example}

\chapter{Tetrads and Newman-Penrose Tetrads}

\begin{tabular}{lll}
We saw & 1+3 & \tabularnewline
 & 1+1+2 & formalisms in congruences\tabularnewline
 & 2+2 & \tabularnewline
 & $\left(e_{\mu}\right)^{a}\bot T^{a}$ & in singularities\tabularnewline
\end{tabular}

\section{Tetrads}

Some vectors can be decomposed on others

\begin{tabular}{ll}
Vectors as components in coordinate basis & $V=V^{a}\partial_{a}$\tabularnewline
can be generalised as tetrads & $V=V^{A}e_{A}$\tabularnewline
\end{tabular}

Following Chandrasekhar \cite{Chandrasekhar:1985kt} ``The mathematical
theory of black holes'', Ashtekar \cite{Ashtekar:2000hw} \textquotedbl{}Isolated
horizons: Hamiltonian evolution and the first law\textquotedbl{} Appendix
B, Trautman  \textquotedbl{}Einstein-Cartan theory\textquotedbl{},
and \href{https://en.wikipedia.org/wiki/Newman-Penrose_formalism}{Newman-Penrose formalism},
\href{https://en.wikipedia.org/wiki/Vaidya_metric}{Vaidya metric}
we explore tetrads.

\begin{tabular}{lll}
In 4D spacetime: & tetrads vector labels  & :$A,B,...:\,e_{A}$\tabularnewline
 & spacetime indices & :$a,b,...:\,\partial_{a}$\tabularnewline
\end{tabular}

so $e_{A}=e_{A}^{\:i}\partial_{i}$ with $\left.\begin{array}{c}
i\\
A
\end{array}\right\} \in\left[0;3\right]$ or $\begin{array}{c}
i\in\left[0;3\right]\\
A\in\left[\Box,I,I\!I,I\!I\!I\right]
\end{array}$

\begin{tabular}{ll}
Also designated & $e_{A}^{\:i}$ with corresponding forms\tabularnewline
 & $e_{A\:i}=g_{ij}e_{A}^{\:j},g_{ij}$ metric ($e_{A}=e_{A\:i}dx^{i})$\tabularnewline
\end{tabular}

Assume $e_{A}$ orthonormal and $e_{\Box}$ timelike, $e_{I},e_{I\!I},e_{I\!I\!I}$
spacelike (or noted $e_{\left(0\right)}$ and $e_{\left(1\right)},e_{\left(2\right)},e_{\left(3\right)}$)
then
\begin{align*}
g_{ij}e_{A}^{\:i}e_{B}^{\:j}= & \eta_{AB}=\left(e_{A},e_{B}\right)
\end{align*}
with $\eta$ Minkowski metric. This can be seen as the line element
in tangent Minkowski space
\begin{align*}
g_{ij}e_{A}^{\:i}e_{B}^{\:j}dx^{A}dx^{B}= & \eta_{AB}dx^{A}dx^{B}=ds_{M^{T}}^{2}=\left(e_{A},e_{B}\right)x^{A}dx^{B}
\end{align*}
Note in general $e_{A}$ is not orthonormal and $\left(e_{A},e_{B}\right)=\eta_{AB}$
metric of general tangent manifold to spacetime.

For most of the rest, we assume $\eta$ constzant Minkowski

$\exists\eta^{AB}$ inverse: $\eta^{AC}\eta_{CB}=\delta_{\,B}^{A}$

$\Rightarrow\eta^{AB}e_{B}=e^{A}=e_{\:i}^{A}dx^{i}$ inverse form
basis, so line element in spacetime can be seen as
\begin{align*}
\eta_{AB}e^{A}e^{B}= & \eta_{AB}e_{\:i}^{A}e_{\:j}^{B}dx^{i}dx^{j}=g_{ij}dx^{i}dx^{j}=ds^{2}\\
= & \eta_{AB}\eta^{AC}e_{C}\eta^{BD}e_{D}\\
= & \delta_{B}^{\:C}\eta^{BD}e_{C}e_{D}=\eta^{CD}e_{C}e_{D}=e^{D}e_{D}
\end{align*}
so $g_{ij}=\eta_{AB}e_{\:i}^{A}e_{\:j}^{B}$

$\exists g^{ij}$ inverse: $g^{ik}g_{kj}=\delta_{\,j}^{i}$ and

$\eta_{AB}e^{B}=\eta_{AB}\eta^{BC}e_{C}=\delta_{A}^{\:C}e_{C}=e_{A}\Rightarrow\left(\eta_{AB}e^{B}\right)^{i}\partial_{i}=e_{A}^{\:i}\partial_{i}=\eta_{AB}e_{\:j}^{B}g^{ij}\partial_{i}$

Samely $e^{A}=\eta^{AB}e_{B}\Rightarrow\left(\eta^{AB}e_{B}\right)^{i}\partial_{i}=e_{\:j}^{A}g^{ij}\partial_{i}$

while 
\begin{align*}
\delta_{\,j}^{i}=g^{ik}g_{kj}= & \eta_{AB}e_{\:k}^{A}e_{\:j}^{B}g^{ik}\\
= & e_{B}^{\:i}e_{\:j}^{B}\,:\,e_{\:i}^{A}\textrm{ spacetime inverse of }e_{A}^{\:j}
\end{align*}
Thus as
\begin{align*}
\eta^{AB}e_{A}^{\:i}e_{B}^{\:k}g_{kj}= & e_{A}^{\:i}e_{\:l}^{A}g^{kl}g_{kj}\\
= & \delta_{\,l}^{i}\delta_{\,j}^{l}=\delta_{\,j}^{i}=g^{ik}g_{kj}
\end{align*}
we also have
\begin{align*}
g^{ij}= & \eta^{AB}e_{A}^{\:i}e_{B}^{\:j}
\end{align*}
as well as
\begin{align*}
\delta_{\,B}^{A}=\eta^{AC}\eta_{CB}= & \eta^{AC}g_{ij}e_{C}^{\:i}e_{B}^{\:j}\\
= & e_{B}^{\:j}\eta^{AC}e_{C}^{\:i}g_{ij}\\
= & e_{B}^{\:j}e_{\:j}^{A}\,:\,e_{\:i}^{A}\textrm{ tangent space inverse of }e_{B}^{\:i}
\end{align*}
and finally $g^{ij}e_{\:i}^{A}e_{\:j}^{B}=\eta^{AC}e_{C}^{\;j}e_{\:j}^{B}=\eta^{AC}\delta_{C}^{\:B}=\eta^{AB}$
\begin{description}
\item [{In~summary}] Basis $e_{A}=e_{A}^{\:i}\partial_{i}$, inverse form
basis $\eta^{AB}e_{B}=e^{A}=e_{\:i}^{A}dx^{i}$\\
\begin{tabular}{ccc}
orthonormalisation &  $\eta_{AB}=g_{ij}e_{A}^{\:i}e_{B}^{\:j}$ & \tabularnewline
inverse: & $\eta^{AB}=g^{ij}e_{\:i}^{A}e_{\:j}^{B}$ & as $\eta^{AC}\eta_{CB}=\delta_{\,B}^{A}$\tabularnewline
\multicolumn{3}{c}{tangent space metric}\tabularnewline
\end{tabular}\\
\begin{tabular}{rll}
spacetime line element  & $ds^{2}=\eta_{AB}e^{A}e^{B}$ & metric $g_{ij}=\eta_{AB}e_{\:i}^{A}e_{\:j}^{B}$\tabularnewline
inverse metric & $g^{ij}=\eta^{AB}e_{A}^{\:i}e_{B}^{\:j}$ & as $g^{ik}g_{kj}=\delta_{\,j}^{i}$\tabularnewline
form basis & $e_{A}=e_{A}^{\:i}g_{ij}dx^{j}$ & vector inverse basis $e^{A}=e_{\:j}^{A}g^{ij}\partial_{i}$\tabularnewline
\end{tabular}\\
so components of basis inverse of components of form basis\\
\begin{tabular}{lll}
right inverse: & $e_{\:j}^{A}e_{A}^{\:i}=\delta_{\,j}^{i}$ & spacetime inverse\tabularnewline
left inverse: & $e_{A}^{\:i}e_{\:i}^{B}=\delta_{A}^{\:B}$ & tangent space inverse\tabularnewline
\end{tabular}\\
\begin{tabular}{lll}
and we can define & form component & $g_{ij}e_{A}^{\:j}=e_{A\:i}$\tabularnewline
 & vector component & $g^{ij}e_{\:j}^{A}=e^{A\:i}$\tabularnewline
\end{tabular}
\end{description}
Since, chosing $e_{\Box}$ timelike,
\begin{align*}
\eta_{AB}=\eta^{AB}= & \left(\begin{array}{cccc}
-1 & 0 & 0 & 0\\
0 & 1 & 0 & 0\\
0 & 0 & 1 & 0\\
0 & 0 & 0 & 1
\end{array}\right)
\end{align*}
$e_{\:i}^{A}=\eta^{AB}e_{B\:i}$ then $e_{\:i}^{\Box}=-e_{\Box\:i}$
and $e_{\:i}^{\left(\mu\right)}=-e_{\left(\mu\right)\:i},\mu\in\left[1;3\right]$
so the metric reads
\begin{align*}
g_{ij}= & e_{\:i}^{\Box}e_{\:j}^{\Box}+e_{\,i}^{I}e_{\,j}^{I}+e_{\,i}^{I\!I}e_{\,j}^{I\!I}+e_{\,i}^{I\!I\!I}e_{\,j}^{I\!I\!I}\\
= & -e_{\Box\:i}e_{\Box\:j}+e_{I\:i}e_{I\:j}+e_{I\!I\:i}e_{I\!I\:j}+e_{I\!I\!I\:i}e_{I\!I\!I\:j}
\end{align*}
(recall 1+1+2)

\subsection{Tetrad tensor components}

Any tensor can be projected onto tetrad frame
\begin{align*}
A_{A}= & e_{A\:i}A^{i}=e_{A}^{\:i}A_{i}\textrm{ for }1\negmedspace-\textrm{tensors}\\
A^{A}= & \eta^{AB}A_{B}=e_{\:i}^{A}A^{i}=e^{A\:i}A_{i}
\end{align*}
with spacetime tensors given by
\begin{align*}
A^{i}= & e_{A}^{\:i}A^{A}=e^{A\:i}A_{A}, & A_{i}= & e_{\:i}^{A}A_{A}=e_{A\:i}A^{A}
\end{align*}
For a 2-tensor
\begin{align*}
T_{AB}= & e_{A}^{\:i}e_{B}^{\:j}T_{ij}=e_{A}^{\:i}T_{iB}
\end{align*}
inverted by
\begin{align*}
T_{ij}= & e_{\:i}^{A}e_{\:j}^{B}T_{AB}=e_{\:i}^{A}T_{Aj}
\end{align*}
Indices are contracted, raised and lowered with respective metrics

\subsubsection{Note on 2-tensors and their matrix representation}

The tensorial product has no order, keeping origin in mind
\begin{align*}
T_{abc}X_{de}= & X_{de}T_{abc}
\end{align*}
same numbers but
\begin{align*}
A_{abcde}\equiv & T_{abc}X_{de}\ne X_{de}T_{abc}
\end{align*}
Contractions only care about index positions
\begin{align*}
T_{ab}X^{bc}= & X^{bc}T_{ab}\ne X^{cb}T_{ab}\textrm{ unless }X\textrm{ symmetric}
\end{align*}
Symmetrisations require indices at same level in general: $T_{[a}^{\;b]}$
has no sense

Trace is contraction with metric $T_{\:a}^{a}=T_{ab}g^{ab}$

\paragraph{Matrix representation of 2-tensors}

\begin{align*}
\left[T_{ab}\right]= & \left[T_{a1},\dots,T_{an}\right]=\left[\begin{array}{c}
T_{1b}\\
\vdots\\
T_{nb}
\end{array}\right]
\end{align*}
Line or column division is selected when using matrix product to represent
tensor contraction. Then first index stands for line number while
secind is column number. Index position is encoded in matrix relative
position for contraction.
\begin{example}
\begin{align*}
A_{ab}= & \left(\begin{array}{cccc}
0 & A_{01} & 0 & 0\\
0 & A_{11} & 0 & 0\\
0 & 0 & 0 & 0\\
0 & 0 & 0 & 0
\end{array}\right)=A_{01}\delta_{a}^{0}\delta_{b}^{1}+A_{11}\delta_{a}^{1}\delta_{b}^{1}\\
B^{ab}= & \left(\begin{array}{cccc}
B^{00} & B^{01} & 0 & 0\\
0 & 0 & 0 & 0\\
0 & 0 & 0 & 0\\
0 & 0 & 0 & 0
\end{array}\right)=B^{00}\delta_{0}^{a}\delta_{0}^{b}+B^{01}\delta_{0}^{a}\delta_{1}^{b}
\end{align*}
same order as matrix product
\begin{align*}
A_{ac}B^{cb}=0= & \left(A_{01}\delta_{a}^{0}\delta_{c}^{1}+A_{11}\delta_{a}^{1}\delta_{c}^{1}\right)\left(B^{00}\delta_{0}^{c}\delta_{0}^{b}+B^{01}\delta_{0}^{c}\delta_{1}^{b}\right)\\
= & A_{01}B^{00}\delta_{a}^{0}\delta_{0}^{b}\underset{0}{\underbrace{\delta_{0}^{1}}}+A_{01}B^{01}\delta_{a}^{0}\delta_{1}^{b}\underset{0}{\underbrace{\delta_{0}^{1}}}+A_{11}B^{00}\delta_{a}^{1}\delta_{0}^{b}\underset{0}{\underbrace{\delta_{0}^{1}}}+A_{11}B^{01}\delta_{a}^{1}\delta_{1}^{b}\underset{0}{\underbrace{\delta_{0}^{1}}}=0
\end{align*}
opposite order
\begin{align*}
A_{cb}B^{ac}=\left[B^{ac}\right]\left[A_{cb}\right]= & \left(\begin{array}{cccc}
0 & A_{01}B^{00}+A_{11}B^{01} & 0 & 0\\
0 & 0 & 0 & 0\\
0 & 0 & 0 & 0\\
0 & 0 & 0 & 0
\end{array}\right)\\
= & \left(A_{01}\delta_{c}^{0}\delta_{b}^{1}+A_{11}\delta_{c}^{1}\delta_{b}^{1}\right)\left(B^{00}\delta_{0}^{a}\delta_{0}^{c}+B^{01}\delta_{0}^{a}\delta_{1}^{c}\right)\\
= & A_{01}B^{00}\delta_{b}^{1}\delta_{0}^{a}\underset{1}{\underbrace{\delta_{0}^{0}}}+A_{01}B^{01}\delta_{b}^{1}\delta_{0}^{a}\underset{0}{\underbrace{\delta_{1}^{0}}}+A_{11}B^{00}\delta_{b}^{1}\delta_{0}^{a}\underset{0}{\underbrace{\delta_{0}^{1}}}+A_{11}B^{01}\delta_{b}^{1}\delta_{0}^{a}\underset{1}{\underbrace{\delta_{1}^{1}}}\\
= & \left(A_{01}B^{00}+A_{11}B^{01}\right)\delta_{0}^{a}\delta_{b}^{1}
\end{align*}
traces: $A=A_{01}g^{01}+A_{11}g^{11}\ne A_{11}$, $B=B^{00}g_{00}+B^{01}g_{01}\ne B^{00}$

Trace = matrix trace only for $A_{a}^{\:b}$ or $B_{\:b}^{a}$!
\end{example}

\subsection{Directional derivatives, connections and Ricci rotation coefficients}

Recall: Vectors are differential operators

$e_{A}=e_{A}^{\:i}\partial_{i}$ generalises with covariant derivative
$e_{A}=e_{A}^{\:i}\nabla_{i}=\nabla_{e_{A}}=\nabla_{A}$

Partial derivatives generalise as ($\phi$ scalar function)
\begin{align*}
\phi_{,A}= & e_{A}^{\:i}\partial_{i}\phi=e_{A}^{\:i}\phi_{,i}=e_{A}^{\:i}\phi_{;i}\textrm{ for scalars}
\end{align*}
Applied to a form
\begin{align*}
A_{A;B}= & e_{B}^{\:i}\nabla_{i}A_{A}=e_{B}^{\:i}\nabla_{i}\left(e_{A}^{\:j}A_{j}\right)=e_{B}^{\:i}\nabla_{\partial_{i}}\left(e_{A}^{\:j}A_{j}\right)\\
= & e_{B}^{\:i}\left[e_{A}^{\:j}A_{j;i}+A_{j}e_{A\,;i}^{\:j}\right]\\
= & e_{A}^{\:j}A_{j;i}e_{B}^{\:i}+e_{B}^{\:i}e_{A\,j;i}e_{C}^{\:j}A^{C}
\end{align*}

\begin{description}
\item [{Recall:~connection}] defined by $\nabla_{b}\partial_{a}=\Gamma_{ab}^{c}\partial_{c}$
\end{description}
\begin{proof}
Connection definition

From directional derivative of a vector $\nabla_{X}Y$ in a basis
$e_{A}$
\begin{align*}
X^{A}\nabla_{e_{A}}\left(Y^{B}e_{B}\right)= & X^{A}Y_{,A}^{B}e_{B}+Y^{B}X^{A}\nabla_{e_{A}}e_{B}
\end{align*}
define connection as decomposition of $\nabla_{e_{A}}e_{B}$ on $e_{C}$
\begin{align*}
\nabla_{e_{A}}e_{B}= & \omega_{BA}^{C}e_{C}\\
\Rightarrow\nabla_{X}Y= & X^{A}\left(Y_{,A}^{B}e_{B}+Y^{B}\omega_{BA}^{C}e_{C}\right)\\
= & X^{A}\left(Y_{,A}^{C}+Y^{B}\omega_{BA}^{C}\right)e_{C}=X^{A}\left(\nabla_{e_{A}}Y\right)^{C}e_{C}\\
\Rightarrow\left(\nabla_{e_{A}}Y\right)^{C}= & Y_{,A}^{C}+Y^{B}\omega_{BA}^{C}
\end{align*}
Choose $e_{A}=\partial_{a},Y=\partial_{a}=\delta_{a}^{b}\partial_{b}$
\begin{align*}
\left(\nabla_{a}\partial_{b}\right)^{c}= & \left(\delta_{b}^{c}\right)_{,a}+\omega_{b^{\prime}a}^{c}\delta_{b}^{b^{\prime}}=\omega_{ba}^{c}=\Gamma_{ba}^{c}
\end{align*}
or
\begin{align*}
\nabla_{a}\partial_{b}= & \Gamma_{ba}^{c}\partial_{c}
\end{align*}
\end{proof}
Here
\begin{align*}
\nabla_{B}e_{A}= & \Gamma_{AB}^{C}e_{C}=\Gamma_{AB}^{C}e_{C}^{\:i}\partial_{i}\\
= & e_{B}^{\:b}\nabla_{b}e_{A}=e_{A\,;b}e_{B}^{\:b}=e_{A\,;b}^{\:i}e_{B}^{\:b}\partial_{i}
\end{align*}
\warningsign $\Gamma_{AB}^{C}$ may not have same symmetries as $\Gamma_{ab}^{c}$!

So connection coefficients read
\begin{align*}
e_{B}^{\:i}e_{A\,;i}^{\;j}=\Gamma_{CAB}\eta^{CD}e_{D}^{\:j}\Rightarrow g_{jk}e_{C}^{\:k}e_{A\,;i}^{\;j}e_{B}^{\:i}= & \Gamma_{FAB}\eta^{FD}e_{D}^{\:j}g_{jk}e_{C}^{\:k}\\
= & \Gamma_{FAB}\eta^{FD}\eta_{DC}\\
= & \Gamma_{FAB}\delta_{\:C}^{F}=\Gamma_{CAB}
\end{align*}

Defining Ricci rotation coefficients (connection)
\begin{align*}
\gamma_{CAB}= & e_{C}^{\:j}e_{A\,j;i}e_{B}^{\:i}=\Gamma_{CAB}\textrm{ we get }\\
A_{A;B}= & e_{A}^{\:j}A_{j;i}e_{B}^{\:i}+\gamma_{CAB}A^{C}
\end{align*}

\subsubsection{Properties of $\gamma$}

\begin{align*}
e_{A\,i;j}= & e_{\:i}^{C}\gamma_{CAB}e_{\:j}^{B}=\delta_{i}^{\,k}e_{A\,k;l}\delta_{j}^{\,l}
\end{align*}
Antisymmetry in first pair: $\gamma_{CAB}=-\gamma_{ACB}$ $\left(\gamma_{\left[CA\right]B}=\gamma_{CAB}\right)$
\begin{proof}
From $0=\eta_{AB;i}=\left[e_{Aj}e_{B}^{\:j}\right]_{;i}$ (not valid
for non-orthonormal basis!)
\begin{align*}
\left[e_{Aj}e_{B}^{\:j}\right]_{;i}= & e_{Aj;i}e_{B}^{\:j}+e_{Aj}e_{B\,;i}^{\:j}\\
= & e_{\:j}^{C}\gamma_{CAD}e_{\:i}^{D}e_{B}^{\:j}+e_{Aj}e^{Cj}\gamma_{CBD}e_{\:i}^{D}\\
= & \gamma_{BAD}e_{\:i}^{D}+\gamma_{ABD}e_{\:i}^{D}=\eta_{AB;i}=0\\
\Rightarrow\gamma_{ABD}= & -\gamma_{BAD}
\end{align*}
\end{proof}
Returning to 1-form directional derivative equation

Totally projected covariant derivative of frame/tetrad defines: intrinsic
derivative (notation)
\begin{align*}
A_{A|B}= & e_{A}^{\:i}A_{i;j}e_{B}^{\:j}\Leftrightarrow A_{i;j}=e_{\:i}^{A}A_{A|B}e_{\:j}^{B}\\
= & A_{A;B}-\gamma_{NAB}\eta^{NM}A_{M}
\end{align*}
(Notation change: $X_{\left[a|b|c\right]}$ to $X_{\left[A\brokenvert B\brokenvert C\right]}=\eta_{BD}X_{\left[A\quad C\right]}^{\quad D}$)

This can be generalised to any tensor (e.g. Riemann)

So Riemann covariant derivative leads to
\begin{align*}
R_{ABCD|F}= & R_{ijkl;m}e_{A}^{\:i}e_{B}^{\:j}e_{C}^{\:k}e_{D}^{\:l}e_{F}^{\:m}\\
\textrm{with }R_{ABCD;F}= & \left[R_{ijkl;m}e_{A}^{\:i}e_{B}^{\:j}e_{C}^{\:k}e_{D}^{\:l}\right]_{;m}e_{F}^{\:m}\\
\Rightarrow R_{ABCD|F}= & R_{ABCD;F}-\eta^{NM}\left[\gamma_{NAF}R_{MBCD}+\gamma_{NBF}R_{AMCD}+\gamma_{NCF}R_{ABMD}+\gamma_{NDF}R_{ABCM}\right]
\end{align*}

\subsubsection{Ricci rotation coefficients as Levi-Civita-like connection}

The $\gamma$ can be computed from mere partial derivatives

Define
\begin{align*}
\lambda_{ABC}= & e_{Bi;j}\left[e_{A}^{\:i}e_{C}^{\:j}-e_{A}^{\:j}e_{C}^{\:i}\right]\\
= & 2e_{Bi;j}e_{A}^{\:[i}e_{C}^{\:j]}\\
= & 2e_{B\left[i;j\right]}e_{A}^{\:i}e_{C}^{\:j}=\left[e_{Bi,j}-e_{Bj,i}\right]e_{A}^{\:i}e_{C}^{\:j}
\end{align*}
as covariant derivative connection $\Gamma$ is symmetric in GR and
\begin{align*}
e_{B\left[i;j\right]}= & e_{B\left[i,j\right]}-\Gamma_{\left[ij\right]}^{k}e_{Bk}=e_{B\left[i,j\right]}
\end{align*}
Moreover, as
\begin{align*}
e_{A}^{\:i}e_{Bi;j}e_{C}^{\:j}= & e_{A}^{\:i}e_{\:i}^{D}\gamma_{DBF}e_{\:j}^{F}e_{C}^{\:j}\\
= & \gamma_{ABC}
\end{align*}
we have
\begin{align*}
\lambda_{ABC}= & \gamma_{ABC}-\gamma_{CBA}
\end{align*}
By circular permutation and using $\gamma$ antisymmetry
\begin{align*}
\lambda_{ABC}-\lambda_{BCA}+\lambda_{CAB}= & \gamma_{ABC}-\gamma_{CBA}\\
 & -\gamma_{BCA}+\gamma_{ACB}\\
 & +\gamma_{CAB}-\gamma_{BAC}\\
= & 2\gamma_{ABC}\\
\Leftrightarrow\gamma_{ABC}= & \frac{1}{2}\left(\lambda_{ABC}-\lambda_{BCA}+\lambda_{CAB}\right)
\end{align*}
looking like a Levi-Civita connection for the metric derivative-like
$\lambda$

By construction $\lambda_{ABC}=-\lambda_{CBA}$

\subsubsection{Commutation and strucutre constants}

The Lie bracket of tetrad yields an algebra defined by its structure
constants
\begin{align*}
\left[e_{A},e_{B}\right]f= & e_{A}^{\:i}\left[e_{B}^{\:j}f_{,j}\right]_{;i}-e_{B}^{\:j}\left[e_{A}^{\:i}f_{,i}\right]_{;j}\\
= & \left[e_{A}^{\:i}e_{B\,;i}^{\:j}-e_{B}^{\:i}e_{A\,;i}^{\:j}\right]f_{,j}\textrm{ as }f_{,i;j}=f_{,j;i}\\
= & \left[e^{Dj}\gamma_{DBF}e_{\:i}^{F}e_{A}^{\:i}-e^{Dj}\gamma_{DAF}e_{\:i}^{F}e_{B}^{\:i}\right]f_{,j}\\
= & \left[\gamma_{DBA}-\gamma_{DAB}\right]e^{Dj}f_{,j}\\
= & \left[-\gamma_{BDA}+\gamma_{ADB}\right]\eta^{DC}e_{C}^{\:j}f_{,j}\\
= & \left(-\gamma_{B\;A}^{\;C}+\gamma_{A\;B}^{\;C}\right)e_{C}^{\:j}f_{,j}=\left(\gamma_{A\;B}^{\;j}-\gamma_{B\;A}^{\;j}\right)f_{,j}\\
\equiv & C_{\;AB}^{C}e_{C}f=C_{\;AB}^{C}e_{C}^{\:j}f_{,j}\\
\Leftrightarrow C_{\;AB}^{C}= & \gamma_{\;BA}^{C}-\gamma_{\;AB}^{C}
\end{align*}
(If basis commute, e.g. $\partial_{a}$, $C_{\;AB}^{C}=0$)

\subsubsection{Ricci and Bianchi identities}

The Ricci identities applied to tetrads yield
\begin{align*}
2e_{A\,i\left[;k;l\right]}= & R_{mikl}e_{A}^{\:m}=R_{Aikl}
\end{align*}
Projecting on the remaining indices
\begin{align*}
R_{ABCD}= & R_{Aikl}e_{B}^{\:i}e_{C}^{\:k}e_{D}^{\:l}\\
= & \left\{ e_{A\,i;k;l}-e_{A\,i;l;k}\right\} e_{B}^{\:i}e_{C}^{\:k}e_{D}^{\:l}\\
= & \left\{ \left[e_{\:i}^{F}\gamma_{FAG}e_{\:k}^{G}\right]_{;l}-\left[e_{\:i}^{F}\gamma_{FAG}e_{\:l}^{G}\right]_{;k}\right\} e_{B}^{\:i}e_{C}^{\:k}e_{D}^{\:l}\\
= & \left\{ e_{\:i;l}^{F}\gamma_{FAC}e_{B}^{\:i}e_{D}^{\:l}+\gamma_{BAC;D}+\gamma_{BAG}e_{\:k;l}^{G}e_{C}^{\:k}e_{D}^{\:l}\right.\\
 & \left.-e_{\:i;k}^{F}\gamma_{FAD}e_{B}^{\:i}e_{C}^{\:k}-\gamma_{BAD;C}-\gamma_{BAG}e_{\:l;k}^{G}e_{C}^{\:k}e_{D}^{\:l}\right\} \\
= & \gamma_{BAC;D}-\gamma_{BAD;C}+\gamma_{FAC}e_{B}^{\:i}e_{\:i}^{G}\gamma_{G\quad H}^{\quad F}e_{\:l}^{H}e_{D}^{\:l}-\gamma_{FAD}e_{B}^{\:i}e_{\:i}^{G}\gamma_{G\quad H}^{\quad F}e_{\:k}^{H}e_{C}^{\:k}\\
 & +\gamma_{BAG}\left[e_{C}^{\:k}e_{\:k}^{F}\gamma_{F\quad H}^{\quad G}e_{\:l}^{H}e_{D}^{\:l}-e_{D}^{\:l}e_{\:l}^{F}\gamma_{F\quad H}^{\quad G}e_{\:k}^{H}e_{C}^{\:k}\right]\\
= & \gamma_{BAC;D}-\gamma_{BAD;C}+\gamma_{FAC}\gamma_{B\quad D}^{\quad F}-\gamma_{FAD}\gamma_{B\quad C}^{\quad G}+\gamma_{BAG}\left[\gamma_{C\quad D}^{\quad G}-\gamma_{D\quad C}^{\quad G}\right]
\end{align*}
The Bianchi identities can also be expressed as combination of Riemann
and Ricci totation coefficients

Bianchi in in coordinates implies, in basis components:
\begin{align*}
R_{ij\left[kl;m\right]}=0\Leftrightarrow & R_{AB\left[CD|F\right]}=0=e_{A}^{\:i}e_{B}^{\:j}e_{[C}^{\:k}e_{D}^{\:l}e_{F]}^{\:m}R_{ij\left[kl;m\right]}
\end{align*}
from Riemann symmetries and antisymmetrisation development
\begin{align*}
R_{ij\left[kl;m\right]}= & \frac{1}{3}\left[R_{ijkl;m}+R_{ijlm;k}+R_{ijmk;l}\right]\\
\Leftrightarrow0= & \frac{1}{3}\left[R_{ABCD|F}+R_{ABDF|C}+R_{ABFC|D}\right]
\end{align*}
and
\begin{align*}
R_{ABCD|F}=R_{ABCD;F} & -e_{A;m}^{\:i}R_{iBCD}e_{F}^{\:m}\\
 & -e_{B;m}^{\:j}R_{AjCD}e_{F}^{\:m}\\
 & -e_{C;m}^{\:k}e_{F}^{\:m}R_{ABkD}\\
 & -e_{D;m}^{\:l}e_{F}^{\:m}R_{ABCl}
\end{align*}
each term built on same model
\begin{align*}
e_{A;m}^{\:i}e_{F}^{\:m}R_{iBCD}= & e^{Gi}\gamma_{GAH}e_{\:m}^{H}e_{F}^{\:m}R_{iBCD}\\
= & \eta^{GN}e_{N}^{\;i}\gamma_{GAF}R_{iBCD}\\
= & \eta^{GN}\gamma_{GAF}R_{NBCD}
\end{align*}
so Bianchi identity reads
\begin{align*}
0= & R_{AB\left[CD;F\right]}-\frac{\eta^{NM}}{3}\left\{ \vphantom{{\color{blue}-}\gamma_{\underset{{\color{blue}AN}}{\underline{{\scriptstyle NA}}}F}R_{MBCD}}\right.\\
 & {\color{blue}-}\gamma_{\underset{{\color{blue}AN}}{\underline{{\scriptstyle NA}}}F}R_{MBCD}+{\color{green}-}\gamma_{\underset{{\color{green}AN}}{\underline{{\scriptstyle NA}}}C}R_{MBDF}+{\color{red}-}\gamma_{\underset{{\color{red}AN}}{\underline{{\scriptstyle NA}}}D}R_{MBFC}+\gamma_{\underset{{\color{blue}BN}}{\underline{{\scriptstyle NB}}}F}R_{\underset{{\color{blue}MA}}{\underline{{\scriptstyle AM}}}CD}+\gamma_{\underset{{\color{green}BN}}{\underline{{\scriptstyle NB}}}C}R_{\underset{{\color{green}MA}}{\underline{AM}}DF}+\gamma_{\underset{{\color{red}BN}}{\underline{{\scriptstyle NB}}}D}R_{\underset{{\color{red}MA}}{\underline{{\scriptstyle AM}}}FC}\\
 & \left.+\gamma_{\underset{{\color{green}CN}}{\underline{{\scriptstyle NC}}}F}R_{AB\underset{{\color{green}DM}}{\underline{{\scriptstyle MD}}}}+\gamma_{\underset{{\color{blue}DN}}{\underline{{\scriptstyle ND}}}C}R_{AB\underset{{\color{blue}FM}}{\underline{{\scriptstyle MF}}}}+\gamma_{\underset{{\color{red}FN}}{\underline{{\scriptstyle NF}}}D}R_{AB\underset{{\color{red}CM}}{\underline{{\scriptstyle MC}}}}+{\color{red}-}\gamma_{\underset{{\color{red}DN}}{\underline{{\scriptstyle ND}}}F}R_{ABCM}+{\color{green}-}\gamma_{\underset{{\color{green}FN}}{\underline{{\scriptstyle NF}}}C}R_{ABDM}{\color{blue}-}\gamma_{\underset{{\color{blue}CN}}{\underline{{\scriptstyle NC}}}D}R_{ABFM}\vphantom{{\color{blue}-}\gamma_{\underset{{\color{blue}AN}}{\underline{{\scriptstyle NA}}}F}R_{MBCD}}\right\} \\
= & R_{AB\left[CD;F\right]}-\frac{\eta^{NM}}{3}\left\{ 2\left[\gamma_{N[B\brokenvert F}R_{\brokenvert A]MCD}+\gamma_{N[B\brokenvert D}R_{\brokenvert A]MFC}+\gamma_{N[B\brokenvert D}R_{\brokenvert A]MFC}\right]\right.\\
 & \left.+\lambda_{FND}R_{ABCM}+\lambda_{CNF}R_{ABDM}+\lambda_{DNC}R_{ABFM}\vphantom{\gamma_{N[B\brokenvert F}R_{\brokenvert A]MCD}}\right\} 
\end{align*}

Application (see Ch.~\ref{chap:Motion-of-a})

\subsection{General non-orthonormal tetrads}

In case $\eta_{AB}$ is not Minkowski the Ricci coefficients are NOT
antisymmetric but from covariant derivative projection, we get
\begin{align*}
\eta_{AB;i}e_{C}^{\;i}= & \eta_{AB;C}=\gamma_{BAC}+\gamma_{ABC}\ne0
\end{align*}
Avoiding use of antisymmetry, expression for structure coefficients
remains
\begin{align*}
C_{\;AB}^{C}= & \gamma_{\;AB}^{C}-\gamma_{\;BA}^{C}
\end{align*}
It turns out that Ricci and Bianchi identities also retain the same
forms

\section{Tetrads from Einstein-Cartan gravity}
\begin{description}
\item [{Nomenclature}] tetrad=vierbein=4D %
\begin{tabular}{|l}
frame\tabularnewline
basis\tabularnewline
\end{tabular}\\
In $n$ dimensions: vielbein=$n$D %
\begin{tabular}{|l}
frame\tabularnewline
basis\tabularnewline
\end{tabular}
\end{description}
One can represent gravity with the frame bundle $FM$: for a spacetime
$M$ of dimension $n$, $FM=P\left(GL\left(n,\mathbb{R}\right),M,\pi\right)$
with connection $\omega$, canonical form $\theta$ and soldering
form $\tilde{\theta}$

Frame bundle: principal bundle with structure group $GL\left(n,\mathbb{R}\right)$
and projection $\pi$ on the base $M$
\begin{defn}
Fiber bundle

Manifold with
\begin{itemize}
\item a base manifold: $M$
\item Fibers, isomorphic to the bundle structure group: $F\sim G$, here
$G=SO_{0}\left(1,3\right)$
\item a projector $\pi$ from the bundle manifold onto the base manifold:
$\pi\left(p\right)=x$
\end{itemize}
\end{defn}
See \cite{Isham:1999qu,Nakahara2003,fecko:2006}
\begin{itemize}
\item %
\begin{tabular}[t]{lll}
In the same way as & a manifold & is locally Minkowski\tabularnewline
 & a bundle & is locally $M\times F$\tabularnewline
\end{tabular}
\item $G$ the structure group glues the fibers together
\end{itemize}
\begin{example}
see \cite[p372]{Nakahara2003}

The cylinder is the trivial bundle $S^{1}\times F$ with %
\begin{tabular}[t]{l}
$S^{1}$, the circle\tabularnewline
$F=\left[-1,1\right]$\tabularnewline
\end{tabular}for the structure group $G=\left\{ Id_{F\to F}\right\} $

while

the Möbius strip is locally $S^{1}\times F$ but for $G=\mathbb{Z}_{2}=\left\{ Id_{F\to F},\,g\right\} $
and $g:\begin{array}[t]{l}
F\to F\\
t\mapsto-t
\end{array}$ (see Fig.~\ref{fig:Cylinder-and-M=0000F6bius})
\begin{figure}
\includegraphics[width=1\columnwidth]{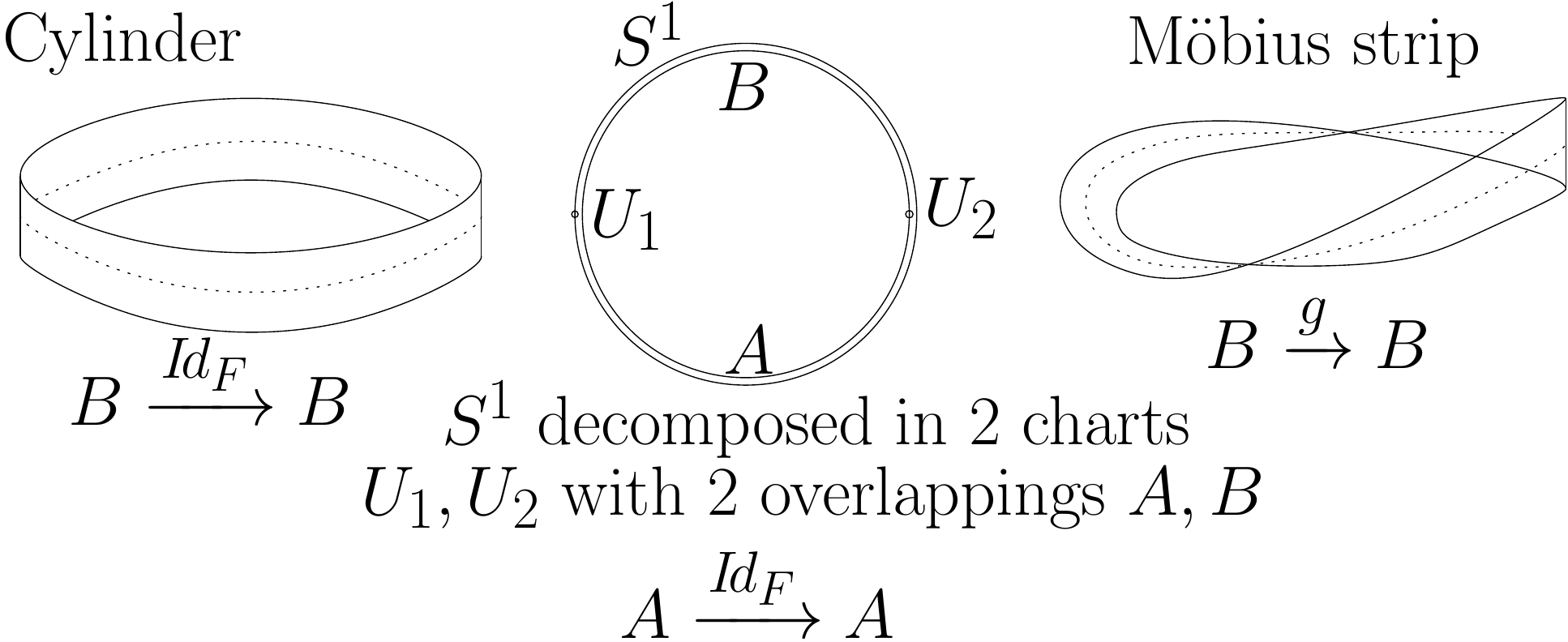}

\caption{\label{fig:Cylinder-and-M=0000F6bius}Cylinder and Möbius strip bundles}

\end{figure}
\end{example}

\subsection{In the frame bundle}

At one point of $M$ one can choose a frame $e_{A}$

The transforms from $e_{A}$ to $e_{A}^{\prime}=ge_{A}$ is done by
linear maps: elements of $GL\left(n,\mathbb{R}\right)$. This inducesa
bijection between $e_{A}$ and $g\in GL\left(n,\mathbb{R}\right)$

$GL\left(n,\mathbb{R}\right)$ implies the fiber to be made of any
vielbein

So $\eta$ is then not Minkowski.

\begin{figure}
\includegraphics[width=1\columnwidth]{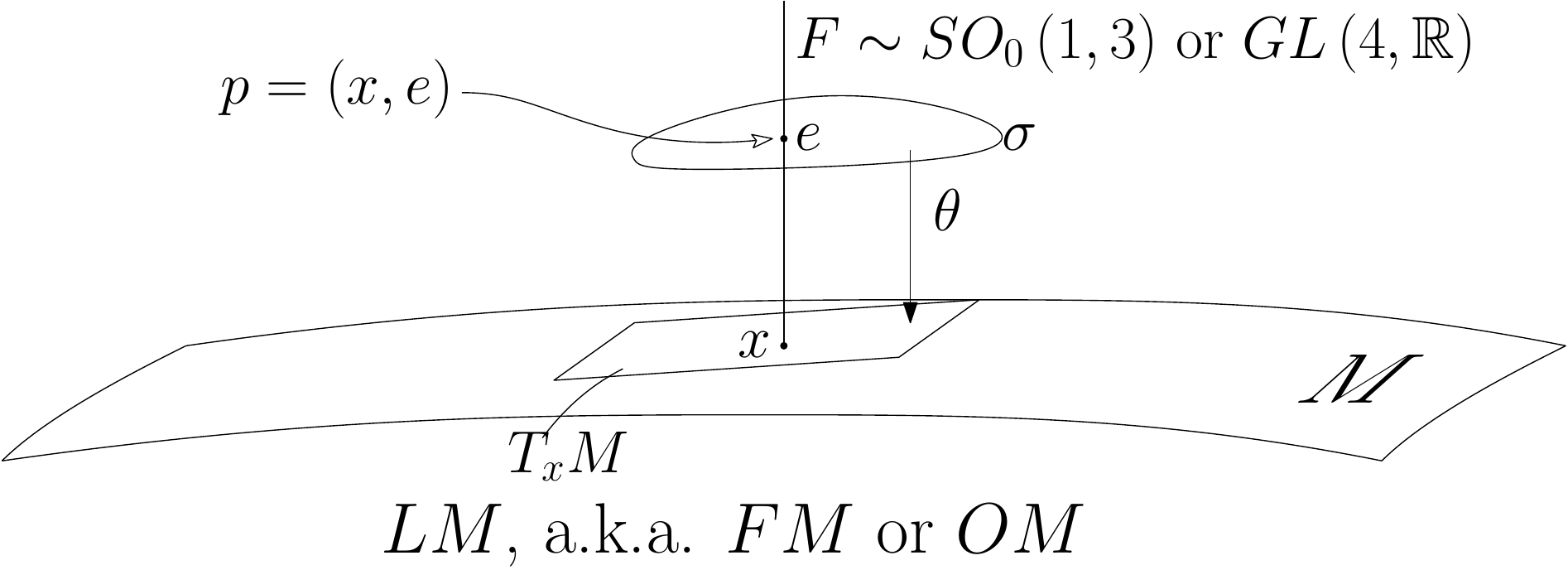}

\caption{\label{fig:Frame-Bundles-representation}Frame Bundles representation}

\end{figure}

The restriction from $GL\left(n,\mathbb{R}\right)$ to the Lorentz
group $SO_{0}\left(1,3\right)\Rightarrow$ orthonormal bases, related
by Lorentz transforms (see Fig.~\ref{fig:Frame-Bundles-representation})

\subsubsection{The canonical form $\theta$}

gives horizontal vector components of bundle vector in a basis
\begin{align*}
\theta\left(e,v\right)= & \left\{ v^{A}\right\}  & \textrm{for }V= & V^{A}e_{A} & \textrm{horizontal part}\\
\theta_{e_{A}}\left(v\right)= & V^{A} &  & +V^{v} & \textrm{vertical part}
\end{align*}
in the tangent to the bundle

and realises

\subsubsection{The soldering form $\tilde{\theta}$}

Displacement along the fiber (here a Lorentz transform) also corresponds
to moving on the base $M$ (Lorentz transform in the spacetime manifold

\subsubsection{The section $\sigma$ in local open set of bundle}

Choose a basis (element of fiber) above a spacetime point neighbourhood 

$\sigma\left(p\right)=e\left(x\right)$: element of fiber group (here
frame) in a neighbourhood

In $LM$, $\sigma^{*}\theta=e$ on $M$: $\sigma^{*}\theta^{A}=e^{A}$

($\sigma^{*}$ equivalent to $\pi$: projection, in this context)

\subsubsection{The connection $\omega$}

Ensures parallel transport/covariant derivative

\paragraph{Parallel transport in bundles:}

offers a different approach

$\gamma$, curve in base $M$, is lifted to $\gamma^{\prime}$

The connection defines the horizontal part of the tangent plane to
the bundle manifold, while the fiber is canonically vertical.

The vertical drift of $\gamma^{\prime}$ measures the bundle curvature

\begin{figure}
\subfloat[\label{fig:Bundle-closed-loop:}Bundle closed loop: curvature]{\includegraphics[width=0.5\textwidth]{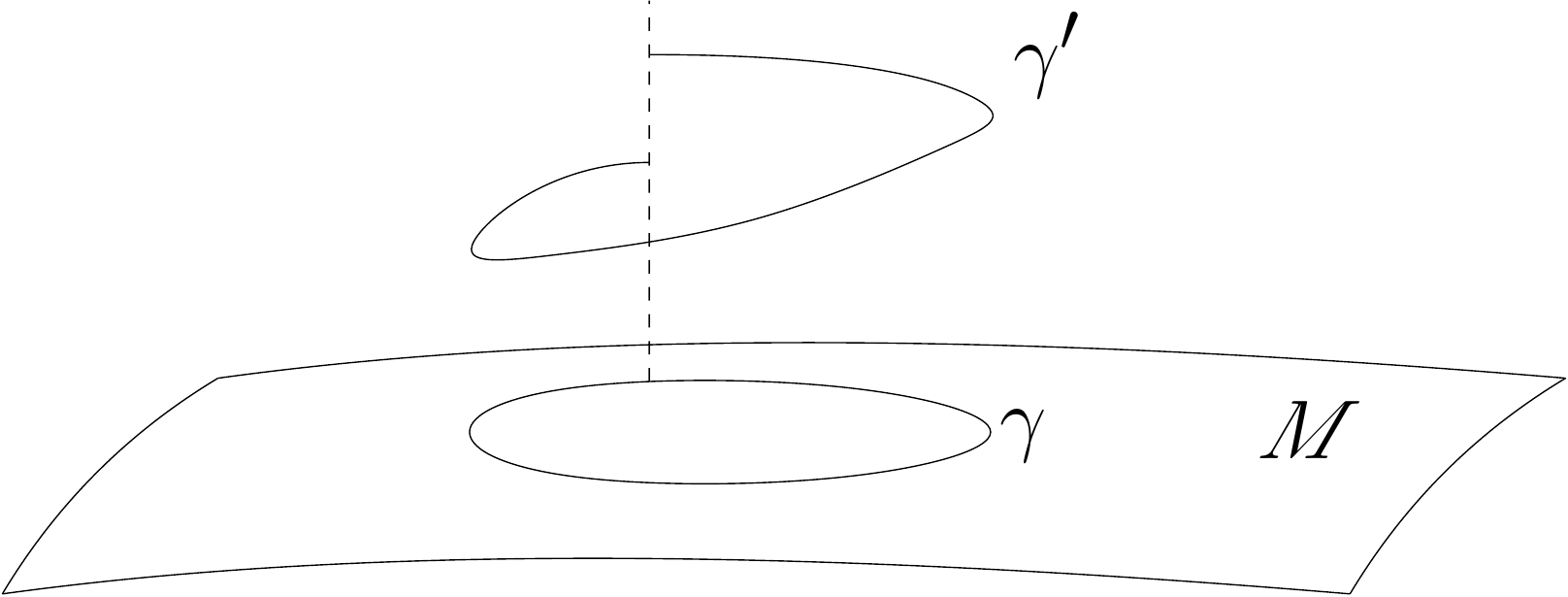}

}\subfloat[\label{fig:Bundle-horizontal-lift}Bundle horizontal lift of a curve]{\includegraphics[width=0.5\textwidth]{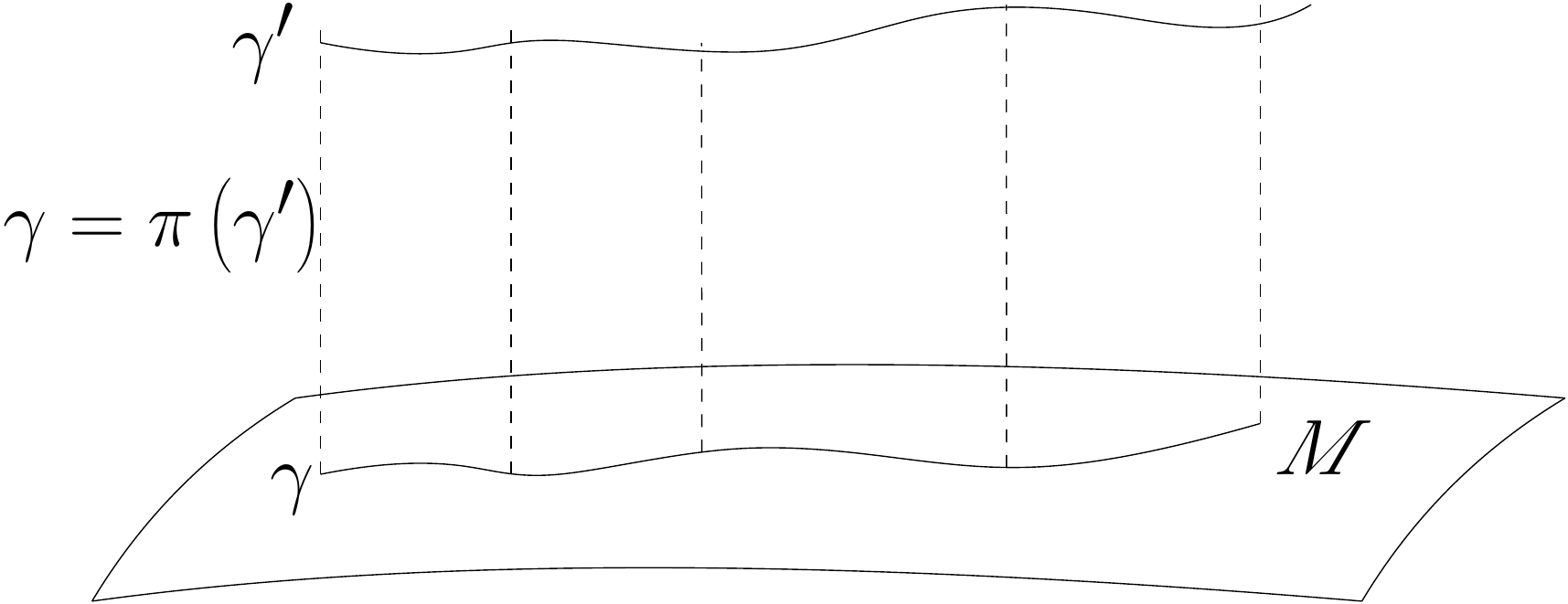}

}

\caption{Bundle Parallel transport}

\end{figure}

For $\gamma$ closed loops, see Fig.~\ref{fig:Bundle-closed-loop:}

In parallel transport, $\dot{\gamma}^{\prime}$ remains horizontal
(geodesic on $M$, see Fig.~\ref{fig:Bundle-horizontal-lift})

$\phi$: quantity transported along $\gamma^{\prime}$
\begin{align*}
\dot{\gamma}\phi= & D\phi\left[\dot{\gamma}^{\prime}\right]\\
\textrm{on }M:\,\dot{\gamma}\pi\phi= & \nabla_{\dot{\gamma}}\pi\phi\left[\dot{\gamma}\right]
\end{align*}
Using the connection
\begin{align*}
\textrm{in frame bundle }\omega= & \omega_{\:B}^{A}\theta_{A}\theta^{B}\\
\textrm{on }M\hfill\sigma^{*}\omega= & \Gamma_{BC}^{A}e_{A}e^{B}e^{C}\\
\textrm{with }\nabla_{e_{A}}e_{B}= & \Gamma_{AB}^{C}e_{C}
\end{align*}

\begin{defn}
Connections

$\omega$ is an Ehresmann connection

$\Gamma$ is a Koszul connection (as in GR)

Connections were developed first by \uline{Lie} to study solutions
of differential equationsas trajectories on manifolds. Lie PhD student
\uline{Cartan} introduced torsion and connections with frames.

While \uline{Riemann geometry} describes manifolds compared with
tangent planes, \uline{Cartan geometry} defines manifolds from
the motion of a mobile frame on them.

Cartan's PhD student \uline{Ehresmann} defined a more general connection
independent of frames, without the need for soldering.

Its abstract use allowed to describe gauge theories with fiber bundles
in which the covariant derivative on the base yields, as a connection,
the gauge field coupling.

Connections have been applied to %
\begin{tabular}[t]{l}
- gravitation, GR,... (e.g. Penrose)\tabularnewline
- particle physics, gauge theory\tabularnewline
\end{tabular}
\end{defn}
Applied to a frame form $e^{A}$ on $M$ ($\mathcal{D}$ exterior
covariant derivative on $M$)
\begin{align*}
\mathcal{D}e^{A}= & de^{A}+\hat{\omega}_{\:B}^{A}\wedge e^{B}
\end{align*}
Advantage: with appropriate representation, spinors can be coupled
to gravity with the Fock-Ivanenko covariant derivative
\begin{align*}
D_{A}\psi= & \partial_{A}\psi-\frac{i}{2}\omega_{A}^{\:bc}S_{bc}\psi
\end{align*}
where $S_{bc}$ is the Fock-Ivanenko coefficient in spinor space and
$\omega_{A}^{\:bc}$ is the connection representation called sin connection

\subsubsection{Curvature and torsion}

are defined as parallel transport of
\begin{itemize}
\item connection: curvature 2-form
\begin{align*}
\Omega_{\;B}^{A}= & d\omega_{\:B}^{A}+\omega_{\:C}^{A}\wedge\omega_{\:B}^{C}
\end{align*}
and
\begin{align*}
\Omega_{\;B}^{A}= & \frac{1}{2}R_{\;BCD}^{A}e^{C}\wedge e^{D}\textrm{ in terms of Riemann}
\end{align*}
\item canonical form: torsion 2-form
\begin{align*}
\Theta^{A}= & d\theta^{A}+\omega_{\:B}^{A}\wedge\theta^{B}
\end{align*}
and
\begin{align*}
= & \frac{1}{2}
\end{align*}
\begin{align*}
\Theta^{A}= & \frac{1}{2}T_{\;BC}^{A}e^{B}\wedge e^{C}\textrm{ in terms of torsion tensor}
\end{align*}
For $de^{A}=0$ (holonomic frames, similar to $\partial_{a}$)
\begin{align*}
T_{\;BC}^{A}= & 2\Gamma_{\left[BC\right]}^{A}
\end{align*}
In GR $\Gamma_{BC}^{A}=\Gamma_{\left(BC\right)}^{A}$ so $T_{\;BC}^{A}=0$
\end{itemize}

\section{Newman-Penrose tetrads}

From a usual tetrad with %
\begin{tabular}[t]{ll}
$e_{\Box}$ & timelike\tabularnewline
$e_{\left(\mu\right)}$ & spacelike\tabularnewline
\end{tabular} orthonormal!

wecan build a fully null basis $z_{\,\,A}^{a}=\left(l^{a},n^{a},m^{a},\overline{m}^{a}\right)$
\begin{align*}
l= & \frac{e_{\Box}+e_{I}}{\sqrt{2}}, & n= & \frac{e_{\Box}-e_{I}}{\sqrt{2}},\textrm{ and} & m= & \frac{e_{I\!I}+i\,e_{I\!I\!I}}{\sqrt{2}} & \Rightarrow\overline{m}= & \frac{e_{I\!I}-i\,e_{I\!I\!I}}{\sqrt{2}}
\end{align*}
Null basis
\begin{align*}
l.l= & \frac{e_{\Box}^{2}+e_{I}^{2}}{2}=0=n.n=m.m=\frac{e_{I\!I}^{2}-e_{I\!I\!I}^{2}}{2}=\overline{m}.\overline{m}
\end{align*}
Orthonormal by blocks
\begin{align*}
l.m= & l.\overline{m}=n.m=n.\overline{m}=0
\end{align*}
Normalisation
\begin{align*}
l.n= & \frac{e_{\Box}^{2}-e_{I}^{2}}{2}=-1\textrm{ and} & m.\overline{m}= & \frac{e_{I\!I}^{2}+e_{I\!I\!I}^{2}}{2}=1
\end{align*}
Then the metric becomes, through normalisation
\begin{align*}
\eta_{AB}=\eta^{AB}= & \left[\begin{array}{cccc}
0 & -1 & 0 & 0\\
-1 & 0 & 0 & 0\\
0 & 0 & 0 & 1\\
0 & 0 & 1 & 0
\end{array}\right]
\end{align*}
which yields the basis and form basis
\begin{align*}
e_{1}= & l, & e_{2}= & n, & e_{3}= & m, & e_{4}= & \overline{m}: & z_{A}= & \left(l,n,m,\overline{m}\right)\\
e^{1}= & \eta^{1B}e_{B}=-n, & e^{2}= & \eta^{2B}e_{B}=-l, & e^{3}= & \overline{m}, & e^{4}= & m: & z^{A}= & \left(-n,-l,\overline{m},m\right)
\end{align*}
and the metric
\begin{align*}
g_{ab}= & \eta_{AB}z_{a}^{A}z_{b}^{B}=-2\left(l_{(a}n_{b)}-m_{(a}\overline{m}_{b)}\right)
\end{align*}

Basis vectors as directional derivatives are given symbols:
\begin{align*}
e_{1}= & e_{1}^{\:a}\partial_{a}\equiv D=l^{a}\nabla_{a}, & e_{2}= & n^{a}\nabla_{a}\equiv\Delta, & e_{3}= & m^{a}\nabla_{a}\equiv\delta, & e_{4}= & \overline{m}^{a}\nabla_{a}\equiv\overline{\delta}
\end{align*}
The connection full information lies in the Ricci rotation coefficients
named Newman-Penrose spin coefficients:
\begin{align*}
\varkappa= & \gamma_{311}=-m^{a}l^{b}\nabla_{b}l_{a} & \pi= & \gamma_{241}=\overline{m}^{a}l^{b}\nabla_{b}n_{a} & \varepsilon= & \frac{1}{2}(\gamma_{211}+\overset{\begin{array}{c}
{\color{green}-\gamma_{431}}\\
{\color{green}\shortparallel}
\end{array}}{\overbrace{\gamma_{341}})}\\
 &  & {\color{green}=} & {\color{green}-\gamma_{421}} & = & \frac{1}{2}\left(\overline{m}^{a}l^{b}\nabla_{b}m_{a}-n^{a}l^{b}\nabla_{b}l_{a}\right)\\
\sigma= & \gamma_{313}=-m^{a}m^{b}\nabla_{b}l_{a} & \mu= & \gamma_{243}=\overline{m}^{a}m^{b}\nabla_{b}n_{a} & \beta= & \frac{1}{2}(\gamma_{213}+\overset{\begin{array}{c}
{\color{green}-\gamma_{433}}\\
{\color{green}\shortparallel}
\end{array}}{\overbrace{\gamma_{343}})}\\
 &  & {\color{green}=} & {\color{green}-\gamma_{423}} & = & \frac{1}{2}\left(\overline{m}^{a}m^{b}\nabla_{b}m_{a}-n^{a}m^{b}\nabla_{b}l_{a}\right)\\
\varrho= & \gamma_{314}=-m^{a}\overline{m}^{b}\nabla_{b}l_{a} & \lambda= & \gamma_{244}=\overline{m}^{a}\overline{m}^{b}\nabla_{b}n_{a} & \alpha= & \frac{1}{2}(\gamma_{214}+\overset{\begin{array}{c}
{\color{green}-\gamma_{434}}\\
{\color{green}\shortparallel}
\end{array}}{\overbrace{\gamma_{344}})}\\
 &  & {\color{green}=} & {\color{green}-\gamma_{424}} & = & \frac{1}{2}\left(\overline{m}^{a}\overline{m}^{b}\nabla_{b}m_{a}-n^{a}\overline{m}^{b}\nabla_{b}l_{a}\right)\\
\tau= & \gamma_{312}=-m^{a}n^{b}\nabla_{b}l_{a} & \nu= & \gamma_{242}=\overline{m}^{a}n^{b}\nabla_{b}n_{a} & \gamma= & \frac{1}{2}(\gamma_{212}+\overset{\begin{array}{c}
{\color{green}-\gamma_{432}}\\
{\color{green}\shortparallel}
\end{array}}{\overbrace{\gamma_{342}})}\\
 &  & {\color{green}=} & {\color{green}-\gamma_{422}} & = & \frac{1}{2}\left(\overline{m}^{a}n^{b}\nabla_{b}m_{a}-n^{a}n^{b}\nabla_{b}l_{a}\right)
\end{align*}
Complex conjugation simply involves exchanging indices $3$ and $4$
or $m$ and $\overline{m}$

\subsection{Transportation equations}

Directional derivatives applied to the NP tetrad yields
\begin{align*}
Dl= & \left(\varepsilon+\overline{\varepsilon}\right)l-\overline{\varkappa}m-\varkappa\overline{m} & Dn= & -\left(\varepsilon+\overline{\varepsilon}\right)n+\pi m+\overline{\pi m}\\
\Delta l= & \left(\gamma+\overline{\gamma}\right)l-\overline{\tau}m-\tau\overline{m} & \Delta n= & -\left(\gamma+\overline{\gamma}\right)n+\nu m+\overline{\nu m}\\
\delta l= & \left(\overline{\alpha}+\beta\right)l-\overline{\varrho}m-\sigma\overline{m} & \delta n= & -\left(\overline{\alpha}+\beta\right)n+\mu m+\overline{\lambda m}\\
Dm= & \overline{\pi}l-\varkappa n+\left(\varepsilon-\overline{\varepsilon}\right)m & \Delta m= & \overline{\nu}l-\tau n+\left(\gamma-\overline{\gamma}\right)m\\
\delta m= & \overline{\lambda}l-\sigma n+\left(\beta-\overline{\alpha}\right)m & \overline{\delta}m= & \overline{\mu}l-\varrho n+\left(\alpha-\overline{\beta}\right)m
\end{align*}
Application:
\begin{itemize}
\item $l$ is tangent to geodesics $\Leftrightarrow\varkappa=0$\\
then $Dl=\left(\varepsilon+\overline{\varepsilon}\right)l$, non affine
geodesics equation
\item $l$ tangent to affinely parameterised geodesics $\Leftrightarrow\varkappa=0$
and $\varepsilon=-\overline{\varepsilon}$ ($\varepsilon=i\mathcal{E}$):
then $Dl=0$
\end{itemize}
Samely
\begin{itemize}
\item $n$ tangent to geodesics $\Leftrightarrow\nu=0$ so $\Delta n=-\left(\gamma+\overline{\gamma}\right)n$
\item $n$ affinely parameterised tangent to geodesics $\Leftrightarrow\nu=0$
and $\gamma=-\overline{\gamma}$ ($\gamma=i\mathcal{G}$) so that
$\Delta n=0$
\end{itemize}

\subsection{Commutators}

The connection torsion freeness is equivalent to the forms of the
directional derivatives commutators: 
\begin{align*}
\left[z_{m},z_{n}\right]= & \nabla_{m}z_{n}-\nabla_{n}z_{m}\\
= & -2\Gamma_{nm}^{k}z_{k}
\end{align*}
Explicitly
\begin{align*}
\left[\Delta,D\right]= & \left(\gamma+\overline{\gamma}\right)D+\left(\varepsilon+\overline{\varepsilon}\right)\Delta-\left(\pi+\overline{\tau}\right)\delta-\left(\tau+\overline{\pi}\right)\overline{\delta}\\
\left[\delta,D\right]= & \left(\overline{\alpha}+\beta-\overline{\pi}\right)D+\varkappa\Delta-\left(\overline{\varrho}+\varepsilon-\overline{\varepsilon}\right)\delta-\sigma\overline{\delta}\\
\left[\delta,\Delta\right]= & -\overline{\nu}D+\left(\tau-\overline{\alpha}-\overline{\beta}\right)\Delta+\left(\mu-\gamma+\overline{\gamma}\right)\delta+\overline{\lambda}\overline{\delta}\\
\left[\overline{\delta},\delta\right]= & \left(\overline{\mu}-\mu\right)D+\left(\overline{\varrho}-\varrho\right)\Delta+\left(\alpha-\overline{\beta}\right)\delta-\left(\overline{\alpha}-\beta\right)\overline{\delta}
\end{align*}
Replace second directional derivative by its vector on the L.H.S.
and the directional derivatives by their vectors on the R.H.S. yields
combinations of transport equations

\subsection{Weyl, Ricci and Riemann representations}

From the Riemann decomposition into Weyl and its trace \cite[Eq:~3.147]{carroll-2004}
the tetrad components of Riemann, Weyl and Ricci can be written as
\begin{align*}
R_{\quad CD}^{AB}= & C_{\quad CD}^{AB}+\frac{4}{n-2}R_{\;[C}^{[A}\eta_{\;D]}^{B]}-\frac{2R}{\left(n-1\right)\left(n-2\right)}\eta_{\;[C}^{[A}\eta_{\;D]}^{B]}
\end{align*}
with the tetrad components of Ricci:
\begin{align*}
R_{AC}= & \eta^{BD}R_{ABCD}\textrm{ and } & R= & \eta^{AB}R_{AB}
\end{align*}

\paragraph{For NP tetrads:}

\begin{align*}
R= & 2\left(-R_{12}+R_{34}\right)
\end{align*}
The tracefreeness of Weyl leads to
\begin{align*}
\eta^{AD}C_{ABCD}= & -C_{1BC2}-C_{2BC1}+C_{3BC4}+C_{4BC3}=0
\end{align*}
and if in addition
\begin{align*}
B=C\Rightarrow C_{1314}= & C_{2324}=C_{1332}=C_{1442}=0
\end{align*}
Moreover, the Riemann symmetries yield
\begin{align*}
R_{a\left[bcd\right]}= & \frac{1}{3}\left(R_{abcd}+R_{acdb}+R_{adbc}\right)=0\\
\Rightarrow & C_{1234}+C_{1342}+C_{1423}=0
\end{align*}
Therefore, in the case $b\ne c$, together with the above cyclic antisymmetry,
Weyl tracelessness gives
\begin{align*}
C_{1231}= & C_{1334}; & C_{1241}= & C_{1443};\,C_{1232}=C_{2343};\,C_{1242}=C_{2434}\\
C_{1212}= & C_{3434}; & C_{1342}= & \frac{1}{2}\left(C_{1212}-C_{1234}\right)=\frac{1}{2}\left(C_{3434}-C_{1234}\right)
\end{align*}
so the Riemann tensor can be decomposed with its symmetries and its
writing in terms of Weyl and Ricci

The 10 independent components of the Weyl tensor can be encoded into
5 complex N-P scalars
\begin{align*}
\psi_{0}= & C_{1313}=C_{abcd}l^{a}m^{b}l^{c}m^{d}\\
\psi_{1}= & C_{1213}=C_{abcd}l^{a}n^{b}l^{c}m^{d}\\
\psi_{2}= & C_{1342}=C_{abcd}l^{a}m^{b}\overline{m}^{c}n^{d}\\
\psi_{3}= & C_{1242}=C_{abcd}l^{a}n^{b}\overline{m}^{c}n^{d}\\
\psi_{4}= & C_{2424}=C_{abcd}n^{a}\overline{m}^{b}n^{c}\overline{m}^{d}
\end{align*}
Encoding the Riemann symmetries into the symbol:
\begin{align*}
\left\{ wxyz\right\} = & 8\left(\left[wx\right]\left[yz\right]\right)=4\left[wx\right]\left[yz\right]+4\left[yz\right]\left[wx\right]
\end{align*}
the Weyl tensor can be decomposed on the NP basis
\begin{align*}
C_{abcd}= & \left(\psi_{2}+\overline{\psi}_{2}\right)\left[\left\{ l_{a}n_{b}l_{c}n_{d}\right\} +\left\{ m_{a}\overline{m}_{b}m_{c}\overline{m}_{d}\right\} \right]+\left(\overline{\psi}_{2}-\psi_{2}\right)\left\{ l_{a}n_{b}m_{c}\overline{m}_{d}\right\} \\
 & \left(\psi_{0}\left\{ n_{a}\overline{m}_{b}n_{c}\overline{m}_{d}\right\} +\psi_{4}\left\{ l_{a}m_{b}l_{c}m_{d}\right\} -\psi_{2}\left\{ l_{a}m_{b}n_{c}\overline{m}_{d}\right\} +\psi_{1}\left[\left\{ l_{a}n_{b}n_{c}\overline{m}_{d}\right\} +\left\{ n_{a}\overline{m}_{b}\overline{m}_{c}m_{d}\right\} \right]\right.\\
 & \left.-\psi_{3}\left[\left\{ l_{a}n_{b}l_{c}m_{d}\right\} -\left\{ l_{a}m_{b}m_{c}\overline{m}_{d}\right\} \right]+\textrm{complex conjugates}\right)
\end{align*}
The 10 independent components of the Ricci tensor can be encoded into
4 real scalars $\left\{ \Phi_{00},\Phi_{11},\Phi_{22},\varLambda\right\} $
and 3 complex scalars $\left\{ \Phi_{01},\Phi_{02},\Phi_{12}\right\} $
\begin{gather*}
\begin{array}{rl}
\Phi_{00}= & \frac{1}{2}R_{ab}l^{a}l^{b}\vphantom{\frac{R_{ab}1}{2R_{ab}}}\\
\Phi_{11}= & \frac{1}{4}R_{ab}\left(l^{a}n^{b}+m_{a}\overline{m}_{b}\right)\vphantom{\frac{R_{ab}1}{2R_{ab}}}\\
\Phi_{22}= & \frac{1}{2}R_{ab}n^{a}n^{b}\vphantom{\frac{R_{ab}1}{2R_{ab}}}\\
\varLambda= & \frac{R}{4!}\vphantom{\frac{R_{ab}1}{2R_{ab}}}
\end{array}\begin{array}{rlrll}
\Phi_{01}= & \frac{1}{2}R_{ab}l^{a}m^{b} & \vphantom{\frac{R_{ab}1}{2R_{ab}}}\Rightarrow\Phi_{10}= & \frac{1}{2}R_{ab}l^{a}\overline{m}_{b} & =\overline{\Phi_{01}}\\
\Phi_{02}= & \frac{1}{2}R_{ab}m^{a}m^{b} & \vphantom{\frac{R_{ab}1}{2R_{ab}}}\Rightarrow\Phi_{20}= & \frac{1}{2}R_{ab}\overline{m}^{a}\overline{m}_{b} & =\overline{\Phi_{02}}\\
\Phi_{12}= & \frac{1}{2}R_{ab}m^{a}n^{b} & \vphantom{\frac{R_{ab}1}{2R_{ab}}}\Rightarrow\Phi_{21}= & \frac{1}{2}R_{ab}\overline{m}^{a}n_{b} & =\overline{\Phi_{12}}
\end{array}
\end{gather*}
Because of N-P orthonormalisation of the N-P tetrads, $R_{ab}$ in
the definitions above can be replaced with
\begin{itemize}
\item the tracefree Ricci $Q_{ab}=R_{ab}-\frac{R}{4}g_{ab}$
\item the Einstein tensor $G_{ab}=R_{ab}-\frac{R}{2}g_{ab}$
\end{itemize}
The Ricci and Bianchi identities can written using the above N-P scalars
and their derivatives into scalar N-P equations (see \cite{Chandrasekhar:1985kt},
or more modernly, \cite{frolov})

\subsection{Maxwell's equations}

In N-P formalism, the Faraday tensor's 6 indepenedent components are
encoded in 3 complex Maxwell-NP scalars
\begin{align*}
\phi_{0}= & F_{13}=F_{ab}l^{a}m^{b}; & \phi_{1}= & \frac{1}{2}\left(F_{12}+F_{43}\right)=\frac{1}{2}F_{ab}\left(l^{a}n^{b}-m_{a}\overline{m}_{b}\right)\\
\phi_{2}= & F_{42}=-F_{ab}n_{a}\overline{m}_{b}
\end{align*}
\begin{tabular}{ll}
and Maxwell's equations & in differential geometric form\tabularnewline
\end{tabular}
\begin{align*}
d\star F= & 0 & dF= & 0
\end{align*}
\begin{tabular}{ll}
\phantom{and Maxwell's equations} & in coordinate form\tabularnewline
\end{tabular}
\begin{align*}
F_{\left[ab;c\right]}= & 0 & g^{ac}F_{ab;c}= & 0
\end{align*}
\begin{tabular}{ll}
\phantom{and Maxwell's equations} & in tetrad form\tabularnewline
\end{tabular}
\begin{align*}
F_{\left[AB|C\right]}= & 0 & \eta^{BC}F_{AB|C}= & 0
\end{align*}
are replaced by (in NP formalism)
\begin{align*}
\phi_{1|1}-\phi_{0|4}= & 0 & \phi_{2|1}-\phi_{1|4}= & 0\\
\phi_{1|3}-\phi_{0|2}= & 0 & \phi_{2|3}-\phi_{1|2}= & 0
\end{align*}
In terms of spin coefficients, we can write, e.g., the first two terms
as
\begin{align*}
\phi_{1|1}= & \frac{1}{2}\left(F_{21|1}+F_{43|1}\right)\\
= & \frac{1}{2}\left[-F_{12;1}+\eta^{nm}\left(\gamma_{n11}F_{m2}+\gamma_{n21}F_{1m}\right)+F_{43;1}-\eta^{nm}\left(\gamma_{n41}F_{m3}+\gamma_{n31}F_{4m}\right)\right]\\
= & \phi_{1;1}-\left(\gamma_{131}F_{42}+\gamma_{241}F_{13}\right)\\
= & D\phi_{1}+\varkappa\phi_{2}-\pi\phi_{0}.
\end{align*}
Samely
\begin{align*}
\phi_{0|4}= & \overline{\delta}\phi_{0}-2\alpha\phi_{0}+2\varrho\phi_{1}
\end{align*}
Applying the same procedure to the remaining of the 8 real Maxwell
equations components leads to
\begin{align*}
D\phi_{1}-\overline{\delta}\phi_{0}= & \left(\pi-2\alpha\right)\phi_{0}+2\varrho\phi_{1}-\varkappa\phi_{2},\\
D\phi_{2}-\overline{\delta}\phi_{1}= & -\lambda\phi_{0}+2\pi\phi_{1}+\left(\varrho-2\varepsilon\right)\phi_{2},\\
\Delta\phi_{0}-\delta\phi_{1}= & \left(2\gamma-\mu\right)\phi_{0}-2\tau\phi_{1}+\sigma\phi_{2},\\
\Delta\phi_{1}-\delta\phi_{2}= & \nu\phi_{0}-2\mu\phi_{1}+\left(2\beta-\tau\right)\phi_{2}.
\end{align*}
The energy-momentum of Faraday is
\begin{align*}
T_{ab}= & g^{cd}F_{ac}F_{bd}-\frac{g_{ab}}{4}F_{ef}F^{ef}\\
\Rightarrow T_{AB}= & \eta^{CD}F_{AC}F_{BD}-\frac{\eta_{AB}}{4}F_{EF}F^{EF}
\end{align*}
In terms of NP scalars
\begin{align*}
-\frac{1}{2}T_{11}= & \phi_{0}\overline{\phi}_{0}, & -\frac{1}{2}T_{13}= & \phi_{0}\overline{\phi}_{1}, & -\frac{1}{4}\left(T_{12}+T_{34}\right)= & \phi_{1}\overline{\phi}_{1},\\
-\frac{1}{2}T_{11}= & \phi_{1}\overline{\phi}_{2}, & -\frac{1}{2}T_{11}= & \phi_{2}\overline{\phi}_{2}, & -\frac{1}{2}T_{11}= & \phi_{0}\overline{\phi}_{2},
\end{align*}
while $T_{\;A}^{A}=F_{AC}F^{AC}-\frac{4}{4}F_{EF}F^{EF}=0$, so EFE
yield (recall $\Phi_{nm}$ can be written with $G_{ab}$) for pure
Faraday sources (only electromagnetic)
\begin{align*}
\varLambda= & 0\textrm{ and } & \Phi_{nm}= & \kappa\phi_{n}\overline{\phi}_{m}
\end{align*}

\subsection{Some applications of N-P formalism}

\subsubsection{Gravitational waves}

Governed exclusively by $\psi_{4}=C_{abcd}n^{a}\overline{m}^{b}n^{c}\overline{m}^{d}$

Choosing at infinity $l^{a}=\frac{\left(\hat{t}+\hat{r}\right)^{a}}{\sqrt{2}},n^{a}=\frac{\left(\hat{t}-\hat{r}\right)^{a}}{\sqrt{2}},m^{a}=\frac{\left(\hat{\theta}+i\hat{\varphi}\right)^{a}}{\sqrt{2}}$

In the transverse traceless gauge, linearised G.W. yield, assuming
propagation in $\hat{r}$ direction
\begin{align*}
-\frac{1}{4}\left(\ddot{h}_{\hat{\theta}\hat{\theta}}-\ddot{h}_{\hat{\varphi}\hat{\varphi}}\right)= & -R_{\hat{t}\hat{\theta}\hat{t}\hat{\theta}}=-R_{\hat{t}\hat{\varphi}\hat{t}\hat{\varphi}}=-R_{\hat{r}\hat{\theta}\hat{r}\hat{\theta}}=-R_{\hat{r}\hat{\varphi}\hat{r}\hat{\varphi}}\\
\frac{1}{2}\ddot{h}_{\hat{\theta}\hat{\varphi}}= & -R_{\hat{t}\hat{\theta}\hat{t}\hat{\varphi}}=-R_{\hat{r}\hat{\theta}\hat{r}\hat{\varphi}}=R_{\hat{t}\hat{\theta}\hat{r}\hat{\varphi}}=R_{\hat{r}\hat{\theta}\hat{t}\hat{\varphi}}.
\end{align*}
Their combination yields
\begin{align*}
\psi_{4}= & \frac{1}{2}\left(\ddot{h}_{\hat{\theta}\hat{\theta}}-\ddot{h}_{\hat{\varphi}\hat{\varphi}}\right)+i\ddot{h}_{\hat{\theta}\hat{\varphi}}=-\ddot{h}_{+}+i\ddot{h}_{\times}
\end{align*}
so $\psi_{4}$ encodes all outgoing GW

\subsubsection{Vaidya metric}

A generalisation of Schwarzschild, Vaidya solution represents spherical
distribution of null dust. Recall the toirtoise radius outside the
horizon (see Sec.~\ref{subsec:Tortoise-coordinates})
\begin{align*}
d\bar{r}= & \frac{dr}{1-\frac{2M}{r}}=dr+\frac{2Mdr}{r\left(1-\frac{2M}{r}\right)}=d\left[r+2M\ln\left(\frac{r}{2M}-1\right)\right]
\end{align*}
to build Minkowski-like null coordinates
\begin{gather*}
\left.\begin{array}{rlrl}
u= & t-\bar{r} & v= & t+\bar{r}\\
\Rightarrow t= & u+\bar{r} & \Rightarrow t= & v-\bar{r}
\end{array}\right\} \begin{array}{rlll}
\Rightarrow dt= & du+d\bar{r} & = & dv-d\bar{r}\\
= & du+\frac{dr}{1-\frac{2M}{r}} & = & dv-\frac{dr}{1-\frac{2M}{r}}
\end{array}
\end{gather*}
Then they are used to obtain the Eddington-Filkenstein 

retarded/outgoing metric (see Sec.~\ref{subsec:Eddington-Finkelstein-coordinate})
\begin{align*}
ds^{2}= & -\left(1-\frac{2M}{r}\right)du^{2}-2dudr+r^{2}d\Omega^{2}
\end{align*}
advanced/ingoing metric
\begin{align*}
ds^{2}= & -\left(1-\frac{2M}{r}\right)dv^{2}+2dvdr+r^{2}d\Omega^{2}
\end{align*}
To obtain the Vaidya metric requires to turn the mass parameter into
a mass distribution ofthe null coordinate. As constant mass corresponds
to constant null coordinate, behaving like dust with radius turned
to null directions (expanding or collapsing at light speed) this is
interpreted as pressureless gas of photons or null particles: null
dust

As in E-F Schwarzschild solution, there are 2 cases

\paragraph{Outgoing null dust}

\begin{align*}
ds^{2}= & -\left(1-\frac{2M\left(u\right)}{r}\right)du^{2}-2dudr+r^{2}d\Omega^{2}\\
= & \frac{2M\left(u\right)}{r}du^{2}+ds_{M}^{2}
\end{align*}

\paragraph{Ingoing null dust}

\begin{align*}
ds^{2}= & -\left(1-\frac{2M\left(v\right)}{r}\right)dv^{2}+2dvdr+r^{2}d\Omega^{2}\\
= & \frac{2M\left(v\right)}{r}dv^{2}+ds_{M}^{2}
\end{align*}
as Minkowski verifies
\begin{align*}
ds_{M}^{2}= & -dt^{2}+dr^{2}+r^{2}d\Omega^{2}\\
= & -du^{2}-2dudr+r^{2}d\Omega^{2}\\
= & -dv^{2}+2dvdr+r^{2}d\Omega^{2}
\end{align*}
Analysis of both cases is simplified in N-P formalism

\paragraph{Emitting Vaidya and NP tetrads}

Because of the extra terms from Minkowski the only non-zero Ricci
is
\begin{align*}
R_{uu}= & -\frac{2M_{,u}\left(u\right)}{r^{2}}
\end{align*}
while $g^{uu}=0\Rightarrow R=R_{ab}g^{ab}=0$ so EFE yield $G_{ab}=R_{ab}=\kappa T_{ab}=8\pi T_{ab}$

$\Rightarrow T_{ab}=-\frac{M_{,u}\left(u\right)}{4\pi r^{2}}l_{a}l_{b}$
with $l_{a}dx^{a}=du$

Choosing $l_{a}=-\partial_{a}u,l^{a}=g^{ab}l_{b}$ we recognise the
energy momentum of pure radiation dust with $\rho=-\frac{M_{,u}\left(u\right)}{4\pi r^{2}}$

From NEC (see Sec.~\ref{subsec:Aside:-Energy-conditions}), $\rho\ge0\Rightarrow M_{,u}<0$

so the central body looses mass by emitting radiations out (photons,
neutrinos,...)

Applying N-P formalism, non-zero scalars are
\begin{align*}
\psi_{2}= & -\frac{M\left(u\right)}{r^{3}}, & \Phi_{22}= & -\frac{M_{,u}}{r^{2}}
\end{align*}

\warningsign Maxwell N-P equations are not satisfied!

Expansions:
\begin{align*}
\Theta_{l}= & -\left(\varrho+\overline{\varrho}\right)=\frac{2}{r}\\
\Theta_{n}= & \mu+\overline{\mu}=\frac{-r+2M\left(u\right)}{r^{2}}
\end{align*}
Building the radial null dust Lagrangian ($\mathcal{L}=0,\dot{\theta}=\dot{\varphi}=0$,
from $\left(\frac{ds}{d\lambda}\right)^{2}=0$)
\begin{align*}
\mathcal{L}=0= & -\left(1-\frac{2M\left(u\right)}{r}\right)\dot{u}^{2}+2\dot{u}\dot{r}
\end{align*}
2 solutions
\begin{itemize}
\item $\dot{u}=0$ outgoing ($r\nearrow$ for $t\nearrow$)
\item $\dot{r}=\frac{1}{2}\left(1-\frac{2M\left(u\right)}{r}\right)\dot{u}$
ingoing ($r\searrow$ for $t\nearrow$)
\end{itemize}
In terms of NP tetrad
\begin{align*}
l^{a}= & \left(\begin{array}{cccc}
0, & 1, & 0, & 0\end{array}\right) & n^{a}= & \left(\begin{array}{cccc}
1, & -\frac{\left(1-\frac{2M}{r}\right)}{2}, & 0, & 0\end{array}\right) & m^{a}= & \frac{1}{\sqrt{2}r}\left(\begin{array}{cccc}
0, & 0, & 1, & \frac{i}{\sin\theta}\end{array}\right)
\end{align*}
and their dual basis
\begin{align*}
l_{a}= & \left(\begin{array}{cccc}
-1, & 0, & 0, & 0\end{array}\right) & n_{a}= & \left(\begin{array}{cccc}
-\frac{\left(1-\frac{2M}{r}\right)}{2}, & -1, & 0, & 0\end{array}\right) & m_{a}= & \frac{r}{\sqrt{2}}\left(\begin{array}{cccc}
0, & 0, & 1, & i\sin\theta\end{array}\right)
\end{align*}
The spin coefficients (connection) verify
\begin{align*}
\varkappa= & \sigma=\tau=\nu=\lambda=\pi=\varepsilon=0\\
\varrho= & -\frac{1}{r},\mu=-\frac{1}{2r}\left(1-\frac{2M}{r}\right),\alpha=-\beta=-\frac{\sqrt{2}\cot\theta}{4r},\gamma=\frac{M\left(u\right)}{2r^{2}}
\end{align*}
The restriction to Schwarzschild is simply obtained from setting $M\left(u\right)=M=cst$:
only $\psi_{2}\ne0$

\paragraph{Pure absorbing Vaidya}

In this case, the non-zero Ricci becomes
\begin{align*}
R_{vv}= & \frac{2M_{,v}}{r^{2}}
\end{align*}
and since again $ $ $g^{vv}=0\Rightarrow R=g^{ab}R_{ab}=0$ which
leads, through EFE to

$T_{ab}=\frac{M_{,v}}{4\pi r^{2}}n_{a}n_{b}$ for $n_{a}dx^{a}=-dv$,
again the EMT of pure radiation dust with $\rho=\frac{M_{,v}}{4\pi r^{2}}$
which once again from NEC has $\rho\ge0\Rightarrow M_{,v}>0$ and
leads to the interpretation of an absorbing central object for ingoing
dust radiation.

The non-zero N-P scalars are
\begin{align*}
\psi_{2}= & -\frac{M\left(v\right)}{r^{3}}, & \Phi_{00}= & \frac{M_{,v}}{r^{2}}
\end{align*}
The expansions are now
\begin{align*}
\Theta_{l}= & -\left(\varrho+\overline{\varrho}\right)=\frac{r-2M\left(v\right)}{r^{2}}\\
\Theta_{n}= & \mu+\overline{\mu}=-\frac{2}{r}
\end{align*}
As ingoing Vaidya is one of the few exact dynamical solutions, it
can be used to explore B.H. physics, like check differences between,
e.g., Event Horizons and quasi-local Trapping Horizons (see Defs.~\ref{def:Event-Horizon}
and \ref{def:Trapping-Horizon}; also refer to Defs.~\ref{def:A-Killing-Horizon}
and \ref{def:Cauchy-Horizon(C.H.)}).

Note $r=2M\left(v\right)$ is always marginally outer Trapped Horizon
($\Theta_{l}=0,\Theta_{n}<0$)

Radial null Vaidya Lagrangian ($\mathcal{L}=0,\dot{\theta}=\dot{\varphi}=0$)
\begin{align*}
\mathcal{L}=0= & -\left(1-\frac{2M}{r}\right)\dot{v}^{2}+2\dot{v}\dot{r}
\end{align*}
also has
\begin{itemize}
\item ingoing $\dot{v}=0$ solution
\item outgoing $\dot{r}=\frac{1}{2}\left(1-\frac{2M}{r}\right)\dot{v}$
solution
\end{itemize}
The adapted N-P tetrad to ingoing null dust analysis can be set as
\begin{align*}
l^{a}= & \left(\begin{array}{cccc}
1, & -\frac{\left(1-\frac{2M}{r}\right)}{2}, & 0, & 0\end{array}\right) & n^{a}= & \left(\begin{array}{cccc}
0, & -1, & 0, & 0\end{array}\right) & m^{a}= & \frac{1}{\sqrt{2}r}\left(\begin{array}{cccc}
0, & 0, & 1, & \frac{i}{\sin\theta}\end{array}\right)
\end{align*}
and their dual basis is
\begin{align*}
l_{a}= & \left(\begin{array}{cccc}
-\frac{\left(1-\frac{2M}{r}\right)}{2}, & -1, & 0, & 0\end{array}\right) & n_{a}= & \left(\begin{array}{cccc}
-1, & 0, & 0, & 0\end{array}\right) & m_{a}= & \frac{r}{\sqrt{2}}\left(\begin{array}{cccc}
0, & 0, & 1, & i\sin\theta\end{array}\right)
\end{align*}
The corresponding spin coefficients are
\begin{align*}
\varkappa= & \sigma=\tau=\nu=\lambda=\pi=\gamma=0\\
\varrho= & -\frac{1}{2r}\left(1-\frac{2M}{r}\right),\mu=-\frac{1}{r},\alpha=-\beta=-\frac{\sqrt{2}\cot\theta}{4r},\varepsilon=\frac{M\left(v\right)}{r^{2}}
\end{align*}
Thus the Weyl and Ricci N-P scalars turn as
\begin{align*}
\psi_{0}= & \psi_{1}=\psi_{3}=\psi_{4}=0, & \psi_{2}= & -\frac{M\left(v\right)}{r^{3}}\\
\Phi_{10}= & \Phi_{20}=\Phi_{11}=\Phi_{12}=\Phi_{22}=\varLambda=0, & \Phi_{00}= & \frac{M_{,v}}{r^{2}}
\end{align*}
Again, restricting to Schwarzschild summarises as $M\left(v\right)=M=cst$
and only $\psi_{2}\ne0$

Note $\psi_{4}=0\Rightarrow$ no radiation (G.W.)

\chapter[Geodesics from Hamilton-Jacobi]{Geodesics from Hamilton-Jacobi Formalism}

(From \cite{Bazanski:1988fs})

\paragraph{Hamilton-Jacobi Formalism (recall)}

In the canonical view, one starts from a Hamiltonian $H$ with canonical
variables $q$ and $p$, and look at canonical transformations of
$H,q,p$ into $\mathscr{H},Q,P$. They obey canonical equations
\begin{gather*}
\left\{ \begin{array}{rl}
\dot{q}= & \frac{\partial H}{\partial p}\\
\dot{p}= & -\frac{\partial H}{\partial q}
\end{array}\right.\textrm{ and \ensuremath{\left\{  \begin{array}{rl}
 \dot{Q}=  &  \frac{\partial\mathscr{H}}{\partial P}\\
 \dot{P}=  &  -\frac{\partial\mathscr{H}}{\partial Q} 
\end{array}\right.}}
\end{gather*}
and their Lagrangian only differs by a differential of a \uline{Generating
function}
\begin{align*}
L-\mathscr{L}= & p.\dot{q}-H-P.\dot{Q}+\mathscr{H}=\frac{dF}{dt}\left(q,p,Q,P,t\right)
\end{align*}
Then Hamilton's \uline{Principal function} $S\left(q,t\right)=\int^{\left(q,t\right)}L\,dt$
(action) which is obtained from the Generating function by a Legendre
transform
\begin{align*}
S\left(q,P\right)= & F+Q.P
\end{align*}
Thus
\begin{align*}
\frac{dS}{dt}= & p.\dot{q}-H-\cancel{P.\dot{Q}}+\mathscr{H}+\cancel{\dot{Q}.P}+Q.\dot{P}=\frac{\partial S}{\partial q}\dot{q}+\frac{\partial S}{\partial P}\dot{P}+\partial_{t}S\\
\Leftrightarrow & \left(p-\partial_{q}S\right)\dot{q}+\left(Q-\partial_{P}S\right)\dot{P}+\left(\mathscr{H}-H-\partial_{t}S\right)=0\\
\Leftrightarrow & \left\{ \begin{array}{rl}
\partial_{q}S= & p\\
\partial_{P}S= & Q\\
\mathscr{H}-H= & \partial_{t}S
\end{array}\right.
\end{align*}
Choose $\mathscr{H}=0$ ($\Rightarrow P,Q=cst$) constants of motion
\begin{gather*}
\boxed{H\left(q,\partial_{q}S,t\right)+\partial_{t}S=\mathscr{H}=0}
\end{gather*}
Now if $H=\frac{1}{2}g_{ab}u^{a}u^{b}$ for geodesics, $q^{a}=x^{a},p_{a}=\frac{dx^{a}}{d\tau}$
$\mathscr{H}=0$ $Q^{a},P_{a}=cst$ then we have
\begin{align*}
L= & p_{a}x^{a}-H\\
\mathscr{L}= & P_{a}Q^{a}-\mathscr{H} & \frac{dF}{d\tau}= & p_{a}\dot{x}^{a}-H-P_{a}\dot{Q}^{a}+\mathscr{H}
\end{align*}
Define $S\left(x^{a},P_{a}\right)=F+Q^{a}P_{a}$ then
\begin{align*}
\frac{dS}{d\tau}= & p_{a}\dot{x}^{a}-H-P_{a}\dot{Q}^{a}+\mathscr{H}+\dot{Q}^{a}P_{a}+Q^{a}\dot{P}_{a}\\
= & \partial_{a}S\dot{x}^{a}+\partial_{P_{a}}S\dot{P}_{a}+\partial_{\tau}S
\end{align*}
and we get
\begin{gather*}
\begin{array}{rl}
\partial_{a}S= & p_{a}\\
\partial_{P_{a}}S= & Q^{a}\\
\mathscr{H}-H= & \partial_{\tau}S
\end{array}
\end{gather*}
so if $\mathscr{H}=0$ ($\Rightarrow P_{a},Q^{a}=cst$ of motion)
$H+\partial_{t}S=\mathscr{H}=0$

Thus
\begin{align*}
\partial_{P_{a}}S=Q^{a}\Rightarrow & S=Q^{a}P_{a}+f\left(\tau,x^{a}\right)\\
\partial_{a}f=\partial_{a}S=p_{a}\Rightarrow & S=Q^{a}P_{a}+p_{a}x^{a}+g\left(\tau\right)\\
d_{\tau}g=\partial_{\tau}S=-H\Rightarrow & S=Q^{a}P_{a}+p_{a}x^{a}-\int\frac{1}{2}g_{ab}u^{a}u^{b}d\tau
\end{align*}
but this is not very enlightening. Use \cite{Bazanski:1988fs}

\section{Original derivation of Hamilton-Jacobi equation}

\begin{figure}
\includegraphics[width=1\columnwidth]{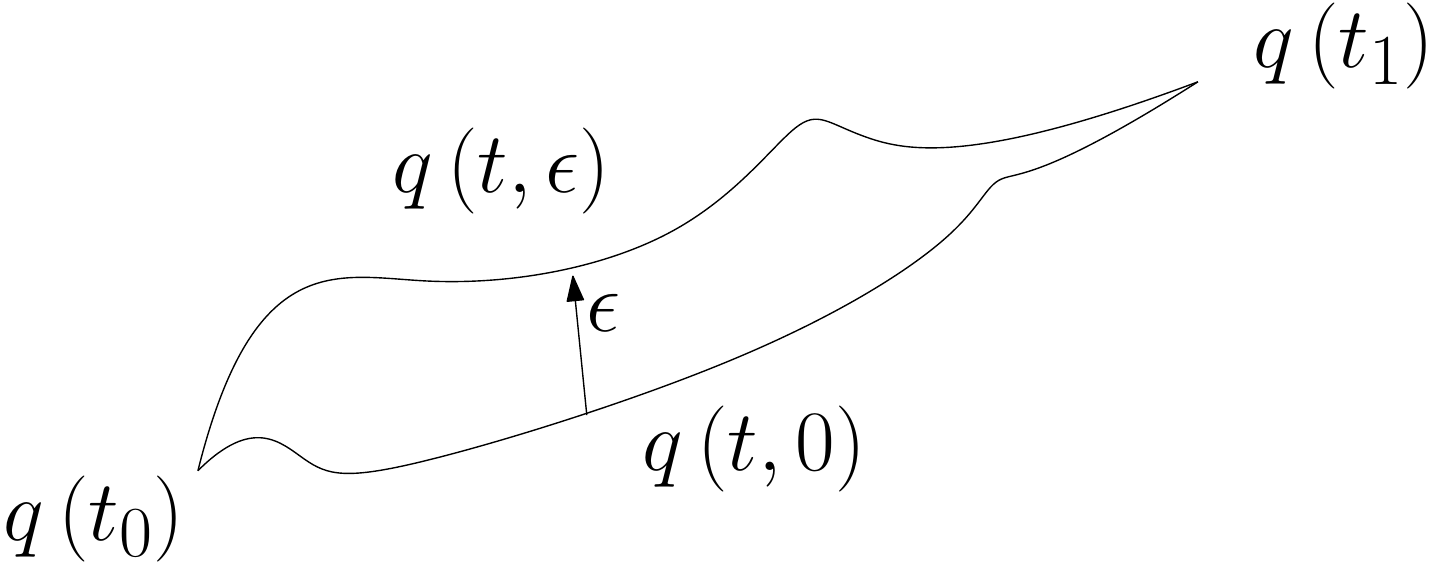}

\caption{\label{fig:Family-of-trajectories}Family of trajectories}

\end{figure}

From Hamilton: denote neighbouring family of trajectories by $\epsilon$
with Lagrangian $\mathscr{L}\left(q,\dot{q},t\right)$ and end points
at $t_{0},t_{1}$:

$q=q\left(t,\epsilon\right)$ trajectory, $q\left(t_{i},\epsilon\right)=q\left(t_{i}\right)\,i=0,1$
(see Fig.~\ref{fig:Family-of-trajectories})

Thus variations yield $\bar{\delta}q^{k}=\partial_{\epsilon}q^{k}\left(t,0\right)\delta\epsilon,\delta t=\frac{dt}{d\epsilon}\left(0\right)\delta\epsilon$

so $\delta q^{k}=\bar{\delta}q^{k}+\dot{q}^{k}\delta t=\left(\partial_{\epsilon}q^{k}+\frac{dt}{d\epsilon}\right)\left(0\right)\delta\epsilon$

The variation of the action yields from $S=\int_{t_{0}}^{t_{1}}\mathscr{L}\left(q,\dot{q},t\right)dt$
\begin{align*}
\delta S= & \left[\mathscr{L}\delta t\right]_{t_{0}}^{t_{1}}+\int_{t_{0}}^{t_{1}}\bar{\delta}\mathscr{L}\,dt
\end{align*}
Adding and substracting to the first term $\partial_{\dot{q}^{k}}\mathscr{L}\,\dot{q}^{k}\frac{dt}{d\epsilon}\delta\epsilon=\partial_{\dot{q}^{k}}\mathscr{L}\,\dot{q}^{k}\delta t$
and expliciting with integration by part for the second term,
\begin{align*}
\int_{t_{0}}^{t_{1}}\partial_{\dot{q}^{k}}\mathscr{L}\,\bar{\delta}\dot{q}^{k}dt= & \left[\partial_{\dot{q}^{k}}\mathscr{L}\,\bar{\delta}q^{k}\right]_{t_{0}}^{t_{1}}-\int_{t_{0}}^{t_{1}}\frac{d}{dt}\left(\partial_{\dot{q}^{k}}\mathscr{L}\right)\bar{\delta}q^{k}dt
\end{align*}
for $\bar{\delta}\dot{q}=\partial_{\epsilon}\dot{q}\delta\epsilon=\frac{d}{dt}\left(\bar{\delta}q\right)$
we get
\begin{gather*}
\boxed{\delta S=\left[\partial_{\dot{q}^{k}}\mathscr{L}\,\delta q^{k}-\left(\partial_{\dot{q}^{k}}\mathscr{L}\,\dot{q}^{k}-\mathscr{L}\right)\delta t\right]_{t_{0}}^{t_{1}}+\int_{t_{0}}^{t_{1}}\left[\partial_{q^{k}}\mathscr{L}-\frac{d}{dt}\left(\partial_{\dot{q}^{k}}\mathscr{L}\right)\right]\bar{\delta}q^{k}dt}
\end{gather*}
\begin{tabular}[t]{ll}
Taking $\delta q=\delta t=0$ at $t_{0},t_{1}$ yields the usual & - Lagrange equations\tabularnewline
 & -Noether identities\tabularnewline
\end{tabular}

Thus the differential of the Principal function $S\left(q,t\right)$
\begin{align*}
dS= & \left[\partial_{\dot{q}}\mathscr{L}.dq-\left(\partial_{\dot{q}}\mathscr{L}.\dot{q}-\mathscr{L}\right)dt\right]_{t_{0}}^{t_{1}}+\int_{t_{0}}^{t_{1}}\left[\partial_{q}\mathscr{L}-\frac{d}{dt}\left(\partial_{\dot{q}}\mathscr{L}\right)\right]dq\,dt\textrm{ on shell}\\
= & \partial_{q}S.dq+\partial_{t}S\,dt
\end{align*}
(with $\partial_{q}\mathscr{L}-\frac{d}{dt}\left(\partial_{\dot{q}}\mathscr{L}\right)=0$)
one gets
\begin{gather*}
\boxed{\begin{array}{rl}
\partial_{q^{k}}S= & \partial_{\dot{q}^{k}}\mathscr{L}=p_{k}\\
-\partial_{t}S= & p.\dot{q}-\mathscr{L}=H\left(q^{k},\partial_{q^{k}}S,t\right)
\end{array}}
\end{gather*}
Thus we have the
\begin{thm}
Jacobi theorem

$S:Q_{n}\times\mathbb{R}^{n+1}\longrightarrow\mathbb{R}$ $S=S\left(q^{k},a^{l},t\right)$
is a complete integral of the Hamilton-Jacobi equation
\begin{align*}
\Rightarrow\partial_{a^{l}}S= & \alpha_{l}
\end{align*}
defines implicitly
\begin{enumerate}
\item $q^{k}=\xi^{k}\left(t,a^{l},\alpha_{m}\right)$ satisfies $\partial_{a^{l}}S=\partial_{q^{k}}S\partial_{a^{l}}\xi^{k}=\alpha_{l}$
\item $p_{k}=\eta_{k}\left(t,a^{l},\alpha_{m}\right)=\partial_{q^{k}}S\left(\xi^{k},a^{l},t\right)=\partial_{q^{k}}S\left(\xi^{k}\left(t,a^{l},\alpha_{m}\right),t\right)$
\end{enumerate}
and $p$ and $q$ are solutions of Hamilton's equations
\end{thm}

\section{Geodesics}

In manifold $\left(M,g_{ab}\right)$, we have $\gamma$ timelike curve
with parameter $\tau\in\left[\tau_{0},\tau_{1}\right],\:x^{a}=\xi^{a}\left(\tau\right)$

The action functional for it to be geodesic writes
\begin{align*}
U\left[\gamma\right]= & \int_{\tau_{0}}^{\tau_{1}}\sqrt{-g_{ab}\dot{\xi}^{a}\dot{\xi}^{b}}d\tau\textrm{ where } & \dot{\xi}^{a}= & \frac{d\xi^{a}}{d\tau}\\
 &  & \bar{\delta}\xi^{a}= & \partial_{\epsilon}\xi^{a}\left(\tau,0\right)\delta\epsilon\\
 &  & \delta\tau= & \partial_{\epsilon}\tau\left(0\right)\delta\epsilon
\end{align*}
so $\delta\xi^{a}=\bar{\delta}\xi^{a}+\dot{\xi}^{a}\delta\tau$

We have
\begin{gather*}
\left\{ \begin{array}{rl}
\partial_{\dot{x}^{a}}\mathscr{L}= & \frac{-g_{ab}\dot{\xi}^{b}}{\sqrt{-g_{ab}\dot{\xi}^{a}\dot{\xi}^{b}}}\\
\partial_{a}\mathscr{L}= & 0
\end{array}\right.\begin{array}{rl}
\Rightarrow\partial_{\dot{x}^{a}}\mathscr{L}\,\dot{\xi}^{a}= & \sqrt{-g_{ab}\dot{\xi}^{a}\dot{\xi}^{b}}=\mathscr{L}\\
\\
\end{array}
\end{gather*}
so applying the same procedure
\begin{align*}
\delta U= & \left[\frac{-g_{ab}\dot{\xi}^{b}}{\sqrt{-\dot{\xi}^{c}\dot{\xi}_{c}}}\,\delta\xi^{a}\right]_{\tau_{0}}^{\tau_{1}}+\int_{\tau_{0}}^{\tau_{1}}d\tau\left[\frac{D}{d\tau}\left(\frac{g_{ab}\dot{\xi}^{b}}{\sqrt{-\dot{\xi}^{c}\dot{\xi}_{c}}}\right)\right]\bar{\delta}\xi^{a}
\end{align*}
Accordingly the Principal function $U\left(x^{a},\tau\right)$ is
constructed so that
\begin{align*}
dU= & \frac{-g_{ab}\dot{\xi}^{b}}{\sqrt{-\dot{\xi}^{c}\dot{\xi}_{c}}}dx^{a}\Rightarrow\raisebox{.8\baselineskip}{\rotatebox{-90}{\ensuremath{\setlength{\arraycolsep}{.5\arraycolsep}\begin{array}{@{}c@{}}
\begin{array}{rr}
\mrot{-\frac{g_{ab}\dot{\xi}^{b}}{\sqrt{-\dot{\xi}^{c}\dot{\xi}_{c}}}} & \mrot{0\textrm{ which is the Hamilton-Jacobi equation}}\end{array}\\
\aunderbrace[lD1r]{\begin{array}{rr}
\mrot{\partial_{a}U=\vphantom{\frac{g_{ab}\dot{\xi}^{b}}{\sqrt{-\dot{\xi}^{c}\dot{\xi}_{c}}}}} & \mrot{\partial_{\epsilon}U=}\end{array}}
\end{array}}}}
\end{align*}
but we also have
\begin{align*}
g^{ab}\partial_{a}U\partial_{b}U= & g^{ab}\times-\frac{g_{ac}\dot{\xi}^{c}}{\sqrt{-\dot{\xi}^{2}}}\times-\frac{g_{bd}\dot{\xi}^{d}}{\sqrt{-\dot{\xi}^{2}}}\\
= & \frac{g_{ac}\dot{\xi}^{c}\dot{\xi}^{a}}{-\dot{\xi}^{d}\dot{\xi}_{d}}=-1
\end{align*}
If we proceed like for the Jacobi theorem, we would break covariance,
so we need a covariant version

\subsection{Jacobi theorem for geodesics}

We consider $g^{ab}\partial_{a}U\partial_{b}U=-1$ a PDE for $U=U\left(x^{a}\right)$
and define a complete integral for it
\begin{defn}
$U:M\times\mathbb{R}^{3}\longrightarrow\mathbb{R}$ $U=U\left(x^{a},a^{k}\right)$
is a complete integral of $g^{ab}\partial_{a}U\partial_{a}U=-1\Leftrightarrow$
\begin{enumerate}
\item $U\left(x^{a},a^{k}\right)$ verifies $\forall a^{l}\in I_{\left(l\right)},g^{ab}\partial_{a}U\partial_{b}U=-1$,
where $I_{\left(l\right)}$ open set of $\mathbb{R}^{3}$
\item $\partial_{a,a^{k}}^{2}U$ is a $4\times3$ matrix of rank $3$
\end{enumerate}
\end{defn}
Note
\begin{itemize}
\item $a^{k}$ are scalars (independent variables of $x^{a}\Leftrightarrow\partial_{a}a^{k}=0$)
\item Since we can Taylor expand
\begin{align*}
\partial_{a}U= & \partial_{a}U\left(x^{b},0\right)+\partial_{a,a^{k}}^{2}U.a^{k}+O\left(\left(a^{k}\right)^{2}\right)
\end{align*}
and $\partial_{a,a^{k}}^{2}U$ $4\times3$ of rank $3$, it can be
inverted\\
and $a^{k}=\left(\partial_{a,a^{k}}^{2}U\right)^{-1}\left(\partial_{a}U-\partial_{a}U_{0}\right)$
so
\begin{align*}
g^{ab}\partial_{a}U\partial_{b}U=-1= & g^{ab}\partial_{a}U\left(x^{c},\left(\partial_{d,a^{k}}^{2}U\right)^{-1}\left(\partial_{d}U-\partial_{d}U_{0}\right)\right)\\
 & \phantom{g^{ab}}\partial_{b}U\left(x^{e},\left(\partial_{f,a^{k}}^{2}U\right)^{-1}\left(\partial_{f}U-\partial_{f}U_{0}\right)\right)\\
\Leftrightarrow\Psi\left(x^{a},\partial_{d}U\right)= & 0
\end{align*}
\end{itemize}
\begin{thm}
\label{thm:Jacobi-theorem-geo}Jacobi theorem for geodesics

$U=U\left(x^{a},a^{k}\right)$ complete integral of \textup{$g^{ab}\partial_{a}U\partial_{b}U=-1\Rightarrow\partial_{a^{k}}U=\alpha_{k},\alpha_{k}\in\mathbb{R}^{3}$
determines $\xi^{a}:\mathbb{R}^{7}\longrightarrow\mathbb{R},x^{a}=\xi^{a}\left(f\left(\tau\right),a^{k},\alpha_{l}\right)$
family of worldlines in $M$, $f$ arbitrary and monotonous, verifying}
\begin{enumerate}
\item \label{enu:jacobiGeo1}$\partial_{a^{k}}U=\alpha_{k}$
\item \label{enu:jacobiGeo2}$\exists\lambda=\lambda\left(\tau\right),\,\lambda\partial_{a}U=g_{ab}\frac{d\xi^{b}}{d\tau}$
\item $x^{a}=\xi^{a}\left(f\left(\tau\right),a^{k},\alpha_{l}\right)$ geodesics
\begin{align*}
\frac{D}{d\tau}\left[g_{ab}\frac{d\xi^{b}}{d\tau}\right]= & \frac{\dot{\lambda}}{\lambda}g_{ab}\frac{d\xi^{b}}{d\tau}
\end{align*}
\item \label{enu:jacobiGeo4}$\frac{df}{d\tau}=\pm\lambda\Leftrightarrow s=f\left(\tau\right)$
proper time along geodesics
\end{enumerate}
\end{thm}
Note if $\tau$ proper time, $f\left(\tau\right)=\tau=s\Rightarrow\frac{df}{d\tau}=1=\pm\lambda\Rightarrow\dot{\lambda}=0$
\begin{align*}
\frac{D}{d\tau}\left[g_{ab}\frac{d\xi^{b}}{d\tau}\right]= & 0
\end{align*}
so we get usual geodesic definition
\begin{proof}
There always exists (simplectic structure) $\varphi\left(x^{a},a^{k}\right)=\tau$
such as $\lefteqn{\underset{\begin{array}{c}
\downarrow\\
{\scriptscriptstyle \textrm{rank 3}}
\end{array}}{\phantom{(\;\partial_{a,a^{k}}^{2}U}}}\left(\partial_{a,a^{k}}^{2}U,\partial_{b}\varphi\right)$ is $4\times4$ of rank 4 (non-singular), and for $f$ monotonous
($f\ne0$)
\begin{gather*}
\left\{ \begin{array}{rll}
\phi\left(x^{a},a^{k}\right)= & f\left(\tau\right) & \textrm{ can supplement}\\
\partial_{a^{k}}U\left(x^{a},a^{k}\right)= & \alpha_{k}
\end{array}\right.
\end{gather*}
and the 4$\times$4 system can be locally inverted into
\begin{gather*}
\boxed{x^{a}=\xi^{a}\left(f\left(\tau\right),a^{k},\alpha_{l}\right)}
\end{gather*}
Plug into algebraic equations, it verifies Condition~\ref{enu:jacobiGeo1}:
$\partial_{a^{k}}U\left(\xi^{a},a^{k}\right)=\alpha_{k}$

As a complete integral, $U$ verifies
\begin{align*}
g^{ab}\partial_{a}U\partial_{b}U= & -1
\end{align*}
so $\partial_{a}U$ is timelike

Differentiating
\begin{align*}
g^{ab}\partial_{a}U\partial_{b,a^{k}}^{2}U= & 0
\end{align*}
so the $\partial_{a,a^{k}}^{2}U$ are orthogonal, thus spacelike and,
from definition, linearly independent

$\Rightarrow\partial_{a}U,\partial_{a,a^{k}}^{2}U$ linearly independent
\begin{align*}
\frac{d}{d\tau}\left(\partial_{a^{k}}U\left(\xi^{a},a^{k}\right)=\alpha_{k}\right)\Leftrightarrow & \partial_{a,a^{k}}^{2}U\frac{d\xi^{a}}{d\tau}=0
\end{align*}
Thus $\frac{d\xi^{a}}{d\tau}$ orthogonal to all three $\partial_{a,a^{k}}^{2}U$,
therefore parallel to $\partial_{a}U$ (in 4D), thus proportional,
and we get $\boxed{g_{ab}\frac{d\xi^{b}}{d\tau}=\lambda\partial_{a}U}$
for $\lambda\left(\tau,x^{a},a^{k},\alpha_{l}\right)$

Using it into $g^{ab}\partial_{a}U\partial_{b}U=-1$, we get
\begin{align*}
g^{ab}\frac{g_{ac}}{\lambda}\frac{d\xi^{c}}{d\tau}\frac{g_{bd}}{\lambda}\frac{d\xi^{d}}{d\tau}= & \frac{g_{ac}}{\lambda^{2}}\frac{d\xi^{c}}{d\tau}\frac{d\xi^{a}}{d\tau}=-1
\end{align*}
Now choose $f\left(\tau\right)=s$ proper time, then $g_{ab}\frac{d\xi^{a}}{ds}\frac{d\xi^{b}}{ds}=-1$ 

Since $\frac{d\xi^{a}}{d\tau}=\frac{d\xi^{a}}{ds}\frac{ds}{d\tau}=\frac{d\xi^{a}}{ds}\frac{df}{d\tau}$
then we have $-1=\frac{g_{ab}}{\lambda^{2}}\frac{d\xi^{a}}{ds}\frac{d\xi^{b}}{ds}\left(\frac{df}{d\tau}\right)^{2}=-\left(\frac{\frac{df}{d\tau}}{\lambda}\right)^{2}$
thus $\boxed{\frac{df}{d\tau}=\pm\lambda\left(\tau\right)}$ which
yields Conditions~\ref{enu:jacobiGeo2} and \ref{enu:jacobiGeo4}

Now from $g_{ab}\frac{d\xi^{b}}{d\tau}=\lambda\partial_{a}U$, we
can write
\begin{align*}
\frac{D}{d\tau}\left[g_{ab}\frac{d\xi^{b}}{d\tau}\right]= & \dot{\lambda}\partial_{a}U+\lambda\left(\partial_{a}U\right)_{;b}\frac{d\xi^{b}}{d\tau}
\end{align*}
and for a scalar $\phi$
\begin{align*}
\phi_{;ab}= & \left(\phi_{,a}\right)_{;b}\\
= & \phi_{,ab}-\Gamma_{ab}^{c}\phi_{,c}\\
= & \phi_{,ba}-\Gamma_{ba}^{c}\phi_{,c}\\
= & \left(\phi_{,b}\right)_{;a}=\phi_{;ba}
\end{align*}
thus 
\begin{align*}
\lboxed{\frac{D}{d\tau}\left[g_{ab}\frac{d\xi^{b}}{d\tau}\right]}= & \dot{\lambda}\partial_{a}U+\lambda^{2}g^{bc}\left(\partial_{a}U\right)_{;b}\partial_{c}U\\
= & \dot{\lambda}\partial_{a}U+\frac{\lambda^{2}}{2}\nabla_{b}\left(g^{bc}\partial_{a}U\partial_{c}U\right)\\
= & \dot{\lambda}\partial_{a}U=\rboxed{\frac{\dot{\lambda}}{\lambda}g_{ab}\frac{d\xi^{b}}{d\tau}}
\end{align*}
\end{proof}
Of course $\partial_{a^{k}}U=\alpha_{k}$ is also telling $\alpha_{k}$
constant of motion of geodesics $\xi^{a}$: $\frac{D}{d\tau}\left[\frac{d\xi_{a}}{d\tau}\right]=\frac{\dot{\lambda}}{\lambda}\frac{d\xi_{a}}{d\tau}$
using the set
\begin{gather*}
\left\{ \begin{array}{rl}
U= & s-s_{0}\\
\partial_{a^{k}}U= & \alpha_{k}
\end{array}\right.
\end{gather*}
instead of an arbitrary function yields a unique solution $x^{a}=\xi^{a}\left(s-s_{0},a^{k},\alpha_{l}\right)$
geodesic for proper time $s$

\section{Geodesic deviation (overview)}

The action functional for evolution of geodesic deviation $r^{a}$
along geodesic $\gamma:x^{a}=\xi^{a}\left(\tau\right)$ reads
\begin{align*}
S\left[\gamma,r\right]= & \int_{\tau_{0}}^{\tau_{1}}d\tau g_{ab}\frac{\dot{\xi}^{a}}{\left(-\dot{\xi}^{c}\dot{\xi}_{c}\right)^{\frac{1}{2}}}\frac{Dr^{b}}{d\tau}
\end{align*}
\begin{tabular}{lll}
Variations with respect to &  $r^{a}$ yield & geodesic equation on $\xi^{a}$\tabularnewline
while & $\xi^{a}$ & geodesic deviation equation on $r^{a}$. As before\tabularnewline
\end{tabular}
\begin{align*}
\bar{\delta}\xi^{a}= & \partial_{\epsilon}\xi^{a}\left(\tau,0\right)\delta\epsilon\\
\bar{\delta}r^{a}= & \partial_{\epsilon}r^{a}\left(\tau,0\right)\delta\epsilon\\
\delta\tau= & \partial_{\epsilon}\tau\left(0\right)\delta\epsilon
\end{align*}
and complete variations are
\begin{align*}
\delta\xi^{a}= & \bar{\delta}\xi^{a}+\dot{\xi}^{a}\delta\tau\\
\delta r^{a}= & \bar{\delta}r^{a}+\dot{r}^{a}\delta\tau
\end{align*}
Thus we get, with $h_{ab}=g_{ab}+\frac{\dot{\xi}_{a}\dot{\xi}_{b}}{\left(-\dot{\xi}^{c}\dot{\xi}_{c}\right)}$
projector on space
\begin{multline*}
\delta S=\left[\frac{1}{\left(-\dot{\xi}^{c}\dot{\xi}_{c}\right)^{\frac{1}{2}}}\left(h_{ab}\frac{Dr^{b}}{d\tau}+\Gamma_{ac}^{b}r^{c}\dot{\xi}_{b}\right)\,\delta\xi^{a}-\frac{1}{\left(-\dot{\xi}^{c}\dot{\xi}_{c}\right)^{\frac{1}{2}}}g_{ab}\dot{\xi}^{a}\delta r^{b}\right]_{\tau_{0}}^{\tau_{1}}\\
-\int_{\tau_{0}}^{\tau_{1}}d\tau\left\{ \left[\frac{D}{d\tau}\left(\frac{h_{ab}}{\left(-\dot{\xi}^{c}\dot{\xi}_{c}\right)^{\frac{1}{2}}}\frac{Dr^{b}}{d\tau}\right)+\frac{R_{abcd}\dot{\xi}^{b}r^{c}\dot{\xi}^{d}}{\left(-\dot{\xi}^{c}\dot{\xi}_{c}\right)^{\frac{1}{2}}}+\Gamma_{ab}^{c}r^{b}\frac{D}{d\tau}\left(\frac{1}{\left(-\dot{\xi}^{c}\dot{\xi}_{c}\right)^{\frac{1}{2}}}\right)\right]\bar{\delta}\xi^{a}\right.\\
\left.+\frac{D}{d\tau}\left(\frac{\dot{\xi}^{a}}{\left(-\dot{\xi}^{c}\dot{\xi}_{c}\right)^{\frac{1}{2}}}\right)\bar{\delta}r^{a}\right\} 
\end{multline*}
When $\delta\epsilon=0,\delta r^{a}$ and $\delta\xi^{a}$ independent
and vanishing at $\tau_{0},\tau_{1}$,
\begin{align*}
\delta S=0\Leftrightarrow & \raisebox{.8\baselineskip}{\rotatebox{-90}{\ensuremath{\setlength{\arraycolsep}{.5\arraycolsep}\begin{array}{@{}c@{}}
\begin{array}{rr}
\mrot{\vphantom{\left(\frac{\dot{\xi}^{a}}{\left(-\dot{\xi}^{c}\dot{\xi}_{c}\right)^{\frac{1}{2}}}\right)}0\textrm{ geodesic equation}} & \mrot{\vphantom{\left(\frac{h_{ab}}{\left(-\dot{\xi}^{c}\dot{\xi}_{c}\right)^{\frac{1}{2}}}\frac{Dr^{b}}{d\tau}\right)}0\textrm{ geodesic deviation equation}}\end{array}\\
\aunderbrace[lD1r]{\begin{array}{rr}
\mrot{\phantom{\frac{R_{abcd}\dot{\xi}^{b}r^{c}\dot{\xi}^{d}}{\left(-\dot{\xi}^{c}\dot{\xi}_{c}\right)^{\frac{1}{2}}}+++}\frac{D}{d\tau}\left(\frac{\dot{\xi}^{a}}{\left(-\dot{\xi}^{c}\dot{\xi}_{c}\right)^{\frac{1}{2}}}\right)=} & \mrot{\frac{D}{d\tau}\left(\frac{h_{ab}}{\left(-\dot{\xi}^{c}\dot{\xi}_{c}\right)^{\frac{1}{2}}}\frac{Dr^{b}}{d\tau}\right)+\frac{R_{abcd}\dot{\xi}^{b}r^{c}\dot{\xi}^{d}}{\left(-\dot{\xi}^{c}\dot{\xi}_{c}\right)^{\frac{1}{2}}}=}\end{array}}
\end{array}}}}
\end{align*}
Solutions: $\xi^{a}\left(\tau\right),r^{a}\left(\tau\right)$ so Principal
function is
\begin{align*}
S\left(\xi^{a},r^{a},\tau\right)= & \int_{\tau_{0}}^{\tau_{1}}d\tau g_{ab}\frac{\dot{\xi}^{a}}{\left(-\dot{\xi}^{c}\dot{\xi}_{c}\right)^{\frac{1}{2}}}\frac{Dr^{b}}{d\tau}
\end{align*}
and differential for solutions $\xi^{a},r^{a}$ yield
\begin{align*}
dS= & \frac{1}{\left(-\dot{\xi}^{c}\dot{\xi}_{c}\right)^{\frac{1}{2}}}\left(h_{ab}\frac{Dr^{b}}{d\tau}+g_{bd}\Gamma_{ac}^{b}r^{c}\dot{\xi}^{d}\right)dx^{a}+g_{ab}\frac{\dot{\xi}^{b}}{\left(-\dot{\xi}^{c}\dot{\xi}_{c}\right)^{\frac{1}{2}}}dr^{a}
\end{align*}
therefore
\begin{align*}
\partial_{\tau}S= & 0\;\begin{array}{rl}
\partial_{r^{a}}S= & g_{ab}\frac{\dot{\xi}^{b}}{\left(-\dot{\xi}^{c}\dot{\xi}_{c}\right)^{\frac{1}{2}}}\end{array}\\
\partial_{a}S= & \frac{1}{\left(-\dot{\xi}^{c}\dot{\xi}_{c}\right)^{\frac{1}{2}}}\left(h_{ab}\frac{Dr^{b}}{d\tau}+g_{bd}\Gamma_{ac}^{b}r^{c}\dot{\xi}^{d}\right)
\end{align*}
and we can combine them into
\begin{itemize}
\item $g^{ab}\partial_{r^{a}}S\partial_{r^{b}}S=-1$ as previously $\frac{\dot{\xi}^{a}\dot{\xi}_{a}}{-\dot{\xi}^{d}\dot{\xi}_{d}}=-1$
\\
and $\partial_{a}S-\Gamma_{ac}^{b}r^{c}\partial_{r^{b}}S=\frac{h_{ab}}{\left(-\dot{\xi}^{c}\dot{\xi}_{c}\right)^{\frac{1}{2}}}\frac{Dr^{b}}{d\tau}$,
so
\item 
\begin{align*}
g^{ab}\partial_{r^{a}}S\left(\partial_{a}S-\Gamma_{ac}^{b}r^{c}\partial_{r^{b}}S\right)= & \frac{\dot{\xi}^{b}}{\left(-\dot{\xi}^{d}\dot{\xi}_{d}\right)^{\frac{1}{2}}}\frac{h_{bc}}{\left(-\dot{\xi}^{e}\dot{\xi}_{e}\right)^{\frac{1}{2}}}\frac{Dr^{c}}{d\tau}\\
= & 0\textrm{ since }h_{bc}\dot{\xi}^{b}=0
\end{align*}
\end{itemize}
Those last two equations are the Hamilton-Jacobi equations for $\left\{ \begin{array}{l}
\textrm{geodesics}\\
\textrm{geodesic deviation}
\end{array}\right.$

Now we will solve them with additional Jacobi-type theorems

\subsection{Theorems on geodesic deviation solutions}

We solve the system
\begin{gather*}
\left\{ \begin{array}{rr}
g^{ab}\partial_{r^{a}}S\partial_{r^{b}}S= & -1\\
g^{ab}\partial_{r^{b}}S\left(\partial_{a}S-\Gamma_{ac}^{b}r^{c}\partial_{r^{b}}S\right)= & 0
\end{array}\right.
\end{gather*}
for $S$ defined in tangent bundle $TM$ and take values in $M$ $\left(x^{a}\right)$
and in $T_{x}M$, tangent space at $x$
\begin{defn}
\label{def:-complete-integral}$S$ complete integral of $\raisebox{.8\baselineskip}{\rotatebox{-90}{\ensuremath{\setlength{\arraycolsep}{.35\arraycolsep}\begin{array}{@{}c@{}}
\begin{array}{rr}
\mrot{-1} & \mrot{\phantom{-}0}\end{array}\\
\aunderbrace[l1D1r]{\begin{array}{rr}
\mrot{\hphantom{+++m+}g\partial_{r}S\partial_{r}S=} & \mrot{g\partial_{r}S\left(\partial S-\Gamma r\partial_{r}S\right)=}\end{array}}
\end{array}}}}$

$S:M\times T_{x}M\times\mathbb{R}^{6}\longrightarrow\mathbb{R}$,
$S=S\left(x^{a},r^{b},a^{k},b^{l}\right)\Leftrightarrow$
\begin{enumerate}
\item $S$ solution of \label{enu:DefCI-solution-of-system} $\forall\left(a^{k},b^{l}\right)\in\mathbb{R}^{6}$
\item 
\begin{align*}
M_{8\times6}= & \left(\begin{array}{cc}
\partial_{a,a^{k}}^{2}S & \partial_{b,b^{l}}^{2}S\\
\partial_{r^{a},a^{k}}^{2}S & \partial_{r^{b},b^{l}}^{2}S
\end{array}\right)
\end{align*}
\label{enu:DefCI-is-rank-6}is rank 6
\end{enumerate}
\end{defn}
Existence of complete integral follows
\begin{thm}
of existence

All complete integral $U\left(x^{a},a^{k}\right)$ of $g^{ab}\partial_{a}U\partial_{b}U=-1$
generates \\
a complete integral $S\left(x^{a},r^{b},a^{k},b^{l}\right)$ of $\raisebox{.8\baselineskip}{\rotatebox{-90}{\ensuremath{\setlength{\arraycolsep}{.35\arraycolsep}\begin{array}{@{}c@{}}
\begin{array}{rr}
\mrot{-1} & \mrot{\phantom{-}0}\end{array}\\
\aunderbrace[l1D1r]{\begin{array}{rr}
\mrot{\hphantom{+++m+}g\partial_{r}S\partial_{r}S=} & \mrot{g\partial_{r}S\left(\partial S-\Gamma r\partial_{r}S\right)=}\end{array}}
\end{array}}}}$ in the form\\
$S=\partial_{a}U\,r^{a}+\partial_{a^{k}}U\,b^{k}$
\end{thm}
From form
\begin{itemize}
\item $\partial_{r^{a}}S=\partial_{a}U$ and $g\partial U\partial U=-1\Rightarrow g\partial_{r}S\partial_{r}S=-1$
\item $\partial_{a}S=\partial_{ab}^{2}U\,r^{b}+\partial_{a,a^{k}}^{2}U\,b^{k}$
\end{itemize}
so
\begin{align*}
g\partial_{r}S\left(\partial S-\Gamma r\partial_{r}S\right)= & g^{ab}\partial_{a}U\left(\partial_{bc}^{2}U\,r^{c}+\partial_{b,a^{k}}^{2}U\,b^{k}-\Gamma_{bc}^{d}r^{c}\partial_{d}U\right)\\
= & g^{ab}\partial_{a}U\left(\partial_{b}U\right)_{;c}r^{c}+g^{ab}\partial_{a}U\partial_{b,a^{k}}^{2}U\,b^{k}\\
= & \lefteqn{\phantom{((}\underbrace{\phantom{\frac{g^{ab}}{2}\partial_{a}U\partial_{b}U}}_{-\frac{1}{2}}}\left(\frac{g^{ab}}{2}\partial_{a}U\partial_{b}U\right)_{;c}+\partial_{a^{k}}\lefteqn{\phantom{((}\underbrace{\phantom{\frac{g^{ab}}{2}\partial_{a}U\partial_{b}U}}_{-\frac{1}{2}}}\left(\frac{g^{ab}}{2}\partial_{a}U\partial_{b}U\right)b^{k}\\
= & 0
\end{align*}
So $S$ verifies definition \ref{def:-complete-integral}-\ref{enu:DefCI-solution-of-system}

From form $S=\partial_{a}U\,r^{a}+\partial_{a^{k}}U\,b^{k}$ we get
\begin{align*}
M_{8\times6}= & \left(\begin{array}{cc}
\partial_{a,a^{k}}^{2}S & \partial_{b,a^{l}}^{2}U\\
\partial_{a,a^{k}}^{2}U & 0
\end{array}\right)
\end{align*}
since $\partial_{b^{l}}\left(\partial_{b}U\right)=0$

As $\partial_{a,a^{k}}^{2}U$ are 3 linearly independent 4-vectors
(rank 3) then the columns of $M_{8\times6}$ are 6 L.I. 8-vectors
and $M_{8\times6}$ is of rank 6 (Def.~\ref{def:-complete-integral}-\ref{enu:DefCI-is-rank-6})
\begin{thm}
\label{thm:Jacobi-theorem-geo-dev}Jacobi theorem on geodesic deviations

$S=\partial_{a}U\,r^{a}+\partial_{a^{k}}U\,b^{k}$ where $U$ complete
integral of $g^{ab}\partial_{a}U\partial_{b}U=-1\Rightarrow$
\begin{enumerate}
\item $\partial_{b^{k}}S=\partial_{a^{k}}U=\alpha_{k}$ determines a family
of geodesics $x^{a}=\xi^{a}\left(f\left(\tau\right),a^{k},\alpha_{l}\right)$
from Jacobi theorem on geodesics (\ref{thm:Jacobi-theorem-geo})
\item \label{enu:jacobiGeoDev2}
\begin{gather*}
\left\{ \begin{array}{rl}
\partial_{a^{k}}S\left(\xi^{a},r^{b},a^{m},b^{l}\right)= & \beta_{k}\\
\partial_{r^{a}}Sr^{a}= & \mu\left(\tau\right)
\end{array}\right.
\end{gather*}
with $\beta_{k}\in\mathbb{R}^{3}$ and $\mu$ arbitrary can be solved
locally for $r^{a}$ and determine $r^{a}=\rho^{a}\left(\tau,a^{k},\alpha_{l},b^{m},\beta_{n},\mu\left(\tau\right)\right)$
which defines for constant $a^{k},\alpha_{l},b^{m},\beta_{n}$ a vector
field along $\xi^{a}$ geodesic
\item $\rho^{a}$ verifies 
\begin{enumerate}
\item \label{enu:jacobiGeoDev3a}$\partial_{r^{a}}S\frac{D\rho^{a}}{d\tau}=\dot{\mu}\left(\tau\right)$
for %
\begin{tabular}[t]{l}
$x^{a}=\xi^{a}$\tabularnewline
$r^{a}=\rho^{a}$\tabularnewline
\end{tabular}and from $\raisebox{.8\baselineskip}{\rotatebox{-90}{\ensuremath{\setlength{\arraycolsep}{.35\arraycolsep}\begin{array}{@{}c@{}}
\begin{array}{rr}
\mrot{g_{ab}\frac{d\xi^{b}}{d\tau}} & \mrot{\partial_{a}U}\end{array}\\
\aunderbrace[l1D1r]{\begin{array}{rr}
\mrot{\lambda\partial_{a}U=} & \mrot{\partial_{r^{a}}S=}\end{array}}
\end{array}}}}$\\
$\lambda\partial_{r^{a}}S\frac{D\rho^{a}}{d\tau}=\lambda\partial_{a}U\frac{D\rho^{a}}{d\tau}=\boxed{g_{ab}\frac{d\xi^{b}}{d\tau}\frac{D\rho^{a}}{d\tau}=\lambda\dot{\mu}}$
\item \label{enu:jacobiGeoDev3b}
\begin{align*}
\frac{D\rho^{a}}{d\tau}= & \left[\lambda g^{ab}\left(\partial_{b}S-\Gamma_{bd}^{c}r^{d}\partial_{r^{c}}S\right)-\dot{\mu}g^{ab}\partial_{r^{b}}S\right]_{\begin{array}{rl}
x= & \xi\\
r= & \rho
\end{array}}\\
= & \lambda g^{ab}\left[\left(\partial_{b}U\right)_{;d}r^{d}+\partial_{b,a^{k}}^{2}U\,b^{k}\right]_{\begin{array}{rl}
x= & \xi\\
r= & \rho
\end{array}}-\left[\dot{\mu}g^{ab}\partial_{b}U\right]_{\begin{array}{rl}
x= & \xi\end{array}}
\end{align*}
\item \label{enu:jacobiGeoDev3c}Geodesic deviation equation
\begin{align*}
\frac{D^{2}\rho^{a}}{d\tau^{2}}+R_{abcd}\frac{d\xi^{b}}{d\tau}\rho^{c}\frac{d\xi^{d}}{d\tau}= & \frac{\dot{\lambda}}{\lambda}\frac{D\rho^{a}}{d\tau}-\frac{d}{d\tau}\left(\frac{\dot{\mu}}{\lambda}\right)\frac{d\xi^{a}}{d\tau}
\end{align*}
\end{enumerate}
\end{enumerate}
\end{thm}
Note as before that $\tau$ proper time and $\mu\left(\tau\right)=\tau$
yields usual geodesic deviation.
\begin{proof}
~
\begin{itemize}
\item From $S=\partial_{a}U\,r^{a}+\partial_{a^{k}}U\,b^{k}$
\begin{itemize}
\item $\partial_{b^{k}}S=\partial_{a^{k}}U=\alpha_{k}$ from Geodesic Jacobi
theorem (\ref{thm:Jacobi-theorem-geo}) and thus \ref{thm:Jacobi-theorem-geo-dev}-\ref{enu:jacobiGeo1}
is fulfilled from it
\item We can rewrite the system
\begin{gather*}
\left\{ \begin{array}{rl}
\partial_{a^{k}}S= & \partial_{a,a^{k}}^{2}U\,r^{a}+\partial_{a^{k}a^{l}}^{2}U\,b^{l}=\beta_{k}\\
\partial_{r^{a}}Sr^{a}= & \partial_{a}U\,r^{a}=\mu
\end{array}\right.\\
\Rightarrow\left\{ \begin{array}{rl}
\partial_{a,a^{k}}^{2}U\,r^{a}= & \beta_{k}-\partial_{a^{k}a^{l}}^{2}U\,b^{l}\\
\partial_{a}U\,r^{a}= & \mu
\end{array}\right.
\end{gather*}
while $\partial_{a}U,\partial_{a,a^{k}}^{2}U$ are linearly independent,
so this invertible linear system ensure existence and unicity of $r^{a}=\rho^{a}\left(\tau,a^{k},\alpha_{l},b^{m},\beta_{n},\mu\right)$
(\ref{thm:Jacobi-theorem-geo-dev}-\ref{enu:jacobiGeoDev2})
\end{itemize}
\item Since $\partial_{a}U\,r^{a}=\mu\Rightarrow\lambda\partial_{a}U\,\rho^{a}=\lambda\mu$
and given $U$ complete integral of $g^{ab}\partial_{a}U\partial_{b}U=-1$,
so
\begin{gather*}
\left\{ \begin{array}{rl}
\lambda\partial_{a}U= & g_{ab}\frac{d\xi^{b}}{d\tau}\\
\frac{D}{d\tau}\left[g_{ab}\frac{d\xi^{b}}{d\tau}\right]= & \frac{\dot{\lambda}}{\lambda}g_{ab}\frac{d\xi^{b}}{d\tau},
\end{array}\right.
\end{gather*}
then $g_{ab}\frac{d\xi^{b}}{d\tau}\rho^{a}=\lambda\mu$. Taking its
$\frac{D}{d\tau}$ we get
\begin{align*}
\frac{D}{d\tau}\left[g_{ab}\frac{d\xi^{b}}{d\tau}\rho^{a}\right]= & \dot{\lambda}\mu+\lambda\dot{\mu}\\
= & \frac{\dot{\lambda}}{\lambda}g_{ab}\frac{d\xi^{b}}{d\tau}\rho^{a}+g_{ab}\frac{d\xi^{b}}{d\tau}\frac{D\rho^{a}}{d\tau}\\
= & \dot{\lambda}\partial_{a}U\rho^{a}+\lambda\partial_{a}U\frac{D\rho^{a}}{d\tau}\\
= & \dot{\lambda}\mu+\lambda\partial_{a}U\frac{D\rho^{a}}{d\tau}\\
\Leftrightarrow\lambda\partial_{a}U\frac{D\rho^{a}}{d\tau}= & \lambda\dot{\mu}\\
\Rightarrow & \boxed{\partial_{r^{a}}S\frac{D\rho^{a}}{d\tau}=\dot{\mu}}
\end{align*}
(\ref{thm:Jacobi-theorem-geo-dev}-\ref{enu:jacobiGeoDev3a})
\item Now from $\partial_{a^{k}}S=\beta_{k}=cst$ we can use again $S=\partial_{a}U\,r^{a}+\partial_{a^{k}}U\,b^{k}$
on shell $\left(\begin{array}{rl}
x= & \xi\\
r= & \rho
\end{array}\right)$ to write
\begin{align*}
0=\frac{D}{d\tau}\beta_{k}=\frac{D}{d\tau}\left(\partial_{a^{k}}S\right)= & \frac{D}{d\tau}\left(\partial_{a,a^{k}}^{2}U\rho^{a}+\partial_{a^{k}a^{l}}^{2}U\,b^{l}\right)\\
= & \partial_{a,a^{k}}^{2}U\frac{D\rho^{a}}{d\tau}+\frac{D}{d\tau}\left(\partial_{a,a^{k}}^{2}U\right)\rho^{a}+\partial_{a,a^{k}a^{l}}^{3}U\frac{d\xi^{a}}{d\tau}b^{l}
\end{align*}
since $b^{l}$ independent parameter\\
2nd term
\begin{align*}
\frac{D}{d\tau}\left(\partial_{a,a^{k}}^{2}U\right)=\left(\partial_{a,a^{k}}^{2}U\right)_{;b}\frac{d\xi^{b}}{d\tau}= & \left(\partial_{a,a^{k}}^{2}U\right)_{;b}\lambda g^{bc}\partial_{c}U & \textrm{from } & \lambda\partial_{a}U=g_{ab}\frac{d\xi^{b}}{d\tau}\textrm{ (}U\textrm{ complete integral)}\\
= & \lambda g^{bc}\left(\partial_{b,a^{k}}^{2}U\right)_{;a}\partial_{c}U & \textrm{from } & \phi_{;ab}=\phi_{;ba}=\left(\phi_{,a}\right)_{;b}\\
= & -\lambda g^{bc}\partial_{b,a^{k}}^{2}U\left(\partial_{c}U\right)_{;a} & \textrm{from } & g^{ab}\partial_{a}U\partial_{b,a^{k}}^{2}U=0\textrm{ (}U\textrm{ complete integral)}
\end{align*}
\\
3rd term from
\begin{align*}
0= & \partial_{a^{k}}\left(g^{ab}\partial_{b}U\partial_{a,a^{l}}^{2}U\right)\\
= & g^{ab}\partial_{a,a^{k}a^{l}}^{3}U\partial_{b}U+g^{ab}\partial_{a,a^{l}}^{2}U\partial_{b,a^{k}}^{2}U
\end{align*}
and with $\lambda\partial_{a}U=g_{ab}\frac{d\xi^{b}}{d\tau}$ we get
\begin{align*}
\partial_{a,a^{k}a^{l}}^{3}U\frac{d\xi^{a}}{d\tau}= & \lambda g^{ab}\partial_{a,a^{k}a^{l}}^{3}U\partial_{b}U=-\lambda g^{ab}\partial_{a,a^{l}}^{2}U\partial_{b,a^{k}}^{2}U
\end{align*}
Thus
\begin{align*}
0= & \partial_{a,a^{l}}^{2}U\left\{ \frac{D\rho^{a}}{d\tau}-\lambda g^{ab}\left[\left(\partial_{b}U\right)_{;c}\rho^{c}+\partial_{b,a^{k}}^{2}U\,b^{l}\right]\right\} 
\end{align*}
Since the vectors $\partial_{a,a^{k}}^{2}U$ are linearly independent
in 4D, bracket is proportional to $g^{ab}\partial_{b}U$:
\begin{align*}
\exists\bar{\mu}\left(\tau\right),\left\{ \right\} =\bar{\mu}g^{ab}\partial_{b}U\Leftrightarrow & \frac{D\rho^{a}}{d\tau}=\lambda g^{ab}\left[\left(\partial_{b}U\right)_{;c}\rho^{c}+\partial_{b,a^{l}}^{2}U\,b^{l}\right]+\bar{\mu}g^{ab}\partial_{b}U
\end{align*}
Contracting with $\partial_{a}U$ we get
\begin{align*}
\partial_{a}U\frac{D\rho^{a}}{d\tau}= & \dot{\mu}\\
= & \lambda\left[g^{ab}\partial_{a}U\left(\partial_{b}U\right)_{;c}\rho^{c}+g^{ab}\partial_{a}U\partial_{b,a^{l}}^{2}U\,b^{l}\right]+\bar{\mu}g^{ab}\partial_{a}U\partial_{b}U
\end{align*}
Since $g^{ab}\partial_{a}U\partial_{b}U=-1$ then any derivatives,
including $g^{ab}\partial_{a}U\left(\partial_{b}U\right)_{;c}=0=g^{ab}\partial_{a}U\partial_{b,a^{k}}^{2}U$
($U$ complete integral)\\
so $\dot{\mu}=-\bar{\mu}$\\
Thus
\begin{gather*}
\boxed{\frac{D\rho^{a}}{d\tau}=\lambda g^{ab}\left[\left(\partial_{b}U\right)_{;c}r^{c}+\partial_{b,a^{k}}^{2}U\,b^{k}\right]_{\begin{array}{rl}
x= & \xi\\
r= & \rho
\end{array}}-\dot{\mu}g^{ab}\left[\partial_{b}U\right]_{\begin{array}{rl}
x= & \xi\end{array}}}
\end{gather*}
and $\left(\partial_{b}U\right)_{;c}=\partial_{c}\left(\partial_{b}U\right)-\Gamma_{bc}^{d}\partial_{d}U$
while $\partial_{a}U=\partial_{r^{a}}S$ and $\partial_{a}S=\partial_{ab}^{2}U\,r^{b}+\partial_{a,a^{k}}^{2}U\,b^{k}$
so
\begin{gather*}
\boxed{\frac{D\rho^{a}}{d\tau}=\left[\lambda g^{ab}\left(\partial_{b}S-\Gamma_{bc}^{d}r^{c}\partial_{r^{d}}S\right)-\dot{\mu}g^{ab}\partial_{r^{b}}S\right]_{\begin{array}{rl}
x= & \xi\\
r= & \rho
\end{array}}}
\end{gather*}
which is (\ref{thm:Jacobi-theorem-geo-dev}-\ref{enu:jacobiGeoDev3b})
\item From ($U$ complete integral) $\lambda g^{ab}\partial_{b}U=\frac{d\xi^{a}}{d\tau}$
the previous equation can write
\begin{align*}
\frac{D\rho^{a}}{d\tau}= & \lambda g^{ab}\left[\left(\partial_{b}U\right)_{;c}\rho^{c}+\partial_{b,a^{k}}^{2}U\,b^{k}\right]-\frac{\dot{\mu}}{\lambda}\frac{d\xi^{a}}{d\tau}
\end{align*}
and taking $\frac{D}{d\tau}$
\begin{align*}
\frac{D^{2}\rho^{a}}{d\tau^{2}}= & \dot{\lambda}g^{ab}\left[\left(\partial_{b}U\right)_{;c}\rho^{c}+\partial_{b,a^{k}}^{2}U\,b^{k}\right]+\lambda\frac{D}{d\tau}\left(g^{ab}\left[\left(\partial_{b}U\right)_{;c}\rho^{c}+\partial_{b,a^{k}}^{2}U\,b^{k}\right]\right)-\frac{d}{d\tau}\left(\frac{\dot{\mu}}{\lambda}\right)\frac{d\xi^{a}}{d\tau}-\frac{\dot{\mu}\dot{\lambda}}{\lambda^{2}}\frac{d\xi^{a}}{d\tau}\\
 & \pushright{\textrm{from }\frac{D}{d\tau}\left[\frac{d\xi^{a}}{d\tau}\right]=\frac{\dot{\lambda}}{\lambda}\frac{d\xi^{a}}{d\tau}}\\
= & \frac{\dot{\lambda}}{\lambda}\frac{D\rho^{a}}{d\tau}+\lambda\frac{D}{d\tau}\left(g^{ab}\left[\left(\partial_{b}U\right)_{;c}\rho^{c}+\partial_{b,a^{k}}^{2}U\,b^{k}\right]\right)-\frac{d}{d\tau}\left(\frac{\dot{\mu}}{\lambda}\right)\lambda g^{ab}\partial_{b}U
\end{align*}
The second term can be decomposed in two, the first of which yields
\begin{align*}
\frac{D}{d\tau}\left[g^{ab}\left(\partial_{b}U\right)_{;c}\rho^{c}\right]= & g^{ab}\left(\partial_{b}U\right)_{;cd}\rho^{c}\frac{d\xi^{d}}{d\tau}+g^{ab}\left(\partial_{b}U\right)_{;c}\frac{D\rho^{c}}{d\tau}\\
 & \omit\hfill\textrm{and commutation of covariant derivatives on scalar}\\
= & g^{ab}\left(\partial_{c}U\right)_{;bd}\rho^{c}\frac{d\xi^{d}}{d\tau}+g^{ab}\left(\partial_{c}U\right)_{;b}\frac{D\rho^{c}}{d\tau}
\end{align*}
while the second gives, from $\left[g^{ab}\partial_{a,a^{k}}^{2}U\partial_{b}U\right]_{;c}=0,\phi_{;ab}=\phi_{;ba}$
and $\frac{d\xi^{a}}{d\tau}=\lambda g^{ab}\partial_{b}U$ (complete
integral), as well as $b^{k}$ independent parameter
\begin{align*}
\frac{D}{d\tau}\left[g^{ab}\partial_{b,a^{k}}^{2}U\,b^{k}\right]= & g^{ab}\left(\partial_{b,a^{k}}^{2}U\right)_{;c}\frac{d\xi^{c}}{d\tau}b^{k}\\
= & \lambda g^{ab}\left(\partial_{c,a^{k}}^{2}U\right)_{;b}g^{cd}\partial_{d}U\,b^{k}\\
= & -\lambda g^{ab}\partial_{c,a^{k}}^{2}U\left(\partial_{d}U\right)_{;b}g^{cd}b^{k}
\end{align*}
Substituting $\frac{D\rho}{d\tau}$ and summing those two terms, the
second one cancels the ``$b^{k}$'' term of the $\frac{D\rho}{d\tau}$
development
\begin{align*}
\frac{D}{d\tau}\left(g^{ab}\left[\left(\partial_{b}U\right)_{;c}\rho^{c}+\partial_{b,a^{k}}^{2}U\,b^{k}\right]\right)= & g^{ab}\left(\partial_{c}U\right)_{;bd}\rho^{c}\frac{d\xi^{d}}{d\tau}+g^{ab}\left(\partial_{d}U\right)_{;b}\left[\lambda g^{dc}\left(\partial_{c}U\right)_{;e}\rho^{e}+\lambda\cancel{g^{dc}\partial_{c,a^{k}}^{2}U\,b^{k}}\right.\\
 & \left.\vphantom{\left(\partial_{c}U\right)_{;e}}-\dot{\mu}g^{dc}\partial_{c}U\right]-\lambda g^{ab}\cancel{g^{cd}\partial_{c,a^{k}}^{2}U\left(\partial_{d}U\right)_{;b}b^{k}}
\end{align*}
so we have (using $g^{cd}\partial_{c}U\left(\partial_{d}U\right)_{;b}=0$)
\begin{align*}
\frac{D^{2}\rho^{a}}{d\tau^{2}}= & \frac{\dot{\lambda}}{\lambda}\frac{D\rho^{a}}{d\tau}-\frac{d}{d\tau}\left(\frac{\dot{\mu}}{\lambda}\right)\frac{d\xi^{a}}{d\tau}+\lambda g^{ab}\left(\partial_{c}U\right)_{;bd}\rho^{c}\frac{d\xi^{d}}{d\tau}+\lambda^{2}g^{dc}g^{ab}\left(\partial_{d}U\right)_{;b}\left(\partial_{c}U\right)_{;e}\rho^{e}
\end{align*}
The last term reads 
\begin{align*}
g^{ab}g^{cd}\left(\partial_{d}U\right)_{;b}\left(\partial_{c}U\right)_{;e}= & g^{ab}\left[\left(\partial_{d}U\right)\left(\partial_{c}U\right)_{;e}g^{cd}\right]_{;b}-g^{ab}\left(\partial_{c}U\right)_{;eb}\partial_{d}Ug^{cd}\\
= & -g^{ab}\left(\partial_{c}U\right)_{;eb}\partial_{d}Ug^{cd}\textrm{ since }g^{cd}\partial_{c}U\partial_{d}U=-1
\end{align*}
Since $U$ complete integral, $\lambda\partial_{d}Ug^{cd}=\frac{d\xi^{c}}{d\tau}$
so
\begin{align*}
\lambda g^{dc}g^{ab}\left(\partial_{d}U\right)_{;b}\left(\partial_{c}U\right)_{;e}= & -g^{ab}\left(\partial_{c}U\right)_{;eb}\frac{d\xi^{c}}{d\tau}
\end{align*}
and the full equation reads, using $\phi_{;ab}=\phi_{;ba}$
\begin{align*}
\frac{D^{2}\rho^{a}}{d\tau^{2}}= & \frac{\dot{\lambda}}{\lambda}\frac{D\rho^{a}}{d\tau}-\frac{d}{d\tau}\left(\frac{\dot{\mu}}{\lambda}\right)\frac{d\xi^{a}}{d\tau}+\lambda g^{ab}\left[\left(\partial_{c}U\right)_{;bd}\rho^{c}\frac{d\xi^{d}}{d\tau}-\left(\partial_{c}U\right)_{;eb}\frac{d\xi^{c}}{d\tau}\rho^{e}\right]\\
= & \frac{\dot{\lambda}}{\lambda}\frac{D\rho^{a}}{d\tau}-\frac{d}{d\tau}\left(\frac{\dot{\mu}}{\lambda}\right)\frac{d\xi^{a}}{d\tau}+\lambda g^{ab}\left[\left(\partial_{c}U\right)_{;bd}\frac{d\xi^{d}}{d\tau}\rho^{c}-\left(\partial_{e}U\right)_{;cb}\frac{d\xi^{c}}{d\tau}\rho^{e}\right]\\
= & \frac{\dot{\lambda}}{\lambda}\frac{D\rho^{a}}{d\tau}-\frac{d}{d\tau}\left(\frac{\dot{\mu}}{\lambda}\right)\frac{d\xi^{a}}{d\tau}+\lambda g^{ab}\left[\left(\partial_{c}U\right)_{;bd}-\left(\partial_{c}U\right)_{;db}\right]\rho^{c}\frac{d\xi^{d}}{d\tau}
\end{align*}
The Ricci identitty here gives, using the Riemann symmetries and $\lambda g^{df}\partial_{f}U=\frac{d\xi^{d}}{d\tau}$
\begin{align*}
\lambda g^{ab}\rho^{c}\frac{d\xi^{d}}{d\tau}2\left(\partial_{c}U\right)_{;\left[bd\right]}= & \lambda g^{ab}\rho^{c}\frac{d\xi^{d}}{d\tau}R_{dbce}g^{ef}\partial_{f}U\\
= & -\lambda R_{\:dce}^{a}g^{ef}\partial_{f}U\rho^{c}\frac{d\xi^{d}}{d\tau}\\
= & -\lambda R_{\:bcd}^{a}g^{df}\partial_{f}U\rho^{c}\frac{d\xi^{b}}{d\tau}\\
= & -R_{\:bcd}^{a}\frac{d\xi^{d}}{d\tau}\rho^{c}\frac{d\xi^{b}}{d\tau}
\end{align*}
which gives (\ref{thm:Jacobi-theorem-geo-dev}-\ref{enu:jacobiGeoDev3c}),
the geodesic deviation equation
\end{itemize}
\end{proof}

\chapter[Motion of a top: M.P.D. equations]{\label{chap:Motion-of-a}Motion of a top in Relativity: the M.P.D.
equations}

\cite{Mathisson:1937zz,Papapetrou:1951pa,Dixon:1970zza,Hojman1975PhDT}

\begin{figure}[t]
\includegraphics[width=1\columnwidth]{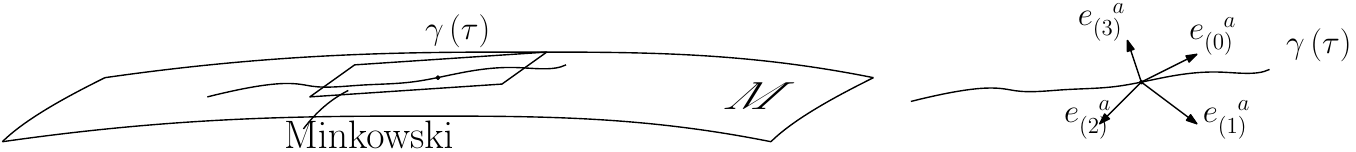}

\caption{\label{fig:Motion-of-a}Motion of a top}

\end{figure}

Describe a top by its worldline $x^{a}\left(\tau\right)$ and the
frame field (orthonormal) attached to it, represented by a tetrad
$e_{\left(A\right)}^{\hspace*{1em}a}\left(\tau\right)$ (see Fig.~\ref{fig:Motion-of-a})
 also known as vierbein (vier=4 in German; $d=n$: vielbein)

The tetrad verifies 10 orthonormal relations
\begin{align*}
e_{A}^{\;a}e_{B}^{\;b}g_{ab}= & \eta_{AB}
\end{align*}
that defines the Minkowski tangent space
\begin{align*}
\eta_{AC}\eta^{CB}= & \delta_{A}^{\;B}
\end{align*}
and the manifold metric and its inverse with
\begin{align*}
e_{A}^{\;a}e_{B}^{\;b}\eta^{AB}= & g^{ab}, & g^{ac}g_{cb}= & \delta_{\:b}^{a}
\end{align*}
Any vectors $V^{a}=V^{A}e_{A}^{\;a}$, $W^{b}=W^{B}e_{B}^{\;b}$ are
defined also in the tangent space and their scalar product is given
with metric in $M$ and in Minkowski tangent space:
\begin{align*}
V^{a}W^{b}g_{ab}= & V^{A}W^{B}e_{A}^{\;a}e_{B}^{\;b}g_{ab}=\eta_{AB}V^{A}W^{B}
\end{align*}
We define the velocity on the top trajectory and the parallel transport
of its tetrad as
\begin{align*}
\frac{dx^{a}}{d\tau}= & \dot{x}^{a}=u^{a}\\
\overset{\circ}{e}_{A}^{\;a}= & \frac{De_{A}^{\;a}}{d\tau}=\frac{de_{A}^{\;a}}{d\tau}+\Gamma_{bc}^{a}e_{A}^{\;b}u^{c}
\end{align*}

\section{Angular velocity tensor}

We define the angular velocity tensor
\begin{gather*}
\boxed{\sigma^{ab}=\eta^{AB}e_{A}^{\;a}\frac{De_{B}^{\;b}}{d\tau}}
\end{gather*}
It is antisymmetric as
\begin{align*}
\frac{D}{d\tau}\left(e_{A}^{\;a}e_{B}^{\;b}\eta^{AB}\right)= & \frac{D}{d\tau}\left(g^{ab}\right)=0\textrm{ since }g^{ab}\textrm{ parallel transported (metricity)}\\
= & \overset{\circ}{e}_{A}^{\;a}e_{B}^{\;b}\eta^{AB}+e_{A}^{\;a}\overset{\circ}{e}_{B}^{\;b}\eta^{AB}\\
\Leftrightarrow\sigma^{ab}= & -\sigma^{ba}
\end{align*}
thus
\begin{gather*}
\boxed{\sigma_{ab}=\sigma_{\left[ab\right]}}
\end{gather*}
It also defines the parallel transport of the tetrad since
\begin{align*}
\frac{D}{d\tau}\left(e_{A}^{\;a}e_{B}^{\;b}g_{ab}\right)= & \overset{\circ}{\eta}_{AB}=0\\
= & \overset{\circ}{e}_{A}^{\;a}e_{B}^{\;b}g_{ab}+e_{A}^{\;a}\overset{\circ}{e}_{B}^{\;b}g_{ab}
\end{align*}
Multiplying by $\eta^{CB}e_{C}^{\;c}$ one gets (after relabelling
dummy indices)
\begin{align*}
\overset{\circ}{e}_{A}^{\;c}e_{C}^{\;a}e_{B}^{\;b}\eta^{CB}g_{cb}= & -e_{A}^{\;c}\underbrace{\eta^{CB}e_{C}^{\;a}\overset{\circ}{e}_{B}^{\;b}}_{{\textstyle \sigma^{ab}}}g_{ab}\\
\Leftrightarrow & \boxed{\overset{\circ}{e}_{A}^{\;a}=-\left(e_{A}\right)_{b}{\textstyle \sigma^{ab}}}
\end{align*}

\section{Construction of top motion Lagrangian}

We build now the Lagrangian from $\sigma^{ab},u^{a}$ and $g_{ab}$
such that it is homogeneous of degree one in the velocities $\left(u,\sigma\right)$
and constructed out of th 4 invariants:
\begin{align*}
a_{1}= & g_{ab}u^{a}u^{b}=u^{2}\\
a_{2}= & g_{ac}g_{bd}\sigma^{ab}\sigma^{cd}=\textrm{tr}\sigma^{2}\\
a_{3}= & g_{ac}g_{bd}g_{ef}u^{a}\sigma^{ce}u^{b}\sigma^{fd}=u\sigma\sigma u\\
a_{4}= & g_{ac}g_{bd}g_{ef}g_{gh}\sigma^{ha}\sigma^{bc}\sigma^{de}\sigma^{fg}=\textrm{tr}\sigma^{4}
\end{align*}
so $\mathscr{L}=\mathscr{L}\left(a_{1},a_{2},a_{3},a_{4}\right)=\mathscr{L}\left(u^{2},\textrm{tr}\sigma^{2},u\sigma\sigma u,\textrm{tr}\sigma^{4}\right)$

To get the equations of motion requires to vary $\mathscr{L}$ with
respect to the 6 independent components of $e_{A}^{\;a}$ and to $x^{a}.$
We obtain them through the antisymmetric tensor $\theta^{ab}$, which
variation is defined as (with 6 non zero components)
\begin{align*}
\delta\theta^{ab}= & \eta^{AB}e_{A}^{\;a}De_{B}^{\;b}=\eta^{AB}e_{A}^{\;a}\left(\delta e_{B}^{\;b}+\Gamma_{cd}^{b}e_{B}^{\;c}\delta x^{d}\right)=-\delta\theta^{ba}
\end{align*}
from variation of the metric

Let us relate now $\delta\sigma^{ab}$ with $\frac{D\delta\theta^{ab}}{d\tau}$
to use $\delta\theta^{ab}$ related to $\sigma_{ab}$ as $\delta x^{a}$
to $u^{a}$

Parallel transport for tensors gives us
\begin{align*}
D\sigma^{ab}= & \delta\sigma^{ab}+\Gamma_{cd}^{a}\sigma^{cb}\delta x^{d}+\Gamma_{cd}^{b}\sigma^{ac}\delta x^{d}
\end{align*}
and
\begin{align*}
\frac{D\delta\theta^{ab}}{d\tau}= & \frac{d}{d\tau}\left(\delta\theta^{ab}\right)+\Gamma_{cd}^{a}\delta\theta^{cb}u^{d}+\Gamma_{cd}^{b}\delta\theta^{ac}u^{d}
\end{align*}
Since equations are tensorial, they are independent of frame.

Choosing a locally flat frame ($\Gamma=0$, which we always can from
the equivalence principle, and see Theorem~\ref{thm:As-the-Christoffel})
this becomes
\begin{gather*}
\left\{ \begin{array}{rlrl}
D\sigma^{ab}= & \delta\sigma^{ab} & \textrm{ and using }\sigma^{ab}= & \eta^{AB}e_{A}^{\;a}\overset{\circ}{e}_{B}^{\;b}\\
\frac{D\delta\theta^{ab}}{d\tau}= & \frac{d}{d\tau}\left(\delta\theta^{ab}\right) & \textrm{ and }\overset{\circ}{e}_{A}^{\;a}= & \dot{e}_{A}^{\;a}+\underbrace{\Gamma_{cd}^{a}}_{=0\textrm{ locally}}e_{A}^{\;c}u^{d}
\end{array}\right.
\end{gather*}
we have
\begin{itemize}
\item 
\begin{align*}
D\sigma^{ab}= & \eta^{AB}\left(\delta e_{A}^{\;a}\overset{\circ}{e}_{B}^{\;b}+e_{A}^{\;a}\delta\left(\overset{\circ}{e}_{B}^{\;b}\right)\right)\\
= & \eta^{AB}\left[\delta e_{A}^{\;a}\dot{e}_{B}^{\;b}+e_{A}^{\;a}\left(\delta\dot{e}_{B}^{\;b}+\Gamma_{cd,e}^{a}e_{B}^{\;c}u^{d}\delta x^{e}\right)\right]
\end{align*}
\item 
\begin{align*}
\frac{D\delta\theta^{ab}}{d\tau}=\frac{d\delta\theta^{ab}}{d\tau}= & \frac{d}{d\tau}\left(\eta^{AB}e_{A}^{\;a}\left(\delta e_{B}^{\;b}+\Gamma_{cd}^{b}e_{B}^{\;c}\delta x^{d}\right)\right)\\
= & \eta^{AB}\left(\dot{e}_{A}^{\;a}\delta e_{B}^{\;b}+e_{A}^{\;a}\left[\delta\dot{e}_{B}^{\;b}+\Gamma_{cd,e}^{b}e_{B}^{\;c}\delta x^{d}u^{e}\right]\right)
\end{align*}
\end{itemize}
Substracting
\begin{align*}
D\sigma^{ab}-\frac{D\delta\theta^{ab}}{d\tau}= & \eta^{AB}\left[\delta e_{A}^{\;a}\dot{e}_{B}^{\;b}-\dot{e}_{A}^{\;a}\delta e_{B}^{\;b}+e_{A}^{\;a}\left(2\Gamma_{c\left[d,e\right]}^{b}e_{B}^{\;c}u^{d}\delta x^{e}\right)\right]
\end{align*}
and we recognise the Riemann tensor for $\Gamma=0$ case in $2\Gamma_{c\left[d,e\right]}^{a}$
from its definition $R_{\:bcd}^{a}=-2\Gamma_{b\left[c,d\right]}^{a}+2\Gamma_{[c|e}^{a}\Gamma_{|d]b}^{ae}$

Moreover for $\Gamma=0$, the definition $\delta\theta^{ab}=\eta^{AB}e_{A}^{\;a}\delta e_{B}^{\;b}$
gives
\begin{align*}
g_{bc}e_{A}^{\;c}\delta\theta^{ba}=\eta^{BC}\left(e_{A}\right)_{b}e_{B}^{\;b}\delta e_{C}^{\;a}=\eta^{BC}\eta_{AB}\delta e_{C}^{\;a}= & \delta e_{A}^{\;a}\\
\textrm{as well as }\overset{\circ}{e}_{A}^{\;a}=\dot{e}_{A}^{\;a}= & -\left(e_{A}\right)_{b}{\textstyle \sigma^{ab}}
\end{align*}
so the difference reads
\begin{align*}
D\sigma^{ab}-\frac{D\delta\theta^{ab}}{d\tau}= & \eta^{AB}\lefteqn{\phantom{_{A}-g_{dc}e_{A}^{\;c}\delta\theta^{da}\left(e_{B}\right)_{e}{\textstyle \sigma^{be}}+\left(e_{A}\right)_{c}{\textstyle \sigma^{ac}}g_{de}e_{B}^{\;e}\delta\theta^{bd}+}\underbrace{\phantom{e_{A}^{\;a}e_{B}^{\;c}}}_{g^{ac}}}\lefteqn{\phantom{\theta-g_{dc}e_{A}^{\;c}\delta\theta^{da}\left(e_{B}\right)_{e}{\textstyle \sigma^{be}}+}\underbrace{\phantom{\left(e_{A}\right)_{c}{\textstyle \sigma^{ac}}g_{de}e_{B}^{\;e}\delta\theta^{bd}}}_{{\textstyle \sigma^{ac}}g_{dc}\delta\theta^{bd}}}\lefteqn{\phantom{\;-g_{dc}e_{A}^{\;c}\delta\theta^{da}\left(e_{B}\right)_{e}{\textstyle \sigma^{be}}+\left(e_{A}\right)_{c}{\textstyle \sigma^{ac}}g_{de}}\overset{\begin{array}{c}
-\delta\theta^{db}\\
\shortparallel
\end{array}}{\phantom{\delta\theta^{bd}}}}\lefteqn{\phantom{\,-}\underbrace{\phantom{g_{dc}e_{A}^{\;c}\delta\theta^{da}\left(e_{B}\right)_{e}}}_{-\delta\theta^{ad}g_{de}}}\lefteqn{\phantom{-\qquad\,\,\,}\overset{\begin{array}{c}
-\delta\theta^{ad}\\
\shortparallel
\end{array}}{\phantom{\delta\theta^{da}}}}\left[-g_{dc}e_{A}^{\;c}\delta\theta^{da}\left(e_{B}\right)_{e}{\textstyle \sigma^{be}}+\left(e_{A}\right)_{c}{\textstyle \sigma^{ac}}g_{de}e_{B}^{\;e}\delta\theta^{bd}+e_{A}^{\;a}e_{B}^{\;c}2\Gamma_{c\left[d,e\right]}^{b}u^{d}\delta x^{e}\right]\\
= & \delta\theta^{ad}\sigma_{\;d}^{b}-\sigma^{ac}\delta\theta_{\:c}^{b}+g^{ac}R_{\:ced}^{b}u^{d}\delta x^{e}\\
\Leftrightarrow & \boxed{D\sigma^{ab}=\frac{D\delta\theta^{ab}}{d\tau}+\sigma^{ac}\delta\theta_{c}^{\;b}-\delta\theta^{ac}\sigma_{c}^{\;b}-g^{ac}R_{\:cde}^{b}u^{d}\delta x^{e}}
\end{align*}
For the variational principle, we use $\delta\sigma^{ab}$ so recall
the parallel transport variations
\begin{align*}
D\sigma^{ab}= & \delta\sigma^{ab}+\Gamma_{cd}^{a}\sigma^{cb}\delta x^{d}+\Gamma_{cd}^{b}\sigma^{ac}\delta x^{d}
\end{align*}
and
\begin{align*}
\frac{D\delta\theta^{ab}}{d\tau}= & \delta\dot{\theta}^{ab}+\Gamma_{cd}^{a}\delta\theta^{cb}u^{d}+\Gamma_{cd}^{b}\delta\theta^{ac}u^{d}
\end{align*}
so we can extract

\noindent\fbox{\begin{minipage}[t]{1\columnwidth - 2\fboxsep - 2\fboxrule}%
\vspace{-0.5cm}
\begin{multline*}
\delta\sigma^{ab}=\delta\dot{\theta}^{ab}+\Gamma_{cd}^{a}\left(\delta\theta^{cb}u^{d}-\sigma^{cb}\delta x^{d}\right)+\Gamma_{cd}^{b}\left(\delta\theta^{ac}u^{d}-\sigma^{ac}\delta x^{d}\right)\\
+\sigma^{ac}\delta\theta_{c}^{\;b}-\delta\theta^{ac}\sigma_{c}^{\;b}-g^{ac}R_{\:cde}^{b}u^{d}\delta x^{e}
\end{multline*}
\end{minipage}}

that should be used in the variational principle. We also need the
canonical momenta

Defining $L_{i}=\frac{\partial\mathscr{L}}{\partial a_{i}}$, we also
need the derivatives of the invariants with respect to the velocities:
\begin{itemize}
\item ~\vspace{-0.99cm}
\begin{flalign*}
\frac{\partial u^{2}}{\partial u_{a}}= & \frac{\partial\left(u_{b}u_{c}g^{bc}\right)}{\partial u_{a}}=\delta_{b}^{a}u_{c}g^{bc}+\delta_{c}^{a}u_{b}g^{bc}=u^{a}+u^{a}=2u^{a}\phantom{\frac{\partial\left(u_{d}\sigma^{ed}g_{ec}\right)}{\partial u_{a}}==}
\end{flalign*}
\item ~\vspace{-0.99cm}
\begin{align*}
\frac{\partial\left(u\sigma\sigma u\right)}{\partial u_{a}}=\frac{\partial\left(\left(u\sigma\right)_{b}\left(u\sigma\right)_{c}g^{bc}\right)}{\partial u_{a}}= & \frac{\partial\left(u\sigma\right)_{b}}{\partial u_{a}}\left(u\sigma\right)_{c}g^{bc}+\frac{\partial\left(u\sigma\right)_{c}}{\partial u_{a}}\left(u\sigma\right)_{b}g^{bc}\\
= & \frac{\partial\left(u_{d}\sigma^{de}g_{eb}\right)}{\partial u_{a}}\left(u\sigma\right)_{c}g^{bc}+\frac{\partial\left(u_{d}\sigma^{ed}g_{ec}\right)}{\partial u_{a}}\left(u\sigma\right)_{b}g^{bc}\\
= & \delta_{d}^{a}\sigma^{de}g_{eb}\sigma^{bf}u_{f}+\delta_{d}^{a}\underbrace{\sigma^{ed}}_{-\sigma^{de}}g_{ec}u_{f}\underbrace{\sigma^{fc}}_{-\sigma^{cf}}\\
= & 2\delta_{d}^{a}\sigma^{de}\sigma_{e}^{\;f}u_{f}\\
= & 2\sigma^{ab}\sigma_{bc}u^{c}\phantom{e}\qquad\qquad\qquad\textrm{and since }\sigma_{ab}=\sigma_{\left[ab\right]}
\end{align*}
\item ~\vspace{-0.99cm}
\begin{align*}
\frac{\partial\left(\textrm{tr}\sigma^{2}\right)}{\partial\sigma_{ab}}=\frac{\partial\left(\sigma_{cd}\sigma_{ef}g^{de}g^{cf}\right)}{\partial\sigma_{ab}}= & \frac{1}{2}\left(\delta_{c}^{a}\delta_{d}^{b}-\delta_{c}^{b}\delta_{d}^{a}\right)\sigma_{ef}g^{de}g^{cf}+\frac{1}{2}\left(\delta_{e}^{a}\delta_{f}^{b}-\delta_{e}^{b}\delta_{f}^{a}\right)\sigma_{cd}g^{de}g^{cf}\\
= & \frac{1}{2}\left(\sigma^{ba}-\sigma^{ab}+\sigma^{ba}-\sigma^{ab}\right)\\
= & -2\sigma^{ab}=2\sigma^{ba}
\end{align*}
\item ~\vspace{-0.99cm}
\begin{align*}
\frac{\partial\left(\textrm{tr}\sigma^{4}\right)}{\partial\sigma_{ab}}= & \frac{\partial\left(\sigma_{cd}\sigma_{ef}\sigma_{gh}\sigma_{ij}g^{cj}g^{de}g^{fg}g^{hi}\right)}{\partial\sigma_{ab}}\\
= & \left[\frac{1}{2}\left(\delta_{c}^{a}\delta_{d}^{b}-\delta_{c}^{b}\delta_{d}^{a}\right)\sigma_{ef}\sigma_{gh}\sigma_{ij}+\frac{1}{2}\left(\delta_{e}^{a}\delta_{f}^{b}-\delta_{e}^{b}\delta_{f}^{a}\right)\sigma_{cd}\sigma_{gh}\sigma_{ij}+\frac{1}{2}\left(\delta_{g}^{a}\delta_{h}^{b}-\delta_{g}^{b}\delta_{h}^{a}\right)\sigma_{cd}\sigma_{ef}\sigma_{ij}\right.\\
 & \left.+\frac{1}{2}\left(\delta_{i}^{a}\delta_{j}^{b}-\delta_{i}^{b}\delta_{j}^{a}\right)\sigma_{cd}\sigma_{ef}\sigma_{gh}\right]g^{cj}g^{de}g^{fg}g^{hi}\\
= & \frac{1}{2}\left[\sigma_{\,f}^{b}\sigma^{fi}\sigma_{i}^{\:a}-\sigma_{\,f}^{a}\sigma^{fi}\sigma_{i}^{\:b}+\sigma_{c}^{\:a}\sigma_{\,h}^{b}\sigma^{ch}-\sigma_{c}^{\:b}\sigma_{\,h}^{a}\sigma^{hc}+\sigma_{cd}\sigma^{da}\sigma^{bc}-\sigma_{cd}\sigma^{db}\sigma^{ac}+\sigma_{\,d}^{b}\sigma^{df}\sigma_{f}^{\:a}\right.\\
 & \left.-\sigma_{\,d}^{a}\sigma^{df}\sigma_{f}^{\:b}\right]\\
= & \frac{1}{2}\left[\sigma^{bc}\sigma_{cd}\sigma^{da}+\sigma^{da}\sigma_{cd}\sigma^{bc}+\sigma^{da}\sigma^{bc}\sigma_{cd}+\sigma^{bc}\sigma^{da}\sigma_{cd}+\sigma^{bc}\sigma_{cd}\sigma^{da}+\sigma_{dc}\sigma^{bd}\sigma^{ca}+\sigma^{bc}\sigma_{cd}\sigma^{da}\right.\\
 & \left.+\sigma^{da}\sigma_{cd}\sigma^{bc}\right]\\
= & 4\sigma^{bc}\sigma_{cd}\sigma^{da}
\end{align*}
\item ~\vspace{-0.99cm}
\begin{align*}
\frac{\partial\left(u\sigma\sigma u\right)}{\partial\sigma_{ab}}= & \frac{\partial\left(u^{c}\sigma_{cd}g^{de}\sigma_{ef}u^{f}\right)}{\partial\sigma_{ab}}\\
= & \frac{u^{c}}{2}\left(\delta_{c}^{a}\delta_{d}^{b}-\delta_{c}^{b}\delta_{d}^{a}\right)g^{de}\sigma_{ef}u^{f}+u^{c}\sigma_{cd}\frac{g^{de}}{2}\left(\delta_{e}^{a}\delta_{f}^{b}-\delta_{e}^{b}\delta_{f}^{a}\right)u^{f}\\
= & \lefteqn{\phantom{\frac{1}{2}\;u^{a}\sigma^{bc}u_{c}-u^{b}\sigma^{ac}u_{c}+u_{c}}\underbrace{\phantom{\sigma^{ca}\!}}_{-\sigma^{ac}}\phantom{u^{b}\:u_{c}}\underbrace{\phantom{\sigma^{cb}}}_{-\sigma^{bc}}}\frac{1}{2}\left(u^{a}\sigma^{bc}u_{c}-u^{b}\sigma^{ac}u_{c}+u_{c}\sigma^{ca}u^{b}-u_{c}\sigma^{cb}u^{a}\right)\\
= & \frac{1}{2}\left(2u^{[a}\sigma^{b]c}u_{c}+2u^{[a}\sigma^{b]c}u_{c}\right)\\
= & 2u^{[a}\sigma^{b]c}u_{c}
\end{align*}
\end{itemize}
We can now write the momenta
\begin{align*}
P^{a}\equiv-\frac{\partial\mathscr{L}}{\partial u_{a}}= & -\frac{\partial u^{2}}{\partial u_{a}}L_{1}-\frac{\partial\left(u\sigma\sigma u\right)}{\partial u_{a}}L_{3}\\
= & -2u^{a}L_{1}-2\sigma^{ab}\sigma_{bc}u^{c}L_{3}\\
S^{ab}\equiv-\frac{\partial\mathscr{L}}{\partial\sigma_{ab}}= & -\frac{\partial\left(\textrm{tr}\sigma^{2}\right)}{\partial\sigma_{ab}}L_{2}-\frac{\partial\left(u\sigma\sigma u\right)}{\partial\sigma_{ab}}L_{3}-\frac{\partial\left(\textrm{tr}\sigma^{4}\right)}{\partial\sigma_{ab}}L_{4}\\
= & -2\sigma^{ba}L_{2}-2u^{[a}\sigma^{b]c}u_{c}L_{3}-4\sigma^{bc}\sigma_{cd}\sigma^{da}L_{4}
\end{align*}
To obtain the variations $\delta x$ we will need to vary $\mathscr{L}$
with $g_{ab}$, and so, for the symmetric tensor $g_{ab}$, we compute
the derivatives of the invariants w.r.t. $g_{ab}$:
\begin{itemize}
\item ~\vspace{-0.99cm}
\begin{align*}
\frac{\partial u^{2}}{\partial g_{ab}}= & \frac{\partial\left(u^{c}u^{d}g_{cd}\right)}{\partial g_{ab}}=\delta_{c}^{(a}\delta_{d}^{b)}u^{c}u^{d}=u^{(a}u^{b)}=u^{a}u^{b}\phantom{+\delta_{c}^{(a}\delta_{f}^{b)}\sigma^{cd}\sigma^{ef}g_{de}}
\end{align*}
\item ~\vspace{-0.99cm}
\begin{align*}
\frac{\partial\left(\textrm{tr}\sigma^{2}\right)}{\partial g_{ab}}=\frac{\partial\left(\sigma^{cd}\sigma^{ef}g_{de}g_{cf}\right)}{\partial g_{ab}}= & \delta_{d}^{(a}\delta_{e}^{b)}\sigma^{cd}\sigma^{ef}g_{cf}+\delta_{c}^{(a}\delta_{f}^{b)}\sigma^{cd}\sigma^{ef}g_{de}\\
= & \sigma^{c(a}\sigma_{\,c}^{b)}+\sigma^{(a|d}\sigma_{d}^{\:|b)}\\
= & \sigma^{(a|c}\sigma_{c}^{\:|b)}+\sigma^{(a|d}\sigma_{d}^{\:|b)}\\
= & 2\sigma^{ac}\sigma_{c}^{\:b}
\end{align*}
\item ~\vspace{-0.99cm}
\begin{align*}
\frac{\partial\left(u\sigma\sigma u\right)}{\partial g_{ab}}= & \frac{\partial}{\partial g_{ab}}\left(u^{c}g_{cd}\sigma^{de}g_{ef}\sigma^{fg}g_{gh}u^{h}\right)\\
= & \delta_{c}^{(a}\delta_{d}^{b)}u^{c}\sigma^{de}\sigma_{ef}u^{f}+u_{d}\sigma^{de}\delta_{e}^{(a}\delta_{f}^{b)}\sigma^{fg}u_{g}+u^{c}\sigma_{cd}\sigma^{dg}\delta_{g}^{(a}\delta_{h}^{b)}u^{h}\\
= & u^{(a}\sigma^{b)c}\sigma_{cd}u^{d}+u^{c}\sigma_{c}^{\;(a}\sigma^{b)d}u_{d}+u^{c}\sigma_{cd}\sigma^{d(a}u^{b)}\\
= & u^{(a}\sigma^{b)c}\sigma_{cd}u^{d}+u_{c}\sigma^{c(a}\sigma^{b)d}u_{d}+u^{(a}\sigma^{b)d}\sigma_{dc}u^{c}\\
= & 2u^{(a}\sigma^{b)c}\sigma_{cd}u^{d}+u_{c}\sigma^{c(a}\sigma^{b)d}u_{d}\\
= & u^{a}\sigma^{bc}\sigma_{cd}u^{d}+u^{d}\sigma_{dc}\sigma^{ca}u^{b}+\frac{1}{2}\left(u_{c}\sigma^{ca}\sigma^{bd}u_{d}+u_{d}\sigma^{da}\sigma^{bc}u_{c}\right)\\
= & u^{a}\sigma^{bc}\sigma_{cd}u^{d}+u_{c}\sigma^{ca}\sigma^{bd}u_{d}+u^{d}\sigma_{dc}\sigma^{ca}u^{b}
\end{align*}
\item ~\vspace{-0.99cm}
\begin{align*}
\frac{\partial\left(\textrm{tr}\sigma^{4}\right)}{\partial g_{ab}}= & \frac{\partial}{\partial g_{ab}}\left(\sigma^{cd}g_{de}\sigma^{ef}g_{fg}\sigma^{gh}g_{hi}\sigma^{ij}g_{jc}\right)\\
= & \sigma^{cd}\delta_{d}^{(a}\delta_{e}^{b)}\sigma^{ef}\sigma_{fg}\sigma_{\;c}^{g}+\sigma^{cd}\sigma_{d}^{\:f}\delta_{f}^{(a}\delta_{g}^{b)}\sigma^{gh}\sigma_{hc}+\sigma^{cd}\sigma_{de}\sigma^{eh}\delta_{h}^{(a}\delta_{i}^{b)}\sigma_{\;c}^{i}+\sigma^{cd}\sigma_{de}\sigma^{ef}\sigma_{f}^{\;j}\delta_{j}^{(a}\delta_{c}^{b)}\\
= & \sigma^{c(a}\sigma^{b)d}\sigma_{de}\sigma_{\;c}^{e}+\sigma^{cd}\sigma_{d}^{\;(a}\sigma^{b)e}\sigma_{ec}+\sigma^{cd}\sigma_{de}\sigma^{e(a}\sigma_{\;c}^{b)}+\sigma^{(a|d}\sigma_{de}\sigma^{ec}\sigma_{c}^{\;|b)}\\
= & \sigma^{c(a}\sigma^{b)d}\sigma_{de}\sigma_{\;c}^{e}+\sigma^{d(a}\sigma^{b)e}\sigma_{ec}\sigma_{\;d}^{c}+\sigma^{e(a}\sigma^{b)c}\sigma_{cd}\sigma_{\;e}^{d}+\sigma^{d(a}\sigma^{b)c}\sigma_{ce}\sigma_{\;d}^{e}\\
= & 2\lefteqn{\phantom{\,\sigma^{ca}\sigma^{bd}\sigma_{de}\sigma_{\;c}^{e}+}\underset{{\scriptscriptstyle \begin{array}{c}
{\scriptscriptstyle \shortparallel}\\
{\scriptscriptstyle \sigma^{da}\sigma^{bc}\sigma_{ce}\sigma_{\;d}^{e}}
\end{array}}}{\phantom{\sigma^{cb}\sigma^{ad}\sigma_{de}\sigma_{\;c}^{e}}}}\left(\sigma^{ca}\sigma^{bd}\sigma_{de}\sigma_{\;c}^{e}+\sigma^{cb}\sigma^{ad}\sigma_{de}\sigma_{\;c}^{e}\right)\\
= & 4\sigma^{ca}\sigma^{bd}\sigma_{de}\sigma_{\;c}^{e}
\end{align*}
\end{itemize}
Defining
\begin{align*}
Q^{ab}\equiv & -\frac{\partial\mathscr{L}}{\partial g_{ab}}=-\frac{\partial u^{2}}{\partial g_{ab}}L_{1}-\frac{\partial\left(\textrm{tr}\sigma^{2}\right)}{\partial g_{ab}}L_{2}-\frac{\partial\left(u\sigma\sigma u\right)}{\partial g_{ab}}L_{3}-\frac{\partial\left(\textrm{tr}\sigma^{4}\right)}{\partial g_{ab}}L_{4}\\
= & -u^{a}u^{b}L_{1}-2\sigma^{ac}\sigma_{c}^{\:b}L_{2}-\left(u^{a}\sigma^{bc}\sigma_{cd}u^{d}+u_{c}\sigma^{ca}\sigma^{bd}u_{d}+u^{d}\sigma_{dc}\sigma^{ca}u^{b}\right)L_{3}-4\sigma^{ca}\sigma^{bd}\sigma_{de}\sigma_{\;c}^{e}L_{4}
\end{align*}
We can proove the relation between $\sigma,S,u$ and $P$ writing
in terms of momenta definitions

First
\begin{align*}
\sigma^{ac}S_{c}^{\:b}-S^{ac}\sigma_{c}^{\:b}= & \sigma_{\;c}^{a}\left(-2\sigma^{bc}L_{2}-2u^{[c}\sigma^{b]d}u_{d}L_{3}-4\sigma^{bd}\sigma_{de}\sigma^{ec}L_{4}\right)\\
 & +\left(2\sigma^{ca}L_{2}+2u^{[a}\sigma^{c]d}u_{d}L_{3}+4\sigma^{cd}\sigma_{de}\sigma^{ea}L_{4}\right)\sigma_{c}^{\:b}\\
= & 2\underbrace{\left(\sigma^{ac}\sigma_{c}^{\:b}-\sigma^{ac}\sigma_{c}^{\:b}\right)}_{0}L_{2}+\left(\left[u^{a}\sigma^{cd}u_{d}-u^{c}\sigma^{ad}u_{d}\right]\sigma_{c}^{\:b}+\sigma_{\;c}^{a}u^{b}\sigma^{cd}u_{d}-\sigma_{\;c}^{a}u^{c}\sigma^{bd}u_{d}\right)L_{3}\\
 & +4\underbrace{\left(\sigma_{\;c}^{a}\sigma^{ce}\sigma_{ed}\sigma^{db}-\sigma^{ae}\sigma_{ed}\sigma^{dc}\sigma_{c}^{\:b}\right)}_{0}L_{4}\\
= & \lefteqn{\phantom{\quad u^{b}\sigma^{ac}\sigma_{cd}u^{d}-u^{a}\sigma^{bc}\sigma_{cd}u^{d}+}\underbrace{\phantom{\sigma^{ad}u_{d}\sigma^{bc}u_{c}-\sigma^{ac}u_{c}\sigma^{bd}u_{d}}}_{0}}\left(u^{b}\sigma^{ac}\sigma_{cd}u^{d}-u^{a}\sigma^{bc}\sigma_{cd}u^{d}+\sigma^{ad}u_{d}\sigma^{bc}u_{c}-\sigma^{ac}u_{c}\sigma^{bd}u_{d}\right)L_{3}\\
= & -2u^{[a}\sigma^{b]c}\sigma_{cd}u^{d}L_{3}
\end{align*}
while
\begin{align*}
u^{a}P^{b}-u^{b}P^{a}= & u^{a}\lefteqn{\phantom{u^{a}-}\underset{\uparrow}{\phantom{L_{1}}}\negthickspace\negthickspace\negthickspace\negmedspace\underset{\begin{array}{c}
\rule{4.3cm}{0.4pt}\\
{\scriptscriptstyle 0}
\end{array}}{\phantom{2u^{b}L_{1}-2\sigma^{bc}\sigma_{cd}u^{d}L_{3}+2u^{a}}}\negthickspace\negthickspace\negthickspace\negmedspace\underset{\uparrow}{\phantom{L_{1}}}}\left(-2u^{b}L_{1}-2\sigma^{bc}\sigma_{cd}u^{d}L_{3}\right)+\left(2u^{a}L_{1}+2\sigma^{ac}\sigma_{cd}u^{d}L_{3}\right)u^{b}\\
= & 2\left(u^{b}\sigma^{ac}\sigma_{cd}u^{d}-u^{a}\sigma^{bc}\sigma_{cd}u^{d}\right)L_{3}\\
= & 2\times-2u^{[a}\sigma^{b]c}\sigma_{cd}u^{d}L_{3}\\
= & 2\left(\sigma^{ac}S_{c}^{\:b}-S^{ac}\sigma_{c}^{\:b}\right)\\
\Leftrightarrow & \boxed{2u^{[a}P^{b]}=4\sigma^{[a|c}S_{c}^{\:|b]}}
\end{align*}
This result can be used in the form
\begin{align*}
2S^{bc}\sigma_{c}^{\;a}= & 2u^{[a}P^{b]}+2S^{ac}\sigma_{c}^{\:b}
\end{align*}
to simplify the writing of $Q^{ab}$ following
\begin{align*}
\lboxed{2u^{(a}P^{b)}-4S^{(a|c}\sigma_{c}^{\:|b)}=} & 2u^{(a}P^{b)}-2\left(S^{ac}\sigma_{c}^{\:b}+S^{bc}\sigma_{c}^{\:a}\right)\\
= & 2\left(u^{(a}P^{b)}-u^{[a}P^{b]}\right)-4S^{ac}\sigma_{c}^{\:b}\\
= & 2P^{a}u^{b}-4S^{ac}\sigma_{c}^{\:b}\\
\textrm{and using definitions }= & 2\left(-2u^{a}L_{1}-2\sigma^{ac}\sigma_{cd}u^{d}L_{3}\right)u^{b}+4\left(2\sigma^{ca}L_{2}+2u^{[a}\sigma^{c]d}u_{d}L_{3}+4\sigma^{cd}\sigma_{de}\sigma^{ea}L_{4}\right)\sigma_{c}^{\:b}\\
= & 4\left(-u^{a}u^{b}L_{1}-2\sigma^{ac}\sigma_{c}^{\:b}L_{2}-\left[u^{b}\sigma^{ac}\sigma_{cd}u^{d}+u^{a}\sigma^{cd}u_{d}\sigma_{\;c}^{b}+u^{c}\sigma^{ad}u_{d}\sigma_{c}^{\:b}\right]L_{3}\right.\\
 & \left.-4\sigma^{ea}\sigma^{bc}\sigma_{cd}\sigma_{\;e}^{d}L_{4}\right)\\
= & 4\left(-u^{a}u^{b}L_{1}-2\sigma^{ac}\sigma_{c}^{\:b}L_{2}-\left[u^{a}\sigma^{bc}\sigma_{cd}u^{d}+u_{d}\sigma^{da}\sigma^{bc}u_{c}+u^{d}\sigma_{dc}\sigma^{ca}u^{b}\right]L_{3}\right.\\
 & \left.-4\sigma^{ea}\sigma^{bc}\sigma_{cd}\sigma_{\;e}^{d}L_{4}\right)\\
= & \rboxed{4Q^{ab}}
\end{align*}
We can now write the variations of the Lagrangian $\mathscr{L}$ with
respect to variations of independent variables $u^{a},x^{a},\sigma^{ab}$
\begin{align*}
\delta\mathscr{L}= & \frac{\partial\mathscr{L}}{\partial u_{a}}\delta u^{a}+\frac{\partial\mathscr{L}}{\partial\sigma_{ab}}\delta\sigma^{ab}+\frac{\partial\mathscr{L}}{\partial g^{bc}}\frac{\partial g^{bc}}{\partial x^{a}}\delta x^{a}\\
= & -P_{a}\delta u^{a}-S_{ab}\delta\sigma^{ab}-Q^{bc}g_{bc,a}\delta x^{a}
\end{align*}
Recall $\delta\sigma^{ab}$ as a function of $\delta\theta^{ab}$
and $\delta x^{a}$, we get
\begin{align*}
\delta\mathscr{L}= & -P_{a}\delta u^{a}-S_{ab}\left\{ \delta\dot{\theta}^{ab}+\sigma^{ac}\delta\theta_{c}^{\;b}-\delta\theta^{ac}\sigma_{c}^{\;b}+\Gamma_{cd}^{a}u^{d}\delta\theta^{cb}+\Gamma_{cd}^{b}u^{d}\delta\theta^{ac}\right.\\
 & \left.\vphantom{\dot{\theta}^{ab}}-\left(\Gamma_{cd}^{a}\sigma^{cb}+\Gamma_{cd}^{b}\sigma^{ac}+g^{ac}R_{\:ced}^{b}u^{e}\right)\delta x^{d}\right\} -\frac{1}{4}\left(2u^{(b}P^{c)}-4S^{(b|d}\sigma_{d}^{\:|c)}\right)g_{bc,a}\delta x^{a}\\
= & -P_{a}\delta u^{a}-S_{ab}\left(\delta\dot{\theta}^{ab}+\sigma^{ac}\delta\theta_{c}^{\;b}-\delta\theta^{ac}\sigma_{c}^{\;b}+\Gamma_{cd}^{a}u^{d}\delta\theta^{cb}+\Gamma_{cd}^{b}u^{d}\delta\theta^{ac}\right)\\
 & +\left[S_{ab}\left(\Gamma_{cd}^{a}\sigma^{cb}+\Gamma_{cd}^{b}\sigma^{ac}+g^{ac}R_{\:ced}^{b}u^{e}\right)-\frac{1}{4}\left(2u^{(b}P^{c)}-4S^{(b|e}\sigma_{e}^{\:|c)}\right)g_{bc,d}\right]\delta x^{d}
\end{align*}
We can simplify the first and last 2 terms of the $\delta x$ factor
using the momenta correspondance in the form $2S^{ce}\sigma_{e}^{\;b}=2S^{be}\sigma_{e}^{\:c}+2u^{[b}P^{c]}$
and relabelling dummy indices
\begin{align*}
\mathscr{J}= & S_{ab}\left(\Gamma_{cd}^{a}\sigma^{cb}+\Gamma_{cd}^{b}\sigma^{ac}\right)-\frac{1}{4}\left(2u^{(b}P^{c)}-4S^{(b|e}\sigma_{e}^{\:|c)}\right)g_{bc,d}\\
= & \Gamma_{cd}^{a}\left(S_{ab}\sigma^{cb}+S_{ba}\sigma^{bc}\right)-\frac{1}{4}\left(2\left[u^{(b}P^{c)}-u^{[b}P^{c]}\right]-2\left[2S^{be}\sigma_{e}^{\:c}\right]\right)g_{bc,d}\\
= & -2S_{ae}\sigma^{ec}\Gamma_{cd}^{a}-\frac{g_{bc,d}}{2}\left(P^{b}u^{c}-2S^{be}\sigma_{e}^{\:c}\right)
\end{align*}
since $S$ and $\sigma$ are antisymmetric. We can also introduce
$g^{ab}$ as a label intermediate
\begin{align*}
\mathscr{J}= & -2S_{ae}\sigma^{ec}\Gamma_{cd}^{a}-\frac{g^{ab}}{2}g_{bc,d}\left(P_{a}u^{c}-2S_{ae}\sigma^{ec}\right)\\
= & S_{ae}\sigma^{ec}\left(\frac{g^{ab}}{2}g_{bc,d}-\Gamma_{cd}^{a}\right)-\frac{g^{ab}}{2}g_{bc,d}P_{a}u^{c}
\end{align*}
Using again the momenta correspondence with
\begin{align*}
2S_{ae}\sigma^{ec}= & 2S^{ce}\sigma_{ea}-u_{a}P^{c}+P_{a}u^{c}
\end{align*}
giving also
\begin{align*}
A^{cb}\equiv2S^{ce}\sigma_{e}^{\:b}-u^{b}P^{c}= & 2S^{be}\sigma_{e}^{\:c}-u^{c}P^{b}\\
= & A^{bc}=A^{\left(cb\right)}
\end{align*}
recalling
\begin{align*}
\Gamma_{cd}^{a}= & \frac{g^{ab}}{2}\left(g_{cb,d}+g_{db,c}-g_{cd,b}\right)
\end{align*}
\begin{align*}
\mathscr{J}= & \left(A_{\;a}^{c}+P_{a}u^{c}\right)\left(\frac{g^{ab}}{2}g_{bc,d}-\Gamma_{cd}^{a}\right)-\frac{g^{ab}}{2}g_{bc,d}P_{a}u^{c}\\
= & A_{\;a}^{c}\frac{g^{ab}}{2}\left(g_{bc,d}-g_{cb,d}-g_{db,c}+g_{cd,b}\right)+P_{a}u^{c}\left(\frac{g^{ab}}{2}g_{bc,d}-\Gamma_{cd}^{a}-\frac{g^{ab}}{2}g_{bc,d}\right)\\
= & \frac{A^{\left(cb\right)}}{2}\left(g_{cd,b}-g_{bd,c}\right)-\Gamma_{cd}^{a}P_{a}u^{c}\\
= & -\Gamma_{cd}^{a}P_{a}u^{c}+A^{\left(cb\right)}g_{[c|d,|b]}=-\Gamma_{cd}^{a}P_{a}u^{c}
\end{align*}
since $A^{\left(cb\right)}B_{\left[cb\right]}=-A^{\left(bc\right)}B_{\left[bc\right]}=-A^{\left(cb\right)}B_{\left[cb\right]}=0$

Thus we can write
\begin{align*}
\delta\mathscr{L}= & -P_{a}\delta u^{a}+\left(S_{ab}g^{ac}R_{\:ced}^{b}u^{e}-\Gamma_{cd}^{a}P_{a}u^{c}\right)\delta x^{d}-S_{ab}\left(\frac{D\delta\theta^{ab}}{d\tau}+\sigma^{ac}\delta\theta_{c}^{\;b}-\delta\theta^{ac}\sigma_{c}^{\;b}\right)\\
= & \frac{\partial\mathscr{L}}{\partial u_{a}}\delta u^{a}+\frac{\partial\mathscr{L}}{\partial x^{a}}\delta x^{a}+\frac{\partial\mathscr{L}}{\partial\theta_{ab}}\delta\theta^{ab}
\end{align*}
and we can proceed to the equations of motion. First noting
\begin{align*}
\int\frac{\partial\mathscr{L}}{\partial x^{a}}\delta x^{a}d\tau= & \int S^{cb}R_{bcea}u^{e}\delta x^{a}d\tau-\int\Gamma_{ca}^{b}P_{b}u^{c}\delta x^{a}d\tau
\end{align*}
We recognise $\Gamma_{ca}^{b}P_{b}u^{c}\delta x^{a}=P_{b}\left(\frac{D\delta x^{b}}{d\tau}-\frac{d\delta x^{b}}{d\tau}\right)$
and proceed with an integration by parts which surface terms, where
conventionally the variations are set to 0, are vanishing:
\begin{align*}
-\int\Gamma_{ca}^{b}P_{b}u^{c}\delta x^{a}d\tau= & \int\left(\frac{DP_{b}}{d\tau}-\frac{dP_{b}}{d\tau}\right)\delta x^{b}d\tau+0
\end{align*}
and the corresponding Euler-Lagrange equation yields
\begin{align*}
-\frac{D}{d\tau}\left(\frac{\partial\mathscr{L}}{\partial u_{a}}\right)= & \frac{DP^{a}}{d\tau}=-\frac{\partial\mathscr{L}}{\partial x_{a}}=-R_{\:ebc}^{a}u^{e}S^{bc}-\frac{DP^{a}}{d\tau}+\frac{dP^{a}}{d\tau}\\
\Leftrightarrow & \boxed{\frac{DP^{a}}{d\tau}=-\frac{1}{2}R_{\:ebc}^{a}u^{e}S^{bc}+\frac{1}{2}\frac{dP^{a}}{d\tau}}
\end{align*}
Similarly
\begin{align*}
\int\frac{\partial\mathscr{L}}{\partial\theta_{ab}}\delta\theta^{ab}d\tau= & \int\left(S_{bc}\sigma_{\;a}^{c}-S_{ac}\sigma_{\;b}^{c}\right)\delta\theta^{ab}d\tau+\int\frac{D\delta\theta^{ab}}{d\tau}S_{ab}d\tau
\end{align*}
and a similar integration by parts with vanishing surface terms reads
\begin{align*}
\int\frac{D\delta\theta^{ab}}{d\tau}S_{ab}d\tau= & -\int\frac{DS_{ab}}{d\tau}\delta\theta^{ab}d\tau+0
\end{align*}
The Euler-Lagrange equation of motion, this time, is
\begin{align*}
-\frac{D}{d\tau}\text{\ensuremath{\left(\frac{\partial\mathscr{L}}{\partial\sigma_{ab}}\right)}}= & \frac{DS^{ab}}{d\tau}=-\frac{\partial\mathscr{L}}{\partial\theta_{ab}}=S^{ac}\sigma_{c}^{\;b}-S^{bc}\sigma_{c}^{\;a}-\frac{DS^{ab}}{d\tau}\\
\Leftrightarrow & \boxed{\frac{DS^{ab}}{d\tau}=\frac{1}{2}\left(S^{ac}\sigma_{c}^{\;b}-S^{bc}\sigma_{c}^{\;a}\right)}
\end{align*}
and the momenta correspondence allows to write
\begin{gather*}
\boxed{\frac{DS^{ab}}{d\tau}=\frac{1}{4}\left(P^{a}u^{b}-P^{b}u^{a}\right)}
\end{gather*}

We can proove that $J^{2}=\frac{1}{2}g_{ab}g_{cd}S^{ac}S^{bd}$ is
a constant og motion: ($S_{ab}$ antisymmetric)
\begin{proof}
\begin{align*}
\frac{DJ^{2}}{d\tau}=\frac{\partial J^{2}}{\partial S^{ab}}\frac{DS^{ab}}{d\tau}= & \frac{1}{2}g_{cd}g_{ef}\left[\frac{1}{2}\left(\delta_{a}^{c}\delta_{b}^{e}-\delta_{a}^{e}\delta_{b}^{c}\right)S^{df}+\frac{1}{2}\left(\delta_{a}^{d}\delta_{b}^{f}-\delta_{a}^{f}\delta_{b}^{d}\right)S^{ce}\right]\frac{DS^{ab}}{d\tau}\\
= & \frac{1}{2}\left[\frac{1}{2}\left(g_{ad}g_{bf}-g_{bd}g_{af}\right)S^{\left[df\right]}+\frac{1}{2}\left(g_{ca}g_{eb}-g_{cb}g_{ea}\right)S^{\left[ce\right]}\right]\frac{1}{2}P^{[a}u^{b]}\\
= & \frac{1}{2}\left[\frac{1}{2}\left(S_{ab}+S_{ab}\right)+\frac{1}{2}\left(S_{ab}+S_{ab}\right)\right]\frac{P^{[a}}{2}u^{b]}\\
= & S_{\left[ab\right]}\frac{P^{[a}u^{b]}}{2}=\frac{1}{2}S_{ab}P^{a}u^{b}=0
\end{align*}
since $S_{ab}u^{b}=0$
\end{proof}
The equations of motions are then
\begin{gather*}
\boxed{\begin{array}{rl}
\frac{DP^{a}}{d\tau}= & -\frac{1}{2}R_{\:ebc}^{a}u^{e}S^{bc}\\
\frac{DS^{ab}}{d\tau}= & \frac{1}{2}P^{[a}u^{b]}
\end{array}}
\end{gather*}
\cite{Hojman1975PhDT} uses a definition of anticommutators from quantum
mechanics $\left[ab\right]_{Hoj}=2\left[ab\right]$, and some ad hoc
and non-consistant modifications in $\delta\mathscr{L}$ ($\frac{1}{2}$
factor in front of his $\frac{\partial\mathscr{L}}{\partial\sigma_{ab}}\delta\sigma^{ab}$)
that leads to $S_{Hoj}^{ab}\equiv2S^{ab}$ and $u^{[a}P^{b]_{Hoj}}=2u^{[a}P^{b]}=4\sigma^{[a|c}S_{c}^{\:|b]}=2\sigma^{[a|c}S_{Hoj\,c}^{\qquad|b]}=\sigma^{[a|c}S_{Hoj\,c}^{\qquad|b]_{Hoj}}$
and would yield
\begin{align*}
\frac{DP^{a}}{d\tau}= & -\frac{1}{4}R_{\:ebc}^{a}u^{e}S_{Hoj}^{bc}\\
\frac{DS_{Hoj}^{ab}}{d\tau}= & S_{Hoj}^{[a|c}\sigma_{c}^{\;|b]}=\frac{1}{2}S_{Hoj}^{[a|c}\sigma_{c}^{\;|b]_{Hoj}}=\frac{1}{2}P^{[a}u^{b]_{Hoj}}=\frac{1}{2}\left(P^{a}u^{b}-P^{b}u^{a}\right)
\end{align*}
However he writes what we will take as a theorem and can be obtained
by ad hoc redefinition of $u_{Hoj}^{a}=\frac{1}{2}u^{a}$
\begin{align*}
\frac{DP^{a}}{d\tau}= & -\frac{1}{2}R_{\:ebc}^{a}u_{Hoj}^{e}S_{Hoj}^{bc}\\
\frac{DS_{Hoj}^{ab}}{d\tau}= & P^{a}u_{Hoj}^{b}-P^{b}u_{Hoj}^{a}
\end{align*}
Dropping the $Hoj$ index, it looks like the MPD equations found in
\cite{Papapetrou:1951pa} in the form
\begin{gather*}
\frac{D}{ds}\left(mu^{a}+\frac{DS^{ab}}{ds}u_{b}\right)=-\frac{1}{2}R_{\:bcd}^{a}u^{b}S^{cd}\\
\frac{DS^{ab}}{ds}+u^{a}\frac{DS^{bc}}{ds}u_{c}-u^{b}\frac{DS^{ac}}{ds}u_{c}=0
\end{gather*}
if one defines
\begin{align*}
P^{a}= & mu^{a}+\frac{DS^{ab}}{ds}u_{b}
\end{align*}
These were not obtained from canonical formalism. Since $S^{ab}$
antisymmetric, it preserves the Papapetrou definition of mass:
\begin{align*}
m= & -u_{a}P^{a}=m-\frac{DS^{\left[ab\right]}}{ds}u_{(a}u_{b)}=m
\end{align*}
which is however not satisfactory as it should be $M=\sqrt{-P^{a}P_{a}}$.

Now
\begin{align*}
P^{a}= & \frac{u^{b}u_{b}}{u^{c}u_{c}}P^{a}+\frac{u^{a}P^{b}u_{b}}{u^{c}u_{c}}-\frac{u^{a}P^{b}u_{b}}{u^{c}u_{c}}\\
= & \frac{\left(P^{b}u_{b}\right)}{u^{c}u_{c}}u^{a}+\frac{P^{a}u^{b}-P^{b}u^{a}}{u^{c}u_{c}}u_{b}\hfill\textrm{ and }P^{a}u^{b}-P^{b}u^{a}=\frac{DS^{ab}}{d\tau}\\
= & \frac{P^{b}u_{b}}{u^{c}u_{c}}u^{a}+\frac{DS^{ab}}{d\tau}\frac{u_{b}}{u^{c}u_{c}}
\end{align*}
Since $P^{a}u_{a}=-m$, choosing $u^{c}u_{c}=-1$ and $s=-\tau$,
weget the Papapetrou from of $P$ but with
\begin{align*}
P_{a}P^{a}= & -M^{2}=-m^{2}-P_{a}\frac{DS^{ab}}{ds}u_{b}
\end{align*}

Papapetrou's $\frac{DP}{d\tau}$ equation actually contains 3rd derivatives
in $s$:
\begin{align*}
\frac{D}{ds}\left(\cdots+\frac{DS^{ab}}{ds}u_{b}\right)\textrm{ while }S^{ab}= & \frac{\partial\mathscr{L}}{\partial\frac{D\theta_{ab}}{ds}}\equiv\frac{D\theta^{ab}}{ds}
\end{align*}
so it does not necessarily conserves spins

Hojman \cite{Hojman1975PhDT} and Papapetrou \cite{Papapetrou:1951pa}
don't agree because of the presence of $S$ in $\frac{DP}{d\tau}$
but coincide in the limit of small spins using the same gauges and
constraints as Hanson and Regge 

For the special relativistic case, we add the choices
\begin{align*}
S^{ab}P_{b}= & 0\\
e_{\left(0\right)}^{a}= & \frac{P^{a}}{M}\\
x^{0}= & \tau\\
P^{a}P_{a}= & -M^{2}=f\left(\frac{1}{2}S^{ab}S_{ab}\right),
\end{align*}
that is $\exists f,-M^{2}=f\left(J^{2}\right)$: $f$ is a Regge trajectory
that allows to construct $\mathscr{L}$

\bibliographystyle{apsrev4-2}
\bibliography{referencias,shortnames}
\printindex{}
\end{document}